%% file: thermo_review.tex
\documentclass[final,3p,times,onecolumn,nopreprintline]{elsarticle}

\usepackage[latin1]{inputenc}
\usepackage[english]{babel}
\usepackage[T1]{fontenc}

\usepackage{graphicx}
\usepackage{color}
\usepackage{bm}
\usepackage{amssymb}
\usepackage{amsmath}
\usepackage{amscd}
\usepackage{eucal}
\usepackage{mathrsfs}
\usepackage{amsbsy}
\usepackage[normalem]{ulem}

 \usepackage{titlesec}
\titleformat{\section} {\normalfont\sffamily\fontsize{14}{15}\bfseries}{\thesection}{1em}{}
\titleformat{\subsection} {\normalfont\sffamily\fontsize{11}{13}\bfseries}{\thesubsection}{1em}{}
\titleformat{\subsubsection} {\normalfont\bfseries}{\thesubsubsection}{1em}{}

\usepackage{showkeys}
\biboptions{numbers,sort&compress}

\newcommand{\dbar}{{\mathchar'26\mkern-11mud}}
\newcommand{\e}{{\rm e}}
\newcommand{\rmd}{{\rm d}}
\newcommand{\rmi}{{\rm i}}

\newcommand{\half}{{\textstyle{\frac{1}{2}}}}

\newcommand{\up}{\uparrow}
\newcommand{\dn}{\downarrow}

\newcommand{\eps}{\epsilon}

\newcommand{\vsig}{\varsigma}
\newcommand{\vrho}{\varrho}

\newcommand{\eminus}{ e^{\operatorname{-}} }

\newcommand{\Jparticle}{J_\rho }
\newcommand{\Jparticlei}{J_{\rho,i} }
\newcommand{\JparticleL}{J_{\rho,{L}} }
\newcommand{\JparticleR}{J_{\rho,{R}}  }

\newcommand{\Jenergy}{J_{u} }
\newcommand{\Jenergyi}{J_{{u},i} }
\newcommand{\JenergyL}{J_{u,L} }
\newcommand{\JenergyR}{J_{u,R} }
\newcommand{\JenergySC}{J_{u,{\rm SC}} }
\newcommand{\JenergyM}{J_{u,M} }
\newcommand{\Jenergyph}{J_{u,{\rm ph}} }

\newcommand{\Jelectrici}{J_{{e},i} }
\newcommand{\JelectricL}{J_{e,L} }
\newcommand{\JelectricR}{J_{e,R} }
\newcommand{\JelectricSC}{J_{e,{\rm SC}} }

\newcommand{\Jelectric}[1]{J_{{e},#1} }

\newcommand{\Jheati}{J_{{h},i} }
\newcommand{\JheatL}{J_{h,L} }
\newcommand{\JheatR}{J_{h,R} }
\newcommand{\JheatSC}{J_{h,{\rm SC}} }
\newcommand{\JheatM}{J_{h,M} }
\newcommand{\Jheatph}{J_{h,{\rm ph}} }
\newcommand{\Jheat}[1]{J_{{h},#1} }

\newcommand{\kB}{k_{\rm B}}
\newcommand{\bigbra}[1]{\big\langle {#1}\big\arrowvert}
\newcommand{\bigket}[1]{\big\arrowvert {#1}\big\rangle}

\newcommand{\DS}{\Delta \mathscr{S}}
\definecolor{DarkGreen}{rgb}{0,0.7,0}

\newcommand{\red}[1]{\textcolor{black}{#1}}
\newcommand{\green}[1]{\textcolor{black}{#1}}
\newcommand{\colorproofs}[1]{\textcolor{black}{#1}}

\makeatletter
\def\ps@pprintTitle{%
 \let\@oddhead\@empty
 \let\@evenhead\@empty
 \def\@oddfoot{\hfill   {\bf July 13, 2017}}     % <== change date here
 \let\@evenfoot\@oddfoot}
\makeatother
\begin{document}

\begin{frontmatter}

\title{Fundamental aspects of steady-state conversion of heat to work at the nanoscale}
%\title{Fundamental aspects of steady state heat to work conversion at the nanoscale}

\author[Como,INFN]{Giuliano Benenti}
\ead{giuliano.benenti@uninsubria.it}
\author[Como,Natal]{Giulio Casati}
\ead{giulio.casati@uninsubria.it}
\author[Yokohama]{Keiji Saito}
\ead{saitoh@rk.phys.keio.ac.jp}
\author[Grenoble]{Robert S.\ Whitney}
\ead{robert.whitney@grenoble.cnrs.fr}

\address[Como]{Center for Nonlinear and Complex Systems,
Dipartimento di Scienza e Alta Tecnologia,\\
Universit\`a degli Studi dell'Insubria, 
Via Valleggio 11, 22100 Como, Italy}
\address[INFN]{Istituto Nazionale di Fisica Nucleare, Sezione di Milano,
via Celoria 16, 20133 Milano, Italy}
\address[Natal]{International Institute of Physics, Federal University of
Rio Grande do Norte, Natal, Brazil}
\address[Yokohama]{Department of Physics, Keio University
3-14-1 Hiyoshi, Kohoku-ku, Yokohama 223-8522, Japan}
\address[Grenoble]{Laboratoire de Physique et Mod\'elisation des 
Milieux Condens\'es (UMR 5493), Universit\'e Grenoble Alpes and CNRS,\\ 
Maison des Magist\`eres, 25 Avenue des Martyrs, BP 166, 38042 Grenoble, France}

\begin{abstract}
In recent years, the study of heat to work conversion has been re-invigorated by nanotechnology.  
Steady-state devices do this conversion without any macroscopic moving parts, through steady-state flows of microscopic particles such as electrons, photons, phonons, etc.   
This review aims to introduce some of the theories used to describe these steady-state flows in a variety of mesoscopic or nanoscale systems.  These theories are introduced in the context of idealized machines which convert heat into electrical power (heat-engines) or convert electrical power into a heat flow (refrigerators).  In this sense, the machines could be categorized as thermoelectrics, 
although this should be understood to include photovoltaics when the heat source is the sun.
As quantum mechanics is important for most such machines, they fall into the field of quantum thermodynamics.
In many cases, the machines we consider have few degrees of freedom, 
however the reservoirs of heat and work that they interact with are assumed to be macroscopic.
This review discusses different theories which can take into account different aspects 
of mesoscopic and nanoscale physics,
such as coherent quantum transport, magnetic-field induced effects (including topological ones such as the quantum Hall effect),  and single electron charging effects. It discusses the efficiency of thermoelectric conversion, and the thermoelectric figure of merit.
More specifically, the theories presented are 
(i) linear response theory with or without magnetic fields, 
(ii) Landauer scattering theory in the linear response regime and far from equilibrium,
(iii) Green-Kubo formula for strongly interacting systems within the
linear response regime,
(iv) rate equation analysis for small quantum machines with or without interaction effects,
(v) stochastic thermodynamic for fluctuating small systems.
In all cases, we place particular emphasis on the fundamental questions about the bounds on ideal machines.
Can magnetic-fields change the bounds on power or efficiency?
What is the relationship between quantum theories of transport and the laws of thermodynamics?
Does quantum mechanics place fundamental bounds on heat to work conversion which are absent in 
the thermodynamics of classical systems?
\end{abstract}


\end{frontmatter}


\tableofcontents

\input{introduction.tex}

\input{basics.tex}

\input{efficiencies.tex}

\input{landauer.tex}

\input{kubo.tex}

\input{master.tex}

\input{other-steady-state.tex}

\input{engines.tex}

\input{conclusions.tex}

%-------------Acknowledgements-----------------------------
\section{Acknowledgements}

We would like to express our gratitude to 
V. Balachandran, 
R. Bosisio,
K. Brandner,
M. B\"uttiker,
S. Chen, 
R. Fazio, 
V. Giovannetti,
C. Goupil, 
F. Haupt, 
M. Horvat, 
Ph. Jacquod, 
F. Mazza, 
C. Mej\'{\i}a-Monasterio, 
H. Ouerdane, 
T. Prosen,
R. S\'anchez, 
U. Seifert,
J. Splettstoesser, 
G. Strini,
F. Taddei, 
S. Valentini,
and 
J. Wang, 
with whom we have had the pleasure of collaborating on the topics discussed in this review paper.
\green{We thank P.~Hofer, B.~Sothmann, R.~Uzdin and an anonymous referee for comments that greatly improved this review.} 
G.B. and G.C. acknowledge the support of the MIUR-PRIN.
G.B. acknowledges the financial support of the INFN through the project ``QUANTUM''.
K.S was supported by JSPS KAKENHI; grant numbers JP25103003 and JP26400404.
R.W. acknowledges the financial support of the COST Action MP1209 ``Thermodynamics in the quantum regime'' and the CNRS PEPS Energie grant ``ICARE''.

\appendix
% Usual Physics Reports style gives appendices labels like  "Appendix A",  "Appendix B", etc.
%  But their is not space for this text in the "Table of Contents".
%  Thus the next command changes the style so appendices just have alphabetic labels, "A", "B", etc.
\renewcommand*{\thesection}{\Alph{section}}

\input{appendix-weakly-nonlinear.tex}

\input{appendix-classical-reservoirs.tex}

\input{appendix-second-to-first-quantization.tex}




%--- appendix

%--- references
\section*{References}
\bibliographystyle{elsarticle-num}
\bibliography{PR-bibliography}


%\begin{thebibliography}{00}
%%% Text of bibliographic item
%
%\bibitem{}
%
%\end{thebibliography}
\end{document}

%% file: introduction.tex
\section{Introduction}
\label{sec:intro}

Since the industrial revolution, the transformation of heat into work has been at the centre of technology.
The earliest examples were steam engines, and current examples range from solar cells 
to nuclear power stations.  
The quest to understand the physics of this transformation led to the theory of thermodynamics.
During the 19th century it became clear that heat and work were simply two different forms of energy 
(the first law of thermodynamics), but that heat is special because it has entropy associated with it,
and no process is allowed to reduce this entropy (the second law of thermodynamics). 
The most concrete prediction of this theory of thermodynamics was that one could never convert 
heat into work with an efficiency exceeding the Carnot efficiency \cite{carnot}, and that this Carnot 
efficiency is always less than one.

A great revolution came with Boltzmann, who made the connection between Newton's deterministic laws of motion
and thermodynamic ideas of the difference between heat and work.  
\green{His work showed} that the laws of thermodynamics emerged at a large scale from a combination of
Newton's laws for each microscopic particle with the statistical uncertainty of our  
knowledge of the positions and velocities of those particles.  
This became known as the theory of statistical mechanics.
It completely changed the status of thermodynamics, which was no longer considered as an underlying theory of nature, but rather an {\it effective} theory that applies to macroscopic systems.
However, this only emphasizes the beauty and power of thermodynamics; it is a simple set of laws for macroscopic 
observables like heat and work, which does not require us to model the microscopic details of the system in question.

The advent of quantum mechanics completely changed the vision of statistical mechanics, forcing us to consider the
quantization of energy levels, the statistics of quantum particles (fermionic or  bosonic), etc.
Yet, this revolution in the microscopic theory did not change the rules of thermodynamics that apply to the macroscopic machines typically used for heat to work conversion. 

In contrast, in recent years we have become increasingly interested in machines that convert heat into 
electrical power at a microscopic level, \green{where quantum mechanics plays a crucial role.  The study of such systems is increasingly becoming known as {\it quantum thermodynamics}.}   
Thermoelectric and photovoltaic devices are some of the simplest examples of this, and it is often said that they
differ from other machines by having no moving parts.
However, it is more accurate to say that they differ from other machines by having no macroscopic moving parts
(i.e. no turbines, pistons, etc).  Instead, they work with steady-state currents of microscopic particles (electrons, photons, phonons, etc) which are all quantum in nature.
Nanotechnology has significantly advanced efforts in this direction, giving us unprecedented control of individual quantum particles.  The questions of how this control can be used for new forms of heat to work conversion has started to be addressed in recent years.
This scientific activity has been boosted by the increasing importance placing on 
sustainable energy for the world's population.  
Most experts expect that small efficient sources of power (heat to work conversion) or refrigeration (work to heat conversion) will be part of the energetic mix of the future.  
We should also not neglect the recovery of the waste heat that is generated in many machines
(from car exhausts to industrial processes). The objective would be to turn waste heat into electrical power 
without impeding the operation of the machine in question.

In spite of the progress made in the last few years, the efficiency of macroscopic thermoelectrics remains rather low \cite{goldsmid,DiSalvo-review,Koumoto2013,Macia2015,dresselhaus,snyder,kanatzidis,majumdarrev,shakouri2009,shakouri2011,dubi, Perroni2016,Pop}. 
%This efficiency depends on physical properties of a given material, namely the electrical conductivity $\sigma$, 
%the thermal conductivity $\kappa$, and the Seebeck coefficient $S$.  It is usually expressed by 
%a dimensionless quantity, the thermoelectric figure of merit,
%$ZT=(\sigma S^2/\kappa)\, T$ \cite{ioffebook}. 
The efficiency is often quantified by the material's dimensionless figure of merit, $ZT$, 
defined in Eq.~(\ref{Eq:ZT-intro}).
High efficiency requires high $ZT$, see Table~\ref{Table:ZT}.  
More than 50 years after Ioffe's discovery that doped semiconductors exhibit relatively large thermoelectric effects
\cite{ioffebook,ioffereview}, and in spite of recent achievements, the most efficient actual devices still operate 
at $ZT$ around 1. 
This corresponds to heat to work conversion with an efficiency which is about a sixth of Carnot efficiency (see Table~\ref{Table:ZT}). While even a small improvement would be most welcome, it is generally accepted that $ZT\sim 3$ (heat to work conversion at about a third of Carnot efficiency) would be necessary for wide-spread industrial and household applications.
For example, one would be able to replaced current pump-based household refrigerators by thermoelectric ones with $ZT\approx 3$. 
However, so far, no clear paths exist to reach that target. 

In such a situation it is useful to investigate an approach which starts from first principles, i.e. from the fundamental microscopic dynamical mechanisms which underline the phenomenological laws of heat and particles transport. 
These methods are particularly suited to study nanoscale systems, which have been considered with interest since
Hicks and Dresselhaus theoretically studied quantum-well
structures in low dimensions and showed that there was a potential to increase the thermoelectric figure
of merit \cite{hd93,hd93b,hhsd96}. 
In this context, enormous achievements in nonlinear dynamical systems and the new tools developed have led to a much better understanding of the statistical behavior of dynamical systems. For example, the question of the derivation of the phenomenological Fourier law of heat conduction from the dynamical equations of motion has been studied in great detail \cite{lepri,dhar,lepri2016}. Theoretical work in this direction have led to the possibility to control the heat current and devise heat diodes,  
transistors, and thermal logic gates
\cite{Terraneo-Payrard-Casati2002,hanggi2012}. Preliminary experimental results have also been 
obtained \cite{majumdar,terasaki}.  We are confident that this theoretical approach, combined with 
sophisticated numerical techniques, may lead to substantial progress on the way of improving the long standing problem of thermoelectric efficiency. 
The study of dynamical complexity of these structures may lead to entirely 
new strategies for developing materials with higher efficiencies of heat to work conversion. 
An additional motivation in favor of steady-state devices, such as thermoelectrics, is that
mechanical engines' efficiencies drop very rapidly at low power output.
This drop is much less significant in thermoelectrics, making them particularly interesting as 
candidates for very small scale (e.g. at micro or nano-scale) power production.

\subsection{The aim of this review}

We believe that a better understanding of the fundamental dynamical mechanisms which control heat and particles transport is desirable. The combined efforts of physicists and mathematicians working in nonlinear dynamical systems and statistical mechanics, condensed matter physicists, and material scientists may prove useful to contribute substantially to the progress in this field of 
\green{importance for our energy supply and its environmental impact.} 
 
The purpose of the present review is to introduce the basic tools and fundamental results on steady-state conversion of heat to work, mainly from a statistical physics and dynamical system's perspective. We hope our review will help bridging the gap among rather diverse communities and research fields, such as \green{non-equilibrium statistical mechanics, mesoscopic physics, mathematical physics of dynamical systems, and strongly correlated many-body systems of condensed matter. }

We start this review with a short overview of non-equilibrium thermodynamics in chapter \ref{sec:nonequilibrium}, where fundamental results on linear response theory and Onsager reciprocity relations are discussed. In chapter \ref{sec:TE}
we then explain basic abstract definitions of thermoelectric heat to work conversion efficiency. 

In chapter \ref{Sect:scattering-theory} we review the microscopic Landauer or Landauer-B\"uttiker scattering theory for systems of non-interacting electrons, and explain how the concept of energy filtering leads to thermoelectric effects which 
convert heat to work in such systems.
Chapter~\ref{sec:landauer} discusses the scattering theory in the linear-response regime, addressing the question of  thermodynamic efficiency and the figure of merit in the context of energy filtering, chiral edge-states, external noise and probe reservoirs.  
Chapter~\ref{Sect:scatter-nonlin} then discusses non-linear scattering theory for systems far from equilibrium,
showing how the laws of thermodynamics emerge naturally from the scattering theory.
It also addresses the treatment of electron-electron interaction at the mean-field level within scattering theory.

One of the most exciting avenues for future investigations of heat to work conversion at the nanoscale 
will be that of systems with strong electron-electron interactions (those for which a mean-field treatment is insufficient).  We set the stage for this in chapters~\ref{sec:interacting}-\ref{Sect:master-examples}.
Chapter \ref{sec:interacting} reviews ideas on the relation between high efficiencies and phase transitions in interacting systems, and introduces the Kubo formalism in this context.
Chapter~\ref{Sect:Qu-Master-Eqn} 
introduces a rate equation method, which is well adapted to describe simple quantum
systems with strong electron-electron interactions, and chapter~\ref{Sect:master-examples} uses this method
to model a variety of quantum thermoelectrics and quantum machines.
Chapter~\ref{Sect:other-steady-state} discusses some other steady-state machines which convert heat to work.

While this review concentrates on steady-state machines, chapter~\ref{sec:CTM}  and chapter~\ref{sec:machines} briefly discuss driven systems, and mentions similarities and differences
compared to steady-state machines.
Chapter~\ref{sec:CTM} treats cyclic machines using methods such as finite-time-thermodynamics.
Chapter~\ref{sec:machines} treats systems modelled by stochastic thermodynamics.
We conclude in chapter~\ref{sec:conclusions} with some 
remarks on future prospects of the field.

\subsection{Further reading: textbooks and reviews}

In this review we will cite numerous works, and apologize for those that we have overlooked.
To help the reader get a complete picture of the field, 
here we give a brief list of useful textbooks and reviews. 
One can start with textbooks on thermodynamics \cite{callen,mazur,kubo} 
and thermoelectricity \cite{ioffebook,goldsmid,rowe}, these focus on bulk systems rather than nanostructures, but present many useful results.   
Useful reviews on thermoelectric effects include Ref.~\cite{DiSalvo-review,shakouri2009,shakouri2011,Pop},
with perspectives for  nano-structured materials given in Ref.~\cite{Koumoto2013,Macia2015}.

Ref.~\cite{sothmannreview} gives a less mathematical overview of the work on quantum dots discussed in this review, 
other related issues are reviewed in Ref.~\cite{Haupt-review}.  The thermodynamics of systems modelled by Markovian
quantum master equations (Lindblad master equations) is reviewed in Ref.~\cite{Kosloff-review}.
Thermoelectric effects in atomic and molecular junctions are reviewed in 
Refs.~\cite{dubi,Ratner2013,Bergfield-Ratner}. 
Ref.~\cite{DiVentra-book} 
addresses theories of transport in nano-systems discussed in this review and beyond.
Refrigeration using superconducting junctions are reviewed in
Ref.~\cite{Pekola-review2006,pekola2012}.  Thermal transport at the nanoscale is reviewed 
in Ref.~\cite{Cahill2003,Cahill2014}.  

On the more theoretical side, the fluctuations of small systems (classical or quantum) can be modelled using stochastic thermodynamics \cite{review-Seifert2007,seifert,review-vandenBroeck,review-Broeck-Esposito}.  
The highly active field of quantum thermodynamics was recently reviewed in Refs.~\cite{Vinjanampathy2016,Millen2016,Goold2016}.

\subsection{Power, efficiency and textbook thermodynamics}
\label{Sect:intro-eff}

To proceed with the introduction it is useful to recall one or two definitions from a first course on thermodynamics.
A heat-engine's conversion of heat into work is typically described by two quantities; the power generated, $P_{\rm gen}$, 
and the efficiency of the converter, $\eta$.  In the case where the power generated is electrical, then
$P_{\rm gen} \ =\ J_{\rm e}\ \Delta V$,
where $J_{\rm e}$ is the electrical current against a voltage difference $\Delta V$. 
The heat engine (eng) efficiency, $\eta_{\rm eng}$, 
is defined as the ratio of the power generated to the heat input, so
\begin{eqnarray}
\eta_{\rm eng} \ = \ { P_{\rm gen}\over J_{\rm H} }\, ,
\label{Eq:eta-eng}
\end{eqnarray}
where $J_{\rm H}$ is the heat flow out of the hot (H) reservoir.
Textbook thermodynamics \cite{callen} tells us that one cannot generate work directly from a heat reservoir, one needs
a pair of reservoirs; hot and cold.  The upper bound on the efficiency of any heat to work conversion 
is that of Carnot,
\begin{eqnarray}
\eta_{\rm eng}^{\rm Carnot} \ = \  1 - {T_{\rm C} \over T_{\rm H}} \, , 
\end{eqnarray}
where $T_{\rm H}$ and $T_{\rm C}$ are the temperatures of the hot (H) and cold (C) reservoirs.
This means the maximum efficiency is always less than one, and becomes very small when
the hot and cold reservoir have similar temperatures.

The conversion of work into heat flow by a refrigerator is also described by two quantities;
the cooling power, and the coefficient of performance (often called COP).
The refrigerator's cooling power is the heat current $J_{\rm C}$ that it sucks out of the cold reservoir  being refrigerated. 
The refrigerator (fri) coefficient of performance, $\eta_{\rm fri}$, is defined as the ratio of cool power to the power absorbed, so 
\begin{eqnarray}
\eta_{\rm fri} \ = \  {J_{\rm C} \over P_{\rm abs}} \,  ,
\label{Eq:eta-fri}
\end{eqnarray}
where $P_{\rm abs}$ is the power absorbed by the refrigerator. If $P_{\rm abs}$ is electrical power,
then it equals $J_{\rm e}\, \Delta V$. However, unlike for the heat-engine, $J_{\rm e}$ is the electrical current in the direction driven by the voltage difference $\Delta V$.
Thermodynamics tells us that we cannot use work to directly cool a reservoir,  
one needs a second reservoir at ambient temperature (which we will call the hot reservoir) 
in which to dump the heat extracted from the cold reservoir.
The upper bound on the coefficient of performance
is that of Carnot,
\begin{eqnarray}
\eta_{\rm fri}^{\rm Carnot} \ = \  \left( {T_{\rm H} \over T_{\rm C}}-1 \right)^{-1} \, ,
\label{Eq:fri-Carnot}
\end{eqnarray}
where $T_{\rm C}$ and $T_{\rm H}$ are the temperatures of reservoirs C and H.
The coefficient of performance is nothing but the efficiency of the refrigerator,
however it has an unusual feature that it can be bigger than one.
The laws of thermodynamics allow the coefficient of performance to be bigger than one for all 
$T_{\rm C} > \half T_{\rm H}$ and it can diverge as $T_{\rm C}$ approaches $T_{\rm H}$, see Eq.~(\ref{Eq:fri-Carnot}).
This is because a good refrigerator (i.e. a refrigerator with an efficiency close to that of Carnot)
can pump many Watts of heat from cold to hot for each Watt of power that consumes, so long as the 
difference in temperature between cold and hot is small.

\subsection{Thermoelectrics : traditional versus quantum}
\label{Sect:Thermoelectrics_traditional_vs_quantum}

%========================================
\begin{figure}
\centerline{\includegraphics[width=0.8\textwidth]{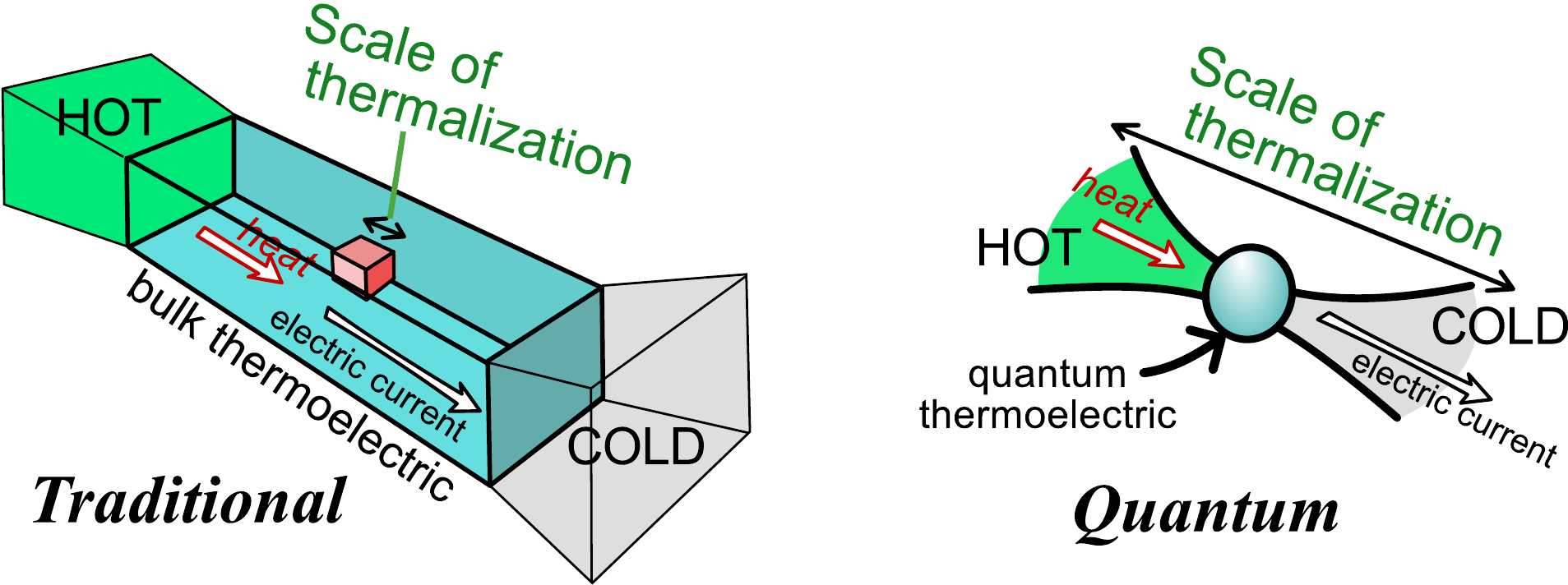}}
\caption{\label{fig:bulk-vs-nano}
A sketch of the qualitative difference between (a) traditional thermoelectrics and (b) quantum thermoelectrics. 
In (a) the distance on which the electrons relax to a local equilibrium is the shortest lengthscale in the system.
Thus, one can treat the electrons inside the thermoelectric structure as being in local thermal equilibrium, with 
a local temperature which varies smoothly across the thermoelectric. The system can then be described by Boltzmann transport equations.  
In contrast, in mesoscopic or nanoscale thermoelectric devices, the thermoelectric structure is of similar size or smaller than the lengthscale on which electrons relax to a local equilibrium.  Then the physics of the system can be much richer, exhibiting intrinsically quantum effects such as interference effects or strong correlation effects,
and one cannot make the approximations necessary to use the standard Boltzmann transport theory.
}
\end{figure}
%========================================

Traditional thermoelectrics have no structure on scales smaller than the 
electronic relaxation length \colorproofs{(except the unit cell which provides their band-structure)}, see Fig.~\ref{fig:bulk-vs-nano}a, where the relaxation length is the distance travelled by an excited electron before inelastic scatterings cause it to relax to thermal equilibrium.
At room temperature, this relaxation length is usually of the order of the mean free path, since electron scattering is typically dominated by inelastic electron-phonon scattering, which
thermalizes the electrons at the same time as causing electrical resistance by relaxing the electrons' momentum.
As such, the relaxation length can be estimated to equal the mean free path extracted from the mobility of the sample in the usual way; it is typically some tens of nanometres.  A thermoelectric with no structure on a scale smaller than this (excepting the unit cell which determines its band structure) is usually well described by Boltzmann transport theory, which assumes a local equilibrium at each point in the thermoelectric, with the temperature and 
electro-chemical potential of this local equilibrium varying smoothly across the thermoelectric.  

Much of the current interest in nanoscale thermoelectrics is because they have structures smaller than this
relaxation length; in many cases the whole thermoelectric is smaller than a relaxation length, see Fig.~\ref{fig:bulk-vs-nano}b.  This is the origin of new physics, such as quantum interference effects or strong correlation effects, for which one cannot use the standard Boltzmann transport theory.  In particular,
the transport properties of the system become non-local on all scales smaller than the relaxation length.
This means that one has to talk in terms of the conductance of the whole system, rather than the conductivity at each point within it.
The objective of this review is to discuss situations of heat to work conversion which are {\it not} described by the 
Boltzmann transport theory.  As such we do not discuss this Boltzmann transport theory, beyond mentioning its similarities to the scattering theory in Section.~\ref{Sect:Boltzmann-Equation}.  The Boltzmann theory for thermoelectrics can be found in 
chapter 3 of Ref.~\cite{goldsmid} or other textbooks.

At low temperatures (typically less than a Kelvin), electron-electron and electron-phonon interactions are rather weak,
as a result the relaxation length can be many microns (or in some cases even a significant fraction of a millimetre \cite{Pfeiffer,Umansky}).  Many system have structures smaller than this, and so will not be described by the usual Boltzmann theory.
At low temperatures, the relaxation length should not be confused with the electron's mean free path.
In such a system, the mean free path is dominated by static disorder which induces only elastic scattering; this 
causes resistance by relaxing the electron's momentum without causing 
thermalization.\footnote{\green{Such elastic scatterings randomize the direction of the electron's momentum without modifying the magnitude of that momentum.  Hence, the average momentum decays, but the kinetic energy of each electron does not change.}} 
Then the relaxation length will be much larger than the mean free path, and so it cannot be experimentally determined from the mobility.  Instead, it must be measured by directly generating an out of equilibrium electron distribution, and studying how it relaxes to a thermal distribution  \cite{Pothier1997,Ritchie-thermometer-2009,Ritchie-thermometer-2013}.  In general the electron-electron scattering length scales with a lower power of temperature than the electron-phonon scattering length, for details see \cite{Pekola-review2006,pekola2012}.  Thus at low enough temperatures, the relaxation length is given by the electron-electron scattering. This means that electrons first thermalize amongst each other via electron-electron interactions, and only afterwards do they thermalize with the phonons
through electron-phonon interactions.
In such cases, if one is driving the electron gas, it may reach a steady-state where electrons thermalize 
(between themselves) at a temperature which is different from that of the phonons in their vicinity.

\subsubsection{\green{Quantum coherence}}
\green{
If the device is small and thermalization is weak, then one may also have quantum coherence over the scale of the device.  Then particles not only maintain their energy as they pass through the device,
they also keep (at least partially) the coherence of their wavefunction.  This means that particles in the device can be in a coherent superposition of different states.  
There can then be interference between the different paths that a particle travels along inside the device.  
It means that quantum correlations can build up between different particles in the device.
Such superposition interference and correlation effects can give rise to numerous effects in quantum thermoelectrics that are absent in classical ones.  Interference can induce thermoelectric effects in systems where they would otherwise be absent, and can modify them in systems which would already have a thermoelectric response (either reducing or increasing them).
These effects are very diverse and we will touch on a number of them in this review.  
}

\green{
A complete picture of the effects of quantum coherence has yet to emerge.
However, we can at least state that
coherence adds another parameter to the system; the quantum mechanical phase. 
When this phase can be manipulated by experimentalists, it can often be adjusted to improve the relevant properties of the system in question and make a better heat engine or refrigerator. One of the earliest examples of this is in 
Ref.~\cite{scully03}, but we cite other examples throughout this review.  Of course, there are usually various classical system parameters that can be manipulated to improve a device's performance.
However one should never underestimate the practical benefit of having one more parameter (the quantum phase) to tune to optimize the machine's performance.
}

\subsection{From linear-response to far-from-equilibrium }

In many applications of thermoelectrics, one is interested in temperature differences which are not small compared to the average temperature.  For example, a proposed application in the automotive industry 
is to generate electricity from the waste heat in a vehicle's exhaust pipe.
The heat reservoir (the exhaust gases) would typically be at 600-700K when the cold reservoir would be the environment at 270-300K.  So the temperature difference is of order the average temperature.
 Thus, one should ask whether one can use a linear response theory to describe this situation. 
 Linear response theory is based on the idea of expanding to linear order about the local equilibrium, and is only expected to work when the temperature difference and bias are small compared to the average temperature.
The answer to the question is very different depending on whether one is considering a traditional bulk thermoelectric sketched in
Fig.~\ref{fig:bulk-vs-nano}a, or a nanoscale thermoelectric sketched in  Fig.~\ref{fig:bulk-vs-nano}b.
The reason is that linear response theory applies when the difference in temperature and chemical potential 
is small (compared to the average temperature)
{\it on the scale of the relaxation length}.  This is very different for the two cases in Fig.~\ref{fig:bulk-vs-nano}.

In the case of a traditional bulk thermoelectric, the temperature drop happens over a few millimetres when the relaxation length is on the scale of tens of nanometres. This means that the temperature drop on the scale of the relaxation length is tiny.  For the above example of an exhaust pipe, the temperature drop
on the scale of a relaxation length would be of order 0.003K (taking a relaxation length of order 10nm), 
which is obviously very much less than the average temperature at any point in the thermoelectric.  
This means that every point in the thermoelectric is extremely close to a local equilibrium, because electrons only fly a distance equal to the relaxation length before relaxing, and so the electrons at a given point all come from neighboring regions with almost the same temperatures and chemical potentials.  In such a situation linear-response theory should work extremely well.

In the case of a nanoscale thermoelectric, the hot and cold reservoirs come within a few nanometres of each other, and the entire temperature drop happens across that nanoscale device which is smaller than a relaxation length.
Thus linear-response theory fails as soon as the temperature difference between the two reservoirs is not much smaller than the average temperature of the two reservoirs.
This makes it extremely clear that if one wanted to use nanoscale thermoelectrics in the above example
of an exhaust pipe, one would absolutely need to describe them with a non-linear theory in which the 
nanostructure is driven far from equilibrium by its coupling to two reservoirs at very different temperatures.

The great simplicity of the linear-response regime is that we can write all thermoelectric properties of the system in terms of four parameters; the electrical conductance, $G$, the thermal conductance, $K$, 
the Seebeck coefficient, $S$, and the Peltier coefficient, $\Pi$.  
While we do not give precise definitions of these quantities until section~\ref{sec:linear_response},
it is important to explain here that this means one can separate the problem in two.  The first problem 
is to find the $G$, $K$, $S$ and $\Pi$ for a given thermoelectric system (either experimentally or 
by modelling).
The second problem is to find the relationship between a \colorproofs{thermoelectric's parameters ($G$, $K$, $S$ and $\Pi$),} and 
\green{the efficiency and power output into a given load that the thermoelectric is attached to.
This second problem is discussed in detail in sections~\ref{sec:stopping_voltage} and \ref{sec:ZT}, 
with the latter section particularly considering the little studied situation where}
a magnetic field breaks the well known symmetry relation between $\Pi=TS$
(a magnetic field allows one to have $\Pi\neq TS$ without violating the Onsager relations).  
In the absence of magnetic field,  we will show in  section~\ref{sec:TE} that there are two principle quantities that are crucial for heat to work conversion.
The first quantity is the dimensionless combination known as the dimensionless figure of merit
\begin{eqnarray}
ZT = {G\, S^2 \, T \over K }\, .
\label{Eq:ZT-intro}
\end{eqnarray}
This gives a measure of the efficiency of the device, via the formulas in section~\ref{sec:ZT},
which will show us that for a heat engine to achieve a given efficiency  we require a given value of $ZT$, 
as indicated in Table \ref{Table:ZT}.
The second quantity is sometimes called the power factor $S^2G$, and is a measure of the maximum power 
such a thermoelectric heat engine can produce, see Eq.~(\ref{Eq:max-power}).
These two quantities act as crucial guides to experimentalists and theorist, since they tell them that a good thermoelectric
requires maximizing $G$ and $S$, while minimizing the thermal conductance $K$.
However, this logic is greatly complicated by the fact that  $G$, $S$ and $K$ are not independent parameters, but instead each of them depends in a different manner on the underlying electronic dynamics, making it hard to optimize the heat to work conversion without a good microscopic theory of these electron dynamics.
One such microscopic model is the scattering theory presented in Chapter~\ref{sec:landauer}.

%%%%%%%%%%%%%%%%%%%%%%%%%%%%%%%%%%%%%%
\begin{table}
\centerline{
\begin{tabular}{|c|c|} 
\hline
Desired efficiency & \ $\phantom{\Big|}$ Necessary $ZT$ \ \\
\hline
$\phantom{\Big|}$ Carnot efficiency &  $\infty$ \\
$\phantom{\Big|}\ 9/10 \ \times\ $ Carnot efficiency &  360\\
$\phantom{\Big|}\ 3/4 \ \times\ $ Carnot efficiency &  48\\
$\phantom{\Big|}\ 1/2 \ \times\ $ Carnot efficiency &  8\\
%\hline
%\end{tabular}
%\hskip 1.2cm
%\begin{tabular}{|c|c|} 
%\hline
%Desired efficiency &  $\phantom{\Big|}$ Necessary $ZT$ \\
%\hline
$\phantom{\Big|}\ 1/3 \ \times\ $ Carnot efficiency & 3 \\
% $\phantom{\Big|}\ 1/4 \ \times\ $ Carnot efficiency & ${16/9} \ \sim \ 1.77$ \\
$\phantom{\Big|}\ 1/6 \ \times\ $ Carnot efficiency & ${24/25} \ \sim \ 1$ \\
$\phantom{\Big|}\ 1/10 \ \times\ $ Carnot efficiency &  ${40/81} \ \sim \ 0.5$ \\
$\phantom{\Big|}\ 1/100 \ \times\ $ Carnot efficiency &  \ ${400/9801}\ \sim \ 0.04$\  \\
\hline
\end{tabular}}
\hskip 5mm
\caption{\label{Table:ZT}
Examples of the dimensionless figure of merit $ZT$ necessary for a desired heat-engine efficiency,
see Eq.~(\ref{etamaxB0}).
This connection between the maximum efficiency and $ZT$ is convenient, as it is easier to
calculate $ZT$ from basic transport measurements than to measure the maximum efficiency directly.
Current bulk semiconductor thermoelectrics have $ZT\sim1$, while a $ZT\sim 3$ would be necessary for
most industrial or household applications.
However the connection between maximum efficiency and $ZT$ only exists in the linear-response regime, as
$ZT$ has no meaning outside the linear-response regime.
}
\end{table}
%%%%%%%%%%%%%%%%%%%%%%%%%%%%%%%%%%%%%%

The situation is much more complicated in the nonlinear regime in which the system is far from equilibrium.
Then one can no longer write the physics in terms of a few constants (like $G$, $S$, etc.) as in the linear-response regime.   Instead, the electrical and heat currents become nonlinear functions of the temperatures and electro-chemical potentials of the reservoirs for which few general statements can be made without explicit considerations of the microscopic dynamics.  It is well known that $ZT$, as defined in Eq.~(\ref{Eq:ZT-intro}),
is no longer a measure of thermodynamic efficiency \cite{shakouri2007,Muralidharan-Grifoni2012,Meair-Jacquod2013,whitney2013-catastrophe,Azema-Lombardo-Dare2014,Crepieux2015}, with it sometimes over-estimating and sometimes 
under-estimating the real efficiency in the nonlinear regime.  Despite efforts in this direction, there is currently no 
nonlinear version of $ZT$. In other words, in the nonlinear regime there is no experimental quantity that acts as a 
short cut to finding the maximum efficiency, the only way to find the maximum efficiency of a given system
is to measure it.
Similarly, there is no short cut to getting a theoretical prediction of a system's maximum efficiency, 
it must be calculated from that system's microscopic dynamics. 

In chapters~\ref{Sect:scatter-nonlin} and \ref{Sect:Qu-Master-Eqn} we consider two different models of such
microscopic dynamics suitable for treating different systems far from equilibrium.  Both methods rely on approximations, either an assumption that  interactions between electrons in the nanostructure
are described by a mean-field theory (the scattering theory in chapter~\ref{Sect:scatter-nonlin}) or 
an assumption that the nanostructure only has a few levels and is weakly coupled to the reservoirs (the rate equations in chapter~\ref{Sect:Qu-Master-Eqn}).

\subsection{Thermocouples, quantum thermocouples and nanoscale photovoltaics}

\begin{figure}
\centerline{\includegraphics[width=0.9\columnwidth]{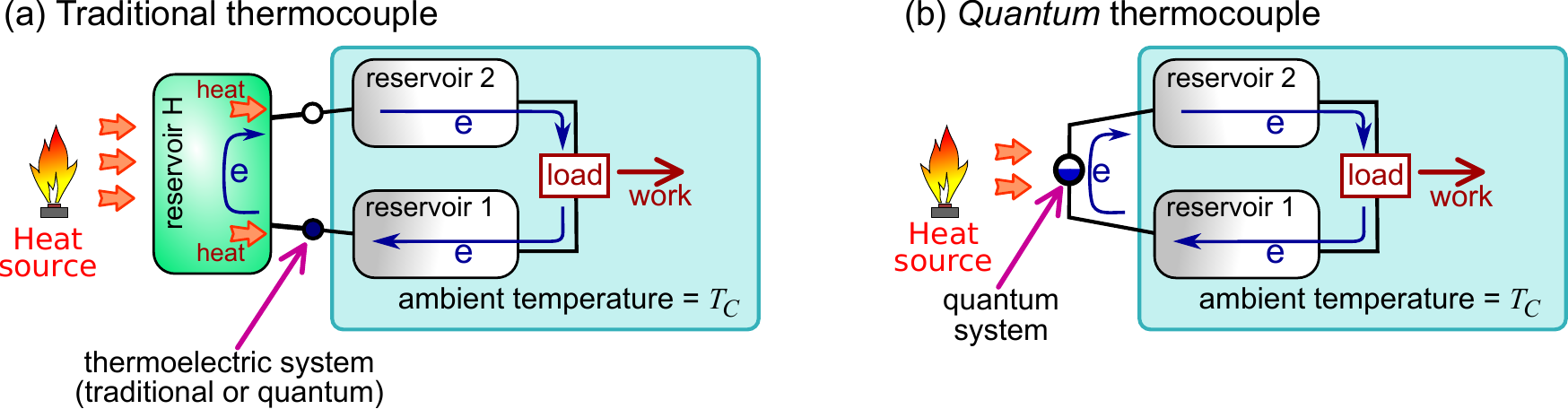}}
\caption{\label{Fig:thermocouple}
In (a) we show a traditional thermocouple made of two different thermoelectrics (open and filled circles)
each coupled to the reservoir being heated (reservoir H) and one of the cold reservoirs (1 or 2).
The heat drives electrons around the circuit from reservoir 1 to reservoir 2, through the load which turns the electrical power into some other kind of work (for example the load could be a motor that generates mechanical work).
In the ideal case, the two thermoelectrics have opposite thermoelectric responses; the one marked by the open circle generates an electrical current in the same direction as the heat flow, while that marked by the filled circle generates an electrical current in the opposite direction to the heat flow.
In (b) we show a new possibility afforded by quantum systems. In this case 
a single quantum system plays the role of the whole thermocouple.
We mark it as a half-filled circle, to indicate that it combines the properties of the two thermoelectrics in (a).
}
\end{figure}

Traditionally to make a thermoelectric heat-engine, one must construct a thermocouple
from two thermoelectric materials (the two ideally having opposite thermoelectric responses),
and heat the region between them (as in Fig.~\ref{Fig:thermocouple}a).
 One assumes this hot reservoir is ideal, meaning it is large and in thermal equilibrium at temperature $T_{\rm H}$.  In this case the thermoelectrics could be traditional or quantum.
The electrons carry heat through them, and thereby generate an electrical current. 
 
However, quantum systems give another possibility.
A suitable quantum system can replace the entire thermocouple (pair of thermoelectrics and hot reservoir),
see Fig.~\ref{Fig:thermocouple}b. 
Heat falls directly on the quantum system in the form of photons, phonons or some other chargeless excitation,
and this drives a steady-state current between the two remaining reservoirs.
We call this a {\it quantum thermocouple}, because it \green{can} exhibit coherent interference effects, quantum correlations 
and non-equilibrium effects across the whole thermocouple.  This makes its physics much richer than a traditional thermocouple.
Many works have considered such systems in recent years, often referring to them as three terminal thermoelectrics,
with one terminal being the one that supplies heat and the other two terminals being those that carry the current.
We discuss such three terminal systems in detail in sections~\ref{sec:probes}-\ref{Sect:Andreev-linear}
and in sections~\ref{Sect:3-term-sys1}-\ref{Sect:cooling-by-heating}. 

It is worth noting that if the heat source is the sun (which is reasonably well approximated by photons emitted from a black-body at 4000K), then the quantum thermocouple can also be thought of as a nanoscale photovoltaic.   
Indeed at a hand-waving level, it works much like a traditional p-n junction photovoltaic.
The electrons at low energies (those in the valence band in a p-n photovoltaic) are coupled to reservoir 1, while those in excited states (the conduction band in a p-n photovoltaic) are coupled to reservoir 2.
Thus when a photon excites an electron from a low to high energy state, that electron flows into reservoir 2.
The empty low energy state is filled by an electron from reservoir 1 (this is often represented as a hole flowing from the system into reservoir 1).  Thus the absorption of a photon causes a net electron flow from reservoir 1 to reservoir 2,
even though reservoir 2 has a higher electro-chemical potential than reservoir 1. It thus converts heat 
into electric work.
The quantum thermocouple systems that
we will discuss in sections~\ref{Sect:3-term-sys1}-\ref{Sect:cooling-by-heating},
are microscopically rather different from a p-n junction photovoltaic, but they still work in the manner outlined here. 

\green{
It is also worth noting that heat gradients may play a role in chemical reactions,
and that these can be considered with donor-acceptor models and reaction path ways 
\cite{Craven-Nitzan2016}, not dissimilar to the models considered in this review.
}

%========================================
%========================================

\subsection{Thermoelectricity as a probe of nanostructures}

Increasingly experimentalists are using the thermoelectric response of nanostructures
as a probe of the physics of those structures.  
It provides complementary information to that extracted from more traditional
transport measurements such as measuring the nanostructure's I-V response.

In this context, the linear-response regime is particularly interesting,
because the linear-response transport properties give us
information about the equilibrium state of the nanostructure.
At a handwaving level, one can say that a standard measure of conductance
(by biasing the sample and measuring the resulting charge current) tells us about the average 
dynamics of those electrons in the nanostructure which have energies close to the Fermi surface.
In contrast, a measurement of the Seebeck coefficient (by applying a temperature difference across the sample
and measuring the resulting potential difference) tells us about the {\it difference} in the dynamics of electrons 
above and below the Fermi surface.  This clearly gives us more information about the sample than the conductance alone.
For example, the sign of the Seebeck coefficient can tell us if the charge carries are electronic excitations (above the Fermi surface) or hole excitations (below the Fermi surface).
This hand waving argument is made quantitative in section~\ref{Sect:scatter-2term-linear}.

The situation is more complicated for the thermoelectric response beyond the linear regime,
just as it is usually harder to understand the nonlinear I-V response of a system than to understand its linear conductance.  In most cases, it is hard to use simple arguments to understand the nonlinear thermoelectric response
of a nanostructure.  Instead, one has to assume a plausible model for the nanostructure, find its thermoelectric response (analytically or numerically) and compare the result with experiments to see how close the model is to the real nanostructure.  The models discussed in chapters~\ref{Sect:scatter-nonlin} and \ref{Sect:Qu-Master-Eqn} 
may provide a good starting point for this sort of analysis.

\subsection{Thermoelectric refrigeration of micron-sized structures to 
extremely low temperatures}

One of the particularly promising applications of nanoscale thermoelectric effects is to refrigerate 
micron-sized (or smaller) structures to unprecedently low temperatures.  Standard cryogenics
typically refrigerate centimetre-sized structures down to 10-100mK, and these are widely used to study
quantum coherent effects in nano-structures, low temperature phase-transitions, etc.
The new idea is to study this type of physics at even lower temperatures by placing
a thermoelectric cooling circuit within the cryogenic refrigerator.  This could
cool a micron-sized region down to a temperature much lower than that of the cryogenic refrigerator.

It is important to note that a standard cryogenic refrigerator cools the lattice of the structure within it, 
that is to say that it cools the structure's phonon gas.  However, at very low temperatures, the coupling between electrons and phonons is very weak.  So it becomes increasingly difficult to cool the electron gas in a structure by cooling the lattice. Instead, one risks to have a cold gas of phonons and a hotter gas of electrons within the structure, with almost no thermal coupling between the two.
Thermoelectric cooling in contrast cools the electrons directly, so may be more efficient for cooling the electron gas than traditional cryogenics.  
In this case, the weak coupling to phonons may be a benefit. If one is interested in the physics of very cold electron gases in metals, quantum hall edge-states, etc.  (say to look for new phases of matter induced by the electron-electron interaction), one may not care if the phonons are hotter than the electrons, so long as they are cold enough that they only couple
weakly to the electron gas.  For this, one could use standard cryogenics to cool the phonons to a few milliKelvin, and then use thermoelectric effects to cool the electrons to much lower temperatures.
Significant experimental progress in cooling has been made using superconducting-normal junctions. 
As these have been well-reviewed elsewhere \cite{Pekola-review2006,pekola2012}, 
we concentrate on other proposed nanoscale refrigeration schemes in this review. 

\green{
A particular application of this nanoscale refrigeration could be the cooling of the environment of solid-state qubits (superconducting circuits, spins in quantum dots, etc.), to maximize the coherence times of such qubits.
This could be crucial for the success of future quantum computers, which require long coherence times.  
However, in this context, we note that an interesting recent work \cite{Splettstoesser2015} shows that it is not only the temperature of the environment which is important in determining the decoherence rate of a superconducting qubit. A temperature difference between different parts of the qubit's environment can lead to a heat current in the vicinity of the qubit which can decohere the qubit more strongly than an environment which is all at the same temperature.  This is a warning that in some cases a non-uniform refrigeration of the environment of qubits may be worse than no refrigeration.
}

\subsection{Phonons and photons as detrimental effects}
\label{Sect:intro-phonons}

%========================================
\begin{figure}
\centerline{\includegraphics[width=0.8\columnwidth]{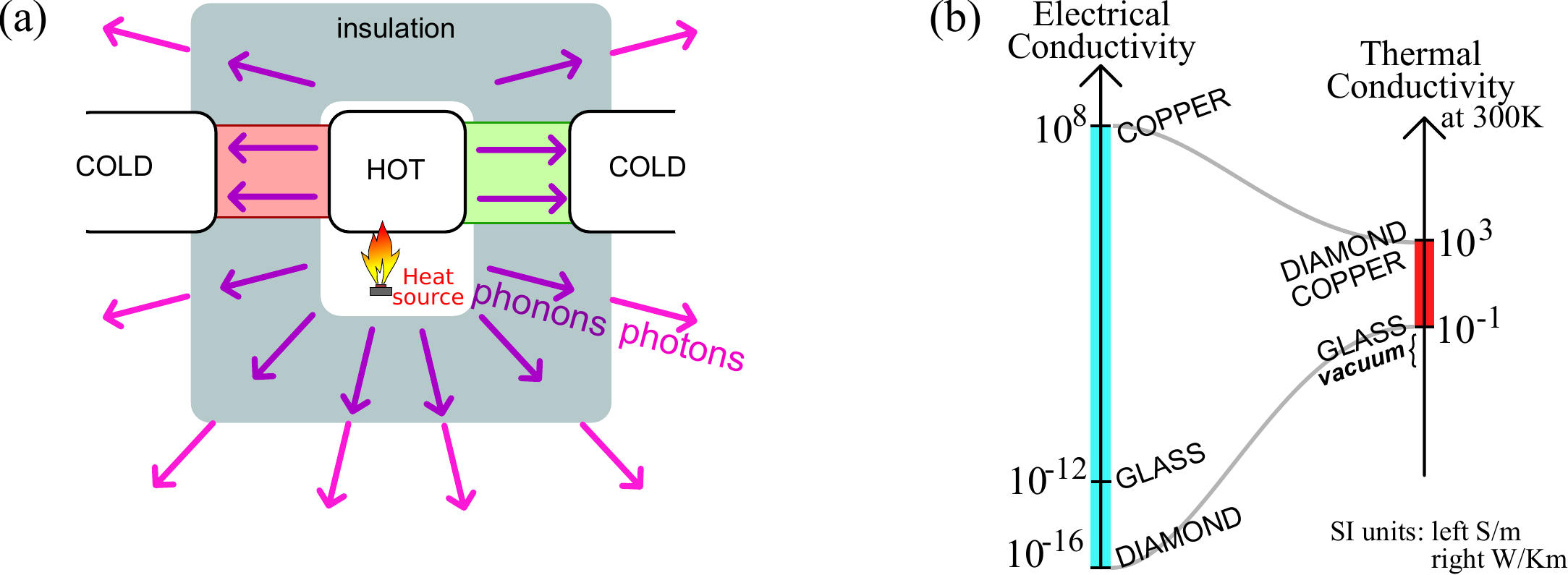}}
\caption{\label{Fig:phonons-intro}
In (a) we sketch of a thermocouple showing how phonons and/or photons carry heat from hot to cold, in parallel with whatever heat the electrons carry through the thermocouple.  It is clear that however efficiently the thermoelectric converts the heat into work, the heat radiated as phonons or photons is lost, greatly reducing the 
overall efficiency of the machine.  Thus it is clear that we want to place an extremely good thermal insulator around
the hot source, and engineer the thermoelectric so the phonon component of its heat conductivity is as small as possible.   The problem is that phonons are uncharged bosons over which we have rather little control.
To illustrate this  (b) shows an adaptation of a figure from Ref.~\cite{conductivity-scales},
it shows the scale of thermal conductivities that exist in nature (spanning about four decades)
compared to electrical conductivities (spanning about 24 decades).  This is a good indication that it is extremely hard to make a good phonon insulator.
}
\end{figure}
%========================================

  Most systems contain charge-less excitation (such as phonons or photons), which
will carry heat from hot to cold in a manner independent of the thermoelectric properties of the system.
The fact these excitation are charge-less makes them much harder to control than electrons,
and as a result we do not have really good thermal insulators.  
Indeed as Fig.~\ref{Fig:phonons-intro}b shows, there is only about a factor of $10^4$ difference in thermal conductivity between the best non-exotic thermal conductors  (such as copper or diamond)  and worst non-exotic thermal conductors (such as glass).  
Even a vacuum has a significant thermal conductivity, because of black-body radiation from hot to cold in the form of photons.
This can be contrasted with factor of $10^{20}$ difference in electrical conductivity between good non-exotic electrical conductors (such as copper) and poor conductors (such as glass).
Thus, while some  phonons and photons will flow through the thermoelectric quantum system, 
and might interact with the electrons there, 
most will flow via other routes, see Fig.~\ref{Fig:phonons-intro}a.
Under such circumstances a thermoelectric heat-engine's efficiency in Eq.~(\ref{Eq:eta-eng}) can be written as
\begin{eqnarray}
\eta_{\rm eng} \ = \ { P_{\rm gen}\over J^{\rm (el)}_{\rm h,H} + J^{\rm (ph)}_{\rm h,H}}
\label{Eq:eta-eng-phonons}
\end{eqnarray}
where $ J^{\rm (el)}_{\rm h,H}$ is the heat carried away from the heat source by the electrons
and $J^{\rm (ph)}_{\rm h,H}$ is the heat carried away from the heat source by phonons, photons, and any other chargeless excitations that may be present.

The heat flow $J^{\rm (ph)}_{\rm h,H}$ cannot contribute to power production,
so we see from Eq.~(\ref{Eq:eta-eng-phonons}) that it only reduces the efficiency.
The efficiency of the power production due to the electronic heat flow, $J^{\rm (el)}_{\rm h,H}$, cannot 
exceed that of Carnot, so $P_{\rm gen}\big/ J^{\rm (el)}_{\rm h,H}  \,\leq\,  \eta_{\rm eng}^{\rm Carnot}$.  
Thus the efficiency in the presence of phonons
must obey
\begin{eqnarray}
\eta_{\rm eng} \ \leq\  \eta_{\rm eng}^{\rm Carnot} \ \times\ {J^{\rm (el)}_{\rm H} \over J^{\rm (el)}_{\rm H} + J^{\rm (ph)}_{\rm H}}
\end{eqnarray}
Thus to achieve a high over all efficiency it is not sufficient to optimize the electronic dynamics to
maximize $ P_{\rm gen}\big/ J^{\rm (el)}_{\rm H}$, one also needs to work to minimize the phonon/photon heat flow.
This requires maximising the insulation around the heat source, and also engineering the thermoelectric's properties
so the phonon heat flow through it is minimal.

The detrimental effect of phonons and photons is even more stark for refrigeration.
Since one is trying to refrigerate the colder of two reservoirs, the phonons and photons will carry heat from
hot to cold, greatly reducing the cooling power.
The cooling power in the presence of phonons will be $J_{\rm h,C}^{\rm (el)} +J_{\rm h,C}^{\rm (ph)}$,
where the phonon or phonon contribution to the heat flow out of the cold reservoir, $J_{\rm h,C}^{\rm (ph)}$,
is {\it negative} and so reduces the cooling power.
Here, $J_{\rm h,C}^{\rm (el)}$ is the cooling power of the electrons alone (defined in section~\ref{Sect:intro-eff}),
and $J_{\rm h,C}^{\rm (ph)}=-J_{\rm h,H}^{\rm (ph)}$ is typically given by a formula of the type in Eq.~(\ref{Eq:J_h,ph}). 
It is the relationship between the electron's cooling power and the phonon's heat flow as a function of $T_{\rm C}$
that will determine the lowest $T_{\rm C}$ that the refrigerator can achieve.
Irrespective of the details, if one reduces $T_{\rm C}$ for fixed $T_{\rm H}$,
then $J_{\rm h,C}^{\rm (ph)}$ will become increasingly negative,
while $J_{\rm h,C}^{\rm (el)}$ will typically decrease and become negative at a given value of $T_{\rm C}$
(in the best case this will happen at $T_{\rm C}=0$).  
Imagine turning on the refrigerator to cool down a cold reservoir, which is initially at the same temperature as the hot one.  Then $J_{\rm h,C}^{\rm (el)}$ is positive and $J_{\rm h,C}^{\rm (ph)}=0$, so heat is sucked out of the
cold reservoir, cooling it down.  As its temperature $T_{\rm C}$ drops,  $J_{\rm h,C}^{\rm (el)}$ drops and  $J_{\rm h,C}^{\rm (ph)}$ becomes increasingly negative. The cooling power $J_{\rm h,C}^{\rm (el)} +J_{\rm h,C}^{\rm (ph)}$ is thus smaller, but cooling continues until $T_{\rm C}$ is such that  
$J_{\rm h,C}^{\rm (el)} +J_{\rm h,C}^{\rm (ph)}=0$.
At this point no further cooling is possible and one has achieved the lowest temperature for the refrigerator.
Thus, we see that minimizing the magnitude of phonon and photon heat flow, $J_{\rm h,C}^{\rm (ph)}$
(for example minimizing $\alpha$ in Eq.~(\ref{Eq:J_h,ph})) is as important as maximizing the electronic cooling power $J_{\rm h,C}^{\rm (el)}$.

It is clear that the efficiency of the refrigerator is reduced by the phonon and photon heat flows in the same manner as the cooling power is, because in the presence of such photons and phonons the numerator in   
Eq.~(\ref{Eq:eta-fri}) becomes the total cooling power $J_{\rm h,C}^{\rm (el)} +J_{\rm h,C}^{\rm (ph)}$,
with $J_{\rm h,C}^{\rm (ph)} <0$ as discussed above.

\subsubsection{Heat carried by phonons and photons}
As a first approximation, the phonon heat flow for the machine is given by that through a typical insulator.
A number of theories for these phonon or photon heat currents 
take the form
\begin{eqnarray}
J_{\rm h,H}^{\rm (ph)}=  \alpha (T_H^\kappa-T_C^\kappa),
\label{Eq:J_h,ph}
\end{eqnarray}
where $J_{\rm h,H}^{\rm (ph)}$ is the heat flow out of the hot (H) reservoir due to phonons or photons,
and both $\alpha$ and $\kappa$ depend on the system in question.
The heat flow out of the cold reservoir, $J_{\rm h,C}^{\rm (ph)}$, is negative and 
equals $-J_{\rm h,H}^{\rm (ph)}$.
The textbook example of such a theory is that of black-body radiation between the two reservoirs,
then $\kappa=4$ and $\alpha$ is the Stefan-Boltzmann constant.
Other examples for phonons in various situations include Refs.~\cite{phonons-Vols,phonons-Zou,Cahill2003,phonons-Wang,phonons-Heron,phonons-Avery,Cahill2014}, while examples for photons in nanostructures 
include Refs.~\cite{photons-Schmitt,photons-Hekking}.
An example relevant to suspended sub-Kelvin nanostructures
is a situation where a finite number $N_{\rm ph}$ of
phonon or photon modes carry heat between the two reservoirs
 \cite{Pendry1983,photons-Hekking,photons-Schmitt,phonons-Heron}.  There, one has
  $\kappa=2$ and $\alpha = t N_{\rm ph}\pi^2 \kB^2/(6h)$, if each mode has the same transmission probability, $t$, with $\kB$ being the Boltzmann constant, and $h$ being the Planck constant.
In many cases, phonons flow diffusively from hot to cold , with regular inelastic scatterings causing thermalization between the phonons,  then the temperature drop on the scale of the thermalization length (inelastic scattering length) is small, and one can apply linear response theory for the phonon heat transport.   If the thermal conductance is approximately temperature independent (on the scale of the temperature difference between hot and cold), then one will 
have a Fourier law for heat flow with $\kappa=1$ and $\alpha$ equalling the phonon thermal conductance.

One of the biggest practical challenges for quantum thermoelectrics is that phonons and photons
will often carry much more heat than the electrons.  This is simply because the hot reservoir
can typically radiate heat in all directions as phonons or photons, while electrons only carry heat 
through the few nanostructures connected to that reservoir.
Thus, in many cases the phonon or photon heat flow will dominate over the electronic one.
However, progress is being made in blocking phonon and photon flow.
One can engineer band gaps in the phonon spectrum by drilling regularly spaced holes in the material to make a phononic crystal (see for example \cite{sound-idea,Phononic-Crystals}).  One can make a highly disordered material known as a phonon glass 
or at least sufficiently disorder to reduce the phonon conduction by a significant factor
(see for example \cite{Phonon-glass1,Phonon-glass2,Phonon-glass3}).
A strategy which makes particular sense for the refrigeration of micron-sized samples to temperatures below that of 
current cryogenics is to suspend the sample being refrigerated. 
This limits its thermal contact with the substrate (which will be at the temperature of the cryostat),
by ensuring that phonons can only flow between the substrate and the micron-sized sample through the 
relatively few phonon modes of the pillars that hold up that sample  \cite{phonons-Heron}.
A typical thermal phonon or photon has a wavelength of $\lambda_{\rm ph} \sim hv_{\rm ph}\big/(\kB T)$,
where $T$ is the temperature and $v_{\rm ph} $ is the velocity of the wave in questions.
For photons in vacuum $v_{\rm ph}=c= 3\times10^8 {\rm ms}^{-1}$, while for photons in solids
$v_{\rm ph}$ varies a lot depending on the material and the type of phonon (longitudinal, transverse, etc.) 
but is typically $10^3$-$10^4 {\rm ms}^{-1}$. 
Thus at cold temperatures (less than one Kelvin) the typical thermal phonon's wavelength is tens of nanometres, 
while the wavelength of thermal photons in vacuum is of the order of a millimetre.
Thus one might imagine that photons will have too long a wavelength to carry heat efficiently into the micron-sized 
island being refrigerated.  This may be true of photons in vacuum, however it is predicted that the metallic wires necessary for the thermoelectric circuit will carry heat via another kind of photons;  these photons are induced by thermal charge fluctuations in the hot part of the circuit, which generate electromagnetic fields that carry heat into
the cold part of the circuit.  In the simplest case the circuit carries a single photon mode 
with a transmission, $t\sim 1$ \cite{photons-Schmitt}, although one can engineer the capacitance and impedance of the circuit to make this transmission much less than one \cite{photons-Hekking}.  
In such cases, the heat flow carried by such circuit photons is given by $J_{\rm h,H}^{\rm (ph)}$ in Eq.~(\ref{Eq:J_h,ph}) with
 $\kappa=2$ and $\alpha = t \pi^2 \kB^2/(6h)$, as mentioned above.

\subsection{The second law of thermodynamics}

Chapters~\ref{Sect:scatter-nonlin} and \ref{Sect:Qu-Master-Eqn} 
will explicitly derive the second law of thermodynamics from the quantum physics of certain systems coupled to reservoirs. 
Since such systems are described by Schr\"odinger's  equation, their dynamics are symmetric under time-reversal.  
However, \green{in these systems,} we will show that the second law of thermodynamics emerges as soon as those systems are 
coupled to macroscopic reservoirs.
What is intriguing is that this result depends very little on the properties of the macroscopic reservoirs,
indeed their dynamics do not even need to be described in detail.  All one needs to get the second law  is that the reservoirs act as equilibrium boundary conditions on the system; that is to say 
all particle arriving at the system from a given reservoir have an equilibrium distribution (defined by the temperature and chemical potential of that reservoir), irrespective of the distribution of particles entering that reservoir from the system. 
\green{
Any reservoir which fulfills  this condition is enough. 
One possibility is that the reservoir is effectively infinite, so that  particles entering it do not leaving it again on the timescale of any experiment.  Another possibility is that the reservoir is large but finite, and contains a weak dissipative process, so that particles entering it are relaxed to the reservoir's equilibrium state before they leave the reservoir again.}
This relaxation process could be due to coupling between the degrees of freedom within the reservoir, or to coupling of that reservoir's degrees of freedom to yet another reservoir.
The microscopic details of these couplings is of no importance, the quantum system will obey
the second law of thermodynamics.

Unfortunately, no proof currently exists that the second law of thermodynamics emerges naturally from the quantum physics of an {\it arbitrary} system coupled to reservoirs, \green{although works in this direction are cited in section~\ref{Sect:Keldysh}}.  The proofs discussed in chapters~\ref{Sect:scatter-nonlin} and \ref{Sect:Qu-Master-Eqn} are special because the systems are particularly simple to treat theoretically, either because they exhibit no non-trivial interactions between particles in quantum system (scattering theory in chapter~\ref{Sect:scatter-nonlin}) or we take the limit of weak coupling between the quantum system and the reservoirs
(rate equations in chapter~\ref{Sect:Qu-Master-Eqn}).  
However, even if we are unable to prove this for an arbitrary system, few experts expect that any such systems will violate the second law.

In contrast, we know that all quantum systems exhibit fluctuations, just as small classical systems exhibit thermal fluctuations.
Thus, even in systems where we know the entropy increases on average, a fluctuation may cause entropy to decrease during a brief period, typically by an amount of order the Boltzmann constant, $k_{\rm B}$.  
These fluctuations average out on longer timescales, ensuring that the entropy does increase on average.  
This means that, if the system is left for a long enough time that its steady-state current 
involves entropy generation much more than $k_{\rm B}$, then it will be extremely unlikely for a fluctuation to
cause the entropy produced during that time to be negative.  
None the less the second law of thermodynamics is only an {\it average} property in such systems. 
There is always a small chance of the entropy being less at the end of the time period during which the quantum machine runs, even if this probability decays exponentially as one increases the time period being considered.
However, we expect that any quantum machine left running long enough to produce a non-microscopic amount of work will have a basically negligible chance of violating the second law.

%% file: basics.tex
\section{Basic thermodynamics of non-equilibrium steady states}
\label{sec:nonequilibrium}

\begin{figure}[b]
\begin{center}
\centerline{\includegraphics[width=0.35\columnwidth]{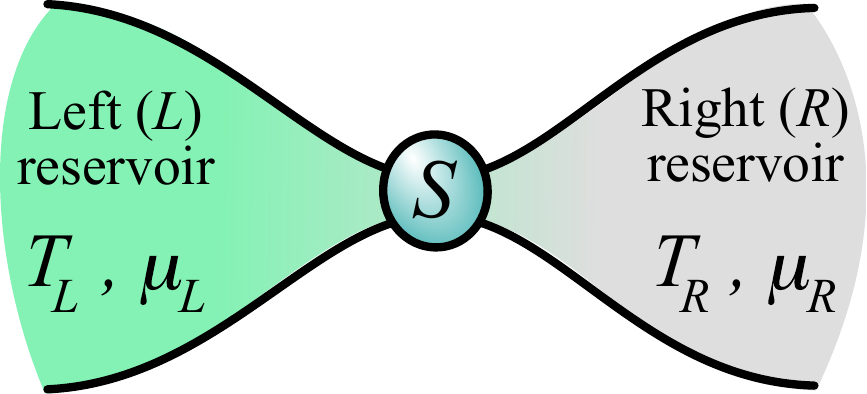}}
\caption{Schematic drawing of steady-state thermoelectric heat to work
conversion. A system $S$ is in touch with two reservoirs at temperatures
$T_L,T_R$ and electrochemical potentials $\mu_L,\mu_R$. 
%We assume $T_L>T_R$ and $\mu_L<\mu_R$.  
Note that, while currents are along the direction connecting
the two reservoirs, the motion of particles 
inside the reservoirs can be two or
three dimensional, and the motion in the system can be one, two or three dimensional.} 
\label{fig:scheme}
\end{center}
\end{figure}

Thermoelectric transport can be conveniently discussed within the 
model sketched in Fig.~\ref{fig:scheme}. Two particle reservoirs of 
respective temperatures $T_L > T_R$ and electrochemical 
potentials $\mu_L < \mu_R$ are connected by a system $S$, 
which allows for the exchange of heat and charged particles. 
We choose the reference values for temperature and electrochemical potential 
to be $T=T_R$ and $\mu=\mu_R$.
As soon as the \emph{steady state} is reached, 
constant \emph{heat and electric currents}, $J_{h}$ and $J_{e}$, 
flow \green{from the left reservoir to the right reservoir}.
We can also write $J_{e}=eJ_\rho$, where $e$ is the electron charge 
and $J_\rho$ the \emph{particle current}, and  
$J_{h}=T J_{\mathscr{S}}$, where $J_{\mathscr{S}}$ is the \emph{entropy current}
($\mathscr{S}$ being the entropy).
Moreover, the heat current is the difference between the 
total \emph{energy current} $J_{u}$ and the electrochemical potential 
energy current $\mu J_\rho$, \green{so that}
$J_{h}=J_{u}-\mu J_\rho=J_{u}-(\mu/e)J_{e}$ \cite{callen}. 
Depending on the sign of the currents, the machine works either 
as a power generator or a refrigerator. 

\subsection{Linear response and Onsager reciprocal relations}
\label{sec:linear_response}

To be in the linear response regime, $\Delta T$ and $\Delta \mu$ must be small.
We assume that both the temperature difference $\Delta T\equiv T_L -T_R>0$ 
and the electrochemical potential difference $\Delta \mu \equiv \mu_L-\mu_R<0$
are small, that is, $|\Delta T|\ll T$ and $|\Delta \mu|\ll k_B T$,
where $k_B$ is the Boltzmann constant.
The \emph{thermodynamic forces} (also known as generalized forces 
or affinities) driving the electric and heat currents are given by 
$\mathcal{F}_e=\Delta V/T$ (where $\Delta V=\Delta \mu/e$ is the applied voltage)
and $\mathcal{F}_h=\Delta T/T^2$ and the relationship
between currents and generalized forces is linear \cite{callen,mazur},
\begin{subequations}
\label{eq:coupledlinear}
\begin{eqnarray}
J_{e}=L_{ee} \mathcal{F}_e + L_{eh} \mathcal{F}_h,
\label{eq:coupledlinear-a}
\\
J_{h}=L_{he} \mathcal{F}_e + L_{hh} \mathcal{F}_h.
\label{eq:coupledlinear-b}
\end{eqnarray}
\end{subequations}
These relations are referred to as phenomenological
\emph{coupled} transport equations or \emph{linear response} equations
or kinetic equations
and the coefficients $L_{ab}$ ($a,b=e,h$) are
known as \emph{Onsager coefficients}.
We will define the matrix of these coefficients as the \emph{Onsager matrix}, ${\bm L}$, 
so 
\begin{eqnarray}
{\bm L} = \left(
\begin{array}{cc}
L_{ee} & L_{eh} \\
L_{he} & L_{hh} 
\end{array}
\right) \ .
\label{Eq:L-matrix}
\end{eqnarray}
%Perhaps we should also stress that we focus on the stationary and steady 
%state situations, where all the forces and responsive currents  
%are independent of time, on average, apart from fluctuations.

The \emph{entropy production rate} accompanying the coupled transport process 
reads \cite{callen,mazur}
\begin{equation}
\dot{\mathscr{S}}=\mathcal{F}_e J_{e} + \mathcal{F}_h J_{h}=
L_{ee} \mathcal{F}_e^2 + L_{hh} \mathcal{F}_h^2 +
(L_{eh}+L_{he}) \mathcal{F}_e \mathcal{F}_h.
\label{eq:sprod}
\end{equation}
The Onsager coefficients are subject to constraints.
\green{
Firstly, if the device is to satisfy the second law of thermodynamics, then  one requires that
 $\dot{\mathscr{S}}\ge 0$ for all  $\mathcal{F}_e$ and $\mathcal{F}_h$.
 It is easy to see from Eq.~(\ref{eq:sprod}) that this requires 
 $L_{ee}\ge 0$ and $L_{hh}\ge 0$, however this alone is not sufficient.
For the entropy production rate to be non-negative for all  $\mathcal{F}_e,\mathcal{F}_h$, 
we need that $\mathcal{F}_e = \mathcal{F}_h = 0$
is a minimum of the function $\dot{\mathscr{S}}$, and not just a saddle-point.
To see when this is the case, we can look at $\dot{\mathscr{S}}$ as a function of $\mathcal{F}_e$ for given  $\mathcal{F}_h$, and see that it is a quadratic function of $\mathcal{F}_e$, with a minimum at 
$\mathcal{F}_e= -\big(L_{eh}+L_{he}\big)\mathcal{F}_h\big/(2L_{ee})$ for which  $\dot{\mathscr{S}}$
takes the value $\big[4L_{ee}L_{hh} - \big(L_{eh}+L_{he}\big)^2 \big]\mathcal{F}_h^2\big/(4L_{ee})$.
This is only non-negative for all $\mathcal{F}_h$ if  $4L_{ee}L_{hh} \ge \big(L_{eh}+L_{he}\big)^2$.
If this inequality were not satisfied, a more little algebra shows the function $\dot{\mathscr{S}}$ would be a saddle-point about $\mathcal{F}_e = \mathcal{F}_h = 0$. 
 Thus the conditions for satisfying the second law of thermodynamics are}
\begin{equation}
L_{ee}\ \ge\  0,
\qquad
\hbox{ and } \ 
L_{hh}\ \geq\  
\frac{(L_{eh}+L_{he})^2}{4 L_{ee}}\ \geq \ 0.
%}
%\end{array}
%\right.
\label{dots}
\end{equation}
Second, assuming the property of time-reversal invariance of the
equations of motion, Onsager derived \cite{onsager} 
fundamental relations, known
as \emph{Onsager reciprocal relations} for the cross coefficients
of the Onsager matrix: $L_{ab}=L_{ba}$.
When an external magnetic field
${\bm B}$ is applied to the system, the laws of physics remain
unchanged if time $t$ is replaced by $-t$, provided that simultaneously
the magnetic field ${\bm B}$ is replaced by $-{\bm B}$. In this
case, the Onsager-Casimir relations \cite{onsager,casimir} read
\begin{equation}
L_{ab}({\bm B})=L_{ba}(-{\bm B}).
\end{equation}
At zero magnetic field, we recover the Onsager reciprocal relations
$L_{ab}=L_{ba}$. Note that only the diagonal coefficients
are bound to be even functions of the magnetic field with
$L_{aa}({\bm B})=L_{aa}(-{\bm B})$. \green{For $a\ne b$,
one has $L_{ab}({\bm B})\ne L_{ab}(-{\bm B})$, so $L_{ab}({\bm B})$ can have any ${\bm B}$ dependence}.

The Onsager coefficients are related to the familiar
transport coefficients. We have 
\begin{eqnarray}
G &=&\left(\frac{J_{e}}{\Delta V}\right)_{\Delta T=0}\ =\ \frac{L_{ee}}{T},
\label{eq:el_conductance}
\\
K &=& \left(\frac{J_{h}}{\Delta T}\right)_{J_{e}=0}
\green{\ =\ \frac{1}{T^2} \left (L_{hh} -{ L_{he}L_{eh} \over L_{ee}}\right)}
\ =\ \frac{1}{T^2}\frac{\det {\bm L}}{L_{ee}},
\label{eq:th_conductance}
\\
S&=&-\left(\frac{\Delta V}{\Delta T}\right)_{J_{e}=0} \ =\ 
\frac{1}{T}\frac{L_{eh}}{L_{ee}},
\label{eq:seebeck}
\end{eqnarray}
where $G$ is the (isothermal) \emph{electric conductance},
$K$ is the \emph{thermal conductance}\footnote{The definition of $K$ can confuse newcomers, 
the idea is that $K$ is given by the heat flow through the sample when it is coupled between two {\it electrically insulating} reservoirs at different temperatures. In such a set-up the reservoirs impose $J_{e}=0$, 
i.e.~the ``open circuit'' condition.
If the sample is a thermoelectric with non-zero $L_{eh}$ then Eq.~(\ref{eq:coupledlinear-a}) implies that 
that a bias will build up across the sample proportional to the temperature difference, $\mathcal{F}_e = -L_{eh} \mathcal{F}_h\big/L_{ee}$. Substituting this into  Eq.~(\ref{eq:coupledlinear-b}) gives Eq.~(\ref{eq:th_conductance}).}, and 
$S$ is the \emph{thermopower} (or \emph{Seebeck coefficient}).
%and ${\bm L}$ denotes the Onsager matrix with matrix elements 
%$L_{ab}$ ($a,b=e,h$).
The \emph{Peltier coefficient}
\begin{equation}
\Pi=\left(\frac{J_{h}}{J_{e}}\right)_{\Delta T=0}
=\frac{L_{he}}{L_{ee}}
\label{eq:Peltier-from-Ls}
\end{equation}
is related to the thermopower $S$ via the Onsager reciprocal 
relation: $ \Pi({\bm B})=TS(-{\bm B})$.
Note that the Onsager-Casimir relations imply 
$G(-{\bm B})=G({\bm B})$ and $K(-{\bm B})=K({\bm B})$,
but in general do not
impose the symmetry of the Seebeck coefficient under the exchange
${\bm B}\to -{\bm B}$.

\green{Inverting the above relations we have 
$L_{ee} = GT$, $L_{eh} = GST^2$, $L_{he}= G\Pi T$ and $L_{hh}=(K+G\Pi S)T^2$. 
Then we see from Eq.~(\ref{dots}) that the system must have 
\begin{eqnarray}
G \ \geq \ 0, \qquad \hbox{ and } \ K \ \geq\ G (ST-\Pi)^2\big/(4T) \ \geq \ 0,
\label{Eq:2nd-law-constraint-on-S-Pi-etc}
\end{eqnarray}
if it is to obey the second-law of thermodynamics.
It is worth noting that the second inequality implies that $K+G\Pi S \geq G(ST +\Pi)^2\big/(4T) \geq 0$.
Taking Eqs.~(\ref{eq:coupledlinear}), we can eliminate} the Onsager matrix elements in favor
of the transport coefficients $G,K,S,\Pi$, thus obtaining
\begin{subequations}
\label{Eq:currents-vs-DeltaV+DeltaT}
\begin{eqnarray}
J_{e} &=& G \Delta V + G S \Delta T,
\label{Eq:Je-vs-DeltaV+DeltaT}
\\
J_{h} &=& G\Pi\Delta V + (K+GS\Pi)\Delta T.
\label{Eq:Jh-vs-DeltaV+DeltaT}
\end{eqnarray}
\end{subequations}
By eliminating $\Delta V$ from these two equations we obtain an interesting
interpretation of the Peltier coefficient. The entropy current reads
\begin{equation}
J_{\mathscr{S}} = \frac{J_{h}}{T}=
\frac{\Pi}{T}\,J_{e}+\frac{K}{T}\Delta T.
%\label{eq:entropycurrent}
\nonumber 
\end{equation}
The first term, $\Pi/T$, can be understood as the entropy transported 
by the electron flow $J_{e}$. Since $J_{e}=e J_\rho$, each 
electron carries an entropy of $e \Pi/T$.
\green{The second term, $K \Delta T/T$, is the entropy generated by a heat flow from hot to cold, in the absence of an electric current.}\footnote{For time-reversal symmetric systems, the same 
interpretation applies to the Seebeck coefficient,
since in this case $S=\Pi/T$.}  
Similarly, the heat flow $J_{h}=TJ_{\mathscr{S}}$ is the sum of two terms,
$\Pi J_{e}$ and $K \Delta T$. 
It is then clear that two distinct processes contribute to 
the thermal transport: the \emph{advective} term $\Pi J_{e}$
is due to the electrical current flow, while the \emph{open-circuit}
term $K \Delta T$ is due to thermal conduction (by both electrons
and phonons) when there is no current flowing.
While the last term is irreversible,
the first one is reversible, that is, it changes sign when 
reversing the direction of the current. 
It can be intuitively understood that efficient energy 
conversion requires to minimize irreversible, dissipative processes 
with respect to reversible processes. 
%Hence, it is desirable to 
%have a large Peltier coefficient and a small heat conductance.

The heat dissipation rate $\dot{Q}$ can be 
computed from the entropy production rate 
in Eq.~(\ref{eq:sprod}),
\begin{equation}
\dot{Q}=T \dot{\mathscr{S}}=
\frac{J_{e}^2}{G}
+\frac{K}{T}(\Delta T)^2
+J_{e}(\Pi-TS)\frac{\Delta T}{T},
\label{eq:qprod}
\end{equation}
where the first term is the \emph{Joule heating}, the 
second term is the heat lost by thermal resistance 
and the last term, which disappears for time-reversal 
symmetric systems, can be negative when 
$J_{e}(\Pi-TS)<0$, thus reducing the dissipated heat. 
It is clear from  
Eq.~(\ref{eq:qprod}) that to minimize dissipative effects
for a given electric current and thermal gradient, we 
need a large electric conductance and low thermal conductance.

We conclude this section with two remarks. 
First, under the assumption of \emph{local equilibrium},
we can write
coupled equations like Eq.~(\ref{eq:coupledlinear}),
connecting the charge and heat current densities $j_e$, $j_h$ 
to local forces, expressed
in terms of gradients $\nabla \mu$, $\nabla T$ rather
than $\Delta \mu$, $\Delta T$ (see, for instance Ref.~\cite{callen}),

\begin{subequations}
\label{eq:coupledlocal}
\begin{eqnarray}
j_e&=&\lambda_{ee} (-\nabla\mu/eT) + \lambda_{eh} \nabla(1/T),
\\
j_h&=&\lambda_{he} (-\nabla\mu/eT) + \lambda_{hh} \nabla(1/T),
\end{eqnarray}
\end{subequations}
with $\lambda_{ab}$ ($a,b=e,h$) elements of the Onsager matrix
${\bm \lambda}$.
In this case,
Eqs.~(\ref{eq:el_conductance}) and (\ref{eq:th_conductance})
can be written with on the left-hand side the
\emph{electric conductivity} $\sigma$ and
the \emph{thermal conductivity} $\kappa$ rather than
the conductances $G$ and $K$ and on the right-hand side the 
kinetic coefficients $\lambda_{ab}$ rather than $L_{ab}$.

As a second remark, notice that 
we can equivalently express the coupled transport equations
in the ``energy representation'' rather than in the ``heat
representation''. That is, we consider the energy 
flow $J_{u}=J_{h}+\mu J_\rho=J_{h}+(\mu/e)J_{e}$ 
instead of the heat flow 
$J_{h}$. In this representation, 
the entropy production rate is given by
$\dot{\mathscr{S}}=\tilde{\mathcal{F}}_e J_{e} 
+ \tilde{\mathcal{F}}_u J_{u}$,
where the thermodynamic forces conjugated to the currents 
$J_{e}$ and $J_{u}$ are 
$\tilde{\mathcal{F}}_e=\Delta(V/T)=\mathcal{F}_e-(\mu/e)\mathcal{F}_h$ 
and $\tilde{\mathcal{F}}_u=\mathcal{F}_h$. 
In the energy representation, the kinetic equations read as follows,
\begin{subequations}
\label{eq:tildeJ1J2}
\begin{eqnarray}
{J}_e &=&\tilde{L}_{ee} \tilde{\mathcal{F}}_e + \tilde{L}_{eu} 
\tilde{\mathcal{F}}_u,
\\
{J}_u &=&\tilde{L}_{ue} \tilde{\mathcal{F}}_e + \tilde{L}_{uu} 
\tilde{\mathcal{F}}_e.
\end{eqnarray}
\end{subequations}
The elements $\tilde{L}_{ab}$ of the Onsager matrix
$\tilde{\bm L}$ are related to the matrix elements 
$L_{ab}$ of ${\bm L}$ as follows,
\green{
\begin{eqnarray}
L_{ee}=\tilde{L}_{ee}, \qquad 
L_{eh}=\tilde{L}_{eu}-(\mu/e) \tilde{L}_{ee}, \qquad 
L_{he}=\tilde{L}_{ue}-(\mu/e) \tilde{L}_{ee}, \qquad 
L_{hh}=\tilde{L}_{uu}- (\mu/e) (\tilde{L}_{eu}+\tilde{L}_{ue})+
(\mu/e)^2\tilde{L}_{ee}. \qquad \nonumber 
\end{eqnarray}
This means that $\det {\bm L}=\det \tilde{{\bm L}}$.}
The Onsager matrix $\tilde{\bm L}$ fulfills
reciprocity relations and obeys the same 
conditions imposed by the positivity of entropy production
as the matrix ${\bm L}$.

\green{Above we defined $T=T_R$, however it is worth noting that 
nothing changes if we take $T$ to be any typical system temperature, i.e. $T_L$, $T_R$ or an average of the two.  Differences due to the choice of $T$ will be at next order in powers of temperature difference, $T_L-T_R$,
when the above analysis is only accurate at lowest order.}

\subsection{\green{Stopping voltage and power versus load resistance}}
\label{sec:stopping_voltage}

\begin{figure}
\centerline{\includegraphics[width=0.5\columnwidth]{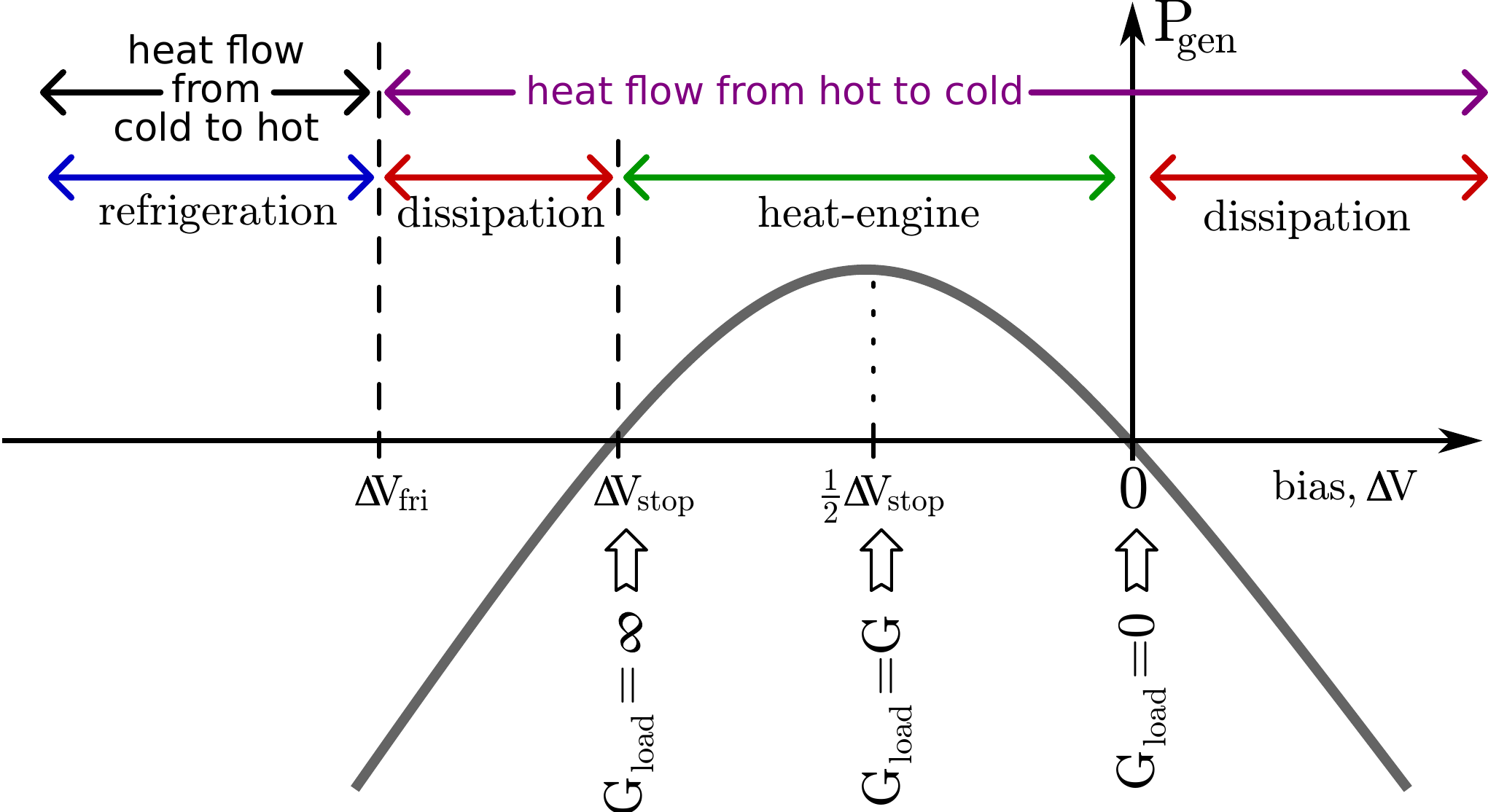}}
\caption{  
\green{A sketch of the dependence of the power generated by a heat-engine, $P_{\rm gen}$,
as a function of the bias, $\Delta V$.  
This parabolic curve comes from Eq.~\ref{Eq:Je-vs-DeltaV+DeltaT} under the assumption that $S>0$,
with $V_{\rm stop}$ given by Eq.~(\ref{Eq:V_stop}).
The curve's maximum is at $\half V_{\rm stop}$, at which $P_{\rm gen}=P^{\rm max}_{\rm gen}$ is given Eq.~(\ref{Eq:P_gen_max}).
The load conductance $G_{\rm load}$ is then given by Eq.~(\ref{Eq:G_load}).
For the bias to be outside the window between $ V_{\rm stop}$ and zero, the bias must be applied to the system via a power supply, see for example Fig.~\ref{Fig:energy-filter-thermocouples}b. For $\Pi >0$, 
the device acts as a refrigerator (using the applied bias to drive heat from cold to hot) if
$\Delta V < \Delta V_{\rm fri}$ given in Eq.~(\ref{eq:V_fri}).  In the regimes marked ``dissipation'' heat flow is from hot to cold and charge flow is from high bias to low bias (like in a resistor), so the system is dissipating 
both heat and work. The cases where $S$ and $P$ are not positive are mentioned in sections~\ref{sec:stopping_voltage} and \ref{sec:stopping_temp}.
}} 
\label{fig:power-vs-bias}
\end{figure}

\green{
The power output of a thermoelectric system (or any other steady-state thermodynamic machine)
depends not just on the machine itself; it also depends on the load it is connected to.
To fix our ideas, imagine a steady-state heat-engine coupled to a load which is an ideal motor,
so it converts into mechanical work all the electrical work supplied by the heat-engine.
The machine sees this load as a resistance, whose resistance determines the relationship 
between the bias across the heat-engine and the electrical current through the heat-engine.
The electrical power generated by the heat-engine and sent into the load will be $P_{\rm gen}=-\Delta V J_e$,
where the minus sign ensures power is generated  when current is driven against a potential difference.
Now if the load's resistance is zero, there will be a current through the thermoelectric but no bias, 
so the power output will be zero.  In contrast, if this resistance is infinite,
there will be a large bias (known as the stopping voltage, discussed below), 
but no current flow, so once again the power output is zero.  
The maximal power output is at a resistance between the two, as sketched in Fig.~\ref{fig:power-vs-bias}.
}

\green{In general, engineers are capable of matching the load resistance to the heat-engine, with the objective of maximizing the power generation.  For example, if the load is an electric motor, changing its resistance may just be a question of adding or removing turns in the coils in the motor.  Thus, what really matters is to calculate the power the heat-engine can generate under optimal conditions; i.e.~when the load resistance is chosen to maximize the power or the efficiency.
To be more quantitative, let us assume the load is electrically in-series with the heat-engine, and its 
resistance is $R_{\rm load} = 1/G_{\rm load}$.  If the heat-engine is described by 
Eq.~(\ref{Eq:Je-vs-DeltaV+DeltaT}),
then current conservation gives
\begin{eqnarray}
J_{\rm e}= -G_{\rm load} \Delta V
\label{Eq:G_load}
\end{eqnarray}
where the minus sign is because the  bias across the load is opposite to across the heat-engine.
Now we know from the above argument that the power the heat-engine gives to the load will be zero at 
$G_{\rm load}=0$ and $G_{\rm load}=\infty$ and will be maximal somewhere between the two, 
see Fig.~\ref{fig:power-vs-bias}. 
In the first case, the bias $\Delta V$ will be zero, when in the second case it will be the 
electrical current  $J_e=0$.  The bias at which $J_e=0$ is called the {\it stopping voltage}, 
because it is the voltage that builds up to stop the current flow; 
from Eq.~(\ref{Eq:Je-vs-DeltaV+DeltaT}) we see that  $J_e=0$ 
occurs at 
\begin{eqnarray}
\Delta V \ =\colorproofs{ \Delta V_{\rm stop}} \ =\ S \Delta T .
\label{Eq:V_stop}
\end{eqnarray}
This is natural, given that the definition of the Seebeck coefficient, $S$, as the ratio between the
voltage and the temperature difference, when the thermoelectric is 
not connected to a circuit (so $J_e=0$).
Thus we know that a heat-engine will generate finite power, $P_{\rm gen}$, will be between
zero and $\Delta V_{\rm stop}$.
}

\green{
Note that we can recast the stopping voltage in terms of thermodynamic forces, 
by defining the stopping force ${\mathcal{F}}_e^{\rm stop}$ 
as the ${\mathcal{F}}_e$ at which the electrical current $J_e$ vanishes for a given ${\mathcal{F}}_h$.
Then from Eq.~(\ref{eq:coupledlinear-a}) we have 
\begin{eqnarray}
{\mathcal{F}}_{e,{\rm stop}} = - {L_{eh} \over L_{ee}} {\mathcal{F}}_h
\label{eq:stopping-force}
\end{eqnarray}
which of course simply means that ${\mathcal{F}}_e^{\rm stop}= \Delta V_{\rm stop}/T$
as it should given Eqs.~(\ref{eq:el_conductance}-\ref{eq:Peltier-from-Ls}).
}

\green{
To find the maximum power generation, $P_{\rm gen}=-\Delta V J_e$, it is convenient to forget Eq.~(\ref{Eq:G_load}) and treat $J_e$ and $\Delta V$ as quantities only related by Eq.~(\ref{Eq:Je-vs-DeltaV+DeltaT}). 
Then, we find the $\Delta V$ which maximizes $P_{\rm gen}$ is 
$\Delta V \ =\  -{1\over 2} S \Delta T  \ \equiv\ {1 \over 2} V_{\rm stop}$.
Hence, the power generated is maximized when the bias is half the stopping voltage
(or when the thermodynamic force ${\mathcal{F}}_e$ is half the stopping force).   
The electric current is then $J_e = \half G S \Delta T$, and so the maximum power is 
\begin{eqnarray}
P^{\rm max}_{\rm gen}\ =\ {1 \over 4} G V_{\rm stop}^2 \ =\ {1 \over 4} G S^2 \, \Delta T^2
\label{Eq:P_gen_max}
\end{eqnarray}
which one can equally write as $P^{\rm max}_{\rm gen} \ =\ {1 \over 4} T L_{ee} \,{\mathcal{F}}_{e,{\rm stop}}^2$ 
with ${\mathcal{F}}_{e,{\rm stop}}$ given by Eq.~(\ref{eq:stopping-force}).
Now using the fact that $\Delta V= -\half S \Delta T$ and  $J_e = \half G S \Delta T$ at maximum power, we can use Eq.~(\ref{Eq:G_load}), if we need to.  The answer is that maximum power is delivered when $G_{\rm load}=G$;
in other words, when the load resistance matches the thermoelectric's resistance.
}

\green{
Note that in this section we assumed $S \geq 0$ (i.e. $L_{eh}\geq0$.
If we change the sign of $S$ (i.e. change the sign of $L_{eh}$), then everything we say here follows through,
if one also changes the sign of the bias.  Thus $V_{\rm stop}$ will be positive, and device will work as a heat engine for positive biases less than $V_{\rm stop}$, with maximum power at $\half V_{\rm stop}$.
This is just like flipping the curve in Fig.~\ref{fig:power-vs-bias} 
about the $y$-axis, so $\Delta V \to -\Delta V$.
}

\subsection{\green{Stopping temperature of a refrigerator}}
\label{sec:stopping_temp}

\green{
Just as a heat-engine obeying Eqs.~(\ref{Eq:currents-vs-DeltaV+DeltaT}) has a stopping voltage, a refrigerator  has a stopping temperature, which is the maximum $\Delta T$ it can support for a given $\Delta V$.
To function as a refrigerator,  Eq.~(\ref{Eq:Jh-vs-DeltaV+DeltaT}) must have $J_h$ with the opposite sign 
from $\Delta T$.  
Here we take a given positive $\Delta T=T_L-T_R$, then refrigeration of reservoir R occurs when $J_h$ is negative.
If the system has positive $S$ and positive $\Pi$ then this requires that
\begin{eqnarray}
\Delta V \ <\ \Delta V_{\rm fri} \ \equiv\ - {K+GS\Pi \over G \Pi} \Delta T\ .
\label{eq:V_fri}
\end{eqnarray}
as shown in Fig.~\ref{fig:power-vs-bias} for $\Delta T > 0$.
We can write $\Delta V_{\rm fri} = \Delta  V_{\rm stop}-K\, \Delta T\big /(G\Pi) $, which means that 
$\Delta V_{\rm fri} < \Delta  V_{\rm stop}$, and so there is always a regime of ``dissipation'' between
the bias at which the system is a heat-engine and the bias at which it is a refrigerator.
We use the term  ``dissipation'' for this regime, because heat flows from hot to cold, and electrical current 
flows from high to low bias (like in a resistor).  This dissipative regime only vanishes in the limit $K \to 0$,
which is the limit which corresponds to  {\it tight-coupling} (see Section~\ref{sec:ZT}).
}

\green{
Inverting Eq.~(\ref{eq:V_fri}) we find that a refrigerator driven by a given negative bias $\Delta V$ will 
not be able to remove heat from the cold reservoir unless
\begin{eqnarray}
\Delta T \ <\ \Delta T_{\rm stop} \ \equiv - {G\Pi \over K+GS\Pi} \Delta V
\label{Eq:T_stop}
\end{eqnarray}
Thus the cold reservoir will get colder as the refrigerator extracts heat from it, 
until the temperature difference approaches tends to $\Delta T_{\rm stop}$, 
at which point the cooling will slow to zero, and the cold reservoir will not get any colder.
}

\green{
If the system has positive $S$, but negative $\Pi$, then its function as a heat-engine is unchanged, but now it works as a refrigerator in a regime of positive $\Delta V $ (as before we take $\Delta T >0$) defined by
\begin{eqnarray}
\Delta V \ >\  \Delta V_{\rm fri} \equiv -{K+GS\Pi \over G \Pi} \Delta T\ .
\end{eqnarray}
where $\Delta V_{\rm fri}$ is now a positive quantity. 
If we replotted  Fig.~\ref{fig:power-vs-bias} for negative $\Pi$ (keeping $S$ positive), the heat-engine regime would be unchanged, but the refrigerator regime would move to positive $\Delta V$ (at $\Delta V > \Delta V_{\rm fri}$), where it would be separated from the heat-engine regime by a dissipation regime 
(at $0<\Delta V < \Delta V_{\rm fri}$).   It is worth noting that $K+GS\Pi$ is always positive (see below Eq.~(\ref{Eq:2nd-law-constraint-on-S-Pi-etc})), but it is smaller when $S$ and $\Pi$ have opposite sign.  Thus, the 
stopping temperature for a given magnitude of the bias is smaller  when $S$ and $\Pi$ have opposite sign.
}

\green{
Finally, we note that the physics for negative $S$ is the same as described above, once we take $S \to -S$, $\Pi \to -\Pi$ and $\Delta V \to -\Delta V$, so the curve in Fig.~\ref{fig:power-vs-bias} is flipped about the y-axis.
}

\subsection{\green{Lowest refrigeration temperature and $ZT$}}

\green{
Section 2.2 of Goldsmid's textbook \cite{goldsmid} gives an elegant argument which says
that the maximum temperature difference that a refrigerator can
achieve is given by its dimensionless figure of merit $ZT$ given in Eq.~(\ref{Eq:ZT-intro}).
This argument is based on the idea that (unlike in Eq.~(\ref{Eq:T_stop})) a large bias is bad for refrigeration, because it generates a lot of Joule heat, and about half of that heat will flow back into the reservoir being cooled. 
For a refrigerator with dimensionless figure of merit $ZT$ given in Eq.~(\ref{Eq:ZT-intro}),
which is cooling a reservoir to a temperature $\Delta T$ below that of the environment (so the environment is at temperature $T$ and the reservoir being cooled is at $T_C = T-\Delta T$).  
The argument leads to the conclusion that $\Delta T \leq \Delta T_{\rm limit}$ 
where $\Delta T_{\rm limit}$ is given by
\begin{eqnarray}
{\Delta T_{\rm limit} \over T}  \ \simeq\   { ZT \over 2}
\label{Eq:T_limit-vs-ZT}
\end{eqnarray}
Thus the refrigerator will never cool a reservoir to a temperature below about $\left(1-\half ZT\right)T$.
Here we outline the argument which leads to this relation, and briefly explain the conditions under which 
it is likely to by broken by a nanoscale refrigerator \cite{whitney2013-catastrophe}.
}

\green{
To arrive at Eq.~(\ref{Eq:T_limit-vs-ZT}), one starts with the linear response equations in Eqs.~(\ref{Eq:currents-vs-DeltaV+DeltaT}) for a refrigerator. One then notes that Eq.~(\ref{Eq:Je-vs-DeltaV+DeltaT}) implies that the system is dissipating electrical work equal to $J_e \Delta V$ as Joule heat (in other words the power generated 
$P_{\rm gen} =- J_e \Delta V <0$), but that this Joule heat does not appear in Eq.~(\ref{Eq:Jh-vs-DeltaV+DeltaT}).  To remedy this, one should add a Joule heating term to Eq.~(\ref{Eq:Jh-vs-DeltaV+DeltaT}), 
this term is non-linear and violated conservation of $J_h$, as such we have to define a heat current  
$J_{h,L}$ from the refrigerator into reservoir L, and a current  $J_{h,R}$ from the refrigerator into reservoir R.
If we were to stay with linear-response and neglect Joule heating, we would have $-J_{h,L}=J_{h,R}=J_h$ where $J_h$ is given by Eq.~(\ref{Eq:Jh-vs-DeltaV+DeltaT}).   However, once we add the Joule heating term
we have $J_{h,L}+J_{h,L} = J_{e}\Delta V$, using Eq.~(\ref{Eq:Je-vs-DeltaV+DeltaT}) this becomes 
$J_{h,L}+J_{h,L} = (G\Delta V + GS\Delta T)\Delta V$.  
Let us assume that the proportion of the Joule heating will go to the cold reservoir (which we take to be reservoir R) is $\alpha$, so the proportion that goes to the 
hot environment (reservoir L) is $1-\alpha$.  Then we have
\begin{eqnarray}
J_{h,R} &=& \ \ \big[G\Pi \Delta V + (K+GS\Pi)\Delta T \big] 
\ +\ \alpha  \, \big[ (G\Delta V + GS\Delta T)\Delta V \big],
\\
J_{h,L} &=& - \big[G\Pi \Delta V + (K+GS\Pi)\Delta T \big] 
\ +\ (1-\alpha)\, \big[ (G\Delta V + GS\Delta T)\Delta V \big],
\end{eqnarray}
where the first square bracket in each expression comes from  Eq.~(\ref{Eq:Jh-vs-DeltaV+DeltaT}),
and the second square-bracket comes from the Joule heating. These equations are an approximation because 
the only non-linear term we consider is that associated with Joule heating, when in reality there are many other non-linear terms coming from the $T$ dependences of $G$, $S$, $\Pi$, and $\Pi$. 
}

\green{
Now, we see that (unlike in Section~\ref{sec:stopping_temp}) making the bias more and more negative 
does not make the heat flow into the cold reservoir $J_{h,R}$ more and more negative.
Instead, the fact the Joule heating term is quadratic in $\Delta V$ means the most negative value of $J_{h,R}$ occurs when $\Delta V= - (\Pi +\alpha S \Delta T)\big/ (2\alpha)$.
Thus the most negative heat current (i.e.~the maximum cooling power) is 
\begin{eqnarray}
J_{h,R} &=& - {G \over 4 \alpha}\big(\Pi +\alpha S \Delta T\big)^2  \ +\ K \Delta T. 
\end{eqnarray}
Now to follow Goldsmid's argument, we assume that the system has $\Pi=ST$, such as in a system with time-reversal symmetry. 
Then we have
\begin{eqnarray}
J_{h,R} &=& K T \left[{ \Delta T \over T} - {ZT \over 4\alpha} \left(1+{ \Delta T \over T}\right)^2 \right]. 
\label{Eq:J_hR-at-max-cooling-power}
\end{eqnarray}
where $ZT$ is given by Eq.~(\ref{Eq:ZT-intro}).
If $J_{h,R}$ is negative for a given $\Delta T$, then the the cold reservoir can be cooled further by the 
refrigerator, this cooling only stops when $\Delta T$ reaches a value where $J_{h,R}$ is no longer negative.
From Eq.~(\ref{Eq:J_hR-at-max-cooling-power}) for $ZT < \alpha$, we see that cooling happens for all
$\Delta T$ down to $\Delta T_{\rm limit}$ where
\begin{eqnarray}
{\Delta T_{\rm limit} \over T} \ =\ {2\alpha \over ZT}  \ \left(\, 1-{ZT \over2\alpha} - \sqrt{1-{ZT \over \alpha}}\,\right)  \qquad \hbox{ for }\ {ZT \over \alpha} < 1
\label{Eq:T_limit-vs-ZT-2}
\end{eqnarray}
As the cooling power can be negative for all $\Delta T \leq  \Delta T_{\rm limit}$, it means that the refrigerator will be able to cool the cold reservoir down to $T_C = T -\Delta T_{\rm in}$ if one waits long enough
(where $T$ is the environment temperature).
}

\green{
In contrast, if  $ZT > \alpha$ then we can see that cooling happens at all $\Delta T$, this implies that the cold reservoir can be cooled to arbitrary low temperatures (even unphysical negative temperatures). This is a clear indication of a deep problem with this argument for large $ZT$.  The problem is that we assumed that we were close enough to linear-response to the linear-response equations and only adding one non-linear term (the term giving Joule heating). This assumption may or may not be reasonable in any given circumstance, however it is clear that it is only self-consistent for systems cooling down to $\left(T-\Delta T_{\rm limit}\right)$ if $\Delta T_{\rm limit}/T $ remains small enough to stay close to the linear-response regime (i.e.~that we can neglect the $\Delta T$ dependence $G$, $K$ and $S$ when $\Delta T \sim \Delta T_{\rm limit}$).  
Thus for a nanoscale system where the linear-response equations fail as soon as $\Delta T/ T$ or 
$e\Delta V\big/(k_{\rm B} T)$ is not small (see Section~\ref{Sect:Thermoelectrics_traditional_vs_quantum}), one sees that Eq.~(\ref{Eq:T_limit-vs-ZT-2}) is only a good estimate of $\Delta T_{\rm limit}$ if $ZT \ll 1$,
in which case we can expand the square-root to get
\begin{eqnarray}
{\Delta T_{\rm limit} \over T} \ =\ {ZT \over 4\alpha} \ +\ {1 \over 8}\left({ZT \over \alpha}\right)^2 
\ +\ {\cal O}\left[ \left({ZT\over \alpha}\right)^3\right] \ .
\end{eqnarray}
Goldsmid's argument made the additional assumption that the Joule heat is approximately equally divided between the hot and cold reservoirs, so $\alpha \simeq 1/2$. Then for $ZT \ll \alpha$ one gets Eq.~(\ref{Eq:T_limit-vs-ZT}).
}

\green{
One can look at Ref.~\cite{whitney2013-catastrophe} to see how the above argument fails for a nanoscale system when $ZT$ is {\it not} small, and so the cooling makes $\Delta T$ large enough that one must take into account all non-linear effects. That work considered a simple non-linear theory of a quantum point-contact (which has $ZT \simeq 1.4$) 
acting as a nanoscale refrigerator. It shows that the  $\Delta T_{\rm limit}$ (and the manner one gets to that limiting temperature) are very different from that discussed above.
}

%% file: efficiencies.tex
\section{Thermodynamic efficiency of steady-state thermal machines}

\label{sec:TE}
One of the pillars of thermodynamics is the existence of an upper bound
on the efficiency of the conversion of heat to work. 
Given any thermal machine operating as a heat engine between
two reservoirs at temperatures $T_L$ and $T_R$ $(T_L>T_R)$,
the efficiency $\eta^{\rm eng}$, defined as the ratio  of the performed work $W$
over the heat $Q_L$ extracted from the high temperature reservoir,
where we use the superscript``eng'' to indicate that it is the efficiency of a heat engine.
This efficiency is bounded by the \emph{Carnot efficiency} $\eta_{\rm C}^{\rm eng}$ \cite{carnot},
\begin{equation}
\eta^{\rm eng} \ =\  \frac{W}{{Q}_L} \leq \eta_{\rm C}^{\rm eng} \ =\ 1-\frac{T_R}{T_L}.
\end{equation}
The ideal Carnot efficiency may be achieved if the conversion process
is \emph{reversible}. Since a thermodynamic reversible transformation
is \emph{quasi-static}, \green{the thermodynamic cycle will take an infinite time.
This is not only impractical, it means the power generated 
(i.e.~the work generated per cycle divided by the cycle's period) is vanishingly small.
Therefore an engine ideally working at the Carnot efficiency would
be useless.  
Of course, the idea is to operate a real machine with a finite cycle time, so the process will not quite be reversible and a small amount of entropy will be generated.  
This will make its efficiency slightly less than  $ \eta_{\rm C}^{\rm eng}$, but it will generate a finite power.  
An important practical question is to quantify how much}
the efficiency deteriorates when heat to work conversion takes place
in a finite time. This is a central question in the field of \emph{finite-time thermodynamics}
(for a review, see \cite{andresen11}).

Hereafter, we focus on steady-state thermal machines, while the
discussion of cyclic thermal machines is postponed to 
Chapter~\ref{sec:CTM}. Owing to the steady-state, we can 
write the efficiencies of heat-engines and refrigerators in terms of heat currents
 and power, as in Section~\ref{Sect:intro-eff}.

 % \begin{equation}
% \eta^{\rm eng} =\frac{P_{\rm gen}}{J_{\rm h,L}},
% \end{equation}
% where the power generated $P_{\rm gen}=\big(\rmd W\big/\rmd t \big)>0$,
% and the rate of heat flow out of
% the hot reservoir $J_{\rm h,L}=\big(\rmd Q_L\big/\rmd t \big)>0$.

% \green{
% We now turn to a refrigerator, which uses work to drive a heat flow against a temperature gradient.
% Its cooling power is defined as the heat current it sucks out of the reservoir being refrigerated.
% If the power the refrigerator absorbs to do the refrigeration is $P_{\rm abs}$, then the 
% refrigerator's efficiency, which is more often called its {\it coefficient of performance} or COP,
% is defined as
% \begin{equation}
% \eta^{\rm fri}=\frac{J_{\rm h,L}}{P_{\rm abs}},
% \end{equation}
% where we assume that it is reservoir L that is being refrigerated, so it is colder than reservoir R, $T_{\rm L} < T_{\rm R}$.
% The upper bound on this COP is the Carnot efficiency, $\eta_{\rm C}^{\rm fri}$,  which can only be achieved if the conversion of work into heat-flow is reversible,  
% \begin{equation}
% \eta^{\rm fri} \ \leq  \ \eta_{\rm C}^{\rm fri} \ =\left(\frac{T_R}{T_L} -1 \right)^{-1}.
% \end{equastion}
% }

\subsection{Figure of merit for thermoelectric efficiency}
\label{sec:ZT}

Within linear response, \green{as given by Eqs.~(\ref{eq:coupledlinear}),}
the efficiency of steady-state conversion of heat to work
reads
\begin{equation}
%\eta=\frac{\dot{W}}{\dot{Q}_L}=\frac{-T\mathcal{F}_e_J_{e}}{J_2}
\eta=\frac{P}{J_{\rm h,L}}=\frac{-(\Delta V) J_{e}}{J_{h}}
=\frac{-T \mathcal{F}_e (L_{ee}\mathcal{F}_e +L_{eh}\mathcal{F}_h)}{L_{he}
\mathcal{F}_e+L_{hh}\mathcal{F}_h},
\label{eq:efficiency}
\end{equation}
where $P=-(\Delta V) \ J_{e}>0$.

The maximum of $\eta$ over $\mathcal{F}_e$, for fixed $\mathcal{F}_h$, 
i.e. over the applied voltage $\Delta V$ for a given temperature 
difference $\Delta T$, is achieved for
\begin{equation}
\mathcal{F}_e=- \frac{L_{hh}}{L_{he}}\left( 
1-\sqrt{\frac{\det{\bm L}}{L_{ee}L_{hh}}}\right)\mathcal{F}_h.
\label{eq:etamaxcondition}
\end{equation}
\green{where we recall that $\det{\bm L}= L_{ee}L_{hh}-L_{eh}L_{he}$.
It is worth doing a bit of algebraic manipulation to write this in term of 
the dimensionless quantity 
\begin{eqnarray}
y= {L_{eh}L_{he}\over \det{\bm L}} \ ,
\end{eqnarray}
where it takes the form
\begin{eqnarray}
\mathcal{F}_e \ =\ \mathcal{F}_{e,{\rm stop}} 
\ {1+y - \sqrt{1+y}
\over y}
\label{Eq:F_e-max-efficency}
\end{eqnarray}
where $\mathcal{F}_{e,{\rm stop}}$ is the stopping force in Eq.~(\ref{eq:stopping-force}).
In systems with $L_{eh}=L_{he}$, such as systems with time-reversal symmetry,
\begin{eqnarray}
y  \ \to\  {L^2_{eh}\over \det{\bm L}} \ = \ {GS^2 T \over K} \ \equiv \ ZT
\end{eqnarray}
is the dimensionless figure of merit introduced in Eq.~(\ref{Eq:ZT-intro}).
Then we can see that a poor thermoelectric, given by the low $ZT$ limit of Eq.~(\ref{Eq:F_e-max-efficency}),
has maximum efficiency at $\mathcal{F}_{e}=\half \mathcal{F}_{e,{\rm stop}}$. This coincides with the
condition for it to have maximum power.  In the opposite limit, an ideal thermoelectric
with $ZT \to \infty$ has a maximum efficiency at $\mathcal{F}_{e}= \mathcal{F}_{e,{\rm stop}}$.
}

\green{Here we continue by considering only systems with $L_{eh}=L_{he}$ (such as those with time-reversal symmetry),
and postpone discussion of systems with $L_{eh}\neq L_{he}$ to section~\ref{sec:efficiencymagnetic}.
Taking Eq.~(\ref{Eq:F_e-max-efficency}) with $y \to ZT$ and substituting it into Eq.~(\ref{eq:efficiency}),
one find that the \emph{maximum efficiency} is  \cite{goldsmid}}
\begin{equation}
\eta_{\rm max}=
\eta_{\rm C}\,
\frac{\sqrt{ZT+1}-1}{\sqrt{ZT+1}+1},
\label{etamaxB0}
\end{equation}
where the Carnot efficiency, $\eta_{\rm C}=1-T_R/T_L$, in the linear-response regime takes the form $\eta_{\rm C} = \Delta T/T= T \mathcal{F}_h$.

\begin{figure}
\begin{center}
\centerline{\includegraphics[width=0.5\columnwidth]{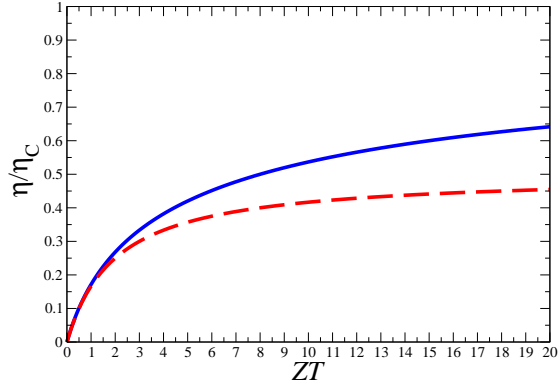}}
\caption{Linear response efficiency for heat to work conversion, 
in units of Carnot efficiency $\eta_{\rm C}$, as a function of the figure
of merit $ZT$. The top and the bottom curve correspond to the 
maximum efficiency $\eta_{\rm max}$ and to the efficiency at 
the maximum power $\eta(P_{\rm max})$, respectively.}
\label{fig:ZT}
\end{center}
\end{figure}

In general, $ZT$ depends on the size of the system, 
since this is the case for $G$, $K$ and $S$.
On the other hand, if we are in the diffusive transport
regime where Ohm's scaling law $G=\sigma A/\Lambda$ and 
Fourier's scaling law $K=\kappa A/\Lambda$ hold, where 
$A$ and $\Lambda$ are the cross section area and length of the material,
$\sigma$ the electric conductivity and 
$\kappa$ the thermal conductivity, then $G/K=\sigma/\kappa$. 
If moreover $S$ is size-independent, then 
the figure of merit can be expressed in terms of 
the material transport coefficients $\sigma$, $\kappa$ and $S$: 
\begin{equation}
ZT=\frac{\sigma S^2}{\kappa}\,T.
\end{equation}
The only restriction imposed by thermodynamics 
(more precisely, by the positivity of the entropy production rate) 
is $ZT\ge 0$, since $G=L_{ee}/T\ge 0$ and $K={\det {\bm L}}/(T^2 L_{ee})\ge 0$.
It is easy to see that $\eta_{\rm max}$ is a monotonous growing function 
of $ZT$, with $\eta_{\rm max}=0$ when $ZT=0$ and 
$\eta_{\rm max}\to \eta_{\rm C}$ when $ZT\to\infty$ (full curve in 
Fig.~\ref{fig:ZT}). 

Note that the divergence of $ZT$ (leading to the Carnot efficiency) 
implies that the condition number
\begin{equation}
{\rm cond}({\bm L})\equiv \frac{[{\rm Tr}({\bm L})]^2}{\det {\bm L}}> ZT
\end{equation}
also diverges. 
As a consequence, the Onsager matrix ${\bm L}$
is ill-conditioned, namely
the ratio  
\begin{equation}
\frac{\lambda_+({\bm L})}{\lambda_-({\bm L})}=
\frac{1+\sqrt{1+\frac{4}{{\rm cond}({\bm L})}}}{1
-\sqrt{1-\frac{4}{{\rm cond}({\bm L})}}}
\end{equation}
diverges; here 
$\lambda_+ ({\bm L})$ and $\lambda_- ({\bm L})$ denote the largest and 
the smallest eigenvalue of ${\bm L}$, respectively.
Therefore, in the limit $ZT\to\infty$ the system
(\ref{eq:coupledlinear}) becomes singular.
That is,
$J_{u}=c J_{e}$, with the proportionality factor $c$ being independent
of the values of the applied thermodynamic forces.
In short, within linear response (and without external magnetic fields
or other effects breaking time-reversal symmetry) the Carnot efficiency
can be obtained only if charge and energy flows are proportional, \green{this is} known as the 
\emph{tight coupling} condition (also sometimes called \emph{strong coupling}).

In most physical systems ${\rm Tr}({\bm L})$ has a finite upper bound, which means the tight coupling condition requires $\det[{\bm L}] \to 0$.  In this case Eqs.~(\ref{eq:el_conductance},\ref{eq:th_conductance}) tell us that the ratio of thermal conductance to electrical conductance vanishes, $K\big/G \to 0$.
Thus a system which achieves large $ZT$ is likely to strongly violate the Wiedemann-Franz law (for an example see
Section~\ref{sec:energyfiltering}).

\subsection{Efficiency at maximum power}
\label{sec:ZTmaxpower}

The output power 
\begin{equation}
P=-(\Delta V) J_{e} 
=-T \mathcal{F}_e (L_{ee}\mathcal{F}_e +L_{eh}\mathcal{F}_h)
\end{equation}
is maximum when 
\begin{equation}
\mathcal{F}_e=-\frac{L_{eh}}{2L_{ee}}\,\mathcal{F}_h
\label{eq:X1max}
\end{equation}
and is given by 
\begin{equation}
P_{\rm max}=\frac{T}{4}\frac{L_{eh}^2}{L_{ee}}\,\mathcal{F}_h^2=
\frac{\eta_{\rm C}}{4}\,\frac{L_{eh}^2}{L_{ee}}\,\mathcal{F}_h.
\label{eq:Pmax}
\end{equation}
Using Eqs.~(\ref{eq:el_conductance}) and (\ref{eq:seebeck})
we can also write
\begin{equation}
P_{\rm max}=\frac{1}{4}\,S^2 G (\Delta T)^2.
\label{Eq:max-power}
\end{equation}
We can see from this last equation that the maximum power
is directly set by the combination $S^2 G$, known 
for this reason as \emph{power factor}. 
Note that $P$ is a quadratic function of $\mathcal{F}_e$ and the
maximum is obtained for the value (\ref{eq:X1max}) 
corresponding to half of the so-called \emph{stopping force} in Eq.~(\ref{eq:stopping-force}).
\begin{equation}
\mathcal{F}_e^{\rm stop}=-\frac{L_{eh}}{L_{ee}}\,\mathcal{F}_h,
\end{equation}
that is, of the value for which the electric current vanishes,
$J_{e}(\mathcal{F}_e^{\rm stop})=0$.
For systems with time reversal symmetry, 
the efficiency at maximum power 
reads \cite{vandenbroeck2005}
\begin{equation}
\eta(P_{\rm max})=\frac{\eta_{\rm C}}{2}\frac{ZT}{ZT+2}.
\label{etawmaxB0}
\end{equation}
This quantity also is a monotonous growing function of 
$ZT$, with $\eta(P_{\rm max})=0$ when $ZT=0$ and
$\eta(P_{\rm max})\to \eta_{\rm C}/2$ when $ZT\to\infty$ (dashed curve in
Fig.~\ref{fig:ZT}). For small $ZT$ we
have $\eta(P_{\rm max})\approx\eta_{\rm max}\approx 
(\eta_{\rm C}/4) ZT$. The difference between 
$\eta(P_{\rm max})$ and $\eta_{\rm max}$ becomes 
relevant only for $ZT>1$. 
\red{
It is useful to point out at this stage that the
bound $\eta_{\rm C}/2$ coincides with the linear response
expansion of the Curzon-Ahlborn efficiency.  This will be discussed
in Section~\ref{sec:CTM} for cyclic thermal machines.}

\subsection{Efficiency versus power}

\begin{figure}
\begin{center}
\centerline{\includegraphics[width=0.5\columnwidth]{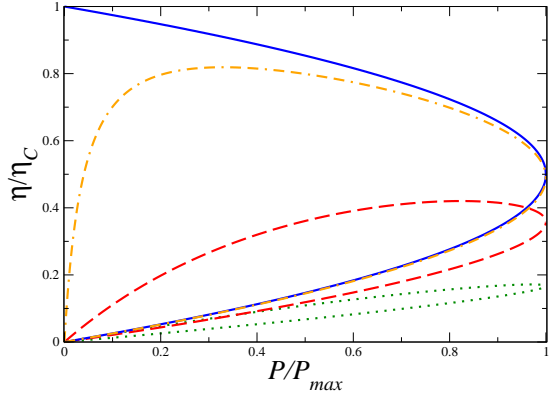}}
\caption{Relative efficiency $\eta/\eta_{\rm C}$ versus normalized 
power $P/P_{\rm max}$. From bottom to top: $ZT=1,5,100$, and $\infty$.
In each curve the lower branch corresponds to a force ratio
$r\le 1/2$, the upper branch to $r\ge 1/2$. Maximum efficiency is 
always achieved on the upper branch.  
\green{
For refrigerators, similar curves can be plotted of efficiency versus cooling power, 
see for example Refs.~\cite{Gordon-Ng-book,Kosloff-Levy2014}
}}
\label{fig:eta_power}
\end{center}
\end{figure}

In this section, we discuss how it is possible to 
establish a linear-response efficiency versus power plot. 
We can express the ratio between the power at a given value of
$\mathcal{F}_e$ and the maximum power as a function of 
the \emph{force ratio} $r=\mathcal{F}_e/\mathcal{F}_e^{\rm stop}$:
\begin{equation}
\frac{P}{P_{\rm max}}=4 r (1-r). 
\end{equation}
This relation can be inverted:
\begin{equation}
r=\frac{1}{2}\,\left[1\pm \sqrt{1-\frac{P}{P_{\rm max}}}\right],
\end{equation}
with the plus sign for $r\ge 1/2$ and the minus sign for $r\le 1/2$.
Inserting this latter relation into Eq.~(\ref{eq:efficiency}) we can 
express the efficiency (normalized to the Carnot efficiency) as 
\begin{equation}
\frac{\eta}{\eta_{\rm C}}=
\frac{\displaystyle{\frac{P}{P_{\rm max}}}}{
\displaystyle{2\left(1+\frac{2}{ZT}\mp \sqrt{
1-\frac{P}{P_{\rm max}}}\right)}},
\end{equation}
where the minus sign corresponds to $r\ge 1/2$, the plus sign to
$r\le 1/2$.
Plots of the normalized efficiency versus the normalized power
are shown in Fig.~\ref{fig:eta_power}, for several values 
of the figure of merit $ZT$. 
Note that, while for low values of $ZT$ the maximum efficiency 
is close to the efficiency at maximum power, for large $ZT$
the difference becomes relevant (see also Fig.~\ref{fig:ZT}).
For $ZT=\infty$ the Carnot efficiency is achieved at $P=0$, i.e. at 
the stopping force: $\mathcal{F}_e=\mathcal{F}_e^{\rm stop}$, namely $r=1$. 

%%%%%%%%%%%%%%%%%%%%%%%%%%%%%%%%%
%%%%%%%%%%%%%%%%%%%%%%%%%%%%%%%%%

\subsection{Coefficient of performance}

When the force ratio exceeds one, $r>1$, the thermoelectric device works as 
a \emph{refrigerator}. In this case the most important benchmark is 
the \emph{coefficient of performance} (COP) 
$\eta^{(r)}=J_{h}/P$ ($J_{h}<0$, $P<0$), given by the ratio of the heat 
current extracted from the cold system over the absorbed power. 
By optimizing this quantity within linear response, we obtain 
\begin{equation}
\eta_{\rm max}^{(r)}=
\eta_{\rm C}^{(r)}\,
\frac{\sqrt{ZT+1}-1}{\sqrt{ZT+1}+1},
\label{etamaxref}
\end{equation}
where $\eta_{\rm C}^{(r)}=T_R/(T_L-T_R)\approx 1/(T \mathcal{F}_h)$ 
is the efficiency
of an ideal, dissipationless refrigerator. 
Since the ratio $\eta_{\rm max}^{(r)}/\eta_{\rm C}^{(r)}$ for refrigeration is equal
to the ratio $\eta_{\rm max}/\eta_{\rm C}$ for thermoelectric power generation,
$ZT$ is the figure of merit for both regimes. 

\subsection{Systems with broken time-reversal symmetry}
\label{sec:efficiencymagnetic}

\begin{figure}
\begin{center}
\centerline{\includegraphics[width=0.5\columnwidth]{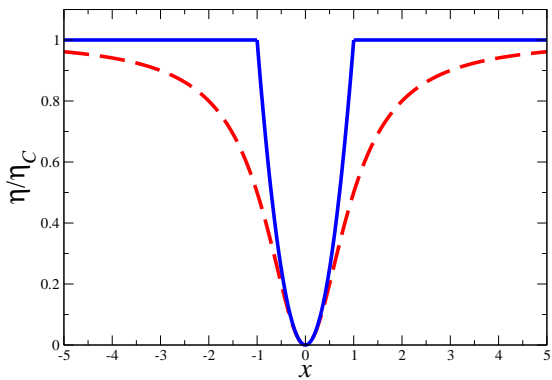}}
\caption{Ratio $\eta/\eta_{\rm C}$ as a function of the asymmetry parameter $x$,
with $\eta=\bar{\eta}(P_{\rm max})$ (dashed curve) and 
$\eta=\bar{\eta}_{\rm max}$
(full curve). For $x=1$, 
$\bar{\eta}(P_{\rm max})=\eta_{\rm C}/2$ and $\bar{\eta}_{\rm max}=\eta_{\rm C}$ 
are obtained for $y(x=1)=ZT=\infty$.}
\label{fig:magnetic}
\end{center}
\end{figure}

The same analysis as above can be repeated 
when time-reversal symmetry is broken, say by a magnetic
field ${\bm B}$ (or by other effects 
such as the Coriolis force). 
The maximum output power is again given by (\ref{eq:Pmax}) 
and the corresponding efficiency at maximum power
\begin{equation}
{\displaystyle
\eta(P_{\rm max}) = \frac{P_{\rm max}}{J_{h}}
= \frac{\eta_{\rm C}}{2}\, \frac{1}{2\frac{L_{ee}L_{hh}}{L_{eh}^2}
-\frac{L_{he}}{L_{eh}}}
}
\end{equation}
%In this case the maximum efficiency and the
%efficiency at maximum power are both determined
%by two parameters \cite{BSC2011}: the asymmetry parameter
is seen to depend on two parameters \cite{BSC2011}: the asymmetry parameter
\begin{equation}
x=\frac{L_{eh}}{L_{he}}=\frac{S({\bm B})}{S(-{\bm B})}
\label{def:x}
\end{equation}
and a generalized ``figure of merit''
\begin{equation}
y=\frac{L_{eh}L_{he}}{\det {\bm L}}=
%\frac{\sigma({\bm B}) S({\bm B})S(-{\bm B})}{\kappa({\bm B})}\,T.
\frac{G({\bm B}) S({\bm B})S(-{\bm B})}{K({\bm B})}\,T \, ,
\end{equation}
\green{where we recall that we have defined $T=T_R$, see the end of section~\ref{sec:linear_response} 
for a discussion of this point.}
Expressed as a function of the parameters $x$ and $y$, 
the efficiency at maximum power reads
\begin{equation}
\eta(P_{\rm max})=
\frac{\eta_{\rm C}}{2}\,\frac{xy}{2+y}.
\label{etawmax}
\end{equation}
The maximum efficiency is again achieved 
when $\mathcal{F}_e$ and $\mathcal{F}_h$ are related as in 
(\ref{eq:etamaxcondition}) and is given by 
\begin{equation}
\eta_{\rm max}= \eta_{\rm C}\,x\,
\frac{\sqrt{y+1}-1}{\sqrt{y+1}+1}.
\label{eq:ZTx}
\end{equation}
In the particular case $x=1$, $y$ reduces to the $ZT$
figure of merit of the
time-symmetric case, Eq.~(\ref{eq:ZTx}) reduces to
Eq.~(\ref{etamaxB0}),
and Eq.~(\ref{etawmax}) to Eq.~(\ref{etawmaxB0}).
While thermodynamics does not impose any restriction on the
attainable values of the asymmetry parameter $x$, the positivity 
of entropy production (Eq.~(\ref{dots})) 
implies $h(x)\le y \le 0$ if $x\le 0$ and
$0\le y \le h(x)$ if $x\ge 0$, where the function $h(x)= 4x/(x-1)^2$.
Note that $\lim_{x\to 1} h(x)=\infty$ and therefore there is no 
upper bound on $y(x=1)=ZT$. For a given value of the asymmetry $x$, 
the maximum (over $y$) 
$\bar{\eta}(P_{\rm max})$ of $\eta(P_{\rm max})$ and the maximum
$\bar{\eta}_{\rm max}$ of $\eta_{\rm max}$ are obtained for $y=h(x)$
and are given by
\begin{equation}
\bar{\eta}(P_{\rm max})=\eta_{\rm C}\frac{x^2}{x^2+1},
\label{eq:boundetapmax}
\end{equation}
\begin{equation}
\bar{\eta}_{\rm max}=
\left\{
\begin{array}{ll}
\eta_{\rm C}\,x^2 & {\rm if}\,\, |x| \le 1,
\\
\\
\eta_{\rm C} & {\rm if}\,\, |x| \ge 1.
\end{array}
\right.
\label{eq:boundetamax}
\end{equation}
The functions $\bar{\eta}(P_{\rm max})(x)$ and 
$\bar{\eta}_{\rm max}(x)$ 
are drawn 
in Fig.~\ref{fig:magnetic}.
In the case $|x|>1$, it is in principle possible to overcome
the \green{Curzon-Ahlborn} limit $\eta_{CA}=\eta_{\rm C}/2$ within linear response and to reach the
Carnot efficiency, for increasingly smaller and smaller figure of merit $y$ as
the asymmetry parameter $x$ increases. The Carnot efficiency is 
obtained for ${\rm det} {\bm L}=(L_{eh}-L_{he})^2/4>0$ when $|x|>1$, 
that is, the tight coupling condition is not fulfilled. 

The output power at maximum efficiency reads
\begin{equation}
P(\bar{\eta}_{\rm max})=\frac{\bar{\eta}_{\rm max}}{4}\frac{|L_{eh}^2-L_{he}^2|}{L_{ee}}\,\mathcal{F}_h.
\end{equation}
Therefore, always within linear response,
it is allowed from thermodynamics 
to have Carnot efficiency and nonzero power
simultaneously when $|x|>1$. 
Such a possibility can be understood on the basis of the 
following argument \cite{BSS2013,BS2013}. We first split each current $J_{i}$
(${i}={e},{h}$) into a 
reversible and an irreversible part, defined by
\begin{equation}
J_{i}^{\rm rev}=\sum_{j=e,h} \frac{L_{ij}-L_{ji}}{2}\,\mathcal{F}_j,
\;\;
J_{i}^{\rm irr}=
\sum_{{j}={e},{h}} \frac{L_{ij}+L_{ji}}{2}\,\mathcal{F}_j.
\label{eq:Jrevirr}
\end{equation}
It is readily seen from Eq.~(\ref{eq:sprod}) and
(\ref{eq:Jrevirr}) that only the irreversible part
of the currents contributes to the entropy production:
\begin{equation}
\dot{\mathscr{S}}=J_{e}^{\rm irr} \mathcal{F}_e + J_{h}^{\rm irr} \mathcal{F}_h.
\end{equation}
The reversible currents $J_{i}^{\rm rev}$ vanish for 
${\bm B}=0$. On the other hand, for broken time-reversal symmetry 
the reversible currents can in principle become arbitrarily large, 
giving rise to the possibility of dissipationless transport.

While in the time-reversal case the linear response
normalized maximum efficiency
$\eta_{\rm max}/\eta_{\rm C}$
and coefficient of performance
$\eta_{\rm max}^{(r)}/\eta_{\rm C}^{(r)}$ 
for power generation and refrigeration
coincide, this is no longer the case with broken
time-reversal symmetry. For refrigeration 
the maximum value of the coefficient of performance reads
\begin{equation}
\eta_{\rm max}^{(r)}=\eta_{\rm C}^{(r)} 
\,\frac{1}{x}\,\frac{\sqrt{y+1}-1}{\sqrt{y+1}+1}.
\label{eq:etarefrigeration}
\end{equation}
For small fields, $x$ will usually be a linear function of the magnetic field,
while $y$ is by construction an even function of the field.
As a consequence, a small external magnetic field either improves
power generation and
worsens refrigeration or vice-versa, while the average
efficiency
\begin{equation}
\frac{1}{2}\left[\frac{\eta_{\rm max}({\bm B})}{\eta_{\rm C}}+
\frac{\eta_{\rm max}^{(r)}({\bm B})}{\eta_{\rm C}^{(r)}}\right]
=\frac{\eta_{\rm max}({\bm 0})}{\eta_{\rm C}}=
\frac{\eta_{\rm max}^{(r)}({\bm 0})}{\eta_{\rm C}^{(r)}},
\end{equation}
up to second order corrections.
Due to the Onsager-Casimir relations, $x(-{\bm B})=1/x({\bm B})$
and therefore by inverting the direction of the magnetic field
one can improve either power generation or refrigeration.

%%%%%%%%%%%%%%%%%%%%%%%%%%%%%%%%
%%%%%%%%%%%%%%%%%%%%%%%%%%%%%%%%

Onsager relations do not impose the symmetry 
$x=1$, i.e., we can have $S({\bm B})\ne S(-{\bm B})$.
However, as discussed in section~\ref{sec:probes} below,
\green{one must have $S(-{\bm B})=S({\bm B})$ for any non-interacting two-terminal system}  
 as a consequence
of the symmetry properties of the scattering matrix \cite{Butcher1990,datta}.
This symmetry
is typically violated when electron-phonon and electron-electron
interactions are taken into account.
While the Seebeck coefficient \green{is usually} found
to be an even function of the magnetic
field in two-terminal purely metallic mesoscopic
systems \cite{lsb98,gbm99},
measurements for certain orientations of 
a bismuth crystal \cite{wolfe1963},
Andreev interferometer experiments \cite{chandrasekhar,
Petrashov03,Chandrasekhar05,Chandrasekhar09}
and theoretical studies \cite{jacquod,SBCP2011,sanchez2011}
have shown that systems
in contact with a superconductor or subject to inelastic
scattering can exhibit non-symmetric thermopower,
i.e., $S(-{\bm B})\ne S({\bm B})$.
So far, investigations of various
classical \cite{horvat2012} and quantum \cite{SBCP2011}
dynamical models have shown arbitrarily large values of the
asymmetry $x$, but correspondingly with low efficiency. 
However, efficiency at maximum power beyond the \green{Curzon-Ahlborn} limit for
$x>1$ has been shown in \cite{BSS2013,vinitha2013,BS2013}
(see section~\ref{sec:probes} below).

There is also current interest in multi-terminal systems with broken time-reversal symmetry,
particularly three-terminal systems in which heat supplied to one terminal drives an electrical current 
between two others.  We discuss a number of such three-terminal systems 
in chapters \ref{Sect:scattering-theory} and \ref{Sect:master-examples}, but mention here  that those in which broken time-reversal symmetry is crucial
to their operation include Aharonov-Bohm rings \cite{entin2012} and quantum hall systems
\cite{
QuNernst2014,sanchez2015a,sanchez2015b,Hofer-Sothmann2015,Vannucci2015,sanchez2016,whitney2016,Samuelsson-Sothmann2016}.

%% file: landauer.tex
\section{Scattering theory for thermoelectric responses}
\label{Sect:scattering-theory}

Landauer's scattering theory is a simple and elegant description of quantum transport.
It is capable of describing the electrical, thermal and thermoelectric properties of 
non-interacting electrons in an arbitrary potential (including arbitrary disorder) 
in terms of the 
probability that the electrons go from one reservoir to another.
These probabilities may be challenging to calculate in complicated structures,
particularly as the electrons propagate as waves which interfere with themselves.
Yet, we can already find out much about such systems' potential for heat-to-work conversion
from the simple fact that the above probabilities  are positive,
and that they reflect electron dynamics which 
obeys time-reversal symmetries (under reversal of any external magnetic field).

This chapter introduces thermoelectric effects within the scattering theory. 
Chapter~\ref{sec:landauer} then discusses
the linear response regime, in particular showing how the structure of the scattering theory leads to
Onsager reciprocal relations and other similar relations, and the relationship between the system's 
scattering properties and its thermoelectric figure of merit $ZT$.
Chapter~\ref{Sect:scatter-nonlin} discusses in detail the nonlinear version of the scattering theory,
and shows that it contains the laws of thermodynamics. This means that no system modelled by scattering theory (in the linear-response regime or the nonlinear regime) can ever violate the first or second law of thermodynamics.
It also shows how Joule heating occurs in systems without a thermoelectric response.

\subsection{Heat-to-work conversion through energy-filtering}

Thermoelectric effects are present whenever the dynamics of the electrons above the Fermi surface are different from the dynamics of electrons below the Fermi surface.
The simplest example of a thermoelectric effect is that of an energy filter.  Scattering theory captures this energy filtering effect in a manner that allows quantitative calculations of currents, efficiencies, etc.  
However, to develop our intuition before launching into quantitative calculations, we first introduce the basic concepts of using energy filtering to perform heat-to-work conversion.

\begin{figure}[t]
\centerline{\includegraphics[width=\textwidth]{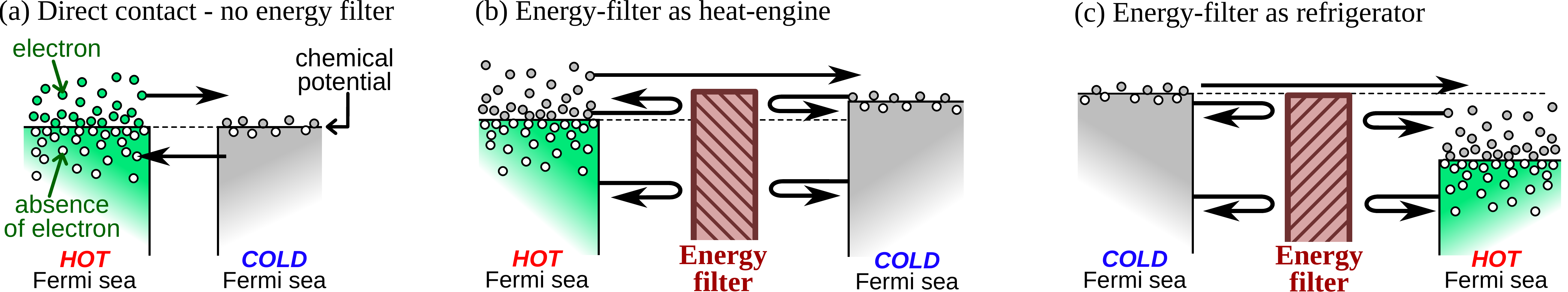}}
\caption{\label{Fig:energy-filter}
The simplest thermoelectric effect to understand is that of an energy filter.
In (a) we show direct connection between two reservoirs of electrons at different
temperatures but the same electrochemical potential in the absence of any energy filter.
Electrons in occupied (shaded) states want to flow into empty (white) states,
crossing from one reservoir to the other to do so.  The resulting flows are marked by the thick black arrows.
In the absence of an energy-filter there is a heat current but no electrical current (the opposite flows of electrons above and below electrochemical potential cancel each other out).
In (b) and (c) we sketch an energy-filter between the hot and cold Fermi seas 
which blocks all particle flow below a certain energy.
In (b) we show how to use it as a heat-engine, it generates power because the temperature difference means that electrons flow from a region of lower electrochemical potential (left) to a region of higher electrochemical potential (right).
In (c) we  show how to use it as a refrigerator, using a potential bias
to ensure that electrons above the Fermi sea can flow out of the cold reservoir, 
cooling it further.
}
\end{figure}

Suppose one has a hot reservoir of electrons and a cold reservoir of electrons, both with the same electrochemical potential (i.e.\ same Fermi energy). 
If we connect them together directly, electrons in full states above 
the Fermi surface of the hot reservoir will flow into empty states in the cold reservoir,
while electrons in full states below the Fermi surface of the cold reservoir will flow into 
empty states in the hot reservoir (see the sketch in Fig.~\ref{Fig:energy-filter}a).
The result is a flow of heat from hot to cold, but no flow of charge, because for every electron above the electrochemical potential flowing one way, there is another electron below the electrochemical potential flowing the other way.  However, if one wants an electrical current,
one simply has to put an energy-filter between the reservoirs that blocks the 
electron flow at certain energies, for example those energies below the electrochemical potential.
Applying this idea in Fig.~\ref{Fig:energy-filter}b, an energy-filter can
allow the high energy electrons on the left to flow to the right (indicated by the upper arrow), while stopping the lower energy electrons
on the right to flow to the left (indicated by the lower arrows).
Thus there is a net electrical current between the reservoirs.
\green{An electrical machine does work by moving charge from a reservoir with lower electrochemical potential to a reservoir with higher electrochemical potential, as would be the case if it were charging up a capacitor plate or a battery.
Thus a a flow of electrons from left to right only generates electrical work if the electrochemical potential is higher on the right, as sketched in Fig.~\ref{Fig:energy-filter}b. There we
show the filter blocking all energies up to certain value,  with the electrochemical potential of the right
reservoir being a bit below this value.}
This system is now converting heat into work, and thus is functioning as a thermodynamic machine.

%========================================
%========================================

One can equally use an energy filter as a refrigerator, to convert electrical power into a heat flow from cold to hot,
in the manner sketched in Fig.~\ref{Fig:energy-filter}c.  The electron states above the electrochemical potential of the cold reservoir have a higher occupation than the states at the same energy in the hot reservoir, because
the hot reservoir 
is biased in such a way that its electrochemical potential is lower than that of the cold reservoir.
Electrons above the cold reservoir's electrochemical potential will escape over the barrier, thereby cooling the cold reservoir further, despite the fact it is colder than the hot reservoir.  These electrons flow from a region of high electrochemical
potential to one of low electrochemical potential, so work is necessary to maintain the potential difference
(supplied by a power source), and ensure that the refrigeration continues.

\begin{figure}[t]
\centerline{\includegraphics[width=\textwidth]{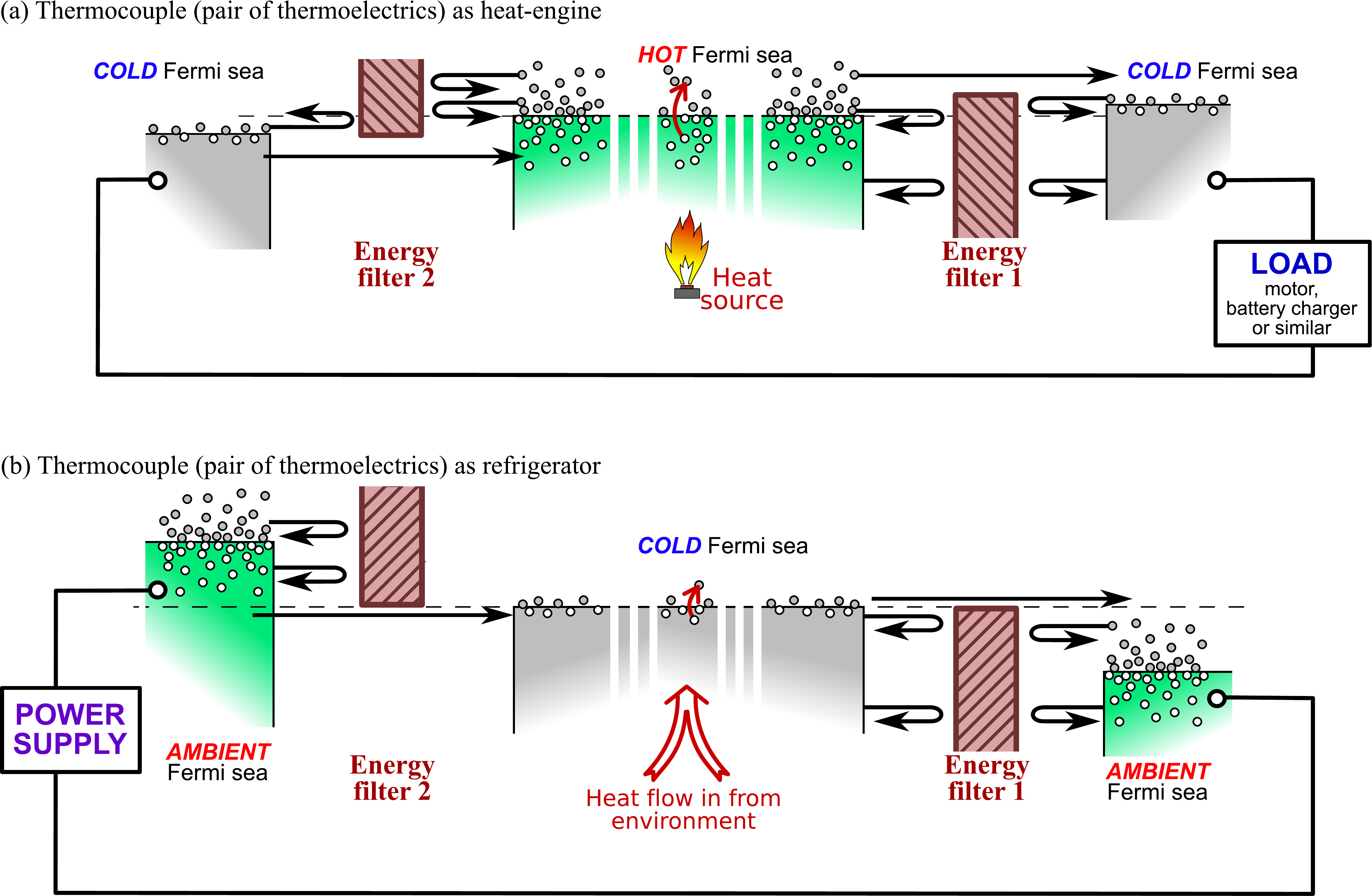}}
\caption{\label{Fig:energy-filter-thermocouples}
Sketch of electron dynamics for a pair of energy filters in a thermocouple geometry \green{such as Fig.~\ref{Fig:thermocouple}a}, 
acting as (a) a power generator or (b) a refrigerator  --- adapted from Refs.~\cite{buttiker2013,buttikerNJP2013}.  
In both cases the energy filter on the right (energy filter 1) only lets through electrons with energies above a certain value (as in Fig.~\ref{Fig:energy-filter}). In contrast the energy filter on the left (energy filter 2) only lets through
electrons {\it below} a certain energy.  In (a) the heat source heats the central region,  inducing a flow of electrons from left to right (against a bias), thereby generating electrical power.  This electrical work can then be converted into another form of work by a suitable load (a \emph{motor} will convert electrical work into mechanical work, a \emph{battery} charger will convert the electrical work into chemical work, etc).
In (b) the power supply generates a bias across the thermocouple and a current from left to right
(flowing due to the bias), so the thermocouple is absorbing work from the power supply.  This flow 
however leads to the refrigeration of the central region.
}
\end{figure}

For steady-state power generation or refrigeration, a single thermoelectric is rarely enough. 
The thermoelectric should carry electrical current, which requires that one form a circuit for this current flow.
The most common way to form a circuit is with a thermocouple, as in Fig.~\ref{Fig:thermocouple}a, 
in which one has two thermoelectrics with different (ideally opposite) thermoelectric responses.  
Fig.~\ref{Fig:energy-filter-thermocouples} shows a sketch of how a thermocouple made of two energy-filters works at the microscopic level.  Filter 1 lets pass electrons with energies above the electrochemical potential of the central region,
while filter 2 lets pass electrons with energies below the electrochemical potential of the central region.
In Fig.~\ref{Fig:energy-filter-thermocouples}a, the heat source maintains the central region at a higher temperature than the rest of the system (cold reservoirs, load, etc.), by exciting electrons (red arrow).  
Electrons flow in from the left (black arrow) below the electrochemical potential of the central region to fill the holes 
in the central region's Fermi sea, even though the electrochemical potential is lower on the left than in the central region. Electrons above the central region's electrochemical potential flow out to the right, even though that means they flow into a region with higher electrochemical potential.  This means the thermocouple is causing an electrical current against a bias.
This means that it can drive electrical current through a load, which converts that electrical work into some other form of work (mechanical, chemical, etc.).

In Fig.~\ref{Fig:energy-filter-thermocouples}b, the central region is being refrigerated by the bias applied to the thermoelectrics by the power supply, so it is colder than the ambient temperature.  In such cases, we cannot rule out a back-flow of heat from the environment in the form of phonons or photons opposing the refrigeration, which excites electrons in the central region (red arrow).  This heat must be removed by the the thermoelectrics.

In both cases, we assume that there is a weak thermalization process in the central region,
which means that any electron entering that region at higher energy 
(or any electron excited by heat arriving from a heat source or back-flow from the environment) 
dissipates that energy to the other electrons in the central region, before arriving at either energy filter.  
Thus electrons arriving at the energy filters from the central region will have a thermal distribution
given by the temperature of the central region.
For this reason, we can calculate the thermoelectric properties of each energy filter separately,
without worrying about how they are connected up or how the temperature difference and bias across each one is generated.

\green{
Above we outlined systems of the type called "traditional thermocouples" in Fig.~\ref{Fig:thermocouple},
which we discuss in more detail in much of chapters~\ref{sec:landauer} and \ref{Sect:scatter-nonlin}. 
Systems of the type called "quantum thermocouples" in Fig.~\ref{Fig:thermocouple} are discussed in sections~\ref{sec:genericmultiterminal}, \ref{Sect:3-term-sys1} and \ref{Sect:3-term-sys2}. }

%========================================
%========================================

\subsection{History of the scattering theory for thermoelectricity}

The literature on Landauer's scattering theory can be divided roughly into two periods.
The first period was that of foundations, it started with Landauer's early publications \cite{Landauer1957,Landauer1970} and  
continuing up to the late 1980s. Papers from this period must be read with great care, because
the theoretical construction of the method was carried out during a time of confusion about 
the experimentally-relevant definition of resistance at the nanoscale.
Once, experiments started to be carried out in the late 1980s 
\cite{q-point-cont:van-Wees1988,q-point-cont:Wharam1988}, 
it became clear how to use the method as a recipe to explain experiments.   This led to the second period, which was its applications to increasingly complex nanostructures.

During the foundational period, the 1981 work of Enquist and Anderson \cite{Enquist-Anderson1981}  laid the foundations for thermal effects, while the 1986 work of Sivan and Imry \cite{Sivan-Imry1986} addressed thermoelectric effects, and by extension heat-to-work conversion.
These two works basically contain all the formalism that we will need, but they must be read 
with caution, because they were written at a time when there was no consensus about whether
the resistance of a perfectly transmitting single channel was zero or finite.
The earliest work suitable for beginners is Ref.~\cite{Butcher1990}, which was written after the consensus was established and develops the formalism for thermoelectric effects further.
Other crucial works are those of Bekenstein \cite{Bekenstein1981,Bekenstein1984} and Pendry \cite{Pendry1983} which use 
scattering theory to show that quantum mechanics places a limit on heat flow, of these works
Ref.~\cite{Pendry1983} is by far the easiest for a modern reader to follow.

\green{Reader who wish to understand the work of 
Enquist and Anderson \cite{Enquist-Anderson1981} 
or the work of Sivan and Imry \cite{Sivan-Imry1986} 
should keep in mind the context in which they were written. At that time}
two formulas for the conductance, $G$, had appeared in the literature: Landauer's original proposition $G \propto T/(1-T)$,
and another proposal $G \propto T$ \cite{Fisher-Lee1981}, where $T$ was the channel's transmission probability.  The former predicts that a perfectly transmitting channel (one with $T=1$) has zero resistance, while the second predicts that it has a finite resistance.
A partial resolution of the confusion is already visible in Ref.~\cite{Enquist-Anderson1981},
which implies that the result depends on the manner in which one measures the voltage.
However, Refs.~\cite{Fisher-Lee1981,Enquist-Anderson1981} add to the confusion
by arguing that the resistance of a perfectly transmitting channel should be zero, and that their results which indicated $G \propto T$ were faulty because the reservoirs were not treated correctly within their theories.  It is now generally agreed that their treatments were not faulty, and results with $G \propto T$ are correct.
B\"uttiker's 1986 work \cite{Buttiker1986}  clarified the situation by 
showing that the $G \propto T$ formula could be generalized to multi-terminal geometries.
This enabled him to show that the conductance measured in two-probe geometries was
$G \propto T$, while that measured in four-probe geometries was more complicated
(because voltage probes acquire a bias such that they carry no current).
Yet, in a certain  limit the four-probe result becomes  
$G \propto T/(1-T)$; this limit being that of weakly-coupled voltage probes in a specific geometry.  
At the same time, Imry \cite{Imry1986,imry} gave a pretty interpretation of this in terms of the idea that a perfectly transmitting channel had zero resistance, but that there is always a contact resistance between
the channel and the bulk leads it is coupled to.  However, this interpretation 
has rather fallen out of favour,
because it is hard to apply in multi-terminal geometries, and it encourages its user to think in terms of summing resistances in series (which is not in-general allowed in phase-coherent conductors).
Crucially, the $G\propto T$ formula fitted the first experiments on point-contacts \cite{q-point-cont:van-Wees1988,q-point-cont:Wharam1988} 
which were of two-terminal type. 
B\"uttiker's multi-terminal version of the $G \propto T$ formula was placed on a more solid theoretical footing by Ref.~\cite{Stone-Szafer1988}, which derived it from the Kubo linear-response formalism.  
Latter B\"uttiker's formula was shown to fit four-terminal experiments \cite{Picciotto2001-four-term-resistance}.

Finally, we mention that Ref.~\cite{beenakker1992} was the first to show a thermoelectric response in a nanostructure (a point-contact), and used scattering theory to explain the experimental observation.
The Bekenstein-Pendry bound on heat-flow was observed experimentally in point-contacts \cite{Molenkamp-Peltier-thermalcond}, and recently verified to high accuracy in quantum Hall edge-states \cite{Jezouin2013}.  

\subsection{The basics of scattering theory}
\label{Sect:scatter-basics}

The scattering theory is based on the idea that one can split the situation under consideration into a small
scattering region coupled to multiple macroscopic reservoirs of free electrons.  The scattering region should then be such that each electron traverses that region from one reservoir to another without exchanging energy with other particles (electrons, phonons, etc).  Thus, an electron that enters the scattering region with energy $E$ from a given reservoir will be a wave with energy $E$ that bounces around elastically (interfering with itself) until it escapes into a reservoir. 
All inelastic processes that could cause dissipation or decoherence are limited to the reservoirs.
Here we follow the less technical route to the scattering theory in Refs.~\cite{imry,datta},
however we mention that one can also derive it using second quantization \cite{Blanter-Buttiker}.

The coupling of the scatterer to each reservoir is written in terms of a set of orthogonal modes in the contact between the scatterer and the reservoir.  Typically one thinks of the connection to the reservoir as a waveguide, so the modes are the transverse modes of this waveguide, although it is sometimes convenient to rotate to another basis of modes, see e.g.~\cite{wj2005,wj2006prb}.
Then, the crucial quantity that encodes this probability for the electron with energy $E$ to go from mode $m$ of reservoir $j$ to mode 
$n$ of reservoir $i$ is the scattering matrix element, ${\cal S}_{in;jm}(E)$.
Since the Hamiltonian for the scatterer is hermitian, 
the scattering matrix must be unitary, so its matrix element ${\cal S}_{in;jm}(E)$ must obey
$\sum_{jm} {\cal S}_{in;jm}(E) {\cal S}^*_{i'n';jm}(E) =\delta_{i'i}\delta_{n'n}$.
The probability to go from mode $m$ of reservoir $j$ to mode 
$n$ of reservoir $i$ is 
\begin{align}
P_{in;jm}(E)\  =\ \big| {\cal S}_{in;jm}(E) \big|^2. 
\label{Eq:scattering-prob}
\end{align}
If we sum this over all modes coupled to reservoirs $i$ and $j$, we get the transmission matrix elements
\begin{align}
{\cal T}_{ij}(E) = \sum_{nm} P_{in;jm} (E);
\label{Eq:def-T_ij}
\end{align}
this can be interpreted as the probability to go from a given mode of reservoir $j$ to any mode of reservoir $i$, summed over all modes of reservoir $j$.  As such, one has
\begin{subequations}
\label{Eq:constraints-T_ij}
\begin{align}
&{\cal T}_{ij}(E) \geq 0 \qquad\hbox{ for all } \ i,\ j,\   E,
\end{align}
while
\begin{align}
&\sum_i {\cal T}_{ij}(E) = N_j(E),
\label{Eq:constraints-T_ij-b} 
\\
&\sum_j {\cal T}_{ij}(E) = N_i(E),
\label{Eq:constraints-T_ij-c}
\end{align}
\end{subequations}
where the $i$ and $j$ sums are over all reservoirs.
Here, $N_j(E)$ is the number of modes in the coupling to reservoir $j$ at energy $E$.
Often authors refer to ${\cal T}_{ii}$ with the symbol ${\cal R}_{ii}$ for ``reflection'', 
since it corresponds to electrons entering the scatterer from reservoir $i$ and being  reflected back into reservoir $i$, 
however we will use  ${\cal T}_{ii}$ here \green{to keep the formulas compact}.

The scatterer has an underlying Hamiltonian which satisfies time-reversal symmetry. This means that
if we reverse the velocity of all particles, 
and reverse the external magnetic field, ${\bm B}$,
then the particles will follow a time-reversed trajectory back to where they came from 
(so incoming electrons become outgoing electrons and vice versa).
Hence, the scattering matrix elements must obey ${\cal S}_{in;jm}(E,-{\bm B}) = {\cal S}^*_{jm;in}(E,{\bm B})$,
which in turn means that the transmission functions obey
\begin{eqnarray}
{\cal T}_{ij}(E,{\bm B})={\cal T}_{ji}(E,-{\bm B})\ .
\label{Eq:T_ij-time-reverse}
\end{eqnarray}
This relation will be fundamental in proving Onsager reciprocal relations for such systems, 
such as the well-known relation between Seebeck and Peltier coefficients.

The Landauer approach tells us that one can write the charge and heat currents out of reservoir  $i$ in terms of ${\cal T}_{ij}(E)$.  The 
charge current $\Jelectrici$ out of reservoir $i$ and into the scatterer is given by counting each electron that crosses the boundary between the scatterer and reservoir $i$.
The number of electrons flowing out of reservoir $i$ and into the scatterer at energy $E$ is proportional to the
number of modes $N_i$ multiplied by the reservoir's occupation at energy $E$, which 
is given by the
Fermi function
\begin{eqnarray}
f_i(E) = \left(1+\exp\left[(E - \mu_i)\big/ (\kB T_i) \right] \right)^{-1},
\label{Eq:f}
\end{eqnarray} 
where $\mu_i=e  V_i$ and $T_i$ are the electrochemical potential and temperature 
of reservoir $i$.  However, there is also a flow of electrons from the scatterer into reservoir $i$.  
\green{The number of electrons that flow into reservoir $i$ at energy $E$ from reservoir $j$ is
proportional to  ${\cal T}_{ij}(E)$ multiplied by reservoir $j$'s occupation $f_j(E)$.}
The total flow of electrons into reservoir $i$ is given by the sum of this over all $j$ (including $j=i$).
The electrical current into the \green{scatterer} from reservoir $i$ is 
\green{then given by the flow of electrons out of the reservoir minus the total flow into it}
\cite{imry,datta},
\begin{eqnarray}
\Jelectrici \! &=& \! \sum_{j} \int_{-\infty}^\infty {{\rm d}E \over h} \ \ e \ 
\, \left[N_i(E)\, \delta_{ij} - {\cal T}_{ij}(E)  \right] \,  f_j (E).  
\label{Eq:I-initial}
\end{eqnarray}
We can make the same argument to define the energy current out of reservoir $i$ into the scatterer, except now each electron carries
an amount of energy $E$ instead of the charge $e $.  
Hence 
\begin{eqnarray}
\Jenergyi \! &=& \! \sum_{j} \int_{-\infty}^\infty {{\rm d}E \over h} \ \  E  \ 
\, \left[ N_i(E)\, \delta_{ij} - {\cal T}_{ij}(E)  \right] \,  f_j (E),  
\label{Eq:I-energy-initial}
\end{eqnarray}
To construct the equivalent formula for the heat current out of a reservoir, 
we must consider the definition of heat in that reservoir.
We take the heat energy in a reservoir's electron gas \green{to be} the total energy of the gas 
minus the energy which that gas would have in its ground-state \green{at the same chemical potential}.
As such, the heat energy can be written as a sum over the energy of all electrons, measured from the reservoir's electrochemical potential.  This means electrons above the electrochemical potential
contribute positively to the heat, while those below the electrochemical potential contribute negatively
to the heat.  The latter can be understood as saying that if one removes an electron below
the electrochemical potential, it increases the heat in the reservoir, because one is pushing the system
further from the zero temperature Fermi distribution (in which all states below the electrochemical potential are filled).
Thus, an electron with energy $E$ leaving reservoir $i$ carries an amount of heat, $\Delta Q_i= E -\mu_i$, out of the reservoir.  The formula for heat current is the same as that for energy current, Eq.~(\ref{Eq:I-energy-initial}), but with $(E-\mu_i)$ in place of $E$.
Hence \green{the heat current into the scatterer from reservoir $i$} \cite{Butcher1990,Nenciu2007,sanchezprb13,Meair-Jacquod2013,Whitney-2ndlaw} is
\begin{eqnarray}
\Jheati \! &=& \! \sum_{j} \int_{-\infty}^\infty {{\rm d}E \over h}\, (E\!-\! \mu_i)\,
\left[ N_i(E)\, \delta_{ij} - {\cal T}_{ij}(E)  \right]  \, f_j (E).\ \ 
\label{Eq:J-initial}
\end{eqnarray}
We note that the heat current obeys
\begin{eqnarray}
\Jheati = \Jenergyi - V_i \Jelectrici,
\end{eqnarray}
where $V_i$ is the electrical bias of reservoir $i$, given by $\mu_i=e  V_i$. 

It is useful to also define $\dot{\mathscr{S}}_i$ as the rate of change of the entropy of reservoir $i$.
Using the Claussius relation that the entropy of a reservoir is its heat divided by its temperature,
and noting that the rate of change of heat in reservoir $i$ is $-\Jheati$,
the rate of change of entropy in reservoir $i$ is 
\begin{align}
\dot{\mathscr{S}}_i = - \Jheati/T_i.
\label{Eq:scatter-dotS}
\end{align}
In the steady-state the entropy of the electrons in the scatterer
does not change with time, thus the rate of change of the total entropy $\dot{\mathscr{S}}$
is simply the sum of the rate of changes in the reservoirs,
\begin{align}
\dot{\mathscr{S}} = - \sum_i \Jheati/T_i.
\label{Eq:scatter-S_total}
\end{align}

Given Eq.~(\ref{Eq:constraints-T_ij-c}), we see that the sum of electrical current $\Jelectrici$, 
or energy current $\Jenergyi$, over all reservoirs $i$ is zero;
\begin{eqnarray}
\sum_i \Jelectrici = \sum_i \Jenergyi = 0\, .
\label{Eq:I-conserve}
\end{eqnarray}
This is nothing but Kirchoff's law of 
current conservation for electrical or energy currents. 
However, we then see that heat-currents \green{into the scatterer} obey
\begin{eqnarray}
\sum_i \Jheati = - \sum_i V_i \, \Jelectrici. 
\label{Eq:scatter-1st-law}
\end{eqnarray}
This means that heat currents are not conserved, since  
the scatterer can be a source or sink for heat.
Section~\ref{Sect:scatter-1st-law} will explain that the right hand side of Eq.~(\ref{Eq:scatter-1st-law}) is 
the electrical power generated by the scatterer (one can already guess this from the fact it is a bias multiplied by an electrical current), \green{which we call $P_{\rm gen}$.}  This means that  Eq.~(\ref{Eq:scatter-1st-law}) 
\green{with $P_{\rm gen}=- \sum_i V_i \, \Jelectrici$} is nothing but the first law of thermodynamics for a steady-state flow.
If the power generated  $P_{\rm gen} >0$,  then the scatterer is absorbing heat from the electronic reservoirs and turning it into electrical power.
In contrast, if $P_{\rm gen} <0$, then the scatterer is absorbing electrical power and emits heat into the electronic reservoirs; one can think of this as Joule heating.

It is important to note that the energy current is conserved, but it is not gauge-independent.  That is to say, the value of the energy current, $\Jenergyi$, depends on our choice of the zero of 
energy. This means that the energy current is not of physical relevance, 
although differences in energy currents may be. 
In contrast, even though they are not conserved, the heat currents are gauge-independent.
Thus they are of physical relevance.

\subsubsection{Scattering theory for two reservoirs}

A common situation is that of only two reservoirs, which we label left ($L$) and 
right ($R$).  Thus, it is worth explicitly considering how the scattering theory simplifies for this situation.
The main specificity of a two reservoir system is that the transmission from right to left must equal that from left to right,
\begin{eqnarray}
{\cal T}_{ LR}(E) \ =\  {\cal T}_{ RL}(E)\ \geq \ 0 \ ,
\label{Eq:LR=RL}
\end{eqnarray}
for any given set of conditions (biases and temperatures) on the reservoirs 
\footnote{At first glance this makes it look like the scattering theory could never predict an asymmetric current-voltage relation, such as that of a diode.  This is not the case, such effects come from the interactions which make the transmission ${\cal T}_{\rm LR}(E)$ depend on the reservoir biases.  A diode would result if ${\cal T}_{\rm LR}(E)$ is 
large when a reservoir is biased positively and small when that reservoir is biased negatively,
even though for any given bias the scatterer respects Eq.~(\ref{Eq:LR=RL}).}.
This can be easily proven by
comparing Eq.~(\ref{Eq:constraints-T_ij-b})  with $j={ L}$ 
with Eq.~(\ref{Eq:constraints-T_ij-c})  with $i={ L}$.
In addition, Eq.~(\ref{Eq:constraints-T_ij-c}) give the following useful results
\begin{eqnarray}
{\cal T}_{ LL}(E)  &=& N_{ L}(E) -  {\cal T}_{ LR}(E) 
\\
{\cal T}_{ RR}(E)  &=& N_{ R}(E) -  {\cal T}_{ LR}(E)
\label{Eq:LLandRR}
\end{eqnarray}

Eqs.~(\ref{Eq:I-initial},\ref{Eq:I-energy-initial}) each have only two terms in the sums over $j$, using 
Eq.~(\ref{Eq:LLandRR}) we get that the currents 
 \footnote{It is interesting
to remark that the transmission-function approach is not limited
to quantum mechanics. For classical non-interacting particles, 
formulas similar to (\ref{Eq:I-initial-two-reservoir}) and (\ref{Eq:I-energy-initial-two-reservoir}) can be
written, where the Boltzmann rather than the Fermi
distribution of injected particles appears, see e.g. \cite{Saito2010}.}
\begin{eqnarray}
\JelectricL = - \JelectricR = \ \int_{-\infty}^\infty {{\rm d}E \over h} \ e  \ 
{\cal T}_{ LR}(E) \, \left[ f_{ L} (E) -f_{ R}(E) \right],  
\label{Eq:I-initial-two-reservoir}
\\
\JenergyL = - \JenergyR = \ \int_{-\infty}^\infty {{\rm d}E \over h} \ E \ 
{\cal T}_{ LR}(E) \, \left[ f_{ L} (E) -f_{ R}(E) \right]. 
\label{Eq:I-energy-initial-two-reservoir}
\end{eqnarray}
For the heat currents we have
\begin{eqnarray}
\JheatL  &=& \int_{-\infty}^\infty {{\rm d}E \over h} \ (E -\mu_{ L})\ 
{\cal T}_{ LR}(E) \, \left[ f_{ L} (E) -f_{ R}(E) \right],  
\nonumber \\
\JheatR  &=& \int_{-\infty}^\infty {{\rm d}E \over h} \ (E -\mu_{ R})\ 
{\cal T}_{ LR}(E) \, \left[ f_{ R} (E) -f_{ L}(E) \right],  
\label{Eq:J-initial-two-reservoir}
\end{eqnarray}
where we recall that $\mu_i= e  V_i$.
Since heat current is not conserved, we expect that $\JheatL \neq -\JheatR$, 
and indeed we have
\begin{eqnarray}
\JheatL +\JheatR = (V_{ R}-V_{ L})\, \JelectricL \ .
\label{Eq:scatter-1st-law-two-reservoir}
\end{eqnarray}
Section~\ref{Sect:scatter-1st-law} 
will explain that this is the first law of thermodynamics for steady-state
state flow in a two reservoir problem.

\subsection{Applicability of scattering theory to given systems}
\label{Sect:scatter-validity}

Scattering theory is a single-electron theory. In other words, the outcome of the scattering for any given electron is assumed to be independent of the outcome of the scattering for any other electron.  This is correct for non-interacting particles, but does it apply to electrons which repel each other rather strongly?  

The generally accepted view is that scattering theory is a quantitatively good model of a system even if a given electron's dynamics is strongly affected by the fluid formed by the other electrons, so long as that electron feels the electrostatic effect of the fluid on average (treated as a fluid 
with a continuous charge distribution given by the modulus-squared of the wavefunctions at each point).  
In more formal language, this is like equivalent to saying that the theory captures mean-field effects of the type
described by a time-independent Hartree approximation.
However, it cannot capture situations where two electrons feel each other's individual dynamics.  For example, it cannot capture the physics of an electron scattering off another one, imparting part of its energy to that electron. Nor can it capture the physics of an electron scattering off the lattice (i.e. electron-phonon scattering) and imparting part of its energy to the lattice.  This is why electrons leave the scatter with the same energy that they entered with, each electron only undergoes elastic scattering from the electrostatic potential due to the lattice and the flow of electrons. The theory also does not capture the correlations induced by interactions between individual electrons. For example, \green{it cannot model a situation in which two electrons which could individually be scattered in either of two direction (say left or right), but where their repulsion means that they are unlikely to go in the same direction as each other.}
Similarly, it does not capture the physics of single-electron interaction effects, such as Coulomb blockade or the Kondo effect.

While the scattering theory can account for the mean-field interactions of each electron with the fluid made of the other electrons, it is hard work to do this for a given realistic situation. In principle, one could start with the bare potential,
determined by the material's chemistry (position of charged ions) and the surrounding electrostatic gates.
One would then add the electron flows from one reservoir to another through the scatterer 
in a manner that is self-consistent.  So the modulus-squared of the wavefunctions of the scattering states 
(integrated over all energies) determine the electrostatic potential at each point in the scatterer, while in turn this potential determines the wavefunctions of the scattering states.  
So if the dynamics are such that electrons have a high probability of spending time in one region of the scatterer
on their way from one reservoir to another, then that will tend to change the potential in the scatterer in such a way to repel electrons from this region.

In practice, one nearly always 
starts by assuming that one knows the electrostatic potential in the scatterer when all reservoirs are
at equilibrium with each other (at the same temperature and electrochemical potential).  Indeed, this assumption is usually more realistic that the assumption that one could ever know the bare electrostatic potential defined by the chemistry and gates in the absence of the conduction electrons.
If one only wishes to treat the linear response regime, as in chapter~\ref{sec:landauer}, then one can calculate the scattering matrix, and hence the transmission functions ${\cal T}_{ij}(E)$ directly from  this equilibrium electrostatic potential.  This is because the small changes in the electrostatic potential \green{of the scatterer that are} induced by applying a small bias (or temperature difference) will not affect the currents
at linear order in the small bias (or temperature difference).

For the nonlinear response, the situation is complicated, even if one knows the equilibrium electrostatic potential.  The reason is that strong biases on a reservoir will deform the electrostatic potential (just as a gate would), and the flow of electrons through the scatterer due to the bias (or temperature difference)
will change the electrostatic potential in the scatterer, thereby changing the scattering properties of the
scatterer.  These effects start making an essential contribution to the system's response as soon as one goes beyond linear response (i.e.~they start contributing at quadratic order in the bias or temperature difference).  We discuss how to treat these effects in the nonlinear response in 
chapter~\ref{Sect:scatter-nonlin}.

However, one should not forget that the difficulty are limited to 
calculating the transmission function, ${\cal T}_{ij}(E)$, for given bias and temperature of 
each reservoir.
If one were given that ${\cal T}_{ij}(E)$, then 
one can simply use the formulas in section~\ref{Sect:scatter-basics} to directly calculate all currents
for that bias and temperature of each reservoir.
Even if one does not have ${\cal T}_{ij}(E)$, 
we know it must obey the relations in section~\ref{Sect:scatter-basics},
as consequences of  time-reversal symmetry, particle conservation, etc.
These results will already be enough to say many things about the 
thermodynamics of a system operating to convert heat into work.
We do this in the linear-response regime in chapter~\ref{sec:landauer}, 
and for nonlinear responses in chapter~\ref{Sect:scatter-nonlin}.

\subsubsection{Transmission function for a point contact}
\label{Sect:scatter-pointcont}

It is instructive to briefly look at examples of nanostructure with simple energy dependent 
transmission function, to get an idea about what can be expected from physical systems.

\begin{figure}
\begin{center}
\centerline{\includegraphics[width=0.95\textwidth]{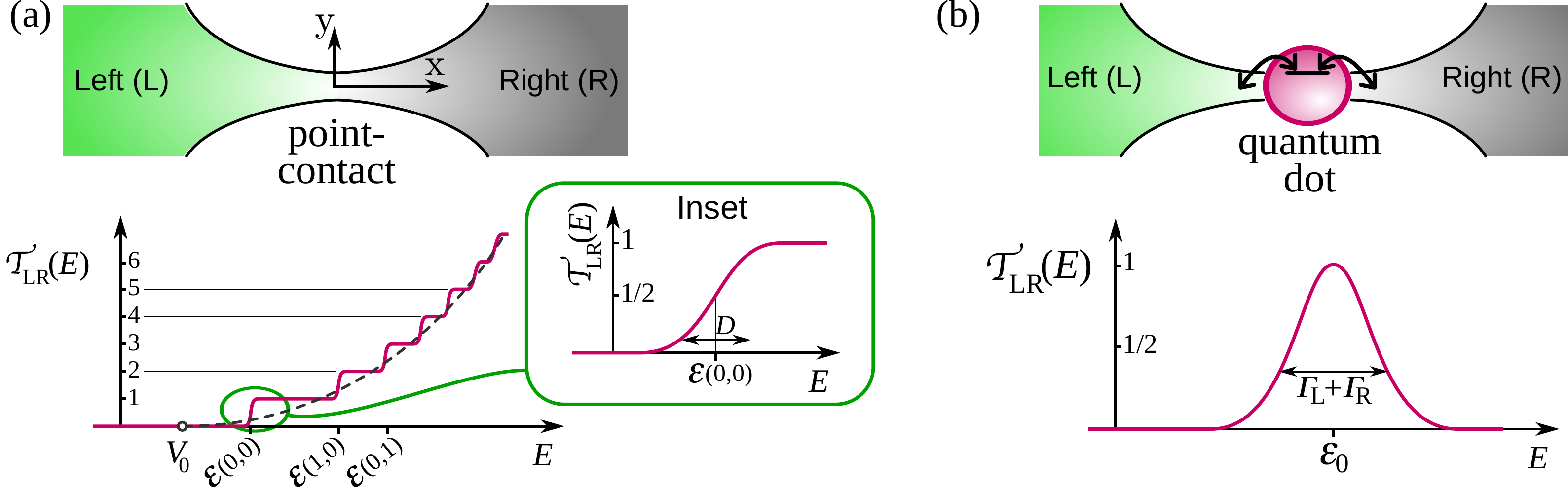}}
\caption{
A sketch of (a) a point contact and (b) a single level quantum dot.  In each case, we sketch the transmission function, ${\cal T}_{ LR}(E)$ as a function of energy $E$, as described in sections~\ref{Sect:scatter-pointcont} and \ref{Sect:scatter-single-level-dot}.  
The sketch of ${\cal T}_{ LR}(E)$ in (a) shows the staircase function which occurs as the number of open modes goes up with increasing $E$. However, most proposal for thermoelectrics involve making the point-contact 
narrow enough that temperature is much less than the distance between the steps, so only the first step is relevant,
as shown in the inset.  In this case there are two parameters to control, the position of this step, $\eps(0,0)$, and the width of the step, $D$. 
A good thermoelectric response occurs if the step is narrower than temperature, $D \ll \kB T$, and positioned within $\kB T$ of the electrochemical potential of the reservoirs.
The sketch ${\cal T}_{ LR}(E)$ in (b) shows the Lorentzian nature of the transmission, 
as typical of a Breit-Wigner form.  Again there are two parameters to control, the Lorentzian's position, $E_0$, and width $\Gamma_{ L}+\Gamma_{ R}$. 
A good thermoelectric response occurs if the Lorentzian is narrower than temperature, 
$\big(\Gamma_{ L}+\Gamma_{ R}\big) \ll \kB T$, and positioned within  $\kB T$ of the electrochemical potential of the reservoirs.
} 
\label{fig:pointcont-qu-dot}
\end{center}
\end{figure}

Firstly we have the point contact, this was the first nanostructure for which a thermoelectric effect was observed experimentally \cite{beenakker1992}.
If we assume it has a smooth enough profile near its narrowest point, its form is approximately parabolic,
it will have the following saddle-point potential \cite{Buttiker-pointcont}, \green{where $x$ is the direction  
along the point contact from one reservoir to the other},
\begin{eqnarray}
V(x,y,z) = V_0 -\half m \omega_x^2 x^2 + \half m \omega_y^2 y^2 + \half m \omega_z^2 z^2 \ .
\end{eqnarray}
Note \cite{Buttiker-pointcont} assumed the point contact was between two two-dimensional electron gases (so there is no z-component to the potential), here we take a three-dimensional problem.
Then it can be shown \cite{Buttiker-pointcont}, that its transmission function is 
\begin{eqnarray}
{\cal T}_{ LR}(E) = \sum_{n_y,n_z} \ {1 \over 1+ \exp \left[- \big(E-\eps(n_y,n_z)\big)\big/D \right]  },
\label{Eq:transmission-pc}
\end{eqnarray}
where for transverse mode $n_y,n_z$ in the point contact one has 
\begin{eqnarray}
\eps(n_y,n_z) = V_0 +  \hbar \omega_y\left( n_y+\half \right) +  \hbar \omega_z\left( n_z+\half \right), \qquad  D \ =\  {\hbar \omega_x \over 2\pi}.
\end{eqnarray}
Thus, the point contact acts as an energy barrier of height $\eps(n_y,n_z) $.
Electrons in the transverse mode $n_y,n_z$, with total energy much less than $\eps(n_y,n_z) $ 
are reflected by this barrier
(${\cal T}_{ LR}(E) \sim 0$), 
while electrons in that mode but with total energy much more than $\eps(n_y,n_z) $ 
pass over the barrier \colorproofs{(see Fig.~\ref{fig:pointcont-qu-dot})}.
Tunnelling through the barrier and reflection above the barrier are significant on energy scales within $D$ of $\eps(n_y,n_z) $, 
and are the physical origin of  ${\cal T}_{ LR}(E)$ switching smoothly from zero to one over a range of energies of order $D$ around $\eps(n_y,n_z)$.

A very long point contact, $\omega_x \to 0$, has  negligible tunneling or over barrier reflection, $D \to 0$.  Thus, such a point-contact's transmission simplifies to the sum of 
Heaviside step-functions, $\sum_n \theta[E-\eps(n_y,n_z) ]$. This gives a transmission function
which takes the form of a staircase, 
with the transmission of the point contact at energy $E$
equalling the  number of steps up to that energy, i.e.~ the number of transverse modes with $\eps(n_y,n_z)  < E$.
For finite values of $\omega_x$ the steps in the staircase become smoothed out, but they will still be clearly distinguishable while $D$ remains less than the energy different between successive steps.

The main regime discussed in the literature \green{on the thermoelectric response of}
 a point contact is for one which is narrow enough ($\omega_y,\omega_z$ are large enough) 
that $\eps(1,0)-\eps(0,0)$ and $\eps(0,1)-\eps(0,0)$ are much larger than temperature or bias.  
This was more or less the case in the first experiments on thermoelectric effects in point contacts \cite{beenakker1992}.
Then, most theory works consider taking $\eps(0,0)$ as close to the reservoir's electrochemical potentials, which means 
only the first step in the staircase plays any role in the physics.  Thus, one
can drop the sums over $n_y,n_z$ in the above expressions, and get a transmission 
which switches from zero at low energies to one at high energies in a manner that gives it the form of a Fermi function
centred at $\eps(0,0)$ with width $D$.  To get significant thermoelectric effects, one want to choose $\omega_x$ such that the width $D$ is of order or less than the reservoir temperatures, so the transmission takes the form of 
a Heaviside $\theta$-function.
Such a point contact gives an energy-filter which is the same those sketched in Fig.~\ref{Fig:energy-filter}, which blocks all electrons below energy $\eps(0,0)$ and lets through electrons above energy $\eps(0,0)$.

A rather different regime was discussed in Refs.~\cite{Brantut-Grenier-et-al2013,Grenier2014},
they considered a case in which temperature was much larger than the distance between steps in the above mentioned staircase.  Then, it is the slope of the staircase as a whole that manners, not the individual steps in the staircase.  If we take the number of $n_y,n_z$ for which $\eps(n_y,n_z)  < E$, grows quadratically with $E$ on scales when we cannot resolve individual steps. Thus in this regime
\begin{eqnarray}
{\cal T}_{LR}(E) \sim \left\{ \begin{array}{ccl}
0 &\qquad \hbox{ for} & E \leq V_0,
\\
{\displaystyle{\pi \,(E-V_0)^2 \over \omega_y \omega_z}} &\qquad \hbox{ for} & E > V_0,
\end{array}\right.
\label{Eq:point-contact-smooth-staircase}
\end{eqnarray}
where we assume the contribution of each individual step is negligible.
A strong thermoelectric effect under situations where the magnitude of 
${\cal T}_{LR}(E)$ changes by a significant proportion within a window of temperature around the electrochemical
potential, this requires that $V_0$ is reasonably close to the electrochemical potentials.
This is currently difficult to do in electronic systems,
it is much easier to pinch-off the point contact (increasing $\omega_y,\omega_z$) than uniformly change the
potential in the vicinity of the point-contact, $V_0$.  Thus the regime in Eq.~(\ref{Eq:point-contact-smooth-staircase}) is little considered in electronic systems, even though it is highly relevant to atomic gases \cite{Brantut-Grenier-et-al2013,Grenier2014}.

%%%%%%%%%%%%%%%%%%%%%%%%%
%%%%%%%%%%%%%%%%%%%%%%%%%%

\subsubsection{Transmission function for  a single-level quantum dot}
\label{Sect:scatter-single-level-dot}

Now, let us turn to the case of a single level quantum dot, or a single-level molecule.  Here we make the assumption that the electrons do not interact with each other, so Coulomb blockade effects (which cannot be treated in the scattering theory) 
are absent.  This would be the case if the quantum dot is so well screened
by surrounding gates, that the electrons in the dot do not feel the presence of each other.
This is a rather drastic assumption, which is rarely satisfied in real experimental systems.
Most real quantum dots have significant Coulomb blockade effects, and so are better modelled by 
another method, such as the rate equation method, see section~\ref{Sect:2-term-sys}. 

However, it is none the less instructive to understand the scattering theory for a quantum dot, before
going on to more sophisticated models.  
Partly, because it is a good introduction to the problem, 
and partly because its results fit rather well 
(perhaps better than one would expect) with the results of more sophisticated calculations \cite{Kennes2013}.

To treat a quantum dot within scattering theory, one can use the following relation
to relate the dot's scattering matrix to its Hamiltonian \cite{Jalabert1992,Alhassid-review},
\begin{eqnarray}
{\cal S}(E) = \hat{1}\ -\ i2\pi  \hat{W}^\dagger \left[{E-\hat{\cal H}_{\rm dot}+i \pi \hat{W}\hat{W}^\dagger}  \right]^{-1}\hat{W},
\label{Eq:S-from-H}
\end{eqnarray}
where $\hat{1}$ is the unit-operator (i.e.\ the unit matrix) in the space of reservoir modes, 
$\hat{\cal H}_{\rm dot}$ is the Hamiltonian of the dot, 
and $\hat{W}$ is the operator coupling the reservoir modes to dot states.
All these operators are most easily written as matrices, in which case $[\cdots]^{-1}$ is simply a matrix inverse.
If the dot only has one state at energy $E_0$, and two reservoirs (each with one mode), which couple
to the dot with strength $w_{ L}$ and $w_{ R}$, then 
\begin{eqnarray}
\hat{W} = \big(w_{ L},w_{ R}\big),  \qquad \hat{W}^\dagger = \left( \begin{array}{c}w^*_{ L} \\ w^*_{ R} \end{array}\right), 
\end{eqnarray}
as a result
\begin{eqnarray}
{\cal S}(E) = \left( \begin{array}{cc} 1 & 0 \\ 0 & 1 \end{array}\right) 
\ -\ {\rmi 2\pi \over E-E_0 +i\pi |w_{ L}|^2 +i\pi  |w_{ R}|^2} 
\left( \begin{array}{cc} 
 |w_{ L}|^2 &  w_{ L}^* w_{ R}\\
w_{ R}^* w_{ L} &  |w_{ R}|^2
\end{array}\right), 
\label{Eq:S-single-level-dot}
\end{eqnarray}
Substituting this into Eqs.~(\ref{Eq:scattering-prob},\ref{Eq:def-T_ij})
and extracting the term corresponding to the transmission from right to left we get
a Lorentzian energy dependence,
\begin{eqnarray}
{\cal T}_{ LR}(E) =  \frac{\Gamma_{ L} \Gamma_{ R} }{\left(E-E_0\right)^2 + \frac{1}{4}\left(\Gamma_{ L} +\Gamma_{ R}\right)^2 },
\label{Eq:Briet-Wigner}
\end{eqnarray}
where we define $\Gamma_i=2\pi |w_ i|^2$ for $i\in { L,R}$, such that $\Gamma_i/\hbar$ is the rate at which 
the dot state decays into reservoir $i$.  Thus, we see that the transmission has  a Breit-Wigner form,  with 
$\Gamma_{ L}+\Gamma_{ R}$ being the energy-broadening of the dot-level due to the coupling to the reservoirs.  The thermoelectric response of systems with this transmission function have been
studied in detail in Refs.~\cite{Paulsson-Datta03,linke2010,elb09,esposito2010,Fahlvik-Svensson2012,Fahlvik-Svensson2013}, and we will refer to this case in sections~\ref{sec:energyfiltering} and \ref{Sect:scatter-nonlin-Carnot}. 

At least one of the more sophisticated methods of treating the same problem \cite{Kennes2013}, 
which includes some amount of Coulomb interaction effects between electrons,
gives basically the same result for the transmission function of a single level quantum dot
(at least in the linear response regime). The Coulomb interaction renormalizes $E_0$, $\Gamma_{ L}$ and 
$\Gamma_{ R}$, but does not change the Lorentzian form of the energy dependence in Eq.~(\ref{Eq:Briet-Wigner}).

 \subsubsection{Transmission functions for more complicated systems}
\label{Sect:transmission-LDA}
 
 \green{
 For more complicated systems, one typically needs to resort to a numerical method to find the 
transmission function.
 The simplest method is to model the quantum system as an $n$ site tight-binding model,
 written in a matrix form (with on-site energies on the diagonal and inter-site couplings for the off diagonal elements), and  substitute this into Eq.~(\ref{Eq:S-from-H}).  The scattering matrix can be found 
 by a numerical diagonalization of the matrix $\left[{E-\hat{\cal H}_{\rm dot}+i \pi \hat{W}\hat{W}^\dagger}  \right]$.
 }
 
 \green{
 A more sophisticated treatment is to use a density functional theory, which treats interaction effects through a local density approximation (LDA), within this approximation one can solve from first principles the problem of the transmission through a molecular structure \cite{car,Pecchia2008,Finch09,Peterfalvi2014,Garcia-Suarez}.  This was used to find the thermoelectric response and figure of merit of various molecules between metallic contacts, and thereby show how to engineer the transmission function; for example by adding side groups to the molecule 
 which introduce Fano resonances at the right energy to generate a strong thermoelectric response
 \cite{Finch09,Peterfalvi2014}.  Other works on using density function theory for thermoelectrics include \cite{vignale,dagosta}.  Recently a powerful software package \cite{Ferrer-GOLLUM14} has been developed
 based on the technique called LDA+U with spectral adjustments for coupled spin, charge and thermal transport.  This package generates the scattering matrix for transport in the presence of non-collinear magnetism, the quantum-Hall effect, Kondo and Coulomb blockade effects, multi-terminal transport, quantum pumps, superconducting nanostructures, etc.  This gives one the information necessary to calculate the thermoelectric response,
 and the system's efficiency as a thermoelectric heat-engine or refrigerator \cite{Ismael,Algharagholy,Sadeghi2017}.
 }

\subsection{Scattering theory with Andreev reflection}
\label{Sect:scatter-Andreev}

Here we assume we are considering a system at low temperatures (typically less than 1 Kelvin), 
coupled to a superconductor whose superconducting gap is much larger than all temperatures or biases in the problem.
Such a superconductor has no electronic states that can contribute to transport, however it does acts as an ``Andreev mirror''.  An electron hitting the superconductor is retro-reflected as a hole (with a Cooper pair going into the superconducting reservoir).
To include this Andreev reflection in the scattering theory \cite{Claughton-Lambert,Beenakker-review}, we  
define the zero of energy as being that of the electrochemical potential of the superconductor.
Then all electron states at negative energies  $E_{\rm electron} < 0$, 
we write in terms of holes (the absence of an electron) 
with positive energy $E=-E_{\rm electron}$.
We thus have two species of particles in the scattering problem at each energy, electrons that we label with $\vsig=+1$, and holes that we label with $\vsig=-1$.
The occupation of reservoir $i$ with electrochemical potential $\mu_i$ and temperature $T_i$, is 
then 
\begin{align}
f_j^\vsig(E) = \left(1+\exp\left[(E - \vsig \mu_j)\big/ (\kB T_j) \right] \right)^{-1}
\quad \hbox{ with } \  E \geq 0.
\label{Eq:f-eh}
\end{align} 
This formula for the electrons ($\vsig=+1$) is identical to Eq.~(\ref{Eq:f}). 
To get this formula for holes ($\vsig=-1$), 
we use the fact that the probability a state at energy $E$ contains a hole is
simply one minus the probability it contains an electron, given by Eq.~(\ref{Eq:f}). 
We then make the observation that $1- (1+\e^x)^{-1} = (1+\e^{-x})^{-1}$, followed by the substitution $E \to -E$ to write the negative electron energies in term of the positive hole energy.

Now we have two species of particles,
the scattering matrix elements are more complicated than in section~\ref{Sect:scatter-basics}: 
they gain the index $\vsig$ which indicates if the 
incoming state is an electron or a hole, and an index $\vrho$ which indicates if the outgoing
state is an electron or a hole.
The probability for a particle $\vsig$ in mode $m$ of reservoir $j$ to scatter into a particle $\vrho$ in mode $n$ of reservoir $i$ is 
\begin{align}
P^{\vrho\vsig}_{in;jm}(E)\  =\ \big| {\cal S}^{\vrho\vsig}_{in;jm}(E) \big|^2. 
\label{Eq:scattering-prob-eh}
\end{align}
If we sum this over all modes coupled to reservoirs $i$ and $j$, we get the transmission matrix elements
\begin{align}
{\cal T}^{\vrho\vsig}_{ij}(E) = \sum_{nm} P^{\vrho\vsig}_{in;jm} (E).
\label{Eq:def-T_ij-SC}
\end{align}
In the absence of the Andreev reflection, incoming electron-states
scatter to outgoing electron-states without changing energy, and incoming hole-states 
scatter to outgoing hole-states without changing energy. In this case, 
 ${\cal S}^{\vrho\vsig}_{in;jm}(E)$ would only be non-zero
for $\vrho=\vsig$ \cite{Beenakker-review}, with
${\cal S}^{+1,+1}_{in;jm}(E \geq 0) = {\cal S}_{in;jm}(E)$
and $
{\cal S}^{-1,-1}_{in;jm}(E \geq 0) = {\cal S}^*_{jm;in}(-E)$,
where ${\cal S}_{in;jm}(E)$ is the scattering matrix above Eq.~(\ref{Eq:scattering-prob}).
However, everything changes in the presence of a superconducting reservoir. 
Every time an electron in the scatterer hits the superconductor, 
it changes into a hole reflected back into the scatterer 
(injecting a Cooper pair into the superconducting reservoir).  
Every time a hole in the scatterer hits the superconductor, it changes into an electron reflected back into scatterer (absorbing a Cooper pair from the superconducting reservoir).
This means that scattering matrix elements for electrons and holes, ${\cal S}^{\vrho\vsig}_{in;jm}(E)$ are no longer zero for $\vrho \neq \vsig$.
Here, $i$ and $j$ label normal (not superconducting) reservoirs, since the superconducting reservoir acts as an Andreev mirror for electrons and holes.
The scattering matrix must still be unitary, because there is conservation of particles in the scatterer (even if the particles make transitions between being electrons and holes).  Since  the scattering matrix is unitary, we have
\begin{eqnarray}
\sum_{j\neq {\rm SC}} \sum_{m\vsig} {\cal S}^{\vrho \vsig}_{in;jm}(E) \ \left[{\cal S}^{\vrho' \vsig }_{i'n';jm}(E)\right]^* 
=\delta_{i'i}\delta_{n'n}\delta_{\vrho'\vrho},
\end{eqnarray}
where $j\neq {\rm SC}$ indicates that the sum is over all reservoirs except the superconducting (SC) reservoir.
This means that 
\begin{subequations}
\label{Eq:constraints-T_ij-eh}
\begin{align}
&{\cal T}^{\vrho\vsig}_{ij} (E) \geq 0 \qquad\hbox{ for all } \ i,j, \vrho, \vsig 
\label{Eq:constraints-T_ij-eh-a}
\end{align}
while
\begin{align}
&\sum_{i\neq {\rm SC}} \sum_\vrho  {\cal T}^{\vrho\vsig}_{ij}(E) = N^\vsig_j(E), 
\qquad
\sum_{j\neq {\rm SC}} \sum_\vsig {\cal T}^{\vrho\vsig}_{ij}(E) = N^\vrho_i(E),
\label{Eq:constraints-T_ij-eh-c}
\end{align}
\end{subequations}
where the $i$ and $j$ sums are over all non-superconducting reservoirs.
As in the absence of Andreev reflection, the dynamics are time-reversed if one reverses all particle velocities
(so incoming particles become outgoing particles and vice versa), and one reverses any external magnetic field, ${\bm B}$,
hence
\begin{eqnarray}
{\cal T}^{\vrho\vsig}_{ij}(E,{\bm B})={\cal T}^{\vsig\vrho}_{ji}(E,-{\bm B})\ .
\label{Eq:T_ij-time-reverse-SC}
\end{eqnarray}

The charge current out of non-superconducting reservoir $i$ is \cite{Beenakker-review} 
\begin{eqnarray}
\Jelectrici = \sum_{j \neq {\rm SC}} \sum_{\vrho\vsig} \int_0^\infty {{\rm d}E \over h} \  \vrho e 
\, \left[N_i^\vrho(E)  \,\delta_{ij}\delta_{\vrho \vsig}-{\cal T}_{ij}^{\vrho \vsig}(E)  \right]
\,  f_j^\vsig (E),  
\label{Eq:I-initial-eh}
\end{eqnarray}
where $j$ is summed over all non-superconducting reservoirs,
while the $\vrho\vsig$-sums are over electrons ($+1$) and  holes ($-1$).
The energy current out of reservoir $i$ is
\begin{eqnarray}
\Jenergyi = \sum_{j \neq {\rm SC}} \sum_{\vrho\vsig} \int_0^\infty {{\rm d}E \over h} E
\, \left[N_i^\vrho(E)  \,\delta_{ij}\delta_{\vrho \vsig} - {\cal T}_{ij}^{\vrho \vsig}(E) \right]
\,  f_j^\vsig (E),  
\label{Eq:I-energy-initial-eh}
\end{eqnarray} 
and the heat current out of reservoir $i$ is \cite{Claughton-Lambert,Whitney-2ndlaw}
\begin{equation}
\Jheati = \sum_{j \neq {\rm SC}} \sum_{\vrho\vsig} \int_0^\infty {{\rm d}E \over h}\, (E\!-\!\vrho \mu_i)
\left[ N_i^\vrho(E)  \,\delta_{ij}\delta_{\vrho \vsig} - {\cal T}_{ij}^{\vrho \vsig}(E) \right]  \, f_j^\vsig (E).\ \ 
\label{Eq:J-initial-eh}
\end{equation}
Comparing these three equations, one can easily see that
\begin{eqnarray}
\Jheati &=& \Jenergyi-V_i \, \Jelectrici. 
\label{Eq:J-to-I-energy-eh}
\end{eqnarray}
It is crucial to recall that throughout this review we use the subscripts ``e'' for electrical current 
and ``h'' for heat current.
In the context of systems with superconductors, this is notation is unfortunate, 
because most of the works we cite use ``e'' for electrons and ``h'' for holes (for which we recall we use ``$\pm 1$'').

If a superconductor is present, then there is a charge-current into it 
(in the form of Cooper pairs), $\JelectricSC$.
In contrast, the heat flow into the superconducting reservoir is zero
as each electron hitting the superconductor 
is Andreev reflected back into the scatterer as a hole with the same energy.
In addition, given that we define the zero of energy as being at the electrochemical potential 
of the superconductor, the energy-current into the superconductor equals the heat-current
into the superconductor, and is also zero.
Thus, since electrical currents and energy currents are conserved, we have
\begin{align}
\JelectricSC &\ =\  - \sum_{i\neq {\rm SC}} \Jelectrici \ ,  
\label{Eq:Isc}
\\ 
0 \ = &\ \JheatSC \ =\ \JenergySC \ =\ -\sum_{i \neq {\rm SC}} \Jenergyi ,
\label{Eq:Jsc}
\end{align}
where again $i$ is summed over the non-superconducting reservoirs.
As the sum of energy currents over all non-superconducting reservoirs is zero, 
we have 
\begin{eqnarray}
\sum_{i\neq {\rm SC}} \Jheati = - \sum_{i\neq {\rm SC}}  V_i \, \Jelectrici \ .
\label{Eq:scatter-1st-law-eh}
\end{eqnarray}
We also note that the entropy of the superconducting reservoir does not change with time,
$\dot{\mathscr{S}}_{\rm SC}=0$, while the rate of change of the entropy in the other reservoirs
is given by Eq.~(\ref{Eq:scatter-dotS}).

Note that the scattering theory presented \green{here} can treat an arbitrary number of superconducting leads with different phases for the superconducting order parameter, but only if all those superconductors
all have the same electrochemical potential.  In cases where there are multiple superconductors with different 
electrochemical potentials one has to use methods beyond this review, such as the methods presented in
Refs.~\cite{KBT-theory1982,KBT-theory1983,Cuevas1999,Melin2011}.

%%%%%%%%%%%%%%%%%%%
%%%%
%%%%%%%%%%%%%%%%%%%

\section{Scattering theory in linear response}
\label{sec:landauer}

Much can be said about the scattering theory of arbitrary systems in the limit where the 
differences in temperature and bias between reservoirs are small on the scale of the average temperature.
In this limit one gets a linear-response theory, where the currents are proportional to the
thermodynamic forces.  This microscopic quantum theory is thus a complement to the 
classical linear-response thermodynamics in chapter~\ref{sec:nonequilibrium},
or indeed a justification for applying classical linear-response thermodynamics to 
such quantum systems.

To get the linear-response version of the scattering theory we 
expand the Fermi functions about a given electrochemical potential and temperature.
For $\kB\Delta T_j$ and $\Delta\mu_j= e V_j$ much less than $\kB T$, we have   
\begin{align}
f_j(E)
&\approx 
f(E) + \frac{\partial f}{\partial T} \,\Delta T_j
 + \frac{\partial f}{\partial \mu} \,\Delta \mu_j
\ =\ 
 f(E)\, - \, f'(E) \,\left[
(E-\mu)\frac{\Delta T_j}{T}+eV_j
\right],
\label{eq:fermilinear}
\end{align}
where we used the fact that $\Delta \mu_i = eV_j$ for a bias $V_j$ on reservoir $j$.
Here we have defined $f'(E)$ as the derivative of the Fermi distribution, so
\begin{eqnarray}
-f'(E) \ \equiv \  -\frac{\partial f}{\partial E} \ = \ 
\frac{1}{4k_BT\cosh^2[(E-\mu)/2k_BT]}\ ,
\label{Eq:fprime}
\end{eqnarray}
is a bell-shaped function centered at $\mu$ and has a width
of the order of $k_B T$.

In the absence of a superconducting reservoir,
we can insert the linear expansion in Eq.~(\ref{eq:fermilinear}) into 
the equations for the currents in section~\ref{Sect:scatter-basics}.
We then note that the terms which are zeroth-order in $\kB\Delta T_j$ and $\Delta\mu_j$ cancel due to 
Eq.~(\ref{Eq:constraints-T_ij}).
This gives us a linear relationship between the currents, $\Jelectrici$ and $\Jheati$, and the thermodynamic forces, $\mathcal{F}_{{ e},i} = V_i/T$ and $\mathcal{F}_{{h},i} = \Delta T_i/T^2$, such that
\begin{equation}
J_{\mu,i}\ =\ \sum_{\nu={ e,h}}\sum_j\
 L_{\mu \nu;ij}\ \mathcal{F}_{\nu,j}, 
\label{Eq:Onsager-matrix-general}
\end{equation}
where $\mu =$e (charge) or h (heat), and $j$ is summed over all reservoirs.
The Onsager coefficients are then given by 
\begin{align}
L_{ee,ij}\ =\ e^2 T I^{(0)}_{ij},\qquad
L_{eh,ij}\ =\ L_{he;ij} \ =\ e T I^{(1)}_{ij},\qquad
L_{hh,ij}\ =\ T I^{(2)}_{ij}.
\label{Eq:Ls-multiterm}
\end{align}
where we define the integral $I^{(n)}_{ij}$ as
\begin{equation}
I^{(n)}_{ij} \equiv \int_{-\infty}^\infty
\frac{dE}{h} (E-\mu)^n \ \left( N_i(E)\delta_{ij} -{\cal T}_{ij}(E)\right) \ \big(- f'(E)\big).
\label{eq:In}
\end{equation}

Given Eq.~(\ref{Eq:constraints-T_ij-c}), one sees that  Eqs.~(\ref{Eq:Ls-multiterm},\ref{eq:In}) 
imply that the linear response heat current is conserved, 
by which we mean it obeys a  Kirchoff's law
\begin{align}
\sum_i \Jheati =0 \ ,
\label{Eq:heat-sum-linear}
\end{align}
where the sum is over all reservoirs.
We warn the reader that this conservation of heat current is 
a specificity of linear-response theory.
In general, we have Eq.~(\ref{Eq:scatter-1st-law}) in place of Eq.~(\ref{Eq:heat-sum-linear}).
If the right hand side of Eq.~(\ref{Eq:scatter-1st-law}) is negative, then the system is consuming electrical power
and producing heat (i.e. the heat flow into the system is less than the heat flow out), in the form of Joule heating. 
In contrast, if the right hand side of Eq.~(\ref{Eq:scatter-1st-law}) is positive, 
then the system is producing electrical power and absorbing heat 
(i.e. the heat flow into the system is more than the heat flow out). 
This is required to conserve energy, and is the origin of the first law of thermodynamics.
So why is it that  Eq.~(\ref{Eq:heat-sum-linear}) suggests that heat flow is conserved (heat flow in equals heat flow out)
irrespective of whether the system is absorbing or producing electrical work?  The reason is that 
Eq.~(\ref{Eq:heat-sum-linear}) is calculated in linear response, which means that it is only accurate to first order
in bias and/or temperature difference.  The power generated or absorbed by the system is {\it quadratic} in these parameters, and so its modification of the heat is not captured by linear response.
To see this, it is sufficient to note that the electrical current is proportional to bias and/or temperature difference, and the power goes like bias times the electrical current, hence the power goes like bias squared and/or bias times temperature difference. 

From Eq.~(\ref{Eq:Ls-multiterm}) we can get various results about the symmetries of the 
matrix of Onsager coefficients. Firstly, we can use the fact that the transmission obeys 
${\cal T}_{ij}(E,{\bm B})={\cal T}_{ji}(E,-{\bm B})$, as described in Eq.~(\ref{Eq:T_ij-time-reverse}) 
to prove that the Onsager coefficients  for an external magnetic field ${\bm B}$ obey the Onsager reciprocal relation
\begin{align}
L_{\mu\nu,ij}({\bm B})=L_{\nu\mu;ji}(-{\bm B})
\label{Eq:Onsager-multiterm}
\end{align}
with  $\mu$ and $\nu$ being either electric ($e$) or heat ($h$).
This is just as Onsager showed in classical thermodynamics. 
However, as Ref.~\cite{Butcher1990} pointed out,  the above microscopic derivation also shows that 
\begin{eqnarray}
L_{eh,ij}({\bm B})=L_{he;ij}({\bm B}).
\label{Eq:Butcher}
\end{eqnarray}
Combining the two above relations means that $L_{eh;ij}({\bm B})=L_{eh;ji}(-{\bm B})$.
Ref.~\cite{Samuelsson-Linke2014} presents experiments that demonstrate such relations,
although the relations found experimentally are rarely perfect for the reasons discussed in that paper.
Of particular importance is the fact that decoherence due to inelastic scattering
leads to a breaking of the equality in Eq.~(\ref{Eq:Butcher}), without affecting the 
equality in Eq.~ (\ref{Eq:Onsager-multiterm}), see section~\ref{sec:probes} 
of this review for more details.  We also note that Andreev reflection from a superconductor 
breaks the equality in Eq.~(\ref{Eq:Butcher}), see section~\ref{Sect:Andreev-linear}.

Symmetries in the underlying system Hamiltonian (such as spin-rotation symmetry, particle-hole symmetry,  or sub-lattice symmetry) lead directly to additional relations between the above Onsager coefficients.
Such relations are given in Ref.~\cite{jwmb}, along with similar relations for the Onsager coefficients
which couple spin transport to charge and heat transport.

\subsection{Onsager matrix}

If the reservoirs are labelled $1,2,3,\cdots,K$, then we can choose to measure all biases and temperatures from those of reservoir 1, so $\mathcal{F}_{{ e},i} = (V_i-V_1)/T_1$ and $\mathcal{F}_{{h},i} = (T_i-T_1)/T_1^2$.
This means that $\mathcal{F}_{{ e},1}= \mathcal{F}_{{h},1}=0$, which 
 simplifies  Eq.~(\ref{Eq:Onsager-matrix-general}).  If Eq.~(\ref{Eq:Onsager-matrix-general}) is written as a matrix equation,  this is equivalent to eliminating two rows and two columns from the matrix,
 resulting in 
\begin{equation}
\left(\begin{array}{c} 
\Jelectric{2} \\
\Jheat{2} \\
\Jelectric{3} \\
\Jheat{3} \\
\vdots \\
\Jelectric{K} \\
\Jheat{K} 
\end{array} \right)
\ \ =\ \
\left(\begin{array}{ccccccc} 
L_{{ee},22} & L_{{eh},22} & L_{{ee},23} & L_{{eh},23} & \cdots &L_{{ee},2K} & L_{{eh},2K} \\
L_{{he},22} & L_{{hh},22} & L_{{he},23} & L_{{hh},23} & \cdots &L_{{he},2K} & L_{{hh},2K}\\
L_{{ee},32} & L_{{eh},32} & L_{{ee},33} & L_{{eh},33} &  \cdots &L_{{ee},3K} & L_{{eh},3K} \\
L_{{he},32} & L_{{hh},32} & L_{{he},33} & L_{{hh},33} &              &L_{{he},3K} & L_{{hh},3K}\\
\vdots & \vdots & \vdots & &  \ddots & & \vdots\\
L_{{ee},K2} & L_{{eh},K2} & L_{{ee},K3} & L_{{eh},K3} &            &L_{{ee},KK} & L_{{eh},KK} \\
L_{{he},K2} & L_{{hh},K2} & L_{{he},K3} & L_{{hh},K3} & \cdots  &L_{{he},KK} & L_{{hh},KK} \\
\end{array} \right)
\ \left(\begin{array}{c} 
\mathcal{F}_{{e},2} \\
\mathcal{F}_{{h},2} \\
\mathcal{F}_{{e},3} \\
\mathcal{F}_{{h},3} \\
\vdots \\
\mathcal{F}_{{e},K} \\
\mathcal{F}_{{h},K} 
\end{array} \right)\ .
\label{Eq:Onsager-matrix}
\end{equation}
We refer to the above matrix as the Onsager matrix, ${\bm L}$, which is $(2K-2)\times(2K-2)$.
This matrix equation does not give the electrical and heat current into reservoir 1.
However, these can be found by using the Kirchoff's laws for  conservation of electrical and heat currents in 
Eqs.~(\ref{Eq:I-conserve},\ref{Eq:heat-sum-linear}), thus
$\Jelectric{1} = -\sum_{i=2}^K \Jelectrici$ and $\Jheat{1}=-\sum_{i=2}^K \Jheati$.

\subsection{Linear-response for two-terminal systems}
\label{Sect:scatter-2term-linear}

The most commonly considered case of Onsager reciprocal relations are for two-terminal systems
which are coupled to two reservoirs; left (L) and right (R).
For two non-superconducting terminals we start with 
Eqs.~(\ref{Eq:I-initial-two-reservoir}-\ref{Eq:J-initial-two-reservoir})
and use Eq.~(\ref{eq:fermilinear}) to expand the Fermi function for reservoir L 
about the electrochemical potential and temperature of reservoir R,
so $e\Delta V = (\mu_{L}-\mu_{R})$ and $\Delta T = T_{L}-T_{R}$.
Then Eq.~(\ref{Eq:Onsager-matrix}) contains only a two-by-two matrix for ${\bm L}$, 
and corresponds to Eq.~(\ref{eq:coupledlinear}) for currents from left to right. 
The Onsager coefficients then read,
\begin{equation}
L_{ee}=e^2 T I_0,\qquad
L_{eh}=L_{he}=e T I_1,\qquad
L_{hh}=T I_2.
\label{Eq:Ls-two-term}
\end{equation}
Here, the integrals $I_n$ have been defined as
\begin{equation}
I_n\equiv \int_{-\infty}^\infty
\frac{dE}{h} \ (E-\mu)^n \ {\cal T}_{LR}(E,{\bm B}) \,\big[\!-\!f'(E)\big],
\label{eq:In-2term}
\end{equation}
for an external magnetic field ${\bm B}$.
It immediately follows 
from Eqs.~(\ref{eq:el_conductance}-\ref{eq:seebeck}) 
that the conductances,
thermopower and Peltier coefficients can all be expressed in terms of the integrals $I_n$:
\begin{equation}
G=e^2 I_0,\qquad
K=\frac{1}{T}\left(I_2-\frac{I_1^2}{I_0}\right),\qquad 
S=\frac{1}{eT}\frac{I_1}{I_0}, \qquad
\Pi = \frac{1}{e}\frac{I_1}{I_0}.
\label{eq:landauer.ex}
\end{equation}

In this two-terminal case, Eq.~(\ref{Eq:Onsager-multiterm}) reduces to
\begin{align}
L_{\mu \nu} (-{\bm B}) \ =\ &  L_{\nu \mu} ({\bm B}) \ ,
\label{Eq:Onsager-2term}
\end{align}
with  $\mu$ and $\nu$ being either electric ($e$) or heat ($h$).
This means that for this quantum system, we recover the famous relation between the Seebeck and Peltier coefficients (see chapter~\ref{sec:nonequilibrium}), 
\begin{equation}
\Pi({\bm B})=TS(-{\bm B}),
\label{Eq:Onsager-2term-2}
\end{equation}
which Onsager proved for systems described
by classical thermodynamics.
However since, we also have $L_{eh}({\bm B})=L_{he}({\bm B})$, this tells us that $L_{\mu \nu} ({\bm B})$ 
is an even function of the external magnetic field ${\bm B}$ for all $\mu, \nu$.
Thus we see that $G({\bm B})$, $K({\bm B})$,  $S({\bm B})$ and $\Pi({\bm B})$ must all be even functions of ${\bm B}$ \cite{Butcher1990}.
However, as will be discussed further in section~\ref{sec:probes},
the relationship $\Pi({\bm B})=TS({\bm B})$ can be broken by decoherence effects or Andreev reflection, 
while the Onsager reciprocal relations ($G({\bm B})=G(-{\bm B})$, $K({\bm B})=K(-{\bm B})$ and  $\Pi({\bm B})=TS(-{\bm B})$) are not.
Thus in realistic systems it is not surprising to see $S({\bm B})$ and $\Pi({\bm B})$ not being even in ${\bm B}$,
while $G({\bm B})$ and $K({\bm B})$ are.

From Eqs.~(\ref{Eq:Ls-two-term}-\ref{eq:landauer.ex}) we see that \cite{mott1},
\begin{equation}
S=
\frac{1}{eT}
\frac{\int_{-\infty}^\infty
dE (E-\mu) {\cal T}_{LR}(E) \,\big[\!-\!f'(E)\big]}{
\int_{-\infty}^\infty
dE {\cal T}_{LR}(E) \,\big[\!-\!f'(E)\big]},
\label{Eq:S-as-ratio-of-integrals}
\end{equation}
with $f'(E)$ being the derivative of the Fermi function in Eq.~(\ref{Eq:fprime}). 
Since $f'(E)$ 
is an even function of $(E-\mu)$, one sees that $S$ vanishes
if ${\cal T}_{LR}(E)$ is symmetric around $\mu$. It is then clear that 
electrons and holes contribute to the thermopower with opposite signs 
and that $S=0$ when there is particle-hole symmetry.
Any system in which the symmetry is broken between the dynamics of electrons above and below the electrochemical will exhibit a finite thermopower.
This occurs when $\mu$ is close to sharp resonances 
of ${\cal T}_{LR}$ \cite{fazio2001,bss10} or close to the mobility edge
of the Anderson metal-insulator phase transition 
\cite{ImryAmir} (the states with energies above the mobility
edge are extended, while those below it are localized), 
where the transmission exhibits a sharp and asymmetric energy 
dependence (the transmission drops exponentially with the 
system size in the insulating regime).

Eq.~(\ref{Eq:S-as-ratio-of-integrals}) also gives a pretty interpretation of the thermopower,
as the following average \cite{mahansofo}
\begin{equation}
S=
\frac{1}{eT} \big\langle E-\mu \big\rangle
\label{Eq:S-as-average}
\end{equation}
Here $\langle E -\mu \rangle$ as the average energy (measured from the electrochemical potential)
of the electrons that are transmitted through the scatterer,
\green{where the average is defined as}
\begin{equation}
\langle \ \cdots\ \rangle \ =\ 
\frac{\int_{-\infty}^\infty
dE \ (\,\cdots\,)\  {\cal T}_{LR}(E) \ \left(-f'(E)\right)  }{
\int_{-\infty}^\infty
dE \ {\cal T}_{LR}(E) \ \left(-f'(E)\right)}.
\label{Eq:defining-average-over-E}
\end{equation}

Eq.~(\ref{Eq:S-as-average}) makes it clear that we can make $S$ as big as we like, by choosing a scatterer which only lets 
through the electrons with very high energy.
Of course, then the flow of electrons through the scatterer will be exponentially small.
This means it will take the system a long time to find the steady state, since if we apply a temperature difference to
a thermoelectric that was previously in equilibrium, the rate at which the bias builds up across the thermoelectric is dependent on the rate at which current flows though it.

A similar analysis tells us that the ratio of thermal to electrical conductance
can be written as
\begin{eqnarray}
{K \over G} \ & = & \frac{ \big\langle (E-\mu)^2  \big\rangle -  \big\langle E-\mu  \big\rangle^2 }{e^2T} 
\label{Eq:K/G}
\end{eqnarray}
We will show in Section~\ref{sec:Sommerfeld-phenomenology} that the right hand side becomes the Lorenz constant, ${\cal L}$, given in Eq.~(\ref{lorenz}), in the limit where the transmission depends only weakly on energy.
So the system obeys the Wiedemann-Franz law in such a limit.  
However, we will also see that for any significant thermoelectric effect, the transmission will be such that the
Eq.~(\ref{Eq:K/G}) will violate the Wiedemann-Franz law,
with section~\ref{sec:energyfiltering} showing that the best thermoelectrics have $K\big/G \to 0$.

\subsubsection{Comparison with the Boltzmann Equation}
\label{Sect:Boltzmann-Equation}

While the Landauer approach describes coherent quantum
transport
\footnote{Note, however, that the Landauer approach
can be useful for understanding transport coefficients in
large conductors as well, by viewing them as a series of elastic 
resistors, connected by reservoirs where  energy is dissipated,
see Ref.~\cite{datta2013}.},
semiclassical transport can be described by means
of the Boltzmann equation. Here we consider transport processes that
occur much slower than the relaxation to local equilibrium
and treat collisions within the relaxation-time
approximation \cite{ashcroftmermin}. That is, collisions drive 
the electronic system to local thermodynamic equilibrium under the
assumption that the distribution of electrons emerging from 
collisions does not depend on the structure of their non-equilibrium 
distribution prior to the collision and that collisions do not
alter local equilibrium.  In this case we can express
\emph{conductivities} and the thermopower in terms of the
integrals
\begin{equation}
K_n\equiv \int_{-\infty}^\infty
dE \ (E-\mu)^n \ \Sigma(E)\  \left(-f'(E)\right).
\label{eq:Kn}
\end{equation}
The form of this function is highly reminiscent of the scattering theory for a two-terminal systems.
Here $\Sigma(E)\approx D(E) t_R(E)\nu(E)^2$ is the transport distribution function, where $D(E)$ is the density of states, $t_R(E)$ the electron
relaxation time, and $\nu(E)$ the electron group velocity.
From the Boltzmann equation one obtains \cite{mahansofo}
\begin{equation}
\sigma = e^2 K_0,\quad
\kappa=\frac{1}{T}\left(K_2-\frac{K_1^2}{K_0}\right),\quad
S=\frac{1}{eT}\frac{K_1}{K_0}.
\label{eq:boltzmann.ex}
\end{equation}
Note that we neglect spin in these results.  If we include spin degeneracy, $\sigma$ and $\kappa$ would be double their above values but $S$ would be unchanged.

More sophisticated treatments of thermoelectric systems include modelling them with density functional theory
methods coupled to a Boltzmann transport theory 
\cite{Arita2008,Held2009,Wissgott2010,Sangiovanni,Knivek2015}.  
Broadly speaking these are the Boltzmann theory equivalent 
of the scattering theory coupled to density functional theory discussed in sect.~\ref{Sect:transmission-LDA}.

\subsubsection{The Sommerfeld expansion for weakly thermoelectric transport}
\label{sec:Sommerfeld-phenomenology}

In macroscopic conductors with weak thermoelectric responses, the transport coefficients are strongly
interdependent. The Wiedemann-Franz (WF) law \cite{wf,kittel} 
relates the electrical and 
thermal conductivities, while Mott's formula \cite{mott1,mott2,mott3}
relates the thermopower to the logarithmic derivative of the conductivity, 
evaluated at the reference electrochemical potential. These phenomenological 
equations can be derived both within the Boltzmann and the Landauer
approaches. In the latter case, conductances rather than conductivities
are considered. 
In both cases, one makes use of a \emph{Sommerfeld expansion} \cite{ashcroftmermin} of integrals
Eq.~(\ref{eq:In}) or Eq.~(\ref{eq:Kn})
to the leading order in $k_B T/E_F$, with $E_F=\mu(T=0)$ being the Fermi energy.
Such expansions are valid when the function ${\cal T}_{LR}(E)$ (for scattering theory)
or the function $\Sigma(E)$ (for the Boltzmann equation) is smooth on the scale of the reservoir
temperatures.  Thus for any given transmission function, going to a sufficiently low temperatures
will take one into a regime where the Sommerfeld expansion is a valid approximation.
In this regime, we expect the energy-filtering effect is weak, and thus expect weak thermoelectric effects.

Hereafter, we focus on the scattering theory approach.
The transmission function is \green{assumed to be slowly varying on the scale of temperature, so it can be
approximated by its Taylor expansion up to first order, }
\begin{equation}
{\cal T}_{LR} (E) \ \approx\  {\cal T}_{LR} (\mu) + \left.\frac{d {\cal T}_{LR} (E)}{dE}\right|_{E=\mu} 
(E-\mu ).
\label{Eq:smooth-T-truncation}
\end{equation}
Inserting this expansion into (\ref{eq:In-2term}), we obtain
the leading order terms of the Sommerfeld expansion of integrals $I_n$:
\begin{equation}
I_0\approx \frac{{\cal T}_{LR}(\mu)}{h},\quad
I_1\approx \frac{\pi^2}{3h}\,(k_B T)^2
\left.\frac{d {\cal T}_{LR} (E)}{dE}\right|_{E=\mu},\quad
I_2\approx  \frac{\pi^2}{3h}\,(k_B T)^2 {\cal T}_{LR}(\mu).
\label{eq:I0I1I2}
\end{equation}
In this derivation, we have used the fact that 
$\partial f/\partial E$ is an even function of 
$(E-\mu)$. Hence, $I_0$ and $I_2$
are determined to the leading order by ${\cal T}_{LR}(\mu)$.  
In contrast, $(E-\mu)\partial f/\partial E$ is an 
odd function of $(E-\mu)$, so that $I_1$ is determined by
the derivative $\left( {d {\cal T}_{LR} (E)}\big/{dE}\right)_{E=\mu}$.
The fact that we assume ${\cal T}_{LR} (E)$ is a smooth enough function of $E$ to 
truncate the expansion in Eq.~(\ref{Eq:smooth-T-truncation}) at leading order in $(E-\mu)$ implies that
we are considering a situation where 
\begin{align}
I_1 \ll \kB T \, I_0
\qquad \hbox{ and } \qquad I_1 \ll I_2\big/ ( \kB T).
\label{Eq:smooth-T-consequences}
\end{align}
There relations 
can equally be written as $L_{eh} \ll L_{ee}/e $ and $L_{eh} \ll eL_{hh}$.
This is equivalent to saying the thermoelectric effects are much weaker than conventional electrical 
and thermal conduction.  More specifically, it implies that $L_{eh}^2 \ll L_{ee}L_{hh}$, hence the figure of
merit $ZT=L_{eh}^2/{\det{\bm L}}\approx L_{eh}^2/L_{ee}L_{hh}\ll 1$, \green{which means the scatterer has} a poor efficiency for heat-to-work conversion.

We then obtain from Eq.~(\ref{eq:landauer.ex})
\begin{equation}
G\approx \frac{ e^2}{h}\,{\cal T}_{LR}(\mu),
\qquad
K\approx \frac{ \pi^2 k_B^2 T}{3h}\,{\cal T}_{LR}(\mu).
\end{equation}
From these relations we find the \emph{Wiedemann-Franz law},
\begin{equation}
\frac{K}{G} \approx {\cal L} T \, ,  \label{wflaw}
\end{equation}
where the constant value 
\begin{equation}
{\cal L} = \frac{\pi^2}{3} \left( \frac{k_B}{e} \right)^2  
\label{lorenz}
\end{equation}
is known as the \emph{Lorenz number}. 
Note that to derive the Wiedemann-Franz law we have considered only the 
leading order term in the Sommerfeld expansion, i.e. we have
neglected in the heat conductance $I_1^2/I_0$ with respect to $I_2$,
as a natural consequence of Eq.~(\ref{Eq:smooth-T-consequences}). 
From Eqs.~(\ref{eq:landauer.ex}) and (\ref{eq:I0I1I2}) we also 
derive \cite{mott1}
\begin{equation}
S \approx
\frac{\pi^2}{3}\left(\frac{k_B}{e}\right) (k_B T)
\left.\frac{d}{dE}\ln {\cal T}_{LR} (E) \right|_{E=\mu} 
%\approx \frac{\pi^2}{3}\left(\frac{k_B}{e}\right) (k_B T)
%\frac{d}{d\mu}\ln G(\mu)
\end{equation}
and consequently \emph{Mott's formula} for the thermopower:
\begin{equation}
S \approx
\frac{\pi^2}{3}\left(\frac{k_B}{e}\right) (k_B T)
\frac{d}{d\mu}\ln G(\mu),
\label{mott}
\end{equation}
where $G (\mu)\approx (e^2/h){\cal T}_{LR}(\mu)$ 
is the electric conductance at electrochemical potential $\mu$.
From this equation we can see that the thermopower vanishes at $T=0$.
We remark that people sometimes use the term ``Mott's formula''  
to refer to Eq.~(\ref{Eq:S-as-ratio-of-integrals}) as well as Eq.~(\ref{mott}),
probably because both formulas appear in Ref.~\cite{mott1}.

Both the Wiedemann-Franz law and the Mott's formula are typically violated at 
higher orders in the Sommerfeld expansion than those considered above,
or when the transmission function is not smoothly varying in the width of order temperature around the electrochemical potential (in which case a Sommerfeld expansion is not possible).
Thus they are typically violated in any system with a large thermoelectric response, or 
large thermoelectric figure of merit, $ZT$.
In particular, the Wiedemann-Franz law has been shown to be violated in strongly interacting systems
\cite{kf96,lo02,Catelani-Aleiner05,dora06,Kubala-Konig-Pekola2008,grsr09,wbxmgh11,Schwiete-Finkelstein2016a,Schwiete-Finkelstein2016b} and in small systems where
transmission can show a significant energy dependence
\cite{fazio2001,stone,shakouri2007b,bss10,bjs10,bbb12,sanchezprb13}.

\subsection{The figure of merit, {\it ZT}, and how to maximize it}
\label{sec:energyfiltering}

Now we turn to large thermoelectric effects, and more particularly large figure of merit, $ZT$,
so we leave behind the Sommerfeld expansion, 
and return to the results in Eqs.~(\ref{Eq:Ls-two-term}-\ref{eq:landauer.ex}).
An interesting question is what is the transmission function
${\cal T}_{LR}(E)$ (or transport distribution function $\Sigma(E)$ in
the Boltzmann approach) that maximizes the
thermodynamic efficiency.
Here, we will reproduce Mahan and Sofo's \cite{mahansofo}
proof that a delta-shaped transmission function leads
to an infinite figure of merit ($ZT\to \infty$) and consequently to Carnot efficiency in the linear response regime.
Their proof was presented in the context of Boltzmann theory, but we will do it in the context of scattering theory.

The first important point is to note that the definition of $ZT$ in Eq.~(\ref{Eq:ZT-intro})
contains the total heat conductance in the denominator. This is typically the sum of the electronic heat conductance, which we will call $K$, and the heat conductance due to other mechanisms (usually phonons and photons), 
which we will call $K_{\rm ph}$.
Thus, in general  
\begin{equation}
ZT=\frac{GS^2}{K+K_{\rm ph}}\,T\, .
\label{Eq:ZT-with-phonons}
\end{equation}
Note that as the phonons and photons are uncharged, they do not contribute to the thermoelectric effects or the charge conductance in the numerator, but do contribute to the denominator.
This makes it clear that phonon/photon heat flow is always detrimental to thermoelectric efficiency, $ZT$.

Thus the first step of Mahan and Sofo's derivation of the maximal efficiency is to assume we could engineer
the system to suppress  phonon and photon heat currents and thus take $K_{\rm ph} \to 0$.
In reality, it is extremely difficult to control phonon and photon heat flows,
although even modest progress in this direction can make a huge difference to $ZT$, 
see section~\ref{Sect:intro-phonons}.
None the less, to find the fundamental upper bound on efficiency, 
it is natural to start by taking  $K_{\rm ph}=0$.
Then substituting Eqs.~(\ref{Eq:Ls-two-term}-\ref{eq:landauer.ex}) 
into Eq.~(\ref{Eq:ZT-with-phonons}), we get
\begin{equation}
ZT=\frac{GS^2}{K}\,T=\frac{I_1^2}{I_0I_2-I_1^2}
\end{equation}
It is revealing to write this in terms of the
averages of the energy (measured from the electrochemical potential)
of the electrons that are transmitted through the scatterer \cite{mahansofo}.
Then we see that
\begin{equation}
ZT= \frac{ \big\langle (E-\mu) \big\rangle^2 }{  \big\langle (E-\mu)^2 \big\rangle- \big\langle (E-\mu) \big\rangle^2}\ \ ,
\label{Eq:ZT-simple-interpretation}
\end{equation}
where we recall that the average over the transmitted electrons is defined in Eq.~(\ref{Eq:defining-average-over-E}).
Eq.~(\ref{Eq:ZT-simple-interpretation}) is crucial, 
because it makes it easy to have a simple physical picture of the value of $ZT$.
Scattering theory (or Boltzmann theory \cite{mahansofo}) tells us that
it is simply the square of the average energy carried by transmitting electrons 
divided by the variance of their energy. 

Thus it is clear to see that one will have $ZT \to \infty$ if one makes the variance of the energies of transmitted electrons vanish.  This is the case if all the transmitted electrons have exactly the same energy $E=E_\star$.
Thus we require that the transmission function ${\cal T}_{LR}(E)$ is shaped like a $\delta$-function, being 
only non-zero in a tiny window around energy $E_\star$, see Fig.~\ref{Fig:slices}b. 
For such an energy filter, that only lets through electrons in the energy window $E_\star$ to $E_\star+\delta E$ 
with $\delta E \to 0$, one has
\begin{equation}
I_n \to \frac{\delta E}{h} \  (E_\star-\mu)^n \ {\cal T}_{LR}(E_\star) \ \left(-f'(E_\star)\right),
\label{eq:In-deltafunct}
\end{equation}
which means $I_n \to(E_\star-\mu)^n I_0$, for all $n$.
One can easily find all Onsager coefficients, thermopower, etc., by substituting this into
Eqs.~(\ref{Eq:Ls-two-term}-\ref{eq:landauer.ex}). 
Intriguingly, the only constraint on the value of $E_\star$
is that the numerator of Eq.~(\ref{Eq:ZT-simple-interpretation}) must not vanish, i.e.~one can take any value of $E_\star \neq \mu$.
Section~\ref{Sect:scatter-nonlin-Carnot} will show that this energy filtering mechanism allows us to achieve the Carnot
efficiency also beyond linear response \cite{linke2002,linke2005}.  However, there we cannot take any $E_\star$, instead in the nonlinear regime one only achieves Carnot efficiency if $E_\star=E^\rightleftharpoons$,
which is related to the bias and temperatures of the two reservoirs by Eq.~(\ref{Eq:eps-reversible}).

It is worth noting that when the variance of the transmitted energies is zero, the ratio
of thermal to electrical conductance vanishes, because of  Eq.~(\ref{Eq:K/G}).  Thus the above mentioned
system with $ZT \to \infty$ violates the Wiedemann-Franz law in the most extreme way, by having
$K\big/G \to 0$.

For experimental evidence of energy filtering, see
\cite{shakouri2006}, where barriers in a
superlattice were used to limit the transport to those electrons
with sufficiently high energy. As a result, a dramatic increase of
the Seebeck coefficient was shown together with a relatively modest
decrease of the electrical conductivity.
Sharp electronic resonances can be found also in molecules
weakly coupled to electrodes and for this reasons molecular
junctions might be efficient for thermoelectric conversion
\cite{majumdar2007,pauly2008,baranger2009,lambert2009,lejinse2010,bss10,dubi,reddy2012,venkataraman2013}.

The above result shows that $ZT\to\infty$ requires that one takes $\delta E \to 0$,
however the current generated is proportional to $\delta E$,
since Eq.~(\ref{eq:In-deltafunct}) scales like $\delta E$.  
Thus for vanishing $\delta E$ one gets 
very high $ZT$ but a vanishing power output.  
High but finite values of $ZT$ can still be achieved if, rather
than delta-shaped transmission function, one considers 
sharply rising \cite{shakouri2004,humphrey2005}
or boxcar-function-shaped \cite{muttalib2013,nikolic,whitney-prl2014,whitney2015} transmission functions.
Such transmission functions allow much greater power outputs than the delta-like function
considered above.

It is crucial to note that the above argument only gives $ZT \to \infty$ because we have assumed 
the phonon contribution to thermal conductivity is vanishingly small, $K_{\rm ph}\to 0$. 
If this conductivity $K_{\rm ph}$ is finite, then everything changes.
In this case, taking $\delta E \to 0$  means $ZT \to 0$, since the denominator of $ZT$ in 
Eq.~(\ref{Eq:ZT-with-phonons}) contains the $K_{\rm ph}$-term, which is independent of $\delta E$, while the numerator goes like $\delta E^2$.
Thus it is crucial to remember that a narrow transmission function is only desirable if one has managed to completely eliminate the phonon and photon heat conductances.

\subsection{Maximizing {\it ZT} for strong phonon heat transport}
\label{Sect:max-ZT-phonons}

The opposite limit to that discussed in the previous section is when phonon heat conductivity, $K_{\rm ph}$, dominates over electron heat conductivity in $ZT$ given by Eq.~(\ref{Eq:ZT-with-phonons}).
Then
one has 
\begin{eqnarray}
ZT \ \simeq\ \frac{G T S^2}{K_{\rm ph}} \ =\  \frac{1}{K_{\rm ph}T}\ I_0 \ \langle E -\mu  \rangle^2   
\label{Eq:def-ZT-phonons}
\end{eqnarray}
where we recall that \green{$ \langle E-\mu \rangle  =I_1/I_0$.}

%%%%%%%%%%%%%%%%%%%%%%%%%%%%%%%%
\begin{figure}
\begin{center}
\centerline{\includegraphics[width=0.75\textwidth]{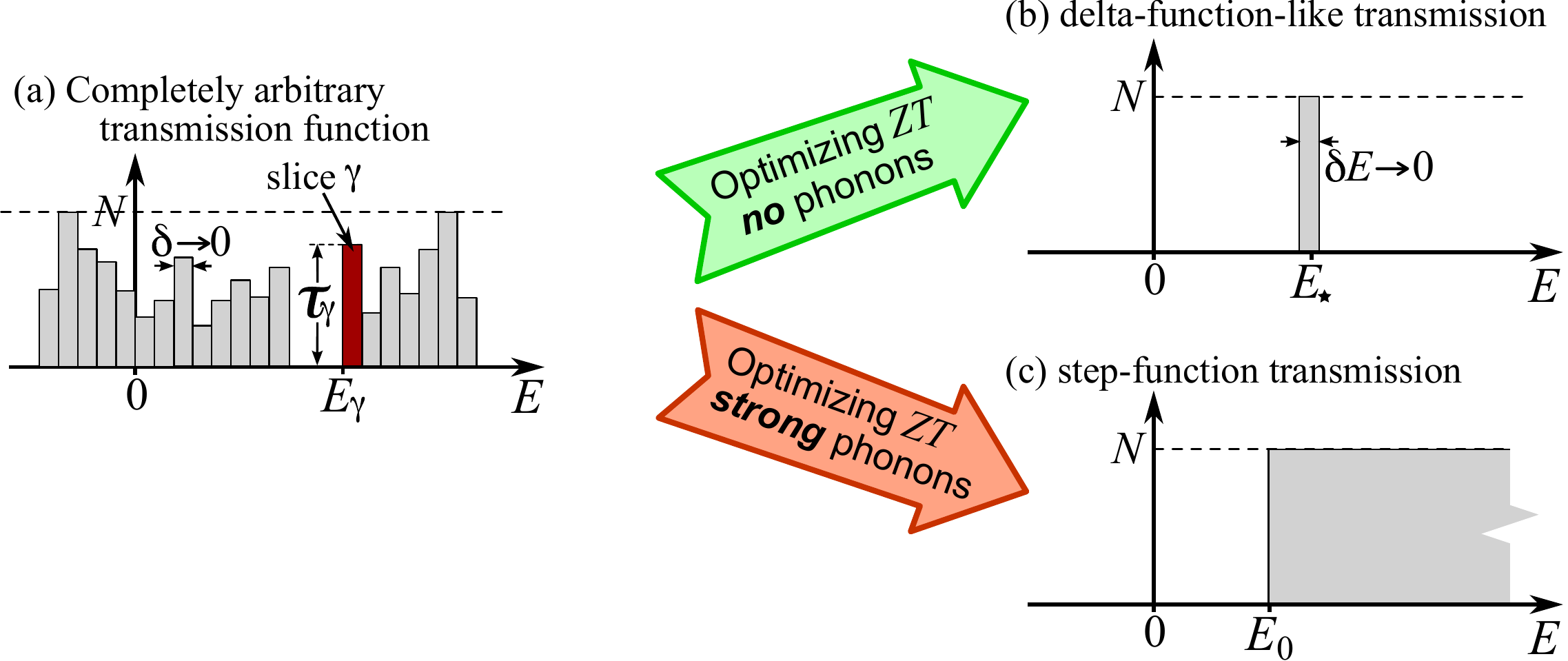}}
\caption{
If one can have any transmission function as in (a), and can change it as desired to maximize $ZT$,
the result is as shown in (b) and (c).  In the absence of phonons carrying heat in parallel with the electrons, shown in (b),
the optimal transmission is the delta-function-like transmission shown in (b) and discussed in 
section~\ref{sec:energyfiltering}.
This gives $ZT \to\infty$, which corresponds to Carnot efficiency. However, if there is {\it any} phonon heat flow in parallel with that of the electrons, such a delta-function-like transmission will give $ZT =0$.
In the limit of very strong phonon heat flow (so the phonons carry more heat than the electrons can), 
the optimal transmission is a step-function (theta-function) shown in (c) and discussed in  
section~\ref{Sect:max-ZT-phonons} 
The optimal transmission for intermediate phonon heat flows is
a boxcar function (band-pass filter)  as discussed in Ref.~\cite{whitney2015}.
 } 
\label{Fig:slices}
\end{center}
\end{figure}
%%%%%%%%%%%%%%%%%%%%%%%%%%%%%%%%%

Our objective here is to find the $E$ dependence of ${\cal T}_{LR}(E)$ which maximizes Eq.~(\ref{Eq:def-ZT-phonons}).   For simplicity in what follows we can measure all energies from the electrochemical potential, 
which is equivalent to saying $\mu=0$.
We consider the case where transmission is dominated by positive energies, so $\langle E\rangle >0$.
If one wishes to consider the case where the transmission is dominated by negative energies 
(which will have the same $ZT$
but equal and opposite $S$), one takes $E \to -E$ is everything that follows.
Continuing with the case dominated by positive energies,
the first thing we note is that transmission at high energies increases both $I_0$ and $\langle E\rangle$, 
so transmission at high energies clearly enhances $ZT$.  Transmission at low energies is more problematic;
allowing electron flow at low energies increase $I_0$, but it reduces $\langle E\rangle$, so its effect on $ZT$ 
is unclear.  To proceed, we follow a similar procedure as for that in Refs.~\cite{whitney-prl2014,whitney2015},
but in this case the algebra is much simpler.
Thus, we start by considering the transmission function as an infinite set of slices each of width $\delta\to 0$, 
as in Fig.~\ref{Fig:slices}a, where we define $\tau_\gamma$ as the transmission of slice $\gamma$, which sits at energy $E_\gamma$. 
Some basic algebra gives 
\begin{eqnarray}
\frac{\rmd I_n}{\rmd \tau_\gamma} = \frac{\delta}{h}\  E_\gamma^n \ \big(-f'(E_\gamma)\big)
\end{eqnarray}
so the rate at which $ZT$ changes with a small increase in $\tau_\gamma$ is 
\begin{eqnarray}
\frac{\rmd (ZT)}{\rmd \tau_\gamma} 
\ =\ \frac{1}{K_{\rm ph}T} \left( \frac{2I_1}{I_0}  
\frac{\rmd I_1}{\rmd \tau_\gamma} - \frac{I_1^2}{I_0^2}  \frac{\rmd I_0}{\rmd \tau_\gamma}  \right)
\ =\  \frac{1}{K_{\rm ph}T}\ \frac{\delta}{h}\ \langle E\rangle \  \big(-f'(E_\gamma)\big)  \ \times \ \big(2E_\gamma -\langle E\rangle \big) \ . 
\end{eqnarray}
This means that increasing $\tau_\gamma$ increases $ZT$ if $E_\gamma > \half \langle E \rangle$, 
but it decreases $ZT$ if $E_\gamma < \half \langle E \rangle$.
As a result, if the scatterer has $N$ transverse modes, 
one can expect that the transmission which maximizes Eq.~(\ref{Eq:def-ZT-phonons})
is the Heaviside theta function shown in Fig.~\ref{Fig:slices}c, 
\begin{eqnarray}
{\cal T}_{LR}(E)=N \  \theta\left(E-E_0\right),
\label{Eq:theta-for-ZT-phonons}
\end{eqnarray} 
with $E_0$ determined by the transcendental equation
$2E_0 = \big\langle E  \big\rangle$.
Here $\big\langle E  \big\rangle$
 is the average energy for the 
transmission function in Eq.~(\ref{Eq:theta-for-ZT-phonons}),
and therefore $\big\langle E  \big\rangle$ depends on $E_0$:
\begin{eqnarray}
\big\langle E  \big\rangle 
\ =\  
\frac{ \int_{E_0}^\infty \rmd E  \ E \ \big(\!-f'(E)\big) }{  \int_{E_0}^\infty \rmd E  \  \big(\!-f'(E)\big) }
\ =\ E_0 \ +\ 
\frac{ \kB T \ln \left[1+\exp[-E_0/(\kB T)] \right]  }{ f(E_0) } \ ,
\end{eqnarray}
where the integrals were evaluated using standard methods, giving the right hand result. 
Hence, the transcendental equation for $E_0$ is
\begin{eqnarray}
\frac{E_0}{\kB T} f(E_0) \ =\  \ln \left[1+\exp[-E_0/(\kB T)] \right] \ .
\label{Eq:transcendental}
\end{eqnarray}
If we define $B_0 = \exp[-E_0/(\kB T)]$, this transcendental equation simplifies to\footnote{\green{
We note that Eqs.~(\ref{Eq:theta-for-ZT-phonons},\ref{Eq:transcendental})
for maximizing $ZT$ in the linear response regime coincide 
with the results for maximizing efficiency in the nonlinear regime
when phonon effects are very strong (so maximizing efficiency requires maximizing the power output),
see section~XIV of Ref.~\cite{whitney2015}.
The transcendental equation given here coincides with that in Eq.~(42) of Ref.~\cite{whitney2015}.}}
 $(1+B_0)\ln[1+B_0] + B_0 \ln[B_0] =0$.
The solution is $B_0= 0.318...$, which means $E_0 = 1.146... \times \kB T$.
Noting that in this case $I_0=(N/h) f(E_0)$ and $\langle E  \rangle=2E_0$, 
one finds that Eq.~(\ref{Eq:def-ZT-phonons}) becomes
\begin{eqnarray}
ZT \ =\  \frac{1}{K_{\rm ph}T} \frac{N E_0^2 f(E_0)}{h}
\ =\  \frac{\kB^2 T}{h K_{\rm ph}}  \ N \ \times \ 0.317...\ .
\label{Eq:def-ZT-phonons-max}
\end{eqnarray}
This is the maximum possible $ZT$ in the case where heat currents are dominated by phonons,
which is the case when $K_{\rm ph} \gg K_{\rm el}$ where the heat conductance due to electrons $K_{\rm el} \sim \kB^2 T N/h$
(this is only an order of magnitude estimate of $K_{\rm el}$).   Thus, the maximum possible $ZT$ is 
of order $K_{\rm el}/K_{\rm ph}$ which is definitely much less than one.
This is expected of course, because phonons carry heat without generating any electrical power.
However, by having a transmission function in the form of a $\theta$-function,  Eq.~(\ref{Eq:theta-for-ZT-phonons}), one gets a finite $ZT$, when the $\delta$-like function in section~\ref{sec:energyfiltering} 
would give strictly zero $ZT$ for any finite $K_{\rm ph}$.

By combining the results of this section with that of the previous one, we conclude that the form of the transmission function  
which optimizes $ZT$ depends on the heat current carried between hot and cold by phonons.  If there are no phonons or they transport  no heat, then the optimal transmission function is very narrow, and gives Carnot efficiency ($ZT \to \infty$).
If phonons transport a lot of heat, a wide transmission function maximizes $ZT$, but this maximum value 
will be rather modest (\green{significantly below} Carnot efficiency).  
In what was presented above, we only considered the two limits (no phonon heat flow and strong phonon heat flow), however section~XIV of  Ref.~\cite{whitney2015} treated the intermediate cases for a nonlinear scattering theory
(where $ZT$ is not meaningful, but efficiency is).  It showed that the maximum efficiency for intermediate phonon heat flows is achieved with a boxcar function (a band pass filter) which transmits all electrons with energies between two energies $E_0$ and $E_1$, and block all electrons outside this energy window, 
see section~\ref{Sect:whitney2015}.

\subsection{Inelastic scattering and probe reservoirs}
\label{sec:probes}

\begin{figure}
\begin{center}
\centerline{\includegraphics[width=0.9\textwidth]{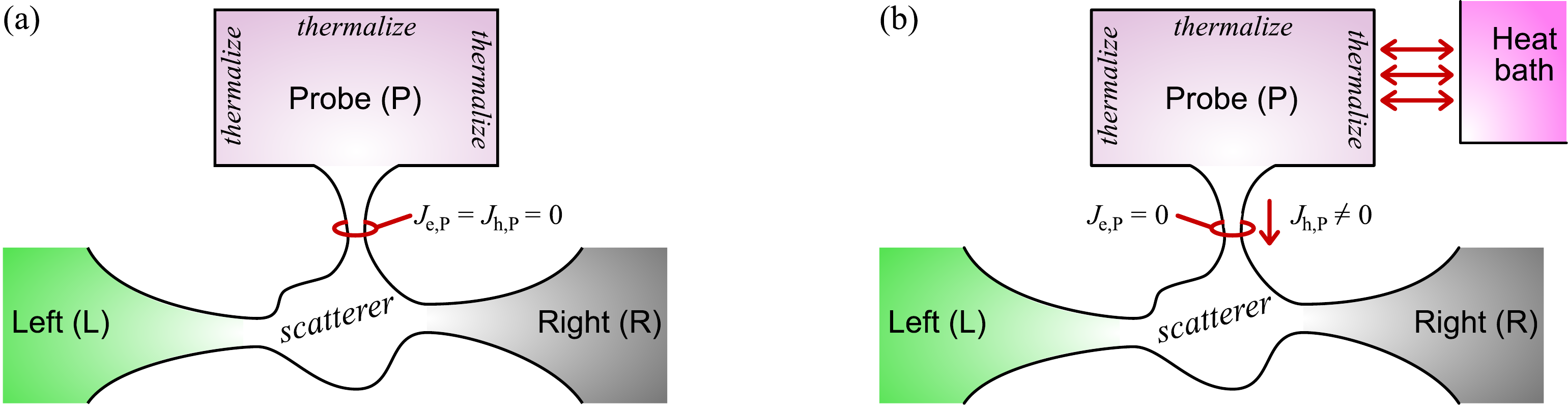}}
\caption{
In (a) we show a sketch of partially-coherent thermoelectric 
transport, with the third terminal acting as a probe reservoir
mimicking inelastic electron-electron scattering. 
Electrons entering the probe reservoir, P, from the scatterer get thermalized at the temperature $T_{P}$
and electrochemical potential $\mu_{P}$, before re-emerging into the scatterer.
The temperature $T_{P}$ and the electrochemical
potential $\mu_{P}$ of the probe reservoir are thus such that the net 
average electric and heat currents into this reservoir vanishes,
$J_{e,P}=J_{h,P}=0$. This setup can be generalized to any number
of probe reservoirs, $k= P1,P2,\cdots$, by setting 
$J_{e,k}=J_{h,k}=0$ for all probes.
In (b) we show a sketch in which the probe reservoir is mimicking inelastic {\it electron-lattice} scattering 
(i.e.\ electron-phonon scattering).  In this case, as the lattice temperature at that point in the nanostructure
will be determined by phonon dynamics, it will not usually be the same as the temperature of the electrons
which scatter from it.  Thus there will be a flow of heat  between the electrons and the lattice, $J_{h,P}\neq 0$, but obviously no flow of charge, $J_{e,P} = 0$. } 
\label{fig:3ter}
\end{center}
\end{figure}

The idea of a probe reservoir was first introduced \cite{Enquist-Anderson1981,Buttiker1986} 
as a simple model of a device for measuring the voltage at a given point in a nano-structure, this is typically known as a {\it voltage probe}.
In the same spirit a number of authors have considered a {\it temperature probe} 
\cite{jacquet2009,Meair2014,Stafford2014,Shastry-Stafford2015,Shastry-Stafforf2016} 
intended to model a device that measures the temperature at a given  point in a nano-structure.
An ideal such probe is a large but finite sized 
reservoir coupled to the 
system, as sketched in Fig.~\ref{fig:3ter}a. \green{The probe reservoir is assumed to be large enough that electrons entering it thermalize before escaping from it. However, it is assumed to be small enough that it will achieve a steady state with respect to the scatterer on an experimentally accessible timescale.
As this probe reservoir is only in contact with the scatterer, its  
temperature and electrochemical potential will build up to their steady-state values, those  
for which the net fluxes of particles and heat into the probe reservoir are zero on average.  We assume the
probe reservoir has achieved this steady state in all the analysis that follows.}
One can then read off the electrochemical potential and temperature of this finite but macroscopic probe reservoir using standard techniques.
It is worth mentioning that it is relatively easy to isolate reservoir electrically,
however it is hard to isolate it thermally from its environment.
Thus one can imagine that many probes will look more like that in Fig.~\ref{fig:3ter}b,
they charge up to a bias 
that ensures the electrical current into it is zero, however they exchange heat with their environment
(marked as ``heat bath'' in  Fig.~\ref{fig:3ter}b) so that the heat current into them is not zero in the steady state.
This is typically fine if one is interested in making a voltage probe \cite{Enquist-Anderson1981,Buttiker1986}
for a system with weak thermoelectric effects. However, a probe of the type in Fig.~\ref{fig:3ter}b obviously changes the heat flow in the system
(as it will typically absorb or emit heat),
which makes it a poor temperature probe. It also makes it a poor voltage probe in a system which exhibits 
a strong thermoelectric responses (because the heat it injects or absorbs will change the electrical currents in the scatterer)
\cite{Meair2014,Stafford2014}.  

%%%%%%%%%%%%%%%%%%%%%%%%%%%%%%%%
%%%%%%%%%%%%%%%%%%%%%%%%%%%%%%%%%

Another reason to consider such probes is that they 
elegantly simulate inelastic scattering in a phenomenological manner \cite{buttiker1988}.
The Landauer scattering approach suffers from the fact it only describes coherent quantum transport,
when real systems often only exhibit partially coherent transport, because there is \emph{inelastic
scattering} due to the interactions of the electrons with phonons,
photons, and other electrons.
One can add probe reservoirs to the model to mimic 
such inelastic scattering.
The advantage of such an approach lies in its simplicity and independence
from microscopic details of inelastic processes.
The probe reservoir is one 
whose parameters (temperature and electrochemical potential)
are chosen self-consistently so that there is no net \emph{average} flux of
particles or heat between this reservoir and the system
(see Fig.~\ref{fig:3ter}).
One can think of the probe as mimicking a small region in the scatterer in which particles relax
to thermal equilibrium.  The only problem with this model is that such equilibration happens completely
(for particles that enter this region) or not at all (for particles that do not enter this region).
This is  rather different from what we expect in more realistic models of electron-electron scattering,
in which we would expect that if a significant number of electrons undergo no inelastic scattering before escaping the
scatterer, then there would also be a significant number which escape after just a single inelastic scattering.
We expect a single such scattering to exchanges energy between the electrons, but not to take them to a perfect thermal distribution (it usually takes many inelastic scatterings for
the electrons arrive at thermal equilibrium).   This model misses this ``partial thermalization'' of electrons,
despite this, it is believed to capture \green{much} of the physics of inelastic scattering in a simple manner. 

As a result, probe reservoirs have been widely used in the literature
and proved to be useful in unveiling nontrivial aspects of
phase-breaking
processes \cite{datta}, heat transport and rectification
\cite{visscher,lebowitz,dhar2007,dhar,lebowitz2009,pereira,segal05,segal,saito06}.
The role of inelastic processes induced by such probes upon thermoelectric responses
was considered in 
\cite{jacquet2009,SBCP2011,sanchez2011,BS2013,BSS2013,segal2013,BS2015,Yamamoto2016}.
Many other works have considered the third probe as a reservoir which supplies heat to the system,
these works will be discussed elsewhere in this review \green{(see sections~\ref{sec:genericmultiterminal}, \ref{Sect:3-term-sys1} and \ref{Sect:3-term-sys2}).}

To model thermalization due to inelastic electron-electron interactions \cite{SBCP2011,sanchez2011}, 
we consider a system with  $K_{P}$ probe reservoirs, and $K_{C}$ current carrying reservoirs,  so Fig.~\ref{fig:3ter}a is an example with $K_{P}=1$ and $K_{C}=2$.
Taking Eq.~(\ref{Eq:Onsager-matrix}), we take reservoirs 1 to $K_{C}$ to be the current carrying reservoirs,
while reservoirs  $K_{C}+1$ to $K_{C}+K_{P}$ are the probe reservoirs.
Then, we can write  Eq.~(\ref{Eq:Onsager-matrix}) as 
\begin{equation}
\left(\begin{array}{c} 
{\bf J}_{C} \\
0 
\end{array} \right)
\ \ =\ \
\left(\begin{array}{cc} 
{\bm L}_{CC} & {\bm L}_{CP}  \\
{\bm L}_{PC} & {\bm L}_{PP}  
\end{array} \right)
\ \left(\begin{array}{c} 
\boldsymbol{{\cal F}}_{{C}} \\
\boldsymbol{{\cal F}}_{{P}} 
\end{array} \right)\ ,
\label{Eq:Onsager-matrix-probes}
\end{equation}
where we label the probe reservoirs as $P$ and the remaining current-carrying reservoirs as $C$.
Thus $\boldsymbol{{\cal F}}_{{C}}$ is a vector of the thermodynamic forces 
(bias and temperature as defined above Eq.~(\ref{Eq:Onsager-matrix})) 
 on the current-carrying reservoirs $2, \cdots, K_{C}$, remember that we already eliminated reservoir 1 in Eq.~(\ref{Eq:Onsager-matrix}).
Similarly $\boldsymbol{{\cal F}}_{{P}}$ is a vector of the forces (bias and temperature) on the probe reservoirs $K_{C}+1, \cdots, K_{C}+K_{P}$.
Then, ${\bf J}_{C}$ is the vector of currents (electrical and heat) in these reservoirs, while there is no current
into the scatterer from the probe reservoirs.
Thus ${\bm L_{PP}}$ is the matrix of the elements of the Onsager matrix ${\bm L}$ which couple probe reservoir forces to probe reservoir currents, so ${\bm L_{ PP}}$ is a $2K_{P}\times 2K_{P}$ matrix.
Similarly,  ${\bm L_{CC}}$ is the matrix of the elements of the Onsager matrix ${\bm L}$ which couple forces on current-carrying reservoir forces to currents on those reservoirs, so ${\bm L_{CC}}$ is a $(2K_{C}-2)\times (2K_{C}-2)$ matrix.
This means ${\bm L_{CP}}$ is a $(2K_{C}-2) \times 2K_{P}$ matrix of the 
coupling between forces on the probe reservoirs and the currents in the charge-carrying reservoirs.

To be concrete for the example in Fig.~\ref{fig:3ter}a, where we identify reservoir 1 with reservoir R, reservoir 2 with reservoir L, and reservoir 3 with the probe reservoir, we have
\begin{eqnarray}
\boldsymbol{{\cal F}}_{{C}} = \left(\begin{array}{c} 
\mathcal{F}_{{e,L}} \\
\mathcal{F}_{{h,L}}
\end{array} \right), 
\qquad
\boldsymbol{{\cal F}}_{{P}} = \left(\begin{array}{c} 
\mathcal{F}_{{e,P}} \\
\mathcal{F}_{{h,P}}
\end{array} \right),
\qquad
{\bf J}_{{C}} = \left(\begin{array}{c} 
\Jelectric{{L}} \\
\Jheat{{L}}
\end{array} \right),
\qquad
\end{eqnarray}
with
\begin{eqnarray}
{\bm L}_{CC} = \left(\begin{array}{cc} 
L_{{ ee,LL}} & L_{{ eh,LL}}\\
L_{{ he,LL}} & L_{{ hh,LL}}
\end{array} \right), 
\;
{\bm L}_{ PP} = \left(\begin{array}{cc} 
L_{{ ee,PP}} & L_{{ eh,PP}}\\
L_{{ he,PP}} & L_{{ hh,PP}}
\end{array} \right), 
\;
{\bm L}_{ CP} = \left(\begin{array}{cc} 
L_{{ ee,LP}} & L_{{ eh,LP}}\\
L_{{ he,LP}} & L_{{ hh,LP}}
\end{array} \right), 
\;
{\bm L}_{ PC} = \left(\begin{array}{cc} 
L_{{ ee,PL}} & L_{{ eh,PL}}\\
L_{{ he,PL}} & L_{{ hh,PL}}
\end{array} \right). 
\end{eqnarray}

Now returning to the general case, we can solve \cite{SBCP2011} 
the second line of Eq.~(\ref{Eq:Onsager-matrix-probes}),
to find that $\boldsymbol{{\cal F}}_{{ P}} = -{\bm L}_{ PP}^{-1}   {\bm L}_{ PC}\boldsymbol{{\cal F}}_{{ C}}$.
Substituting this into the first line of Eq.~(\ref{Eq:Onsager-matrix-probes}), allows us to 
eliminate the probes from the problem, and retrieve a relation for the currents in terms of the forces on the 
current-carrying reservoirs alone.
This gives us ${\bf J}_{{ C}} \ =\ {\bm L}' \ \boldsymbol{{\cal F}}_{{ C}}$, with
\begin{eqnarray}
% {\bf J}_{{ C}} \ =\ {\bm L}' \ \boldsymbol{{\cal F}}_{{ C}} \, ,
%\qquad \  \hbox { with } \ \ 
{\bm L}' \ =\ {\bm L}_{ CC} \ -\   {\bm L}_{ CP} \,{\bm L}_{ PP}^{-1}  \, {\bm L}_{ PC} \ .
\label{Eq:Lprimed}
\end{eqnarray}
This is the transport relation in the presence of probe reservoirs which mimic inelastic scattering within the scatterer.
So the presence of inelastic effects leads one  to replace the ${\bm L}$ matrix for the current-carrying reservoirs (referred to here as ${\bm L}_{ CC}$) by ${\bm L}'$.
The first term in ${\bm L}'$ gives the elastic component of the scattering, so \green{it} is proportional to the probability that electrons traverse the scatterer without any inelastic scattering. The second term gives the inelastic part
and is proportional to the probability that electrons undergo inelastic scattering during the time they traverse the scatterer.

We can show that the inelastic effect do not affect the Onsager reciprocal relation, by noting that
the Onsager relation in Eq.~(\ref{Eq:Onsager-multiterm}) means that  
${\bm L}_{ CC}({\bm B})={\bm L}_{ CC}^{\rm T}(-{\bm B})$,
 ${\bm L}_{ PP}({\bm B})={\bm L}_{ PP}^{\rm T}(-{\bm B})$, and 
 ${\bm L}_{PC}({\bm B})={\bm L}_{ CP}^{\rm T}(-{\bm B})$, where T indicate the matrix transpose.
 Thus it is easy to see that ${\bm L}' ({\bm B}) = [{\bm L}' (-{\bm B})]^{\rm T}$, which is the same as saying
 \begin{eqnarray}
L_{\mu\nu,ij}'({\bm B})=L_{\nu\mu;ji}'(-{\bm B}),
\end{eqnarray}
in the presence or absence of inelastic effects.

In contrast, now that we have inelastic effects, we find that there is no reason for ${\bm L}'$ to obey the relation in 
Eq.~(\ref{Eq:Butcher}).   Even though Eq.~(\ref{Eq:Butcher}) means that each matrix on the right hand side of 
Eq.~(\ref{Eq:Lprimed}) is equal to itself under a transpose within the two-by-two block for each reservoir, this does not mean that the second term in Eq.~(\ref{Eq:Lprimed}) is equal to itself under a transpose within the two-by-two block for each reservoir.  As a result, one can expect that most systems with inelastic scattering will have 
\begin{eqnarray}
L_{\mu\nu,ij}'({\bm B}) \ \neq \ L_{\nu\mu;ij}'({\bm B}),
\end{eqnarray}
because of that inelastic scattering.
It is interesting to note that $L_{\mu\nu,ij}'({\bm B}) - L_{\nu\mu;ij}'({\bm B})$ is proportional to
the probability electrons undergo inelastic scattering as they traverse the scatterer, 
\green{although} the constant of proportionality is likely to be highly system dependent.

Recasting the above general results in terms of the system in Fig.~\ref{fig:3ter}a with two current-carry reservoirs and a single probe reservoir, we get 
\begin{equation}
  \left( \begin{array}{c}
  J_{e,L} \\
  J_{h,L}
\end{array}
\right) 
\ =\   {\bm L}'  \ \left( \begin{array}{cc}
              \mathcal{F}_{e,L} \\
             \mathcal{F}_{h,L}
            \end{array} \right)
\ \ \equiv\  \ \left( \begin{array}{cc}
              L'_{ee} &  L'_{eh} \\
              L'_{he} & L'_{hh}
            \end{array} \right) \,\left( \begin{array}{cc}
              \mathcal{F}_{e,L} \\
             \mathcal{F}_{h,L}
            \end{array} \right),
\label{eq:Lred}
\end{equation}
where  $J_{e,L} = - J_{e,R}$ and  $J_{h,R} = - J_{h,L}$.
The above arguments for the general case mean that the matrix elements $L'_{\mu\nu}$ obey  the Onsager reciprocal relation,
$L'_{\mu\nu}({\bm B})=L'_{\nu\mu}(-{\bm B})$, but do not obey the relation $L'_{\mu\nu}({\bm B})=L'_{\nu\mu}({\bm B})$.
Exactly the same is true if we consider an arbitrary number of probe reservoirs.
As a result, any two-terminal system with inelastic scattering will obey the two-terminal Onsager reciprocal relations
$G({\bm B})=G(-{\bm B})$, $K({\bm B})=K(-{\bm B})$, and $\Pi({\bm B})=TS(-{\bm B})$.
However, the inelastic scattering means that we should not expect either $S({\bm B})$ or $\Pi({\bm B})$ to be even functions of ${\bm B}$.

The thermodynamic efficiencies for this case can
be computed by means of the standard two-terminal formulas
(\ref{eq:ZTx}) and (\ref{etawmax}), with the factors of ${\bm L}$ replaced by the above factors of ${\bm L}'$.
Arbitrarily large values of the asymmetry 
parameter $x=S({\bm B})/S(-{\bm B})=L'_{eh}/L'_{he}$
were obtained in \cite{SBCP2011,sanchez2011} by means
of a three-dot Aharonov-Bohm interferometer model.
The asymmetry was found also for 
chaotic cavities,
ballistic microjunctions \cite{sanchez2011}, and
random Hamiltonians drawn from the Gaussian unitary
ensemble \cite{vinitha2013}.
In \cite{sanchez2011} it was shown that the asymmetry is
a higher-order effect in the Sommerfeld expansion and therefore 
disappears in the low temperature limit.
The asymmetry was demonstrated also in the framework
of classical physics, for a three-terminal deterministic
railway switch transport model \cite{horvat2012}.
In such model, only the values
zero and one are allowed for the transmission functions
${\cal T}_{ji}(E)$, i.e., ${\cal T}_{ji}(E)=1$
if particles injected from terminal $i$ with energy $E$ 
go to terminal $j$ and ${\cal T}_{ji}(E)=0$ is such
particles go to a terminal other than $j$. The transmissions
${\cal T}_{ji}(E)$ are piecewise constant
in the intervals $[E_k,E_{k+1}]$, $(k=1,2,\cdots)$, with switching
${\cal T}_{ji}=1\to 0$ or vice-versa possible at the  
threshold energies $E_k$, with the constraints (\ref{Eq:constraints-T_ij}) 
always fulfilled. 

In all the above instances, no systems were found which had both
large values of the asymmetry parameter, Eq.~(\ref{def:x}), and high thermoelectric efficiency.  
Such a failure was explained by \cite{BSS2013} and is generic
for non-interacting three-terminal systems. In the case where ${\bm B}\ne 0$, current conservation (which is
mathematically expressed by unitarity of the scattering matrix
${\cal S}$) imposes bounds on the Onsager matrix stronger than 
those derived from positivity of entropy production. \green{It takes the form}
\begin{equation}
L_{ee}L_{hh}
-\frac{1}{4}\,(L_{eh}+L_{he})^2 \ge 
\frac{3}{4}\,(L_{eh}-L_{he})^2.
\label{eq:3terbound}
\end{equation}
This only reduces to the third inequality of 
Eq.~(\ref{dots}) in the time-symmetric case 
$L_{eh}=L_{he}$, while it is in general a stronger inequality, 
since the right-hand side of Eq.~(\ref{eq:3terbound}) 
is strictly positive when $L_{eh}\ne L_{he}$. 
As a consequence, Carnot efficiency can be achieved 
in the three-terminal setup only in the time-symmetric case
${\bm B}=0$. On the other hand, the Curzon-Ahlborn linear response
bound $\eta_{CA}=\eta_C/2$ for the efficiency at maximum power 
can be overcome for moderate asymmetries, $1<x<2$, with a
maximum of $4\eta_C/7$ at $x=4/3$. The bounds obtained by 
\cite{BSS2013} are in practice saturated in a 
quantum transmission model reminiscent of the 
railway switch model \cite{vinitha2013}.
Multi-terminal cases with more than three terminals were also discussed for
noninteracting electronic transport \cite{BS2013}.
By increasing the number $K_P$ of probe terminals, 
the constraint from current conservation on the maximum efficiency
and the efficiency at maximum power becomes weaker than that
imposed by Eq.~(\ref{eq:3terbound}). However, the bounds 
Eqs.~(\ref{eq:boundetapmax},\ref{eq:boundetamax})
from the second law of
thermodynamics are saturated only in the limit $K_P\to\infty$. 
Moreover, numerical evidence suggests that the power
vanishes when the maximum efficiency is approached \cite{BS2015}.
It is an interesting open question whether 
similar bounds on efficiency, tighter
than those imposed by the positivity of entropy production,
exist in more general transport models for interacting systems.

\subsection{Generic multi-terminal setups}
\label{sec:genericmultiterminal}

There is increasing interest in systems with more than just two reservoirs carrying currents.
For instance, one may have a third reservoir which is a
phonon heat bath connected to the electrons in the nanostructure.
This can be used as a model of electron thermalization within the nanostructure
due to inelastic electron-lattice (electron-phonon) interactions, as in Fig.~\ref{fig:3ter}b.
Note that in typical nanostructures, phonons in the scatterer are rather strongly coupled with the bath of phonons in the reservoirs, substrate, etc.  Thus the bath of phonons can absorb or supply heat to the electrons,
which mean that $\Jheat{{ ph}} \neq 0$ while  $\Jelectric{{ ph}} = 0$.  

One can imagine using this to supply heat to the nanostructure through the probe reservoir,
by ensuring it is coupled to a reservoir of phonons (or photons) which is hotter than the other reservoirs.
It has been shown that such setups can be favorable for thermoelectric
energy conversion \cite{imry2012}. 
There are many proposals for such devices, for some of them broken time-reversal symmetry 
(via an external magnetic field) is crucial to their operation; these include Aharonov-Bohm rings \cite{entin2012} and quantum Hall systems \cite{sanchez2015a,sanchez2015b,sanchez2016,whitney2016}.

The setup can also act as a 
refrigerator for the local phonon system (modelled here as the probe reservoir). 
The \emph{cooling by heating} phenomenon can also 
be interpreted in terms of a third, photonic terminal powering refrigeration: 
In the proposal by Pekola and Hekking \cite{pekola2007} 
(see also \cite{pekola2011,pekola2012,vandenbroeck2006})
the photons
emitted by a hot resistor can extract 
heat from a cold metal, providing the energy needed to 
electrons to tunnel to a superconductor (separated from the metal
by a thin insulating junction; no voltage is applied over the
junction). If the temperature of the resistor
is suitably set, only the high energy electrons are removed
from the metal, thus cooling it. 
%Such \emph{``Brownian refrigerator''}
%is still to be experimentally demonstrated.
Similar mechanisms have been discussed for cooling 
a metallic lead, connected to another, higher temperature
lead by means of two adjoining quantum dots \cite{vandenbroeck2012}
or for cooling an optomechanical system \cite{eisert2012}.
In both cases, refrigeration is powered by absorption of photons.
Many of these situations can be treated in terms of a multi-terminal scattering theory.
However, a number of them are more naturally treated in terms of the rate equation technique,
so they are discussed in chapter~\ref{Sect:master-examples}.

In a multi-terminal device
all terminals should be treated on equal footing, without
necessarily declaring some of them as probes.
For a linear response approach, the transport coefficients 
must be generalized \cite{Mazza2014}. A generalization of the 
thermopower to the multi-terminal scenario can be obtained by 
introducing the matrix of elements
\begin{equation}
 \label{eq:thermopower}
 S_{ij} = -\left(\frac{\Delta V_i}{\Delta T_j} \right)_{\mbox{\tiny{$\begin{array}{l}
 J_{e,k} = 0, \; \forall k, \\
  \Delta T_{k} = 0,\;  \forall k\neq j
  \end{array}$}} },
 \end{equation}
where $\Delta V_i\equiv \Delta \mu_i/e$ 
is the voltage developed between reservoir 
$i$ and (reference) reservoir $1$. 
In this definition we have imposed that the charge currents in all the leads are 
zero (the voltages are measured at open circuits) and that all but one 
temperature differences are zero
%(of course this  last condition is not required in a two-terminal model). 
While \emph{local} thermopowers correspond to $i=j$, 
\emph{nonlocal} thermopowers are obtained when $i\ne j$, i.e. a temperature 
difference between two reservoirs ($j$ and $1$) induces a voltage also 
between other reservoirs ($i$ and $1$ for $S_{ij}$) at the same 
temperature \footnote{It is worth observing  that 
Eq.~(\ref{eq:thermopower}) differs from other definitions proposed in the literature
For example in Ref.~\cite{Belzig2013} a generalization  of the two-terminal 
thermopower to  a three-terminal system, was proposed
by setting to zero one voltage instead of the corresponding particle current.
While operationally well defined, this choice does not allow one to easily recover 
the thermopower of the two-terminal case.}. 
Generalizations of the electrical and thermal conductances and of the 
Peltier coefficient to the multi-terminal case are provided by the following matrices:
\begin{equation}
 G_{ij} = \left(\frac{J_{e,i}}{\Delta V_j} \right)_{\mbox{\tiny{$\begin{array}{l}
 \Delta T_k = 0, \; \forall k, \\
  \Delta V_{k} = 0,\;  \forall k\neq j
  \end{array}$}} },
\quad
K_{ij} = \left(\frac{J_{h,i}}{\Delta T_j} \right)_{\mbox{\tiny{$\begin{array}{l}
 J_{e,k} = 0, \; \forall k, \\
  \Delta T_{k} = 0,\;  \forall k\neq j
  \end{array}$}} },
\quad
\Pi_{ij} = \left( \frac{J_{h,i}}{J_{e,j}} \right)_{\mbox{\tiny{$\begin{array}{l}
 \Delta T_k = 0, \; \forall k, \\
  \Delta V_{k} = 0,\;  \forall k\neq j
  \end{array}$}} }.
\end{equation}
The Peltier matrix is related to the the thermopower matrix (\ref{eq:thermopower})
through the Onsager reciprocal relations, implying 
$\Pi_{ij}({\bm B})=T S_{ji}(-{\bm B})$. 

The steady-state heat to work conversion efficiency for a multi-terminal 
system is defined as  the power $P$  generated by the machine
(which equals to the sum of all the heat currents {\it exchanged}
between the system and the reservoirs),
divided by the sum of the heat currents {\it absorbed} by the system, i.e. \cite{Mazza2014}
\begin{equation}
\label{effmulti}
\eta \ =\ \frac{P}{\sum_{k_+} J_{h,k}}\ =\   
\frac{\sum_{k=1}^n J_{h,k}}{\sum_{k_+} J_{h,k}}\ =\  
 \frac{ -\sum_{k=1}^n (\mu_k/e)J_{e,k}}{\sum_{k_+} J_{h,k}} \ =\  
\frac{-\sum_{k=2}^n \Delta V_k J_{e,k}}{\sum_{k_+} J_{h,k}},
\end{equation}
where the symbol $\sum_{k_+}$ in the denominator indicates that the sum  is restricted 
to positive heat currents only, and where to derive 
the last two expressions we used 
the charge and energy conservation laws in section~\ref{Sect:scatter-basics}
\footnote{
\green{We have excluded $k=1$ in the last sum of 
Eq.~(\ref{effmulti}) because we have $\Delta V_1=0$ due to our choice of reservoir $1$ as the 
reference.}
}.
The definition (\ref{effmulti}) applies only to the case in which $P$ is positive.
Since the signs of the heat currents $J_{h,k}$ are not known a priori 
(they actually depend on the details of the system), the expression of 
the efficiency depends on which heat currents are positive.
For instance, if for the three-terminal case we set $T_1>T_2>T_3$ 
and focus on those situations
where $J_{h,3}$ is negative 
(positive values of $J_{h,3}$ being associated with regimes where the 
machine effectively works as a refrigerator which extract heat from the 
coldest reservoir of the system), we obtain
$\eta=P/(J_{h,1}+J_{h,3})$ where the heat currents from 
reservoirs $1$ and $2$ are both positive or 
$\eta=P/J_{h,k}$ for $k=1$ or $2$
where only $J_{h,k}$ is positive.  

For a generic multi-terminal setup, the Carnot efficiency is obtained 
by imposing the condition of zero entropy production, 
namely $\dot{\mathscr{S}} = \sum_{k=1}^n J_{h,k}/T_k = 0$. 
In particular, 
for $n=2$ terminals kept at temperatures $T_1$ and $T_3$ (with $T_1>T_3$), 
from the condition $\dot{\mathscr{S}}=0$ and 
the definition of the efficiency, Eq.~(\ref{effmulti}), 
one gets the usual two-terminal Carnot efficiency $\eta_C^{II}=1-T_3/T_1$.
%The Carnot efficiency for a multi-terminal thermal machine is obtained 
%analogously by imposing the condition of zero entropy production.
It is worth noticing that, as shown below for the 
three-terminal case and in contrast to the two-terminal case, 
in general the Carnot efficiency cannot be written in terms of the 
temperatures only, but it depends on the details of the system.
Several instances must be considered separately, already with 
$n=3$ terminals (a reservoir at an intermediate temperature $T_2$ is added)
where we have, as discussed above, three possibilities.
If $J_{h,1}$ only is positive, we obtain
\begin{equation}\label{eq:carnot3t_a}
\eta_{C}=1-\frac{T_3}{T_1}+\frac{J_{h,2}}{J_{h,1}}(1-\zeta_{32})=
\eta_C^{II}+\frac{J_{h,2}}{J_{h,1}}(1-\zeta_{32}),
\end{equation}
where $\zeta_{ij} \equiv T_i/T_j$. 
Note that Eq.~(\ref{eq:carnot3t_a})
is the sum of the two-terminal Carnot efficiency $\eta_C^{II}$ and a term 
whose sign is determined by $(1-\zeta_{32})$.
Since $J_{h,1}>0$, $J_{h,2}<0$ and $\zeta_{32}<1$, it follows that  
$\eta_{C}$ is always \emph{reduced} with respect to its two-terminals 
counterpart $\eta_C^{II}$.
Analogously if $J_{h,2}$ only is positive, we obtain
\begin{equation}\label{eq:carnot3t_c}
\eta_{C}=\eta_C^{II} - \frac{T_3}{T_1}
\left[ \frac{J_{h,1}}{J_{h,2}} (1-\zeta_{13})-(1-\zeta_{12}) \right] ,
\end{equation}
which again can be shown to be reduced with respect to $\eta_C^{II}$, 
since $J_{h,1}<0$, $J_{h,2}>0$, $\zeta_{12}>1$, and $\zeta_{13}>1$.
We notice that this is a hybrid configuration (not a heat engine, neither a 
refrigerator): the hottest reservoir absorbs heat, while the intermediate-temperature 
reservoir releases heat.
However, the heat to work conversion efficiency is legitimately defined since 
generation of power ($P>0$) can occur in this situation.
Finally, if both $J_{h,1}$ and $J_{h,2}$ are positive we obtain
\begin{equation}\label{eq:carnot3t_b}
\eta_{C}=1-\frac{T_3}{T_1} 
\left( 1 + \frac{\zeta_{12} -1}{ 1+ \frac{J_{h,1}}{J_{h,2}} } \right) =
\eta_C^{II} - \frac{T_3}{T_1} \frac{\zeta_{12} -1}{ 1+ \frac{J_{h,1}}{J_{h,2}} }.
\end{equation}
Since $T_1>T_2>T_3$, the term that multiplies $T_3/T_1$ is positive so that 
$\eta_{C}$ is reduced with respect to the two-terminal case.
It can be expected that given a system that works between $T_1$ and $T_3$ 
(with $T_1>T_3$) and adding an arbitrary number of terminals at 
intermediate temperatures will in general lead to Carnot bounds smaller 
than $\eta_C^{II}$ \footnote{Of course, adding terminals at higher (or colder) 
temperatures than $T_1$ and $T_3$ will make $\eta_C$ increase.}.

Within linear response and for the time-reversal symmetric case,
analytical expressions for the efficiency at maximum power,
written in terms of generalized figures of merit, have been derived 
for the three-terminal case \cite{Mazza2014}. 
It turns out that the efficiency at maximum power is always upper
bounded by half of the associated Carnot efficiency, which in turn,
as shown above, is upper bounded by $\eta_C^{II}$.
On the other hand, as shown in Ref.~\cite{Mazza2014} 
in the examples of single and double dot
systems, for two-terminal efficiencies at maximum power 
lower than the CA upper bound, a third terminal can be useful to improve 
both the efficiency at maximum power and the output power. 
Moreover, a multi-terminal device offers enhanced \emph{flexibility} 
that might be useful to improve thermoelectric performances. 
For instance, with three terminals 
one can separate the currents,
with charge and heat flowing to different reservoirs \cite{Mazza2015}.

\subsection{Andreev reflection from superconductors}
\label{Sect:Andreev-linear}

%%%%%%%%%%%%%%%%%%%%%%%%%%
\begin{figure}
\begin{center}
\centerline{\includegraphics[width=0.9\textwidth]{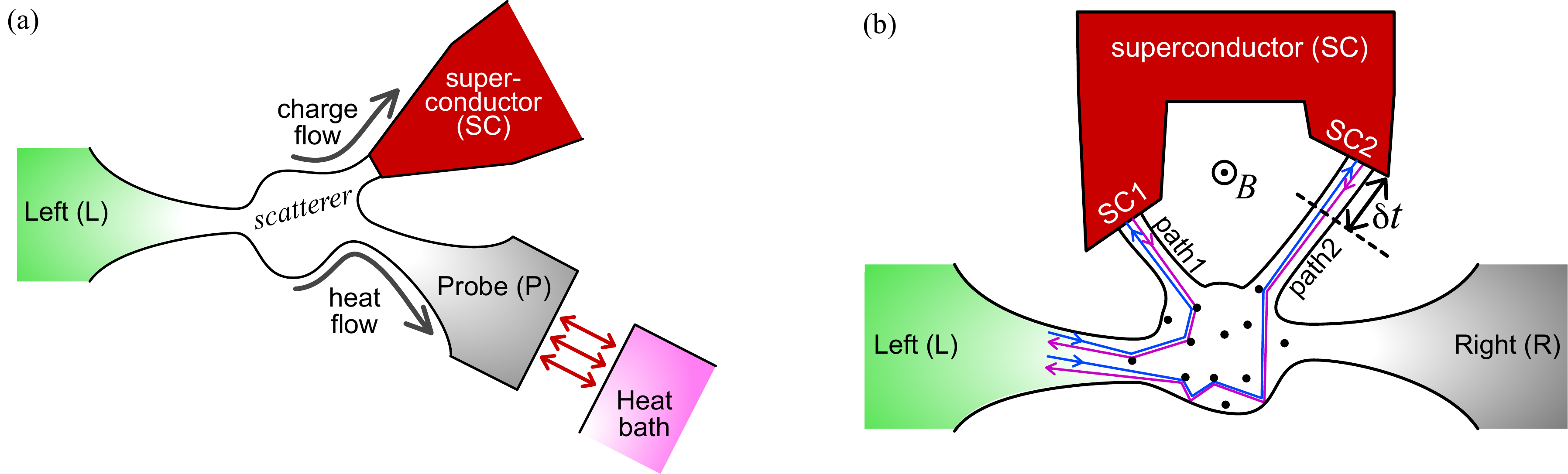}}
\caption{
\label{fig:3term-SC}
Systems which exhibit interesting effects due to  Andreev reflection from a superconductor.
(a) A sketch of the geometry used for heat-charge separation discussed in section~\ref{Sect:heat-charge-separation}.
Charge cannot flow into the probe and heat cannot flow into the superconductor.
(b) A sketch of an Andreev interferometer which generates a thermopower which is an odd function of the applied magnetic field, ${\bm B}$, as discussed in section~\ref{Sect:odd-B}.
}
\end{center}
\end{figure}
%%%%%%%%%%%%%%%%%%%%%%%%%%

In the presence of a superconducting reservoir which induces Andreev reflection, 
we must consider the more complicated expressions for the currents in section~\ref{Sect:scatter-Andreev}.
Performing a linear expansion of the Fermi function, $f^\vsig_j(E)$ in Eq.~(\ref{Eq:f-eh}),
in a similar manner to  Eq.~(\ref{eq:fermilinear}), we recover a relation of the form in Eq.~(\ref{Eq:Onsager-matrix})
for currents into non-superconducting reservoirs,
with all biases from the electrochemical potential of the superconductor.
In this case, the Onsager coefficients  are
\begin{align}
L_{ee,ij} \ =\  e^2T \ I_{ij}^{(0,0)},  \qquad
L_{eh,ij} \ =\  eT \ I_{ij}^{(0,1)},  \qquad
L_{he,ij}\ =\    eT \ I_{ij}^{(1,0)}, \qquad
L_{hh,ij}\ =\  T \ I_{ij}^{(1,1)},
\label{Eq:Andreev-linear1}
\end{align}
where we define the integral $ I_{ij}^{(n,m)}$ as
\begin{align}
I_{ij}^{(n,m)} \ =  \ &
\sum_{\vrho\vsig} \int_0^\infty {{\rm d}E \over h} \  \vrho^{1-n} \,\vsig^{1-m} \,E^{n+m} 
\  
\left[N_i^\vrho(E)  \,\delta_{ij}\delta_{\vrho \vsig}-{\cal T}_{ij}^{\vrho \vsig}(E)  \right]
\,\big[\!-\!f'(E)\big],
\label{Eq:Andreev-linear2}
\end{align}
in which the energy $E$ is measured from the electrochemical potential of the superconductor.
The $\vrho$ and $\vsig$ sums are over over $+1$ for electrons and $-1$ for holes.
We recall that throughout this review we use the subscripts \green{``e'' for electrical current 
and ``h'' for heat current},
when most of the works on systems with superconductors use ``e'' for electrons and ``h'' for holes (for which we use ``$\pm 1$'').

One can easily use Eq.~(\ref{Eq:T_ij-time-reverse-SC}) to show that the system obeys the Onsager reciprocal relation in  Eq.~(\ref{Eq:Onsager-multiterm}).
However, in general one no longer has the equality in Eq.~(\ref{Eq:Butcher}).
This is because $L_{eh,ij}$ is sensitive to the charge carried by particles when they leave the scatterer
(an extra factor of $\vrho$ in the integrand),
while $L_{he,ij}$ is sensitive to the charge carried by particles when they enter the scatterer
(an extra factor of $\vsig$ in the integrand).
The two are the same in the absence of the superconductor (when ${\cal T}_{ij}^{\vrho \vsig}$ is only non-zero for $\vrho=\vsig$), 
however the fact that the Andreev reflection from the superconductor turns electrons into holes (and vice-versa)
means that in general $L_{eh,ij}(B)$ will not equal $L_{he,ij}(B)$ \cite{jwmb}.

\subsubsection{Heat-charge separation}
\label{Sect:heat-charge-separation}
Heat-charge separation can be obtained in a device called ``SPN'', which is 
composed of a generic conductor connected to a superconducting reservoir (S),
a normal metal reservoir (N) and a second normal reservoir whose electrochemical 
potential is set to inhibit the flow of electrical current, thus 
acting as a voltage probe (P).
This set-up naturally realizes heat-charge current separation.
A voltage probe exchanges (on average) by definition only 
heat (energy) with the system, whereas the superconductor, being a poor heat
conductor for temperatures below the gap, can exchange only charges. 
This way, the heat and charge currents, flowing together out
of the normal metal reservoir (N), are split and driven either 
towards the voltage probe (heat), or towards the superconducting 
reservoir (charge).
As a result, it is possible to violate in a controlled fashion the
Wiedemann-Franz law,
greatly enhancing (at low temperatures, i.e. where the 
Sommerfeld expansion holds)
both the efficiency and the power factor
with respect to a standard two-terminal system \cite{Mazza2015}.

\subsubsection{Thermopower as odd-function of external magnetic-field }
\label{Sect:odd-B}

In section~\ref{Sect:scatter-2term-linear}, we explained that the thermopower of a phase coherent scatterer
coupled to two-reservoirs, $S({\bm B})$, was always an even function of the external ${\bm B}$-field; i.e. $S({\bm B})=S(-{\bm B})$.
In section~\ref{sec:probes}, we showed that phase breaking effects (modeled as a probe reservoir)
give the thermopower an indeterminate symmetry under ${\bm B} \to -{\bm B}$.
In that case, any given system's thermopower will contain terms that are even in ${\bm B}$ and others that are odd in ${\bm B}$,
and we can make no general statement about which will be larger.
Here we show that a so-called \emph{Andreev interferometer} has a thermoelectric response that is {\it systematically} odd in the external field; 
i.e. $S({\bm B})=-S(-{\bm B})$.
As such it is a ideal test case for the theories discussed in section~\ref{sec:efficiencymagnetic} since it has an asymmetry parameter $x=-1$.

An Andreev interferometer is a superconducting island formed in a horse-shoe shape so it can be 
coupled to the scatterer at two points, as in Fig.~\ref{fig:3term-SC}b, with a magnetic field, ${\bm B}$, 
through the resulting loop.
Andreev  interferometers have been extensively studied experimentally \cite{chandrasekhar,Petrashov03,Chandrasekhar05,Chandrasekhar09},
and nearly all samples clearly show that $S({\bm B})$ is an odd function of ${\bm B}$.
While those experimental systems had $x=-1$, their thermoelectric responses corresponded
to a generalized figure of merit, $y\ll 1$, 
which means they could not have been useful for applications such as power production
or refrigeration. The theory that we will discuss here captures the basic physics of these systems, but only works in the regime where $y \ll 1$.  One may hope 
that one could get larger figures of merit by a suitable tuning of the parameters of the experimental system, 
even if a different theory is necessary to model such systems. 

This system was modelled using scattering theory in Ref.~\cite{jacquod}, which showed that the thermoelectric
effect could be seen by considering interference between paths that undergo Andreev reflection from the superconductor.
There are many paths that contribute to the transmission \cite{jacquod}, however the basic physics can be understood by considering the two paths shown in Fig.~\ref{fig:3term-SC}b.
In both cases an electron comes from the left reservoir and Andreev reflects back along the same path to
return to reservoir L as a hole, this process removes a charge of $2e$ out of reservoir L.
The difference between the paths is that path 1 reflects off arm SC1 of the superconducting island, while 
path 2 reflects off arm SC2.  The phase acquired by along path 1 is $2E t_1 +\phi/2$
if the electron initially has energy $E$ above the Fermi surface, while the phase acquired by along path 2 is
$2E t_2-\phi/2$ for the same initial electron energy,
where $t_n$ is the time taken to follow path $n$.
The factor of $\phi$ is the external field ${\bm B}$ multiplied by the area of the loop (formed by the 
two arms of the superconductor) measured in units of the magnetic flux quantum, $h/(2e)$.
Thus the superconducting phase is $\phi/2$ at SC1, while it is $-\phi/2$ at SC2,
and this phase is acquired by the wavepacket every time an electron reflects as a hole.
There are many paths going to SC1 with different $t_1$, and many paths going to SC2 with different $t_2$,
however the asymmetric geometry of the two arms means that on average $t_2-t_1 =\delta t$.
Thus we can conclude that the average contribution to the current due to 
interference between path 1 and path 2 is $2eA \cos(2E\delta t -\phi)$.
Here $A$ is a constant related to the probability to follow the paths.

However, if reservoir L is hot but unbiased, for every electron flowing into the scatterer there is a hole flowing into the scatterer.  Thus we must also consider the same paths, but with electron and hole interchanged,
so a hole with energy $\eps$ is injected to be Andreev reflected as an electron from SC1 or SC2 back along the same path, this process removes a charge of  $-2e$ from reservoir L.  
In this case, the average contribution to the current due to 
interference between path 1 and path 2 is $-2eA \cos(2E\delta t +\phi)$.
The sign changes in front of $\phi$ because the phase acquired in the transition hole$\to$electron is opposite from that acquired in the transition electron$\to$hole.  The sum of processes electron$\to$hole and hole$\to$electron gives
an average interference contribution to the current out of reservoir L equalling
\begin{eqnarray}
2eA \left[ \cos (2E\delta t -\phi) - \cos (2E\delta t +\phi) \right] \ = \ 4eA \sin (2E \delta t) \ \sin \phi 
\label{Eq:Andreev-intererence}
\end{eqnarray}
The various other contributions to thermoelectric transport considered in Ref.~\cite{jacquod} have the same $\phi$ dependence.
Thus, we see that heating reservoir L leads to an electrical current in reservoir L that is an odd function of $\phi$,
and hence an odd function of the external field ${\bm B}$.  Similar argument tell us that the electrical conductivity is even in 
$\phi$, which mean that the Seebeck coefficient (which is the ratio of the two) is an odd function of the external magnetic field, $S({\bm B})=-S(-{\bm B})$.

We make a few technical notes about calculating the Onsager matrix for such systems.
Since the superconductor is an island, the average current flow out of it must be zero 
in the steady-state, $\JelectricSC=0$ (we recall that one always has the heat flow $\JheatSC=0$).
Combining this with current conservation, means we expect that
$\JelectricL=-\JelectricR$, while heat conservation in linear response means that $\JheatL=-\JheatR$.
To write an equation of the form in Eq.~(\ref{Eq:Onsager-matrix}), we must chose reservoir 1 as the superconductor,
because we must measure all energies and biases with respect to the electrochemical potential of the 
superconductor.  The temperature of the superconductor is irrelevant, so it is convenient to measure temperatures
from that of reservoir R, which means $\Delta T_{ L} = T_{ L}-T_{ R}$.
Then we get \cite{Claughton-Lambert}
\begin{eqnarray}
\left(\begin{array}{c}
\JelectricL \\
\JelectricR \\
\JheatR
\end{array}\right) \ = \ 
\left(\begin{array}{ccc}
L_{ ee;L-} & L_{ eh;LL}& L_{ ee;L+}  \\
L_{ ee;R-}& L_{ eh;RL}&  L_{ ee;R+} \\
L_{ he;R-} & L_{ hh;RL}&  L_{ he;R+} \\
\end{array}\right)
\left(\begin{array}{c}
\mathcal{F}_{ e,-} \\
\mathcal{F}_{ h,L} \\
\mathcal{F}_{ e,+}
\end{array}\right),
\label{Eq:Onsager-matrix-2term+SC}
\end{eqnarray}
where we write the thermodynamic forces in terms of the sum and difference of biases $V_{ L}$ and $V_{ R}$,
such that $\mathcal{F}_{ e,\pm} =  (V_{ L} \pm V_{ R})\big/ T_{ R}$.
This means that the matrix elements 
$L_{\mu { e};i\pm} = 
\half \left(L_{{ \mu e};i{ L}}
\pm L_{\mu { e};i{ R}}\right)$
for $\mu \in { e,h}$ and $i \in { L,R}$.
Now for a superconducting island, the electrochemical potential of the superconductor must be adjusted to ensure that the condition 
$\JelectricL=-\JelectricR$ is fulfilled,
much as we did with the probe in section~\ref{sec:probes}.  
However, as we measure all energies from the
superconductor's electrochemical potential, this means that we actually adjust 
$\mathcal{F}_{ e,+}$ to ensure that $\JelectricL=-\JelectricR$, 
while the thermodynamic force associated with a bias between left and right is 
$\mathcal{F}_{ e,-}$.  
\green{The condition that $\JelectricL=-\JelectricR$ means that}
\begin{eqnarray}
\mathcal{F}_{ e,+} \ =\  - \frac{ L_{ ee;+-} \mathcal{F}_{ e,-} \ + \  L_{ eh;+L}\mathcal{F}_{ h,L}}{L_{ ee;++}} 
\end{eqnarray}
where for compactness we define $L_{{ e}\nu;\pm j} = \half \left(L_{{ e\nu;L} j}\pm L_{{ e\nu;R} j} \right)$
for $\nu \in { e,h}$ and $j \in { L,-,+}$.
Substituting this into Eq.~(\ref{Eq:Onsager-matrix-2term+SC}), and doing some basic algebra
gives us the two-terminal relations of a system coupled to a superconducting island \cite{Claughton-Lambert},
\begin{eqnarray}
\left(\begin{array}{c}
\JelectricR \\
\JheatR
\end{array}\right) \ = \ 
\left(\begin{array}{ccc}
\tilde{L}_{ ee;R-}
& \tilde{L}_{ eh;RL} \\
\tilde{L}_{ he;R-} & \tilde{L}_{ hh;RL}\\
\end{array}\right)
\left(\begin{array}{c}
\mathcal{F}_{ e,-} \\
\mathcal{F}_{ h,L} 
\end{array}\right)
\qquad \hbox{ with } 
\ \ 
\tilde{L}_{\mu\nu;ij} = L_{\mu\nu;ij}- \frac{L_{{ \mu e};i+}L_{{ e\nu};+j}}{ L_{ ee;++} }\ ,
\label{Eq:Onsager-matrix-2term+SCisland}
\end{eqnarray}
where we recall that $\JelectricL =-\JelectricR$ and $\JheatL=-\JheatR$.
We can then use  Eq.~(\ref{Eq:Andreev-linear1},\ref{Eq:Andreev-linear2}) to get $L_{\mu\nu;ij}$ from the transmission matrix elements.

The handwaving argument that led to Eq.~(\ref{Eq:Andreev-intererence}) applies to 
$L_{\mu\nu;ij}$ rather than  $\tilde{L}_{\mu\nu;ij}$.  However, Ref.~\cite{jacquod} summed all such
contributions to $ L_{\mu\nu;ij}$, inserted them into Eq.~(\ref{Eq:Onsager-matrix-2term+SCisland}),
and found that the asymmetry under $\phi \to -\phi$ 
does indeed carry over into the final result in the case \green{when the coupling to the superconducting reservoir is weaker than the coupling to the other reservoirs.}
For this, they considered a scatterer connected by $N_{\rm SC1}$ and $N_{\rm SC2}$ modes to the two parts of the superconductor, and 
by $N_{ L}$ and $N_{ R}$ modes to the the left and right reservoirs, in the limit where $1 \ll (N_{\rm SC1}+N_{\rm SC2}) \ll (N_{ L}+N_{ R})$.
Under these conditions, they found that
\begin{eqnarray}
S(\phi) \ =\ \frac{\tilde{L}_{ eh;RL}}{\tilde{L}_{ ee;R-}}
\ = \ 
{4 \kB \over e} {N_{\rm SC1}\ N_{\rm SC2} \over (N_{ L}+N_{ R})^2} \ I_{\rm b}(T) \sin \phi.
\end{eqnarray}
The form of the dimensionless factor $ I_{\rm b}(T)$ can be found in Ref.~\cite{jacquod}.
Under the same conditions, they also showed that the Wiedemann-Franz law is violated since
\begin{eqnarray}
{K  \over GT} =   {\cal L} \ \left( 
1 - F(T)  {N_{\rm SC1}^2+N_{\rm SC2}^2 + 2  N_{\rm SC1}N_{\rm SC2}\cos \phi
\over 4(N_{ L}+N_{ R})(N_{\rm SC1}+N_{\rm SC2})} 
\right),
\end{eqnarray}
where ${\cal L} $ is the Lorenz number in Eq.~(\ref{lorenz}), and $F(T)$ is a thermal damping factor 
with  $F(0)=1$.  This violation is much bigger than that found in the absence of a superconducting island
\cite{lsb98},
but remains small since their calculation assumes $N_{\rm SC1}+N_{\rm SC2} \ll N_{ L}+N_{ R}$.
Thus $K \big/(GT)$ is still of order ${\cal L}$ which means that the figure of merit $ZT$ is of order
${\cal L} \ S^2$ which is clearly much less than one in the regime for which their calculation is valid.

It would be a good idea to do theory 
for cases when $(N_{\rm SC1}+N_{\rm SC2})$ is of order $ (N_{ L}+N_{ R})$, 
as $ZT$ should be much larger there.  For that one would have to treat the difficult problem
of multiple scattering from the superconductor; this could be done by treating the scatterer as a random matrix
\cite{Beenakker-review}, or by switching to the Usadel approach \cite{Sev00,Vir04,Vir07a,Vir07b,Tit08}.  
However, we are not aware of 
any works that explore how to maximize $ZT$ in such systems.

\subsection{Mesoscopic fluctuations inducing thermoelectric effects in quantum dots}

The systems discussed in most of this review have their parameters chosen to have simple and strong thermoelectric responses.
However, this is often not the case in real nanoscale systems, since uncontrolled disorder in the system (impurities, dislocations, grain boundaries, etc.)
tends to change and randomize the system parameters.  In such cases thermoelectrics effects  may change significantly from one sample (with one distribution of the microscopic disorder) 
to another superficially identical sample (with a different distribution of disorder).
In macroscopic systems such microscopic effects usually average out across the system, so such fluctuations are of little relevance.  However, nanoscale quantum systems are known as {\it mesoscopic}, because the fluctuations do not average out.
The  {\it universal conductance fluctuations} are the most famous example of such an effect.
They can be understood within the context of scattering theory as variations in the transmission with energy 
due to quantum interference between electron paths that scatter from the disorder in multiple ways.

The energy dependence of transmission leads to thermoelectric effects, as was noted 
in Ref.~\cite{lsb98}.
However, the disorder varies from sample to sample, and thus so does the energy-dependence of the transmission.
If we average over samples, we find that the average transmission is energy independent.  Thus
there is no thermoelectric effect on average.  None the less, the samples have a distribution of Seebeck coefficients centered around zero, so some samples will have positive $S$ while others will have negative $S$. 

Scattering theory has been used to find the typical magnitude of $S$ for a large quantum dot well coupled to the reservoirs \cite{lsb98}, with the dot's level spacing being $\Delta$, and the level broadening being of order $N\Delta$,
where $N$ is the number of modes in the contacts to the reservoirs.
We do not reproduce the calculations in Ref.~\cite{lsb98}, 
but note that they use the Sommerfeld expansion to treat the problem, which relies on 
the transmission function being a smooth function energy on the scale of temperature.
Since the transmission varies on the scale of the level broadening, the Sommerfeld expansion is only valid for $\kB T \ll N \Delta$.
In this regime, Ref.~\cite{lsb98} found that the magnitude of the Seebeck coefficient 
is 
\begin{eqnarray}
S_{\rm typical} \ =  \ \sqrt{ {\rm var}(S) } \ =\  \frac{\pi^3  \ \kB^2  T}{3e \ \beta \ N^2 \Delta}, 
\end{eqnarray}
where $\beta$ is the integer telling us if the system respects time-reversal symmetry in random-matrix theory (time-reversal symmetry means $\beta=1$, while broken time-reversal symmetry due to an external magnetic fields means $\beta=2$).
As one is in the regime given by the Sommerfeld expansion, it is reasonable to assume the 
Wiedemann-Franz law is approximately satisfied (although there will be small mesoscopic oscillations in the ratio $K/G$),
which means the figure of merit 
\begin{eqnarray}
(ZT)_{\rm typical} \ \sim\ {\cal L} \,S_{\rm typical}^2  \ =\  \frac{\pi^4  }{3 \beta \ N^2} \ 
\left(\frac{\kB T}{N\Delta}\right)^2
\end{eqnarray}
where ${\cal L}$ is the Lorenz number in Eq.~(\ref{lorenz}). 
We recall the calculation is valid for $\kB T \ll N\Delta$, which means that it gives $ZT \ll 1$. 
Thus, while these fluctuations are interesting and give us information about the sample, they are too small to 
be useful for heat engines or refrigerators.
Ref.~\cite{jacquod} pointed out that $S_{\rm typical}$ obeys the symmetry discussed below
Eq.~(\ref{Eq:Onsager-2term-2}),
and so could provide an explanation for the previously unexplained even-${\bm B}$ dependence of $S({\bm B})$ for the experimental samples called ``house-geometry'' in 
Ref~\cite{chandrasekhar}.
Similar mesoscopic fluctuations in the context of the nonlinear scattering theory were studied
in Ref.~\cite{sanchez2011},
they were also studied in different regimes of systems with strong Coulomb blockade in
Refs.~\cite{vonOppen2004,Stone-Alhassid2010,Vasenko2015}.

\subsection{Thermoelectricity in disordered systems near the mobility edge}

There have been many works on  thermoelectric effects associated with the mobility-edge in a bulk disordered
semiconductor \cite{Fritzsche71}, and this idea was extended to nanostructures in Ref.~\cite{Sivan-Imry1986}.  The idea is that electronic states below an energy $E_{\rm loc}$ are localized by the disorder, and so cannot flow from hot to cold,
 while those above $E_{\rm loc}$ are delocalized, and so free to flow. 
 At a hand-waving level, one can guess that this will mean the transmission of the disordered system is
 very small below $E_{\rm loc}$ and close to one above $E_{\rm loc}$ (much like the transmission of the point-contact discussed in section~\ref{Sect:scatter-pointcont}).
 Thus, one can immediately see from Eq.~(\ref{Eq:S-as-average}) that the system will
have a significant Seebeck coefficient, $S$.
This simple argument captures the basic idea of the coherent transport regime \cite{pichard2014a} that occurs at low temperatures,
but at higher temperatures activated hopping start to dominate \cite{pichard2014b,Pichard2016}. 
Then, the electrons flow from hot to cold with the aid of thermal activation by phonons,
giving a more complicated (but no less interesting) thermoelectric effect  \cite{pichard2014b,Pichard2016}.
This physics should be visible in disordered semiconductor nanowires, where 
a back-gate could be used to tune $E_{\rm loc}$, and thereby tune the Seebeck coefficient.
This would allow field control of the heat exchange
between the phonons and the electrons at submicron scales in electronic circuits. It could be also
used for cooling hot spots \cite{pichard2015}.

The hopping regime can also be used to make a three-terminal thermoelectric heat-engine \cite{imry2012,imry2013}, 
by heating the phonon gas that activates the transport.  There is has been argued that the physics is dominated 
by the boundary between the nanoscale disordered region being heated and the bulk electronic reservoir
that carry the current generated \cite{imry2013}.

 \subsection{Aharonov-Bohm, quantum Hall and other chiral systems}
\label{sect:qu-hall}
 
 \green{
 There are many proposals for nanoscale heat-engines and refrigerators which require an external  magnetic field for their operation,  these include Aharonov-Bohm rings \cite{entin2012} and quantum Hall systems \cite{
QuNernst2014,sanchez2015a,sanchez2015b,Hofer-Sothmann2015,Vannucci2015,sanchez2016,whitney2016,Samuelsson-Sothmann2016}.
In these cases the external magnetic field does not provide heat or work to the system, but does change the systems dynamics, allowing its dynamics to break time-reversal symmetry.  
The external magnet that generates this magnetic field can the thought of as a catalyst;
it is a resource that changes the system's behavior without being modified itself.  
Of course, if the external magnet is a resistive coil, then it takes work to drive it,
but this is not the case if it is a permanent magnet or a superconducting magnet. 
}

\green{
 Quantum Hall systems are particularly intriguing because electrons flow in chiral edge-states; so for example electrons can flow clockwise around the edge of the system, but cannot flow counter-clockwise.
This makes them the most extreme example of time-reversal symmetry breaking,
since it is not only that the time-reversed state of a given electron has different dynamics.
Instead, here the time-reversed state of an electron going clockwise would be one going counter-clockwise, and this state does not exist in the system.
}

\green{
Thus one can have a quantum-Hall system coupled to three reservoirs (hot, left and right) that has electrons
flowing from hot to left, but no electrons flowing from left to hot.  Superficially, it looks like one can use this
to engineer unphysically good thermoelectric machines, for example one whose power output is independent of the temperature of the left reservoir. However, a more careful analysis requires taking into account the flow of electrons from left to right and right to hot \cite{sanchez2015a,sanchez2015b,sanchez2016,whitney2016}, 
when this is done one recovers the predictions similar to those in section~\ref{sec:efficiencymagnetic}.  Thus  the efficiency must be equal to or less than Carnot efficiency, while the power output is of a similar order of magnitude to systems without magnetic fields \cite{whitney2016}.  
}

\green{
This does not mean that quantum Hall systems are without interest, quite the contrary.  
To build a good thermoelectric nanostructure, it is critical to have a very high degree of control over the electrons, and currently quantum Hall systems are better for this than almost any other electronic system.
For example it is the only context in which one can build a solid state Mach-Zehnder interferometer for electrons
\cite{Heiblum2003,Heiblum2007,Roulleau2008a,Roulleau2009,Strunk2010,Portier2012}.
Such a system has recently been proposed as a powerful 
and efficient heat engines \cite{Hofer-Sothmann2015,Samuelsson-Sothmann2016}.
}

\green{
Turning the situation around, one can use heat and charge transport properties as a probe of physics of a nanostructure.  In this context, recent measurements of the heat and charge transport 
of a point contact in the fractional quantum Hall regime have given a great deal of information about the 
chiral edge-states (and their reconstruction) in these systems \cite{Mitali-Banerjee}.
}

\green{Finally, we mention that works are starting to appear on other topological systems such as 
topological insulators \cite{Ronetti2016}. }
\red{
Due to their bulk properties, many currently known topological insulators (for instance 
bismuth telluride, Bi$_2$Te$_3$) are also excellent       
thermoelectric materials, with applications in power sources for space exploration 
\cite{Schmidt-deep-space}. The possibility that the topologically protected conducting 
channels act as energy filters \cite{Chang2014} and the nontrivial interplay between edge 
(or surface) and bulk states \cite{Xu2014} offer new opportunities to improve thermoelectric
efficiency. Moreover, topological protection against nonmagnetic impurities can ensure 
a good electrical conductivity while phonon conductivity is suppressed by the 
impurities \cite{Chang2014}.
}

 \subsection{\green{Noise in heat and charge currents}}

 \green{ 
Scattering theory has long been used to calculate the noise in the charge current through
a quantum system \cite{Blanter-Buttiker}.  Experimentally this noise can give us much more information than the average current alone, such as the charge of the current carriers (which one could only get from the average current alone if one had a perfect knowledge of the system's transmission probability). This is often summarized with the famous phrase of Landauer that ``the noise is the signal'' \cite{Landauer-noise-is-signal}.  A perfect example of this is that noise measurements were  used to prove that the charge carriers in fractional quantum Hall states are fractionally charged \cite{frac-qu-hall1,frac-qu-hall2}.
}

\green{
One can do the same for the noise in heat or energy currents, both theoretically 
\cite{Sergi2011,Zhan2011,Sanchez2012,Sanchez2013,Azema-Lombardo-Dare2014,Crepieux2015,Battista2014a,Battista2014b}
and experimentally \cite{Jezouin2013,Mitali-Banerjee}, 
and this again should give more information about the system than 
the average currents alone.    In particular, one can look at cross-correlations between noise in the charge current and that in the heat current, which can give information about whether each charge carriers carry
positive or negative amounts of heat \cite{Azema-Lombardo-Dare2014,Crepieux2015}.
The noise in the heat currents into a finite size reservoir will lead to fluctuations in the energy in that reservoir. 
If electron-electron interactions cause the electrons in the reservoir to relax to a Fermi distribution 
(with a well-defined temperature) faster than any process which couples that reservoir to its environment, then
these energy fluctuations can be considered as fluctuations of the effective temperature of the finite-size reservoir \cite{Van-den-Berg2015}.
} 
   
%%%%%%%%%%%%%%%%%%%
%%%%
%%%%%%%%%%%%%%%%%%%

\section{Nonlinear scattering theory and the thermodynamic laws}
\label{Sect:scatter-nonlin}

The central objective of this chapter is to prove that the scattering theory introduced 
in chapter~\ref{Sect:scattering-theory}
contains the laws of thermodynamics.   This will allow us to say that  the laws of thermodynamics
are not violated by any system modelled by the scattering theory.
As a practical consequence, no system modelled by the scattering theory can exceed Carnot efficiency.

We will carry out these proofs in the context of the nonlinear scattering theory 
(which of course means it also applies to the 
linear response scattering theory in chapter~\ref{sec:landauer}).  The reason for this is twofold.
Firstly, the proofs are not more difficult in the nonlinear scattering theory than in the linear response theory. 
Secondly, there is an ambiguity in the linear response theory with respect to the first law of thermodynamics (energy conservation), which is absent in the full nonlinear theory. 
This ambiguity will be made clearer when we address the first law below, however its origins
can already be seen in the discussion below Eq.~(\ref{Eq:heat-sum-linear}).  In linear response, the heat flow into the scatterer equals the heat flow out, even when the scatterer is producing electrical power.  
Superficially, this looks like a violation of the first law of thermodynamics, which says that one cannot generate work without absorbing heat (or vice versa).  However, 
in the linear response regime
the work produced is quadratic in the applied bias, and thus so is the associated reduction in heat.
This means the absorption of heat associated with the work generation is beyond the linear-response theory.  
Hence, linear-response is ambiguous about the first law;
in other words it is hard to tell if a theory is violating the first law or not by only studying its linear-response regime.

\subsection{Calculating transmission in the nonlinear regime}
\label{Sect:scatter-validity2}

Here we return to the question of calculating the transmission function for a given system, that was 
initially discussed in section~\ref{Sect:scatter-validity}, considering  in more detail the case of large biases or temperature differences.
This section is only necessary reading if one wishes to calculate the transmission function for a given nanostructure.
It can be skipped if one is more interested in knowing how an arbitrary nanostructure obeys the laws of thermodynamics, for the reasons outlined at the end of section~\ref{Sect:scatter-validity}.

The objective is to find the electrostatic potential in the scatter for the desired biases and temperature differences between the reservoirs, taking into account the electron flows that occur because the reservoirs are no longer in equilibrium with each other.
There are two main methods to proceed in a manner that make the problem tractable, both found in 
Refs.~\cite{Christen-Buttiker1996a,Christen-Buttiker1996b}.
The first method is to treat the effect of the biases and temperature differences in simple phenomenological models
which enable one to consider situations deep in the nonlinear regime.
The second method is the weakly nonlinear theory \cite{Christen-Buttiker1996a,Christen-Buttiker1996b,Sanchez-Buttiker2005,Buttiker-Sanchez2005}, which involves doing a perturbation expansion about the equilibrium state, by treating the biases and temperature differences as small.
This second method is microscopic, in the sense that it is a recipe which can be used to calculate the
transport properties of a given system from that system's Hamiltonian.  However, such a calculation would require numerical simulation in all but the simplest model situations.

\begin{figure}[t]
\centerline{\includegraphics[width=\textwidth]{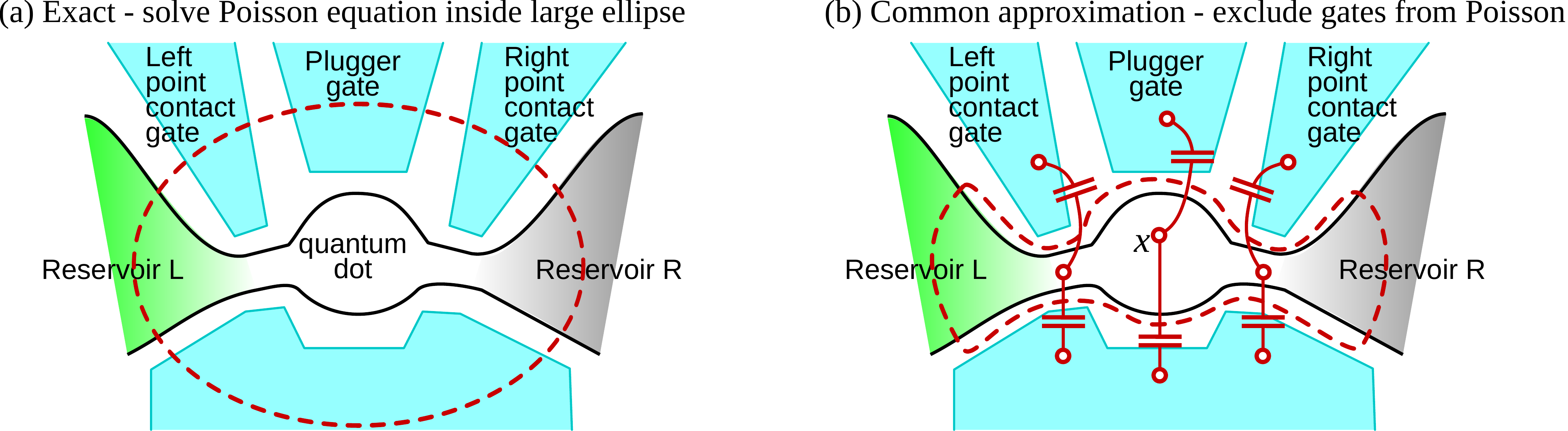}}
\caption{\label{Fig:nonlinear-example}
Example of a two-terminal quantum dot system (reservoir L, quantum dot and reservoir R) defined in a two-dimensional electron gas by
the set of top gates (shown in blue).  Four top gates are necessary to independently control the
size of the quantum dot and the width of the point contacts to the left and right.
When modelling the system using the nonlinear scattering theory, one cannot consider the two-dimensional
electron gas alone, one must also treat its entire electro-static environment, determined by the top-gates
(and the back-gate if present). Thus, in general, one must solve the Poisson equation given by Eq.~(\ref{Eq:Poisson1}) 
within a region significantly bigger than the scatterer (in this case the quantum dot), such as that marked by the dashed red ellipse in (a).  If the gates are all close to ideal (i.e.~good metals), then it may be reasonable to make the approximation indicated in (b).  There the Poisson equation is solved only in the quantum dot and in the nearby parts of its reservoirs (within the dashed red loop), but the gates are taken into account via a capacitive coupling to each point $x$ inside the dashed red loop.  For clarity we sketch each gate as only having a capacitive effect on the nearest part of the nanostructure.  However, experiments show that a gate's capacitive effect applies on a range as large as the nanostructure, so it would be more correct to sketch every point $x$ in the nanostructure as being 
connected capacitively to all gates, with an inverse capacitance which decays smoothly with distance from the gate in question.
}
\end{figure}

Irrespective of which method one uses to calculate the scattering matrix of the system in question, at the end one has a scattering matrix which depends not only on energy $E$, but also on the electrochemical potential and temperature of all the reservoirs in the vicinity of the scatter.  
Note that this requires a change of perspective compared with the linear-response regime.
In the linear response regime, we only cared about those reservoirs which exchanged electrons with the scatterer.
Here we must also consider all those reservoir (gates) which are electrostatically coupled to the scatterer,
e.g. everything within the dashed red ellipse in Fig.~\ref{Fig:nonlinear-example}a.

However, despite this complexity, we know that for any given set of electrochemical potentials and temperatures, there is a unitary scattering matrix which gives the transmission matrix as in Eq.~(\ref{Eq:def-T_ij}). 
This, in turn, enables us to calculate the currents in Eqs.~(\ref{Eq:I-initial},\ref{Eq:J-initial}).
The fact that the scattering matrix is unitary, irrespective of how one calculates that scattering matrix, 
means that the system will always satisfy 
Eqs.~(\ref{Eq:constraints-T_ij}).  This is crucial to sections~\ref{Sect:zeroth-law-scattering}-\ref{Sect:second-law-scattering}, as 
Eq.~(\ref{Eq:constraints-T_ij}) will be the only requirement in our proof 
that an arbitrary system modelled by the scattering theory will obey the laws of thermodynamics.

%========================================
%========================================

\subsubsection{Phenomenological treatment of strong nonlinearities}
\label{Sect:nonlin-phenomen}

The simplest phenomenological model of nonlinear situations is to take the transmission function
for the linear response problem and allow its parameters to depend on the bias and temperature of the reservoirs and gates.  For example, one could take the point-contact discussed in section~\ref{Sect:scatter-pointcont},
and assume the two parameters $\eps(n_y,n_z)$ and $D$ in Eq.~(\ref{Eq:transmission-pc}) depend on the bias and temperatures of reservoirs and gates.
Similarly, one could take the single-level quantum dot discussed in section~\ref{Sect:scatter-single-level-dot},
and assume that the three parameters $\Gamma_{ L}$, $\Gamma_{ R}$ and $\eps_0$ 
in Eq.~(\ref{Eq:Briet-Wigner}) depend on the bias and temperatures of reservoirs and gates.

In principle, a system could have almost any dependence of these parameters on the bias and temperatures of reservoirs and gates. However, one should remember that the physics should be \emph{gauge-invariant}, by which we mean  the physics depends on energy differences, but not on the absolute value of energy.  Thus, a uniform shift of the bias on all reservoirs and gates by $V$ should simply shift the transmission function in energy by $e V$ in such a way that all heat and charge currents are invariant under the uniform shift of the bias.

\begin{figure}
\centerline{\includegraphics[width=0.9\columnwidth]{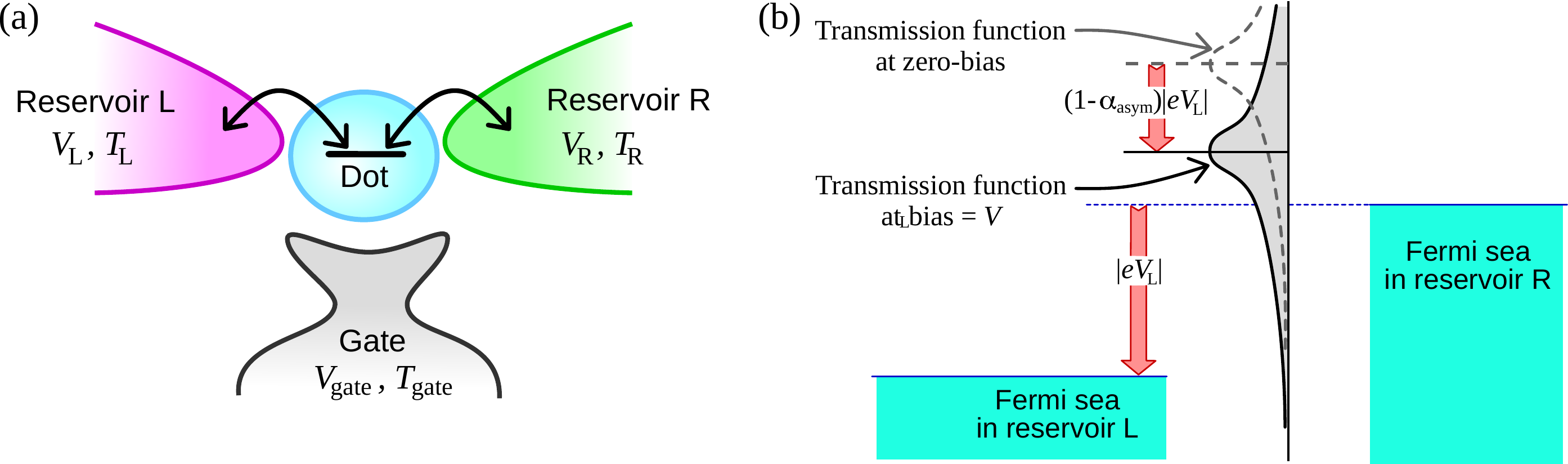}}
\caption{\label{Fig:minimal} 
\green{(a) A simple two terminal system in which the scatterer is a single level quantum dot 
with a tunnel coupling to reservoirs L and R, and a capacitive coupling to a gate (in addition to 
an inevitable capacitative coupling to reservoirs L and R).
(b)} Sketch of how electro-static coupling between the quantum system and the leads modify the 
system's transmission function when a bias is applied across the system.
The sketch is for our minimal mean-field charging approximation, for an un-gated quantum system, so  $\alpha_{\rm gate}=0$ 
in Eq.~(\ref{Eq:T-electrostatics2}).
If the dashed curve corresponds to the transmission when the system is unbiased, the solid (shaded) curve
gives the transmission when the L lead is biased by $V_L$.
It is simply the unbiased transmission curve shifted down by $(1-\alpha_{\rm asym})|\eminus V_L|$.
}
\end{figure}

The simplest possible example of such a phenomenological model is one in which the transmission function is shifted by  
the charge build up around the scatterer caused by the bias, 
without significantly changing the shape of the transmission function.  
In other words, if the system acts as a single-level quantum dot when at zero bias, it still acts as a single-level quantum dot at finite bias.
\green{Let us consider a system coupled to two reservoirs (L and R) and one gate, such as in Fig.~\ref{Fig:minimal}a.}
If ${\cal T}^0_{L,{\rm isl}}(E)$ is the transmission function
when all reservoirs are at the same bias and same temperature ($V_0,T_0$), then
the simplest model is to assume that the transmission function at other biases and temperatures is given by
\begin{eqnarray}
{\cal T}_{L,{\rm isl}}( E)
\simeq \ 
{\cal T}^0_{L,{\rm isl}}\Big( E -\kappa_{ e}\big(
\mathcal{F}_{{ e,L}},\mathcal{F}_{{ e,R}},\mathcal{F}_{{ e, {\rm gate}}}\big) 
-\kappa_{ h}\big(
\mathcal{F}_{{ h,L}},\mathcal{F}_{{ h,R}},\mathcal{F}_{{ h, gate}}\big) \Big),
\label{Eq:T-electrostatics1}
\end{eqnarray}
where our notation means that
$\mathcal{F}_{{ e},i} = (V_i-V_0)\big/T_0$ and $\mathcal{F}_{{ h},i} = (T_i-T_0)\big/T_0^2$,
and we define 
\begin{eqnarray}
\kappa_{ e} (
\mathcal{F}_{{e,L}},\mathcal{F}_{{ e,R}},\mathcal{F}_{{ e, {\rm gate}}}) &=& 
\left(1-\alpha^{ (e)}_{\rm gate}\right)
\left[{e T_0  \over 2}\left(1-\alpha^{(e)}_{\rm asym}\right) \, \mathcal{F}_{{ e,L}}+
{e T_0  \over 2}\left(1+\alpha^{ (e)}_{\rm asym}\right) \, \mathcal{F}_{{ e,R}}
\right] + \alpha^{ (e)}_{\rm gate}\, eT_0 \mathcal{F}_{{e,{\rm gate}}} ,
\label{Eq:T-electrostatics2}
\\
\kappa_{h} (
\mathcal{F}_{{ h,L}},\mathcal{F}_{{ h,R}},\mathcal{F}_{{ h, {\rm gate}}}) &=& 
\alpha^{(h)}_{ L}\ \mathcal{F}_{{ h,L}}
\ + \ \alpha^{ (h)}_{ R}\ \mathcal{F}_{{ h,R}}
\ +\ \alpha^{ (h)}_{\rm gate}\ \mathcal{F}_{{h,{\rm gate}}}  .
\label{Eq:T-electrostatics3}
\end{eqnarray}
The $\alpha^{ (\mu)}_i$ are phenomenological parameters 
which describe the electrostatic environment of the quantum system, where
$0 \leq \alpha^{(e)}_{\rm gate} \leq 1$ and $-1<\alpha^{(e)}_{\rm asym}\leq 1$.
The form of Eq.~(\ref{Eq:T-electrostatics2}) is chosen to ensure that it is gauge-invariant (as discussed above) 
for any value of $\alpha^{(e)}_{\rm gate}$ and $\alpha^{(e)}_{\rm asym}$.
In contrast,  gauge-invariance places no constraint on $\kappa_{h}$, so the parameters $\alpha^{(h)}_i$
can take any value.  Although, in many situations the shift of energy due to temperature effects can be expected to
be less than that due to a bias, in which case the magnitude of the $\alpha^{ (h)}$s are typically smaller
than the magnitudes of the $\alpha^{(e)}$s.
If the gate is absent then $\alpha^{(e)}_{\rm gate}=\alpha^{(h)}_{\rm gate}=0$.
In this case,  $\alpha_{\rm asym}=0$ means that the system feels the same changing effects from the reservoir to its left and its right.
Positive $\alpha_{\rm asym}$ means that charging effects 
are dominated by the charge in the reservoir to the right of the quantum system.
Negative $\alpha_{\rm asym}$ means that charging effects 
are dominated by the charge in the reservoir L.
 If, in contrast, the gate dominates the charge felt by the island then $\alpha_{\rm gate}=1$ and 
$\alpha_{\rm asym}$ is irrelevant.
Fig.~\ref{Fig:minimal}b  is a sketch of the effect of $\kappa_{e}$ for a single-level quantum dot with no gate 
(so $\alpha^{\rm (\mu)}_{\rm gate}=0$), for the case where the $\alpha^{(h)}$s are small enough to be neglected.

%%%%%%%%%%%%%%%%%%%%%%%%%%%%
%%%%%%%%%%%%%%%%%%%%%%%%%%

This approach is simple, easy to understand, and can be used to treat  highly nonlinear situations.
It could be easily extended to the other parameters in the transmission function,
such as making the couplings to the left and right reservoirs in Eqs.~(\ref{Eq:S-from-H}-\ref{Eq:Briet-Wigner}) dependent on the bias on those reservoirs.
It can also be extended by replacing Eqs.~(\ref{Eq:T-electrostatics2},\ref{Eq:T-electrostatics3}) 
by nonlinear functions of the biases and temperatures.
However, whatever one does the model remains phenomenological, and the number of phenomenological parameters increases rather rapidly as one makes the model more sophisticated.  This makes it hard to guess 
what version of the model (and what value of the parameters) to use to predict the properties of a given nanostructure.
However for simple geometries, such as a point-contact, this model may none the less help understand the physics.  It was used in this context to model thermoelectric refrigeration \cite{whitney2013-catastrophe}, in the limit where the gates dominated ($\alpha_{\rm gate} \to 1$).

\subsubsection{Microscopic treatment of weak nonlinearities}
\label{Sect:weakly-nonlinear-microscopic}

In linear response (linear order in biases and temperature differences), 
the electron flows are small enough that the potential in the scatterer remains that of the equilibrium state.
Thus, the scattering matrix is directly given by the dynamics under this unmodified electrostatic potential.
If one goes to one order higher (quadratic order in biases and temperature differences),
then one has to take into account the effect of the linear-response particle flow on the potential in the scatterer.
We expand about the chemical potential and temperature at equilibrium $(V_0,T_0)$, so we expand in powers of 
$\mathcal{F}_{\mu,i}$, where we recall our notation means that
$\mathcal{F}_{{ e},i} = (V_i-V_0)/T_0$ and $\mathcal{F}_{{h},i} = (T_i-T_0)/T_0^2$.
This gives \cite{sanchezprb13,Meair-Jacquod2013,Sanchez-Lopez2013,Lopez-Sanchez2014}, 
\begin{eqnarray}
J_{\mu ,i} = \sum_{\nu= { e,h}} \sum_j L_{\mu\nu,ij} \, \mathcal{F}_{\nu,j} \ +\  
 \sum_{\nu,\kappa= { e,h}} \sum_{j,k} {\cal L}_{\mu\nu\kappa,ijk} \, \mathcal{F}_{\nu,j} \mathcal{F}_{\kappa,k} 
 \label{Eq:J-scatter-2ndorder-expansion}
\end{eqnarray}
where the first term on the right is the linear response as in Eq.~(\ref{Eq:Onsager-matrix-general}),
and the remaining terms are the leading nonlinear corrections.
Note that while the sum over $j$ above is over all reservoirs which exchange electrons with the scatterer, the sum over $k$ is over all reservoirs in the electro-static environment of the scatterer, including gates.
That electrostatic environment is indicated in Fig.~\ref{Fig:nonlinear-example}a by everything inside the 
dashed red ellipse.  In the case were the gates are good enough metals that the charge on their surface is entirely determined by their bias, we can treat the gates in the electrostatic environment as capacitances, as in Fig.~\ref{Fig:nonlinear-example}b, however we cannot avoid a more sophisticated treatment of the electrostatic environment generated by the scatterer itself (the region inside the dashed loop in Fig.~\ref{Fig:nonlinear-example}b).

The linear-response Onsager coefficients, $L_{\mu\nu,ij}$, are given by Eq.~(\ref{Eq:Ls-multiterm}), 
while the nonlinear coefficients are given by the following second derivative of $J_{\mu,i}$;
\begin{eqnarray}
{\cal L}_{\mu\nu\kappa,ijk} ={1 \over 2} \left. {\rmd^2 \,J_{\mu,i} \over \rmd  \mathcal{F}_{\nu,j}  \rmd  \mathcal{F}_{\kappa,k}} \right|_{\mathcal{F}\to 0}
\label{Eq:nonlinear-L-double-derivative}
\end{eqnarray}
where $\mathcal{F}\to 0$ indicates that we take $ \mathcal{F}_{\mu',i'}\to 0$ for all $\mu'$ and $i'$. 
Carefully evaluating these derivative gives results consisting of terms containing zeroth, first and second derivatives of the Fermi function, $f(E)$. However, the fact that 
$\sum_j N_i \delta_{ij}- {\cal T}_{ij}(E)=0$ means that terms containing the zeroth derivative of $f(E)$
do not contribute.
Hence, these second-order coefficients contain two type of terms.
The first type of term is the second derivative of the scattering theory equations 
for $J_{{ e},i}$ and $J_{{ h},i}$ in Eqs.~(\ref{Eq:I-initial},\ref{Eq:J-initial}) which one would have if one assumed that the transmission functions are fixed (does not change with bias or temperature differences).  
The second type of term takes into account the fact that the bias and temperature differences, as given by the set of 
$ \mathcal{F}_{\mu,i}$s will affect the transmission functions, $\big(N_i(E)\delta_{ij} -{\cal T}_{ij}(E)\big)$;  
these terms looks exactly like the linear response terms but with an additional derivative with respect to $ \mathcal{F}_{\mu,i}$ acting on $\big(N_i(E)\delta_{ij} -{\cal T}_{ij}(E)\big)$.

The double-derivative in Eq.~(\ref{Eq:nonlinear-L-double-derivative}) are ugly.
They are slightly simpler if one write everything in terms of $f'(E)$, where the primed indicates $(\rmd /\rmd E)$,
by using $\int \rmd E \,a(E) \,f''(E) = - \int \rmd E \,a'(E) \, f'(E)$.   
For compactness, we follow Refs.~\cite{Christen-Buttiker1996a,Sanchez-Buttiker2005,Buttiker-Sanchez2005} in defining ${\cal A}_{ij} \equiv N_i(E)\delta_{ij}-{\cal T}_{ij}(E)$,
then the coefficients \green{(containing both types of terms discussed above)} are
\begin{subequations}
\label{Eq:nonlinearLs}
\begin{eqnarray}
 {\cal L}_{{eee},ijk} \!&=& \! {e^3T_0 \over 2} \int {\rmd E \over h}  \ 
 \left[ {1 \over e} {\rmd {\cal A}_{ij}\over \rmd \mathcal{F}_{{ e},k}} 
      +  {1 \over e}{\rmd {\cal A}_{ik}\over \rmd  \mathcal{F}_{{ e},j}} 
      + T_0 \delta_{jk} {\cal A}'_{ij}  \right]\  \Big(- \!f'\Big) \ ,
 \\
 {\cal L}_{{ eeh},ijk}  \!&=&\! {e^2 T_0 \over 2} \int {\rmd E \over h}\   
\left[ {\rmd {\cal A}_{ij}\over \rmd  \mathcal{F}_{{ h},k}} 
      + (E-\mu_1)\left( {1 \over e}{\rmd {\cal A}_{ik}\over \rmd \mathcal{F}_{{ e},j}} +T_0 \delta_{jk}  {\cal A}'_{ij}
      \right) \right]\ \Big(-\!f'\Big)\ , 
\\
 {\cal L}_{{ ehh},ijk} \! &=& \! {eT_0 \over 2} \int {\rmd E \over h}\  (E-\mu_1) \
 \left[ {\rmd {\cal A}_{ij}\over \rmd  \mathcal{F}_{{ h},k}} 
      + {\rmd {\cal A}_{ik}\over \rmd  \mathcal{F}_{{ h},j}} + T_0(E-\mu_1) \delta_{jk} {\cal A}'_{ij}
      \right]\ \Big(-\!f'\Big)\ ,
 \\
 {\cal L}_{{ hee},ijk} \!&=&\! {e^2T_0  \over 2} \int {\rmd E \over h} 
 \left[ (E-\mu_1)
 \left(  {1 \over e}{\rmd {\cal A}_{ij}\over \rmd   \mathcal{F}_{{ e},k}} 
      +  {1 \over e}{\rmd {\cal A}_{ik}\over \rmd  \mathcal{F}_{{ e},j}} 
      + T_0\delta_{jk}{\cal A}'_{ij}  \right) 
      +T_0 \left(\delta_{jk}{\cal A}_{ij}-\delta_{ij}{\cal A}_{ik}-\delta_{ik}{\cal A}_{ij} \right)
      \right] \ \Big(-\!f'\Big) \ ,  \qquad
 \\
 {\cal L}_{{ heh},ijk} \!&=&\! {e T_0   \over 2} \int {\rmd E \over h} \ 
(E-\mu_1) \ \left[ {\rmd {\cal A}_{ij}\over \rmd  \mathcal{F}_{{ h},k}} 
      + (E-\mu_1)\left(  {1 \over e}{\rmd {\cal A}_{ik}\over \rmd \mathcal{F}_{{ e},j}} 
      +T_0\delta_{jk}{\cal A}'_{ij}\right) + T_0 \big(\delta_{jk}{\cal A}_{ij} -\delta_{ij}  {\cal A}_{ik}\big)
      \right] \ \Big(-\!f'\Big) \ ,
\\
 {\cal L}_{{ hhh},ijk} \!&=&\!  {T_0  \over 2} \int {\rmd E \over h} 
\ (E-\mu_1)^2 \left[ {\rmd {\cal A}_{ij}\over \rmd  \mathcal{F}_{{ h},k}} 
      + {\rmd {\cal A}_{ik}\over \rmd  \mathcal{F}_{{ h},j}} 
      +T_0(E-\mu_1)\delta_{jk}{\cal A}'_{ij}+T_0\delta_{jk}{\cal A}_{ij}
      \right] \ \Big(-\!f'\Big) , 
\end{eqnarray}
\end{subequations}
where all quantities are evaluated in the limit where $\mathcal{F}_{\mu',i'}\to 0$ for all $\mu'$ and $i'$. 
It is easy to see that $L_{{ ehe};ijk}= L_{{ eeh};ikj}$ and $L_{{ hhe};ijk}= L_{{ heh};ikj}$.
Expanding both sides of Eq.~(\ref{Eq:scatter-1st-law}) up to second order in $\mathcal{F}$ as above, 
one can see that there must be the following relations between certain nonlinear coefficients and certain linear coefficients \cite{Meair-Jacquod2013},
\begin{eqnarray}
\sum_i{\cal L}_{hee,ijk} = -\half T_0 \left( L_{ee,jk} + L_{ee,kj} \right)  
\qquad \hbox{ and } \qquad 
\sum_i  {\cal L}_{heh,ijk} = -T_0 L_{ee,jk}
\end{eqnarray}
It is fairly easy to see that the above expressions satisfy these relations, once one notes that 
Eq.~(\ref{Eq:constraints-T_ij-b}) means that $\sum_i {\cal A}_{ij}=0$.  This means the theory conserves energy up to 
second order, ensuring it obeys the first law of thermodynamics (see section.~\ref{Sect:scatter-1st-law}).

To evaluate Eqs.~(\ref{Eq:nonlinearLs}), we need the derivatives of ${\cal A}_{ij}$ with respect to the $\mathcal{F}$.  
To get these, one defines the so-called characteristic potentials as
\begin{eqnarray}
u_{\nu,k} (x) = \left( {\rmd U(x)\over \rmd  \mathcal{F}_{\nu,k}}\right)_{ \mathcal{F}\to 0}\ .
\label{Eq:u_characteristic}
\end{eqnarray}
These correspond to the change in electrochemical potential at point $x$ in the nanostructure due to 
the change of the thermodynamics potential $\mathcal{F}_{\nu,k}$  in reservoir $k$.
\green{Physically this characteristic potential contains two effects which lead to a change in the electron density at the point $x$ inside and near the scatterer (i) the extra charge injected into each region of the scatterer and (ii) the polarization of the existing charges in the region $x$.}
We use the characteristic potential to write
\begin{eqnarray}
\left({\rmd {\cal A}_{ij}\over \rmd  \mathcal{F}_{\mu,k}} \right)_{ \mathcal{F}\to 0}
= \int \rmd^d x\ {\rmd{\cal A}_{ij}\over \rmd U(x)}\  u_{\mu,k} (x)  &\equiv& \sum_{n} {\rmd {\cal A}_{ij}\over \rmd U_n} u_{\mu,k} (x_n).
\label{Eq:dA/dF-in-terms-of-characteristic-potential}
\end{eqnarray}
Here, ${\rmd (\cdots) \big/ \rmd U(x)}$ is formally a functional derivative, but we assume it is defined
by discretizing space on a grid, and taking the spacing of the grid to zero. 
Then $U_n \equiv U(x_n)$ is the potential at the position $x_n$ of the $n$th site on the grid.
In practice, the grids do not need to be infinitesimally fine, it is sufficient that it is small enough that 
$\big(\rmd {\cal A}_{ij} \big/ U(x)\big)(\cdots)$ varies little between neighbouring sites.  In contrast, the grid must extend far enough into 
each reservoir to capture the fact that a change in $U(x)$ at a point $x$ in the reservoir close to the nanostructure
may change ${\cal A}_{ij}$.
If we substitute Eq.~(\ref{Eq:dA/dF-in-terms-of-characteristic-potential}) into Eqs.~(\ref{Eq:nonlinearLs}), 
we split the problem of calculating $\left({\rmd {\cal A}_{ij}\big/ \rmd  \mathcal{F}_{\mu,k}} \right)$
into two parts; the calculation of the transmission functions dependence on small changes of 
the potential within the scatterer (see Appendix \ref{Sect:T-as-function-of-U}), and the calculation of the characteristic potential (see Appendix \ref{Sect:Characteristic-potentials}).
As these Appendices show, the majority of the work is to calculate the characteristic potential, since this requires solving the Poisson equation to get the potential at a given point in the system from the charge distribution in and around that system.

\subsubsection{Weak nonlinearities for a simple model of a quantum dot}
\label{Sect:microsopic-nonlinear-simple}

 It is instructive to consider a simple model treated in the weakly nonlinear regime; this is the model
 of a  single-level dot \cite{sanchezprb13,Meair-Jacquod2013,Sanchez-Lopez2013,Lopez-Sanchez2014}.
We briefly outline the assumptions that allow us to derive it from the general case, 
discussed in Appendix \ref{Appendix:weakly-nonlinear}, however it is not necessary to follow the details in that appendix  to get a feeling
for the physics of the model.

The main assumption is to treat the Poisson equation as a single site problem (the site being the quantum dot)
with a capacitive coupling to an external gate;
so there is only a single site in the grid discussed in the context of 
Eq.~(\ref{Eq:dA/dF-in-terms-of-characteristic-potential}). 
Then the fact there is only one site means there is only one value of $\left({\rmd {\cal A}_{ij} \big/ \rmd U(x)}\right)$
to calculate, and what is more in such a situation the gauge-invariance discussed in 
Appendix~\ref{Sect:gauge-invariance} 
enables us to replace  $\left({\rmd {\cal A}_{ij} \big/ \rmd U(x)}\right)$ with  ${\cal A}'_{ij}$.
In fact the same argument allows us to replace $\rmd  (\cdots) \big/ \rmd U(x)$ by $-\rmd  (\cdots) \big/ \rmd E$
in all quantities, such as $\nu_k (E,x)$ in Eq.~(\ref{Eq:nu_k}).
The function ${\cal A}_{ij} \equiv N_i(E)\delta_{ij}-{\cal T}_{ij}(E)$ for the single level quantum dot is given by Eqs.~(\ref{Eq:S-single-level-dot},\ref{Eq:Briet-Wigner}).

Suppressing all site labels (since there is only one site), we get
the {\it injectivities}, $D_{\mu,k}$, defined in Eq.~(\ref{Eq:qinj}), which corresponds to the extra charge injected from reservoir $k$ when its bias ($\mu={ e}$) or temperature ($\mu={ h}$) is slightly changed.
They take the form
\begin{eqnarray}
D_{{ e},k}  \ =\ {e^2T \over \eps_0} \int \rmd E \ \big(\!-f'(E)\big)\ \nu_k (E) \ ,
\qquad
D_{{ h},k} \ =\ {eT \over \eps_0} \int \rmd E \ \big(\!-f'(E)\big)\   (E-\mu_1)\ \nu_k (E) \ ,
\end{eqnarray}
for $k={ L,R}$.
Here $\eps_0$ is the permittivity of free space,
$\nu_k(E)$ is the partial density of states associated with particles coming from reservoir $k$,
combining Eq.~(\ref{Eq:S-single-level-dot}) with Eq.~(\ref{Eq:nu_k}), they find that
\begin{eqnarray}
\nu_k (E) = {1 \over 2\pi} \ \frac{\Gamma_k}{\left(E-E_0\right)^2 + \frac{1}{4}\left(\Gamma_{ L} +\Gamma_{ R}\right)^2 } 
\end{eqnarray}
The works in question \cite{sanchezprb13,Meair-Jacquod2013,Sanchez-Lopez2013,Lopez-Sanchez2014}
argue that the discretized Lindhard screening function $\Pi$ for the single site is given by a local function $e^2\Pi=D_{ e,L}+D_{ e,R}$ in the limit of good screening within the scatterer. 
Then Eq.~(\ref{Eq:C-relations}) reduces to $C=-\tilde{C}_{\rm gate}$.
After this we get the characteristic potentials
\begin{eqnarray}
u_{\mu, { L}} = \frac{D_{\mu,{ L}} }{C+D_{ e,L}+D_{ e,R}}\,,
\qquad
u_{\mu, { R}} =\frac{D_{\mu,{ R}} }{C+D_{ e,L}+D_{ e,R}}\,,
\qquad
u_{\mu,{\rm  gate}} = \delta_{\mu, { e}} \ \frac{ C }{C+D_{ e,L}+D_{ e,R}}\,.
\label{Eq:characteristic-potentials-for-qu-dot-model}
\end{eqnarray}
Substituting all the above results into Eqs.~(\ref{Eq:nonlinearLs}) gives us all the nonlinear coefficients,
and so we can get charge and heat currents up to second-order in the thermodynamic forces (biases and temperature differences), from which we can extract power outputs, efficiencies, etc.

With a little effort one can see that this is equivalent to the phenomenological treatment of the single-level quantum system, as discussed in section~\ref{Sect:nonlin-phenomen}, 
in which the bias and temperature difference result is a shift of the 
energy-level of the quantum dot.  However, here the phenomenological functions
in Eqs.~(\ref{Eq:T-electrostatics2},\ref{Eq:T-electrostatics3})
can now be calculated from the microscopic properties of the system.
If we take Eq.~(\ref{Eq:T-electrostatics1}), we see that it implies that
\begin{eqnarray}
\left({\rmd {\cal A}_{ij}\over \rmd  \mathcal{F}_{\mu,k}} \right)_{ \mathcal{F}\to 0} = - 
\left({\cal A}'_{ij} 
\ {\rmd \kappa_\mu \over \rmd  \mathcal{F}_{\mu,k}} \right)_{ \mathcal{F}\to 0}\ .
\end{eqnarray}
We can compare this with Eq.~(\ref{Eq:dA/dF-in-terms-of-characteristic-potential}) for the case where there is only one
site in the sum over $n$, and in which we have used Eq.~(\ref{Eq:Aprimed-to-functional-derivative}) to replace 
$\left({\rmd {\cal A}_{ij}\big/ \rmd  \mathcal{F}_{\mu,k}} \right)$ by $-{\cal A}'_{ij}$.
Then we see that that the the above analysis corresponds to the phenomenological model
in section~\ref{Sect:nonlin-phenomen}, \green{but now the $\kappa$s in 
Eqs.(\ref{Eq:T-electrostatics1}-\ref{Eq:T-electrostatics3}) need not be taken as phenomenological constants,
they can be extracted from the microscopic theory via}
\begin{eqnarray}
\kappa_\mu = u_{\mu,L}   \mathcal{F}_{\mu,L} +  u_{\mu,R}   \mathcal{F}_{\mu,R} +  u_{\mu,{\rm gate}}   \mathcal{F}_{\mu,{\rm gate}},  
\end{eqnarray}
\green{where the recipe above (and in Appendix \ref{Appendix:weakly-nonlinear}) tells us how to calculate the  $u_{\mu,L}$, $ u_{\mu,R}$ and $ u_{\mu,{\rm gate}}$ for a given system.}

Finally, we note that Ref.~\cite{Meair-Jacquod2013} did a similar calculation for a simple model of a point-contact in the weakly nonlinear regime. 

\subsubsection{Consequences for the nonlinear regime}

The central results of Refs.~\cite{sanchezprb13,Meair-Jacquod2013,Sanchez-Lopez2013,Lopez-Sanchez2014} show the strong effect of nonlinear contributions on the thermoelectric response of the system.
This sets in whenever the temperature difference or bias is large enough that the 
nonlinear term in Eq.~(\ref{Eq:J-scatter-2ndorder-expansion}) becomes of similar order to the linear term.
This can be clearly seen in the rectification of charge and heat currents, by which we mean that the sign of the currents 
do not reverse when the sign of the thermodynamic forces are reversed.
Another crucial difference from linear response is that the heat-current is not conserved,
the heat current into the quantum dot is not the same as that which flows out, since the system must obey 
Eq.~(\ref{Eq:scatter-1st-law-two-reservoir}).

Ref.~\cite{sanchezprb13} gives a nonlinear analogue of the Wiedemann-Franz, 
defined as the ratio of heat current to temperature difference ${\Delta T}$ (with no bias) 
divided by the ratio of charge current to bias $V$ (with no temperature difference),
\begin{eqnarray}
\Lambda = {J_{ h,L}(\Delta T, V=0)\big/ \Delta T \over J_{ e,L}(\Delta T=0,V)\big/ V }.
\label{Eq:WF-ratio-nonlin}
\end{eqnarray}
They show that it is given by the usual linear-response Wiedemann-Franz ratio {\it plus} nonlinear corrections 
proportional to  the nonlinear ${\cal L}$ coefficients 
multiplied by $V$ or $\Delta T$ and divided by the linear-response $L$ coefficients.  
However, since heat is not conserved in the nonlinear terms, the value of the ratio is not unique
for a given nanostructure; that is to say it will be different 
 if the currents are measured at the right reservoir instead of the left reservoir (i.e. taking $L\to R$ in Eq.~(\ref{Eq:WF-ratio-nonlin})). 

Ref.~\cite{Meair-Jacquod2013} used the same method to show that 
the efficiency of a heat engine (or the coefficient of performance of a refrigerator) 
is no longer given by its figure of merit $ZT$, as calculated from the linear-response coefficients in 
Eqs.~(\ref{Eq:ZT-intro}). The efficiency can be large or smaller than one would predict from
the linear-response $ZT$.  This can be see phenomenologically from section~\ref{Sect:nonlin-phenomen}; 
the bias will shift the peak in transmission function, $E_0$, in a manner that
depends on the nature of the microscopic parameters.  This can either shift $E_0$ towards or away from the value which optimizes the efficiency.  If it moves $E_0$ towards its optimal value, then the efficiency will be larger than that predicted by the linear-response $ZT$.
In contrast, if it moves $E_0$ away from its optimal value, then the efficiency will be smaller than that predicted by the linear-response $ZT$.
\green{
Since  the bias is typically opposite when a system is used as a refrigerator from when it is used as a heat-engine
(see Fig.~\ref{Fig:energy-filter}), if the bias in the heat-engine configuration pushes a given system's $E_0$ away from its optimal value, then the bias in the refrigeration configuration will push $E_0$ towards its optimal value. 
Thus a system that is a worse heat-engine than expected in the non-linear regime (i.e.\ its efficiency is less than that one would predict from its linear-response $ZT$), will be a better refrigerator in the non-linear regime, 
and vice versa.}

%====================================
\subsection{Equilibrium and the zeroth law of thermodynamics}
\label{Sect:zeroth-law-scattering}

Two systems are said to be in equilibrium, if there is no particle or heat current between them when they are linked by a contact which can carry  particle and heat currents independently.
 \green{That is to say that the contact should not be a "tight-coupling" contact }that lets through particles with only one energy $E_\star$ (i.e.~${\cal T}_{ij}(E)$ should not be $\delta$-function-like in $E$), 
because such a contact always has a heat current equal to $E_\star$ times the particle current, which means 
the particle and heat currents are not independent.
\green{In its dynamic form,} the zeroth law of thermodynamics is the statement that if two systems are in thermodynamic equilibrium with a third system, then they are in thermodynamic equilibrium with each other.
Hence, if the three systems in question are reservoirs of non-interacting electrons,
then there will be no currents between them, irrespective of the nature of the scatterer that 
connects them.

It is trivial to show that the scattering theory obeys the zeroth law, in the sense that if reservoirs are in equilibrium with each other, then there are no particle or heat currents between them however they are connected. 
For any quantum scatterer placed between any number of reservoirs in equilibrium with each other, so  $T_j=T_0$ and $\mu_j=\mu_0$ for all $j$.
The fact that the scattering matrix ${\cal S}$ is unitary implies Eq.~(\ref{Eq:constraints-T_ij-c}).
When we substitute this into Eqs.~(\ref{Eq:I-initial}-\ref{Eq:I-energy-initial}), 
one sees that charge, heat and energy currents are zero.
The same is true in the case with Andreev reflection from a superconductor.
Remembering that we take the electrochemical potential of the superconductor as zero of energy ($\mu_{\rm SC}=0$), reservoir $i$ will be in equilibrium with the superconductor and with the other non-superconducting  reservoirs for $\mu_i=0$ and $T_i=T_0$.
Combining this with Eq.~(\ref{Eq:constraints-T_ij-eh-c}), one sees from Eqs.~(\ref{Eq:I-initial-eh}-\ref{Eq:I-energy-initial-eh}) that the charge, heat and energy currents 
into the non-superconducting reservoirs are all zero.  Eqs.~(\ref{Eq:Isc},\ref{Eq:Jsc}) mean that all the currents out of the superconductor are also zero. 

It is more difficult to show that equilibrium is the {\it only} condition under which there is no 
particle current nor heat current through arbitrary contacts (assuming that they carry particle and heat currents independently, as discussed above).  While it seems natural that this is the case, we do not know of a rigorous proof
for arbitrary ${\cal T}_{ij}(E)$.

\subsection{Work and the first law of thermodynamics}
\label{Sect:scatter-1st-law}

The first law of thermodynamics states that the sum of heat and work remains constant.
Once it was realized that heat and work are just different forms of energy, this
is simply a consequence of energy conservation.
In the context of a steady-state machine the first law can be cast in terms of currents,
in which case it states that the power output of the system must equal the total heat current into it.

In a thermoelectric system the work takes the form of electrical power;
a heat engine converts heat into electrical power, 
while a refrigerator uses electrical power to move heat from a colder reservoir to a hotter one.
For such electrical circuits, adding an electron to a region increases the work in that region by an amount equal to the electrochemical potential of that region. 
Removing that electron reduces the work there by the same amount.
Thus, an electron from reservoir $1$ moving to reservoir $2$ generates a total change in work equal to $(\mu_2-\mu_1)= e (V_2-V_1)$.  In terms of currents, this means the power generated in reservoir $i$ equals
$P_{{\rm gen};i} = - V_i \Jelectrici$,
which means the total power generated is
\begin{eqnarray}
P_{{\rm gen}} = \sum_i P_{{\rm gen};i} = - \sum_i  V_i \Jelectrici \, ,
\label{Eq:P_gen}
\end{eqnarray}
This can be understood by thinking that each reservoir coupled to the system could be an ideal battery (for example a very large ideal capacitor), whose other terminal is coupled to earth 
(where we take earth to be a reservoir with electrochemical potential $\mu=0$).
If the current into the scatterer from the reservoir, $\Jelectrici$, is negative, while the voltage on that reservoir, $V_i$,  is positive  (with respect to ground), then one  is charging up the battery.
Whenever $\Jelectrici$ and $V_i$ have opposite signs, 
the work in the battery is increasing at a rate $-V_i\Jelectrici >0$.
In contrast, if the current $\Jelectrici$ and bias $V_i$ have the same sign, then the current is discharging  the battery; the rate of change of work in the battery is $-V_i\Jelectrici < 0$.
Note, that the definition of the power generated in each reservoir is gauge-dependent,
that is to say that it depends on our choice of the zero of energy, which defines the electrochemical
potential of the earth reservoir.  However, the power generated summed over all reservoirs, 
Eq.~(\ref{Eq:P_gen}), is independent of this choice of the zero of energy, 
and so is gauge-independent.  This can be seen by noting that Eq.~(\ref{Eq:I-conserve})
implies that we can shift all biases by the same arbitrary amount 
without changing $P_{\rm gen}$.

The scattering theory explicitly conserves energy, as each particle leaves the scatterer with the same energy that it entered, with this conservation being apparent in 
Eq.~(\ref{Eq:I-energy-initial}).
Hence, it should be no great surprise that the theory satisfies the first law of thermodynamics.
Indeed, given the above discussion of work and electrical power, it is obvious that 
Eq.~(\ref{Eq:scatter-1st-law}) is the first law of thermodynamics; 
the left hand side is the heat absorbed by the scatterer, while
the right hand side is the power generated by the scatterer, Eq.~(\ref{Eq:P_gen}).  
Thus we have the {\it first law of thermodynamics} as 
\begin{eqnarray}
P_{\rm gen} = \sum_i \Jheati.
\label{Eq:scatter-1st-law-explicit}
\end{eqnarray}
For the case of a two reservoir system, the first law is even simpler,
and the same logic shows that it is that given in Eq.~(\ref{Eq:scatter-1st-law-two-reservoir}).
It is worth noting that
there is no special relation between the heat current out of reservoir $i$, and the power generated in reservoir $i$.  It is only when one sums over all reservoirs , that one finds the equality between
heat input and power output given by the first law of thermodynamics.

Of course, in reality no reservoir is an ideal battery.  In the worst case, reservoir $i$ could be coupled to ground through a resistor, for which one always has $\Jelectrici$ of the same sign as $V_i$.
Such a resistor would simply dissipate the power it absorbs, equal to $-P_{{\rm gen};i}=V_i\Jelectrici$,
as the electricity flowing into that resistor gets dissipated as heat which is lost into the environment.  
However, we would interpret this situation
as the scatterer turning \green{heat into work, which is injected into the reservoir, and then turned back into heat by the fact the reservoir is not ideal, i.e.\ the reservoir contains resistances.}

If there is a superconducting reservoir inducing Andreev reflection, then
the situation changes very little.
Since we have chosen to take the gauge where the zero of energy is the electrochemical potential of the superconductor, the power generated in the SC reservoir is zero.
Thus, the power generated takes a similar form to the case without the superconductor, 
Eq.~(\ref{Eq:P_gen}), except that now $i$ sum being over all non-superconducting reservoirs.
This means  that Eq.~(\ref{Eq:scatter-1st-law-eh}) is the first law of thermodynamics for situations with a superconducting reservoir,
with its left hand side being the heat flow into the scatterer and its right hand side
being the power generated by the scatterer.

\subsubsection{Two reservoir systems without thermoelectric effects: Joule heating, etc.}
\label{Sect:two-reservoir}

Although this review is mostly about using thermoelectric effects to actively convert heat into work,
or convert work into a heat flow (refrigeration), it is worth looking in more detail at the \emph{passive}
\green{work to heat} conversion known as Joule heating.  
If a scatterer has an energy independent transmission function,
then it will exhibit no thermoelectric effects, and will instead act as a resistance.
If we apply a bias across this scatterer, a current will flow, but as the scatterer has a resistance,
we know that the energy used to make the current flow is dissipated as Joule heating.

\green{In scattering theory there is no coupling between the electrons and phonons, so 
the Joule heating takes the form of an electronic heat current from the scatterer into the reservoirs.
This is given by Eq.~(\ref{Eq:scatter-1st-law-explicit})  with negative $P_{\rm gen}$.}

An interesting special case is when the the scatterer is coupled between two reservoirs at the same temperature, but with a bias $V$ between them. This bias generates a current $I$ through the scatterer.
Ref.~\cite{Gurevich1996-equidistribution-Joule-heat} showed that the scatterer generates a Joule heating, $VI$, with exactly half this heat going into \green{the electrons in} each reservoir. 
Remarkably, this result is independent of the details of the scatterer;
even if the scattering region is a dot with weak single-mode coupling to one reservoir and strong many-mode coupling to the other, the Joule heat flow into each reservoir will be the same.

The easiest way to prove this result is to take two reservoirs L and R, and choose that
their electrochemical potentials are $\mu_{ L}= \mu/2$ and $\mu_{ R}=-\mu/2$, respectively. Then, taking Eq.~(\ref{Eq:J-initial-two-reservoir}) 
and splitting the term $(E \pm \mu/2)$ into two separate integrals, 
one has
\begin{eqnarray}
\JheatL &=& {\cal T}_{ LR} \ \ \left( 
\int_{-\infty}^\infty {\rmd E \over h} \ E  
\ \left[ f_{ L}(E) - f_{ R}(E) \right] \ -\ {\mu \over 2}
\int_{-\infty}^\infty {\rmd E \over h} \   
\ \left[ f_{ L}(E) - f_{ R}(E) \right] \right)
\label{Eq:J-energy-independent}
\\
\JheatR &=& {\cal T}_{ LR} \ \ \left( 
\int_{-\infty}^\infty {\rmd E \over h} \ E  
\ \left[ f_{ R}(E) - f_{ L}(E) \right] \ +\ {\mu \over 2}
\int_{-\infty}^\infty {\rmd E \over h} \   
\ \left[ f_{ R}(E) - f_{ L}(E) \right] \right)
\end{eqnarray}
Next we note that $\left[ f_{ L}(E) - f_{ R}(E) \right]$ is an even function of $E$
for this choice of electrochemical potentials (remembering also that there is no temperature difference,
$T_{ L}=T_{ R}$).
This means that the first integral in $\JheatL$ and $\JheatR$ vanishes,
while the second integral equals $\mu/2 \times \JelectricL/e $ in both cases, 
where $\JelectricL$ is given by Eq.~(\ref{Eq:I-initial-two-reservoir}).
Thus, remembering that $\mu = e  V$, we arrive at Ref~\cite{Gurevich1996-equidistribution-Joule-heat}'s observation that the Joule heat radiated into each of the two reservoirs is the same, and equals $V\JelectricL/2$.
In other words the heat currents into the scatterer from the reservoirs are negative, and equal
\begin{eqnarray}
\JheatL = \JheatR = -{V \JelectricL \over 2}
\label{Eq:equidistribution-Joule-heat}
\end{eqnarray}
for any scatterer with an energy-independent transmission, when $T_{ L}=T_{ R}$.

In fact, if ${\cal T}_{LR}$ is $E$-independent, we
can also get simple expression for $\JelectricL$ and $\JheatL$ for $T_{ L}\neq T_{ R}$. 
We start by evaluating the $E$ integrals in $\JelectricL$ given by Eq.~(\ref{Eq:I-initial-two-reservoir}).
For this, we note that 
\begin{eqnarray}
f_i(E) 
&=& \theta (E-\mu_i)  + {{\rm sign}(E-\mu_i)  
\over 1+\exp\big[|E-\mu_i| \big/ (\kB T_i)\big]} \  , 
\end{eqnarray}
% \begin{eqnarray}
% \int_{-\infty}^\infty  \rmd E \, f_i(E) 
% &=& \int_{-\infty}^\infty \rmd E \ \theta (E-\mu_i) 
% \nonumber \\
% & & \quad + \int_{-\infty}^\infty {\rmd E \ {\rm sign}(E-\mu_i)  
% \over 1+\exp\big[|E-\mu_i| \big/ (\kB T_i)\big]} \  , 
% \end{eqnarray}
where the first term on the right is a Heaviside $\theta$-function.
Since the second term on the right hand side of this equation 
is an odd function for $E-\mu_i$, it cancels when we integrate from
$-\infty$ to $\infty$.
In this case, the integrand in Eq.~(\ref{Eq:I-initial-two-reservoir}) 
reduces to the difference of two $\theta$-functions, in which neither $T_{ L}$ nor $T_{ R}$ appears.
Performing this trivial integral one gets
\begin{align}
\JelectricL=-\JelectricR = {e^2 \over h} \ {\cal T}_{ LR} \ V
\label{Eq:I-eps-independent}
\end{align}
for any $T_{ L}$ and $T_{ R}$.
Note that ${\cal T}_{ LR}$ will typically depend on $V$, $T_{ L}$ and $T_{ R}$, as discussed 
at length in section~\ref{Sect:scatter-validity}, so the current may be a very nonlinear 
function of bias and temperature.
If $T_{ L}=T_{ R}$, 
we can then use Eq.~(\ref{Eq:equidistribution-Joule-heat}) 
directly to get $\JheatL$ and $\JheatR$.
For $T_{ L} \neq T_{ R}$, it is easiest to build upon the result for
$T_{ L} = T_{ R}$ in Eq.~(\ref{Eq:equidistribution-Joule-heat}).
From Eq.~(\ref{Eq:J-energy-independent}), we see that the difference between the result
when  $T_{ L} \neq T_{ R}$ from when  $T_{ L} \to T_{ R}$
is
\begin{align}
\JheatL - \JheatL(T_{ L}  \to T_{ R})  
\ =\ \ & {\cal T}_{ LR} \int_{-\infty}^\infty {{\rm d}E \over h} \ (E -\mu_{ L})\  
\left[ f_{ L} (E) - f_{ L}\big(E;T_{ L}  \to T_{ R}\big) \right],  
\end{align}
where $f_{ L}\big(E;T_{ L}  \to T_{ R}\big)$ 
is Eq.~(\ref{Eq:f}) with $\mu_i=\mu_{ L}$ but $T_i =T_{ R}$.
We change variables in the integrals to $\tilde E =E-\mu_{ L}$,
after which the integrand takes the form 
$g(\tilde E) = \tilde E 
\left[\left(1+\exp\left[\tilde E/(\kB T_L)\right]\right)^{-1} -\left(1+\exp\left[\tilde E/(\kB T_R)\right]\right)^{-1} \right] $. We can use the relation 
$\left(1+\e^{-x}\right)^{-1}= 1 - \left(1+\e^{x}\right)^{-1}$ to prove that   
$g(\tilde  E)$ is an even function of $\tilde E$.
This means 
$\int_{-\infty}^\infty \rmd \tilde E g(\tilde E) = 2 \int_0^{\infty} \rmd \tilde E 
\, g(\tilde E)$, and as a result all the integrals take the form
$\int_0^\infty \rmd x x\big/(1+\e^x) = \pi^2/12$.  
This gives us an algebraic result for 
$\JheatL - \JheatL(T_{ L}  \to T_{ R}) $, which 
we add to  $\JheatL(T_{ L}  \to T_{ R}) $
 in Eq.~(\ref{Eq:equidistribution-Joule-heat}).
 Thus shows that  an arbitrary two reservoir system with energy-independent transmission,  has
\begin{eqnarray}
\JheatL &=& {\cal T}_{ LR} \ \left({\pi^2 \over 6h} \left((\kB T_{ L})^2 -(\kB T_{ R})^2\right) - {1\over 2h} (e V)^2 \right).
\end{eqnarray}  
Eq.~(\ref{Eq:scatter-1st-law-two-reservoir}) with $V_{ R}-V_{ L}=V$, then tells us that the expression
for $\JheatR$ equals that for $\JheatL$ with $T_{ L}$ and $T_{ R}$ interchanged. 
Thus, in such systems, the heat current into any reservoir is simply the sum of two terms;
the first term is a conservative flow (it has opposite signs for $\JheatL$ and $\JheatR$)
given by the temperature difference in the absence of an electrical bias, and the second 
term is half the Joule heating induced by the bias (with the same sign for both reservoirs).

While the results in this section are simple, and pretty, 
they do not apply to the thermoelectric systems which we are interested in using to convert between heat and work.  
To be a thermoelectric, the system must have a ${\cal T}_{ LR}$ which depends on $E$. 
Then, in general, none of the expressions presented in this section will apply.

\subsubsection{Two reservoirs when one is a superconductor: Joule heating, etc.}

Another interesting example is that of a two-reservoir system, when reservoirs R is 
a superconductor which induces Andreev reflection.
This is the simplest example of a system coupled to
a superconductor which induces Andreev reflection.
This system is uninteresting from the point of view of heat-to-work conversion,
but it is worth studying  because it clearly shows the effect of Andreev reflection on Joule heating.

Let us say that the right reservoir is a superconductor, while the left (L) reservoir is not.
In this case, there are only terms with $j=i=L$ in Eqs.~(\ref{Eq:constraints-T_ij-eh}).
With a little care, one can rewrite these constraints as
\begin{align}
{\cal T}^{(-1,1)}_{LL}(E) \ =& \ {\cal T}^{(1,-1)}_{LL}(E)\, ,
\\
 {\cal T}^{(1,1)}_{LL}(E) \ =& \ N^{(1)}_{ L}(E) -{\cal T}^{(1,-1)}_{LL}(E)\, ,
\\
 {\cal T}^{(-1,-1)}_{LL}(E) \ =& \ N^{(-1)}_{ L}(E) -{\cal T}^{(1,-1)}_{LL}(E)\, ,
 \end{align}
where we recall that the indices in the superscripts refer to electrons ($1$) or holes ($-1$).
As we take the electrochemical potential of the superconductor as our zero of energy,
a bias of $V_{ L}$ across the scatterer corresponds to taking reservoir L's electrochemical potential
$\mu_{ L}=e V_{ L}$.
Using the above results, Eqs.~(\ref{Eq:I-initial-eh}-\ref{Eq:J-initial-eh}) reduce to
\begin{eqnarray}
\JelectricL \!\!&=& \!\! -\JelectricSC = \ 2e \!\! \int_0^\infty \!\! \rmd E \ \  {\cal T}_{LL}^{(1,-1)}(E)
\,  \left[ f_{ L}^{(1)} (E) - f_{ L}^{(-1)} (E) \right],  \qquad
\label{Eq:I-initial-eh-two-reservoirs}
\\
\JenergyL \!\!\!\! &=& \!\! -\JenergySC \ =\ \JheatSC \ =\  0 \ ,
\\
\JheatL \!\! &=& - V_{ L} \JelectricL \ .
\label{Eq:J-initial-eh-two-reservoirs}
\end{eqnarray}
Eq.~(\ref{Eq:I-initial-eh-two-reservoirs}) means that each electron from reservoir L 
reflected as a hole carries
a current of $2e$ from reservoir L into the superconductor, in the form of a Cooper pair. 
In contrast, each electron reflected as an electron
(either because it never hit the superconductor, or because it Andreev reflected from 
the superconductor an even number of times) carries no current from reservoir L into the superconductor.

The function $\left[ f_{ L}^{(1)} (E) - f_{ L}^{(-1)} (E) \right]$ is an odd function of $V_{ L}$,
always taking the opposite sign to $V_{ L}$.
Thus the current, $\JelectricL$, out of reservoir $L$ is also an odd function of $V_{ L}$,
but takes the same sign as $V_{ L}$ (remember $e$ is negative).
This means the current always flows from the reservoir with higher electrochemical potential to the one with lower electrochemical potential, and so the power generated $P_{\rm gen}= -V_{ L}\JelectricL$ is always negative (although it vanishes at $V_{ L}=0$). 
Hence the scatterer can only dissipate power as a classical resistance would, as Joule heat.
Eq.~(\ref{Eq:J-initial-eh-two-reservoirs}) 
shows that {\it all} of this Joule heat flows into the non-superconducting reservoir (L), 
as none can go into the superconducting reservoir.

Note that despite the fact we allow for an energy-dependent transmission, there is no thermoelectric effect in a two reservoir system, when one of those reservoirs is a superconductor that induces Andreev reflection.
The only role of temperature, which enters through the Fermi functions, 
$f_{ L}^{(\pm1)} (E)$, is to determine the resistance of the scatterer, 
which will typically be a nonlinear function of both $V_{ L}$ and $T_{ L}$.

\subsection{Second law of thermodynamics}
\label{Sect:second-law-scattering}

The process of entropy production is that which we usually call dissipation
in this context,
however dissipation  is treated very lightly in the scattering theory.  
It is simply assumed that dissipation occurs
when the electrons relax to a thermal state in the reservoirs.
The theory contains no microscopic model for this dissipation,
thus it is natural to wonder if this is sufficient for the theory to capture 
the physics of entropy production. In particular, it is natural to wonder whether 
the scattering theory contains the second law of thermodynamics or not.

As we will see below, despite the simplicity of its treatment of entropy production,
the scattering theory {\it does} contain the second law of thermodynamics.
This makes it clear that the second-law of thermodynamics is not reliant 
on the microscopic details of the relaxation process. It is sufficient simply that some such process exists in the reservoirs, and that it induces relaxation on a suitable timescale.

The timescale for relaxation in the scattering theory should be long compared with that of the scattering, 
only then can we treat electrons in the reservoir as non-interacting when they arrive at and when they leave the scatterer.
However, this timescale should also be short enough that electrons injected into a reservoir relax completely to a local thermal distribution \green{(a Fermi distribution determined by that reservoir's temperature and electrochemical potential)} before coming back to the scatterer,
so we can assume that electron's arriving at the scatterer all come from a Fermi distribution with the temperature and electrochemical potential
of the reservoir in question.
However, beyond these assumptions, any relaxation rate or process is acceptable for the scattering theory to work, and for that scattering theory to contain the second law of thermodynamics.

\subsubsection{Second law for a scatterer between two reservoirs}
\label{Sect:scatter-2ndlaw-two-reservoirs}

In the case of an arbitrary system with only two reservoirs, 
the proof that the scattering theory  
contains the second law is rather straightforward, \colorproofs{and has been rediscovered multiple times \cite{Nenciu2007,Bruneau2012,Whitney-2ndlaw,Yamamoto2015}}.
Defining these two reservoirs as left (L) and right (R),
the rate of change of total entropy, Eq.~(\ref{Eq:scatter-S_total}), 
is
\begin{eqnarray}
\dot{\mathscr{S}}\ = \ - {\JheatL \over T_{ L}} -{\JheatR \over T_{ R}} \ .
\label{Eq:dotS-two-term}
\end{eqnarray} 
If we substitute in Eq.~(\ref{Eq:J-initial-two-reservoir}), we get
\begin{eqnarray}
{\dot{\mathscr{S}} \over \kB} = 
- \int_0^\infty {{\rm d}E \over h}\, \big[\xi_{ L}-\xi_{ R}\big]
\ {\cal T}_{ LR}(E)   \ \big[ f(\xi_{ L}) - f(\xi_{ R}) \big], 
\label{Eq:scatter-dotS-equation-two-reservoir}
\end{eqnarray}
where one defines 
\begin{eqnarray}
\xi_i= (E - \mu_i)/(\kB T_i),
\end{eqnarray} 
and one takes $f_i(E)= f(\xi_i)$ for 
$f(\xi)= (1+\exp[\xi])^{-1}$.
Now, since $f(\xi)$ is a monotonically decaying function of $\xi$, 
the product of the two square-brackets in Eq.~(\ref{Eq:scatter-dotS-equation-two-reservoir}) cannot be positive.
Taking this together with the positivity of ${\cal T}_{ LR}(E)$ 
in Eq.~(\ref{Eq:LR=RL}), one concludes that the integrand in 
Eq.~(\ref{Eq:scatter-dotS-equation-two-reservoir}) is not positive at any energy $E$.
Thus, whatever the details of the integral over $E$, one can see that
\begin{eqnarray}
\dot{\mathscr{S}} \geq 0.
\end{eqnarray} 
Thus any two-reservoir system that obeys the scattering theory will automatically
satisfy the second-law of thermodynamics.

\subsubsection{Carnot efficiency for a scatterer between two reservoirs}
\label{Sect:scatter-nonlin-Carnot}

We can use Eq.~(\ref{Eq:scatter-dotS-equation-two-reservoir}) to show the conditions under which Carnot efficiency can be achieved \cite{linke2002}.
We will see that the conditions for achieving Carnot efficiency in the nonlinear regime are a bit stricter than
those for achieving $ZT \to \infty$ in the linear response regime \green{(see section \ref{sec:energyfiltering}).}

By examining the integrand in Eq.~(\ref{Eq:scatter-dotS-equation-two-reservoir}), 
we see that the only energy $E$ at which the transmission 
of an electron from left to right (or right to left) does not generate entropy is the $E$ for which $\xi_{ L}=\xi_{ R}$.
We define this energy as $E^\rightleftharpoons$, 
it obeys
\begin{eqnarray}
E^\rightleftharpoons-e V_{ L} &=& {e (V_{ R}-V_{ L}) \over 1-T_{ R}\big/T_{ L}} \ .
\label{Eq:eps-reversible}
\end{eqnarray}
Physically, $E^\rightleftharpoons$ is the energy at which the Fermi functions of reservoirs L and R are the same, $f(\xi_{ L})=f(\xi_{ R})$, which 
means that the flow of particles from left to right is the same as the flow from right to left
(which is why we give it the symbol $\rightleftharpoons$).
Thus, if particles only flow at this energy, then the flow is ``reversible'' in the thermodynamic sense.

For a system to be Carnot efficient, we require that it is reversible.
In other words, for Carnot efficiency, 
we require that ${\cal T}_{ LR}(E)$ is only non-zero
for $E=E^\rightleftharpoons$. To achieve this one usually considers 
${\cal T}_{ LR}(E)$ to be a Lorentzian or boxcar function centred on
$ E^\rightleftharpoons$, whose width is taken to zero.
Then, since each electron that flows carries the same charge, $e$, 
and the same energy, $E^\rightleftharpoons$, one has a trivial relationship between
$\JelectricL$ and $\JenergyL$, 
\begin{eqnarray}
\JenergyL\ =\ {E^\rightleftharpoons \over e}\, \JelectricL.
\end{eqnarray}
Thus, remembering that $\JheatL=\JenergyL-V_{ L}\JelectricL$,
 a reversible system has
\begin{eqnarray}
{\JelectricL \ (V_{ R}-V_{ L}) \over \JheatL} 
\ =\  
{e (V_{ R}-V_{ L}) \over 
E^\rightleftharpoons-e V_{ L} } 
\ =\ 1 - {T_{ R} \over T_{ L}},
\label{Eq:IV-versus-J-reversible}
\end{eqnarray}
where we have used Eq.~(\ref{Eq:eps-reversible}) to get the second equality.

Let us consider a thermoelectric heat-engine, which is using the heat flow out of a hot reservoir L, $\JheatL$, to generate electrical power $\JelectricL V$, by driving an electrical current $\JelectricL$ against a bias of $V$ (i.e.~the electrons flows from the reservoir with a lower electrochemical potential to the one with a high electrochemical potential). 
If we take reservoir L to be at hot (H) temperature $T_{ H}$, 
and reservoir R to be at cold (C) temperature $T_{ C}$,
a heat-engine made from the above reversible system, Eq.~(\ref{Eq:IV-versus-J-reversible}),  has efficiency
\begin{eqnarray}
\eta_{\rm eng} \equiv  {\JelectricL \ (V_{ R}-V_{ L}) \over \JheatL}  
\ =\ 1 - {T_{ C} \over T_{ H}}.
\end{eqnarray} 
This is the Carnot efficiency for a heat-engine, so the thermodynamically reversible system has Carnot efficiency, as expected.

Similarly, we can consider a thermoelectric refrigerator, which extracts a heat current 
$\JheatL$ from a cold reservoir L, by absorbing the electrical power $\JelectricL V$. 
The absorbed electrical power must come from an electrical current $\JelectricL$, driven by a bias $V$ (note that here the current flows in the direction of the bias, when the flow was against the bias for the heat-engine). 
If we take reservoir L to be at cold (C) temperature $T_{ L}$, 
and reservoir R to be at ambient (0) temperature $T_0$,
a refrigerator made from the above reversible system, Eq.~(\ref{Eq:IV-versus-J-reversible}),  has a coefficient of performance (COP),
\begin{eqnarray}
\eta_{\rm fri} \equiv  {\JheatL \over \JelectricL (V_{ L}-V_{ R})} 
\ =\ {1 \over T_0 \big/ T_{ C} \ -\ 1}.
\end{eqnarray} 
This is the Carnot efficiency for a refrigerator, 
so again the thermodynamically reversible system has Carnot efficiency, as expected.

\subsubsection{Second law in presence of any number of reservoirs}
\label{Sect:scatter-2ndlaw-general}

One can also prove that the second law follows from the scattering theory for an arbitrary scatterer coupled to an arbitrary number of reservoirs at arbitrary temperatures and biases \cite{Nenciu2007}.  
The prove relies only on the structure of the
scattering theory and the positivity of the transmission functions (which in turn comes from the unitarity of the scattering matrix).  It is a little more technical than that given above for two terminal systems.
%because here one cannot assume that  ${\cal T}_{ji}^{\vsig\vrho} \neq {\cal T}_{ij}^{\vrho\vsig}$.  
However, this proof applies even if one of the reservoirs is a superconductor inducing Andreev reflection
(or multiple reservoirs are superconductors, if they are all at the same chemical potential) \cite{Whitney-2ndlaw}.
Here we reproduce Nenciu proof \cite{Nenciu2007}, which is much more elegant than Ref.~\cite{Whitney-2ndlaw}'s proof.

Taking the rate of change of total entropy in Eq.~(\ref{Eq:scatter-S_total})
with the heat currents given by Eq.~(\ref{Eq:J-initial-eh}), 
we can write
\begin{eqnarray}
\dot{\mathscr{S}} \ =\ {\kB \over h}\int_0^\infty  \rmd E \ Z (E), 
\label{Eq:dotS_and_Z}
\end{eqnarray} 
with the integrand
\begin{eqnarray}
Z(E) \ =\  
- \sum_{ij \neq {\rm SC}} \sum_{\vrho\vsig}\ \xi_{i\vrho} \ 
{\cal A}_{ij}^{\vrho\vsig}(E)  \ f(\xi_{j\vsig}), \qquad
\label{Eq:Z}
\end{eqnarray} 
where $i$ and $j$ are summed over all the non-superconducting reservoirs.
Here  we have defined
\begin{eqnarray}
\xi_{i\vrho} \ =\ {E - \vrho e V_i \over \kB T_i},
\qquad \qquad
{\cal A}_{ij}^{\vrho\vsig}(E)  \ =\ N_i^\vrho(E) \delta_{ij}\delta_{\vrho\vsig}-{\cal T}_{ij}^{\vrho\vsig}(E),
\end{eqnarray}
and
$f(\xi) = \left(1+\exp\left[\xi \right] \right)^{-1}$, 
as in section~\ref{Sect:scatter-2ndlaw-two-reservoirs}.  
Our objective is to prove that $Z (E) \geq 0$ for all $E$,
irrespective of the nature of the scatterer,
an immediate consequence of this will be that the second law is satisfied.

The first step is to note that the quantity similar to $Z(E)$ but with $\xi_{i\vrho}$ replaced by $\xi_{j\vsig}$,
obeys 
\begin{eqnarray}
\sum_{ij \neq {\rm SC}} \sum_{\vrho\vsig}\ \xi_{j\vsig} \ 
{\cal A}_{ij}^{\vrho\vsig}(E)  \ f(\xi_{j\vsig}) \ = \ 0
\end{eqnarray} 
as can be seen by using Eq.~(\ref{Eq:constraints-T_ij-eh-c}) to evaluate the sums over $i$ and $\vrho$.
Adding this term to Eq.~(\ref{Eq:Z}) gives
\begin{eqnarray}
Z(E) \ =\  
\sum_{ij \neq {\rm SC}} \sum_{\vrho\vsig}\ 
\left(-{\cal A}_{ij}^{\vrho\vsig}(E) \right)  \ \left(\xi_{i\vrho}-\xi_{j\vsig}\right)\ f(\xi_{j\vsig}). \qquad
\label{Eq:Z2}
\end{eqnarray} 
Now we will use a mathematical trick, for which we need to define a function $F(x) = \int^x \rmd \xi f(\xi)$. Since we know that $f(\xi)$ is a monotonically decaying function of $\xi$, we know that $F(x)$ is a concave function of $x$.
A known inequality for concave functions is
that 
\begin{eqnarray}
F\big(x\big) - F\big(x_0\big) &\leq& (x-x_0 ) \,f(x_0)\ .
\end{eqnarray}
This inequality can be understood as saying the value of a concave function at $x$ is always less than the value of that function's linear Taylor expansion about $x_0$ (this is easy to see graphically).
We identify $x$ with $\xi_{i\vrho}$ and  $x_0$ with $\xi_{j\vsig}$, and substitute this inequality into the right hand side of Eq.~(\ref{Eq:Z2}), and note that Eq~(\ref{Eq:constraints-T_ij-eh-a}) 
means that  ${\cal A}_{ij}^{\vrho\vsig}(E)$ is negative
for all non-zero contributions to the sums; i.e.\ ${\cal A}_{ij}^{\vrho\vsig}(E)$ is only positive when $i=j$ and $\vrho=\vsig$, but the factor of $(\xi_{i\vrho}- \xi_{j\vsig})$ means that such terms make no contribution.
This gives
\begin{eqnarray}
Z(E) \ \geq\  
\sum_{ij \neq {\rm SC}} \sum_{\vrho\vsig} \ 
\left(-{\cal A}_{ij}^{\vrho\vsig}(E) \right)  \ \left(F(\xi_{i\vrho})-F(\xi_{j\vsig}) \right). \qquad
\label{Eq:Z-inequality}
\end{eqnarray} 
Then, Eq.~(\ref{Eq:constraints-T_ij-eh-c}) tells us that the first term in the sum gives zero when summed over $j$ and $\vsig$, and the second term in the sum gives zero when summed over $i$ and $\vrho$.
Thus we immediately have $Z(E) \geq 0$ for any $E$.
Substituting this result into Eq.~(\ref{Eq:dotS_and_Z}), 
we can state that any system modelled by the scattering theory will obey,
\begin{eqnarray}
\dot{\mathscr{S}}  \ =\ - \sum_i {\Jheati \over T_i}\ \geq \ 0 \ ,
\label{Eq:scatter-2nd-law}
\end{eqnarray}
which is the second-law of thermodynamics for such a system.

We recall that this scattering theory does not capture all physical processes in quantum transport.
In particular, it does not capture interaction effects beyond the mean-field level, and it cannot handle multiple superconducting reservoirs at different chemical potentials.  So more general derivations of the second-law would be 
worthwhile.

\subsubsection{Consequence of the second law for Joule heating}

Here we note that the second law has a strong consequence for Joule heating.
For arbitrary temperature differences between reservoirs, a scatterer may absorb heat (generating power)
or create heat (absorbing electrical power).  

However if all reservoirs are at the same temperature, the
system cannot generate electrical power.  As a result it behaves like a resistance, absorbing electrical power 
whenever there are electrical currents through it, and turning that power into Joule heating. 
To see that this is always the case we just have to note  that Eq.~(\ref{Eq:scatter-2nd-law}) in the case where all reservoirs have the same temperature reduces to $\sum_i \Jheati  \leq 0$. Combining this with the first law,
Eq.~(\ref{Eq:scatter-1st-law-explicit}),
give us
\begin{eqnarray}
P_{\rm gen} \ =\ \sum_i \Jheati \ \leq \ 0 \ .
\end{eqnarray}
Thus the scatterer absorbs electrical power (negative $P_{\rm gen}$, and generates generates Joule heat
(sum of heat flows into scatterer is negative).

 \subsection{Upper bound on heat flow and upper bounds on efficiency at given power output}
 
 Having shown that systems modeled by scattering theory always obey the bounds given by the laws of thermodynamics, we now show that quantum mechanics places different bounds on such systems.
 The best known is the Bekenstein-Pendry upper bound on heat flow, which we discuss in the following section.
Sections~\ref{Sect:max-power-and-theta-funct} and \ref{Sect:max-eta-given-P} then discusses the consequences of this bound (or more strictly the aspects of scattering theory which lead to this bound)
for the efficiency of a heat-engine or refrigerator.   In particular, section~\ref{Sect:max-eta-given-P} shows that quantum mechanics can place a stricter upper bound on efficiency than classical thermodynamics alone.

 \subsubsection{The Bekenstein-Pendry bound on heat flow and Nernst's unattainability principle}
 \label{Sect:Nernst}
 
Bekenstein \cite{Bekenstein1981,Bekenstein1984} and Pendry  \cite{Pendry1983} independently noted that there is an upper bound on the heat current  through a single transverse mode. \green{This bound is most easily derived within scattering theory \cite{Pendry1983,Whitney-2ndlaw}; it comes from the quantization of thermal conductance, 
combined with the fact that zero temperature is special. 
Ref.~\cite{Pendry1983} found the maximum heat carried away from reservoir $i$ at temperature $T_i$ by the flow of electrons through a constriction carrying $N$ transverse modes.  Ref.~\cite{Whitney-2ndlaw} made the straightforward generalization to include finite biases and a superconducting reservoir.}
The heat flow out of reservoir $i$ is maximal when that reservoir is coupled to another reservoir at zero temperature \green{(and at the same electrochemical potential)} via  a constriction which lets particles flow at all energies, ${\cal T}(E) = N$.  From Eq.~(\ref{Eq:J-initial-two-reservoir}),  
it then follows that the Bekenstein-Pendry limit on heat flow,
\begin{eqnarray}
J^{\rm max}_{{ h},i} \ =\   {2 \over h} \ N \ \kB^2 T_i^2 \int_0^\infty {x \ \rmd x  \over 1+e^x} 
\ =\ 
{\pi^2 \over 6 h}\  N \ \kB ^2 T_i^2 \ .
\label{Eq:Jmax}
\end{eqnarray}
The number of transverse modes, $N$, is given by  the cross-section in units of the Fermi wavelength of the electrons. 
This means that the maximum rate of entropy flow out of reservoir $i$ is 
\begin{eqnarray}
\dot{\mathscr{S}}^{\rm max}_{{ h},i} \ =\ 
{\pi^2 \over 6 h}\  N \ \kB ^2 T_i \ .
\end{eqnarray}
One has a stricter bound if one restricts to heat flow through a scatterer between a reservoir at temperature $T_{ L}$ and another at $T_{ R}$, without performing any work on reservoirs or scatterer.
\green{This means the reservoirs must be at the same electro-chemical potential to ensure that
neither of them can perform work, and the scattering potential must be time-independent to ensure that it cannot perform work.}
In this case, the maximum heat current out of reservoir L is  
\begin{eqnarray}
J^{\rm max\,(no\,work\, input)}_{{ h,L}} \ =\ \left\{ \begin{array}{ccl} 
{\displaystyle {\pi^2 \over 6 h}}\  N \ \kB ^2 \left(T_{ L}^2 - T_{ R}^2 \right) & \qquad & \hbox{ for } T_{ L} > T_{ R},
\\
0 \phantom{\Big|}  & \qquad & \hbox{ for } T_{ L} < T_{ R}.
\end{array}\right.
\label{Eq:Jmax-no-work}
\end{eqnarray}
\green{For $T_L>T_R$, the bound is reached when the scatterer transmits particles at all energies.
For $T_L <T_R$, the bound is reached when the scatterer reflects particles at all energies.}

The upper bounds in Eqs.(\ref{Eq:Jmax},\ref{Eq:Jmax-no-work}) are of quantum origin;
if we take a naive classical limit for a system with a given cross-section by taking the wavelength to zero, we see that $N\to \infty$, and there ceases to be an upper bound on the heat flow.  However, in reality $N$ is always finite,
and $J^{\rm max}_{{ h},i}$ is 0.5 pW per transverse mode per Kelvin-squared.  To get a better feeling of what this means, suppose we are trying to an object at $T_{ L}=$600K (the typical temperature of an exhaust pipe of a car) by connecting it to the filament of a lightbulb ($10^{-5}$m is diameter with a Fermi wavelength of $10^{-10}$m) the other end of which is connected to a lump of metal at ambient temperature $T_{ R}=$300K.  Then, it can only carry heat out of a metallic reservoir at a rate less than 1350W.  While this is a very large heat flow of such a narrow wire, it is not many orders of magnitude above the heat typically carried by lightbulb filaments.  Thus, the upper bound imposed by quantum mechanics  is not so irrelevantly large as to only be of academic interest.

These Bekenstein-Pendry bound was observed experimentally in point contacts \cite{Molenkamp-Peltier-thermalcond}, and recently verified to high accuracy in quantum Hall edge states \cite{Jezouin2013}.  
Remarkably, if one does the same scattering theory calculation for phonons or photons (replacing the Fermi functions in the scattering theory by Bose function), one arrives at exactly the same result \cite{Pendry1983,Maynard-Akkermans1985}.
\green{In the case of photons, one can make a direct connection to the  Stefan-Boltzmann law  for black-body radiation in the limit of an aperture much larger than the typical photon wavelength. In this limit, the number of traverse photon modes in the aperture, $N$, scales like the temperature-squared, so Eq.~(\ref{Eq:Jmax}) scales like $T^4$, giving  the  Stefan-Boltzmann law.  The difference for electrons in metallic samples is that the number of traverse modes, $N$, is determined principally by the wavelength at the Fermi energy, so $N$ is only weakly dependent on temperature.}

Remarkably the bound in Eq.~(\ref{Eq:Jmax-no-work}) means that there is an {\it upper} bound 
on the rate of entropy production of any two-terminal system to which we do not supply work, such as a heat-engine.    \green{We all know that the second law of thermodynamics gives a lower bound on entropy production, but now we see that quantum mechanics also places an upper bound on the entropy production.  }
The maximum heat flow out of reservoir L is given by 
Eq.~(\ref{Eq:Jmax-no-work}), while the heat flow out of reservoir R is $P_{\rm gen} - J_{ L}$, where the power generated by the system $P_{\rm gen} \geq 0$ (since we are not supplying power to the system), 
but is clearly smaller than $J_{ L}$.  Thus, 
Eq.~(\ref{Eq:dotS-two-term}) is maximal when  $P_{\rm gen} \to 0$, which means that the upper bound on 
the rate of entropy production is
\begin{eqnarray}
\dot{\mathscr{S}}_{\rm max\,(no\,work\, input)}  \ = \  {\pi^2 \over 6 h}\  N \ \kB ^2 \left(T_{ L} - T_{ R} \right)^2 \ \left({T_{L}+T_{R} \over T_{L} T_{R}} \right) .
\label{Eq:max-dotS-no-work}
\end{eqnarray}
Of course, this bound does not apply if we supply work to the system (for which $P_{\rm gen}$ would be negative), because all the work we supply may be converted into heat in the reservoirs, and hence increases their entropy.

We now turn to the case in which we supply work to the system.
As show in Fig.~\ref{Fig:energy-filter}c,  one can move heat from reservoir L to reservoir R even if $ T_{ L} < T_{ R}$, so long as one supplies work to the system in the form of a bias between reservoirs L and R.
 However, no matter how much work one does, the upper bound on the heat extracted \cite{whitney-prl2014,whitney2015} is exactly half $J^{\rm max}_{{ h},L}$ in Eq.~(\ref{Eq:Jmax}).
 \green{This bound is achieved by a two terminal system, in which the scatterer transmits all electrons
 with energies $\eps > \mu_L$ and reflects all electrons with $\eps < \mu_L$
(as in Fig.~\ref{Fig:energy-filter}c) in the limit where $\mu_R$ is very far below $\mu_L$, such that $(\mu_L-\mu_R)/T_L \to \infty$. 
The same bound also applies to three terminal refrigerators. \cite{whitney2016} in which heat is extracted from the reservoir being cooled, without extracting charge from it, by forcing an electrical current to flow between two other electronic reservoirs\footnote{In this case the upper bound on refrigeration is half 
$J^{\rm max}_{h,i}$, where $N$ is the sum of the number of modes on the two leads that carry the electrical current.}}

\green{
We will follow Ref.~\cite{Whitney-2ndlaw} in showing that this leads directly to Nernst's unattainability 
principle,  sometime called the third law of thermodynamics,
which states that it is impossible to cool a system to absolute zero in a finite time.
The temperature change of a reservoir, $\rmd T$, associated with extracting a given amount of heat, $\rmd Q$, from it is determined by that reservoir's heat capacity $C(T) = \left(\rmd Q \big/ \rmd T\right)$.  Thus the rate of change of temperature of reservoir $i$ when heat is extracted at a rate $J_i(T_i)$ is 
\begin{eqnarray}
{\rmd T_i \over \rmd t} \ =\ -{ J_i(T_i) \over C_i(T_i)}\ ,
\nonumber
\end{eqnarray}
where $C_i(T_i)$ is the heat capacity of reservoir $i$ at temperature $T_i$.
The heat capacity of a reservoir of
free electrons is $C_i \propto T_i$, 
so if the maximum heat flow is  $J^{\rm max}_{{ h},i} \propto T_i^2$, then it is easy to see that
the temperature decay is given by}
\begin{eqnarray}
{\rmd T_i \over \rmd t} \  \propto\  -T_i\ .
\label{Eq:weak-Nernst}
\end{eqnarray}
This means that in the ideal case the temperature $T_i$ can decay exponentially towards zero, but can never reach $T_i=0$ on any finite timescale. For non-ideal cases the temperature drop will be slower, and usually stops at a finite temperature.  
Note that if $J^{\rm max}_{{ h},i} \propto T_i^\zeta$ for any exponent $\zeta < 2$, then one can show that the temperature would reach zero in a finite time, 
\green{while a system with $\zeta > 2$ will never reach zero temperature.  
A system that achieves the Bekenstein-Pendry bound on the cooling of a reservoir of free electrons
(and so has $\zeta=2$) is critical. By causing the reservoir temperature to drop exponentially with time, 
it obeys the unattainability principle in its weakest possible form; the reservoir temperature never reaches zero even if it gets exponentially close to zero on a finite timescale.
}

\green{
Note that in interacting quantum systems there has been recent controversy about whether the Nernst unattainability principle is valid.  There were indications that it might not always be valid 
\cite{kolvar2012-violate-Nernst,Levy2012a,Levy2012b,vandenbroeck2012}, 
followed by a number of claims that it is valid 
\cite{Allahverdyan2012-comment,Levy2012-comment,Cleuren-reply2012,Entin-Imry2014-comment,Masanes2014,Freitas2016}. 
It has also been shown \cite{BS15} that quantum mechanics imposes a fundamental 
limitation for cooling by cyclic engines of the type discussed in section~\ref{sec:CTM},
}  
\red{
whose origin is rooted in the dynamical Casimir effect (DCE) 
(for a review of this and other quantum vacuum amplification phenomena see 
\cite{Nation12}).
The DCE concerns the generation of photons from the vacuum due to time-dependent 
boundary conditions or more generally to the change of some parameters of a system. 
Ref.~\cite{BS15} considered a reciprocating refrigerator, 
operating by means of a working medium (a single mode of the electromagnetic field, that is, 
a harmonic oscillator, with a time-dependent frequency), shuttling heat from a cold 
finite-size ``bath'' (a single qubit) to a hot bath. The working medium undergoes 
a four-stroke Otto cycle. Even assuming the ideal case in \green{an} isochore
stroke of the cycle the qubit and the oscillator are prepared in their ground state,
due to the DCE both the oscillator and the qubit are excited, so that at the end of the 
isochore stroke the qubit is left in a state at a nonzero temperature
(note that in this case the change of system's parameters is the switching on/off
of the qubit-oscillator coupling at the beginning/end of the isochore stroke).
As a consequence, for finite-time Otto cycles, the qubit does not attain the
\green{absolute} zero of temperature, even in the limit of an infinite number of cycles. 
% In contrast the qubit thermalizes to a nonzero temperature. 
This fundamental limitation for cooling imposed by the DCE has been recently
confirmed in a more general setup, where a linear and periodically driven quantum
system is coupled with bosonic reservoirs \cite{Freitas2016}.
}

 \subsubsection{Maximum power output of a heat-engine}
 \label{Sect:max-power-and-theta-funct}
 
Refs.~\cite{whitney-prl2014,whitney2015} pointed out that the Bekenstein-Pendry upper bound on heat flow, must place a similar upper bound on the power generated by a heat-engine (since the efficiency is always finite). 
A very quick over-estimate of this upper bound can be made by noting that if one does not provide work to the system,
then the maximum heat-flow between reservoir L and R is given by the bound in Eq.~(\ref{Eq:Jmax-no-work}), and the efficiency must be less than Carnot's efficiency.
This means that the power output for a machine with $N$ transverse modes must be less than
${\pi^2 \over 6h} N \kB^2 (1+T_{ R}/T_{ L}) (T_{ L}-T_{ R})^2$.
However, there is a clear competition between maximizing heat flow (which requires allowing electrons to flow at all energies) and maximizing the efficiency (which involves blocking electron flow at all energies except $E^\rightleftharpoons$, see section~\ref{Sect:scatter-nonlin-Carnot}), which makes this bound unattainable.

Using a method of optimization analogous to the one that we presented in section~\ref{Sect:max-ZT-phonons},
Refs.~\cite{whitney-prl2014,whitney2015} found the following strict upper-bound on the power generated;
\begin{eqnarray}
P_{\rm gen}^{\rm max} \,\equiv\, 
 A_0\, {\pi^2 \over h} \ N\  \kB^2 \big(T_L-T_R\big)^2, 
 \label{Eq:Pmax}
\end{eqnarray}
where $A_0 \simeq 0.0321$. 
This bound is strict in the sense that it is never exceeded, but is achieved by 
a system with a transmission function in the form of a Heaviside $\theta$-function (i.e. a high-pass filter)
which lets through all particles with $E \geq E^\rightleftharpoons$ as defined in Section~\ref{Sect:scatter-nonlin-Carnot}) when one takes 
$e V = 1.146 \,\kB (T_L-T_R)$.

\subsubsection{Maximum efficiency at given power output}
\label{Sect:max-eta-given-P}

%========================================
\begin{figure}
\centerline{\includegraphics[width=0.6\textwidth]{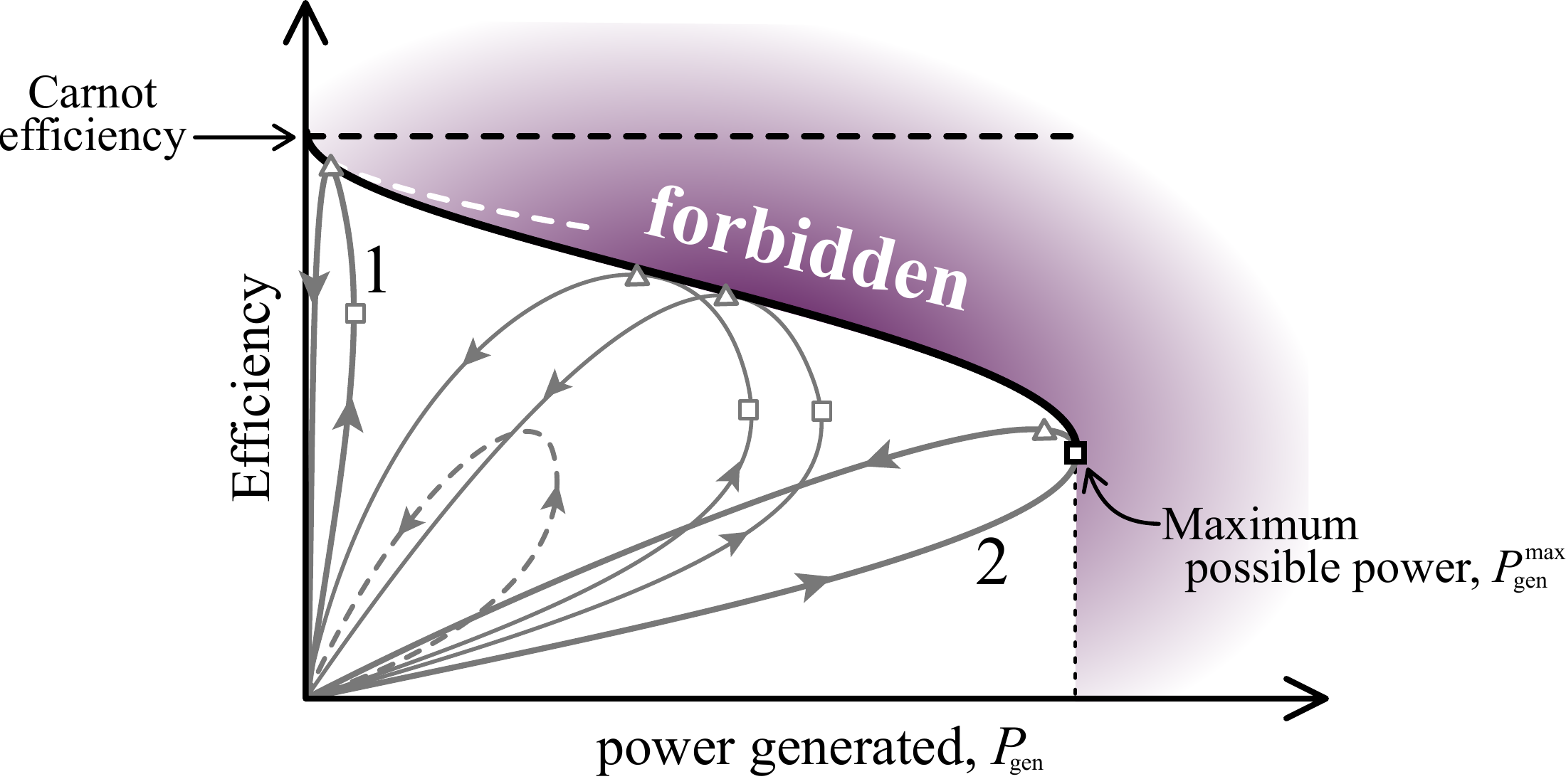}}
\caption{\label{Fig:Eff-max-vs-Pgen}
Here we take given temperatures $T_{ L}$ and $T_{ R}$, 
and sketch
power-efficiency curves (the grey loops) for systems with different transmission functions.
The thick black curve is the envelope of all the loops; it separates the region accessible by systems with
suitably chosen transmission functions, from the region of efficiencies and powers that no system can achieve.  
For small power generation this envelope tends to the dashed white curve given by Eq.~(\ref{Eq:eta-eng-small-Pgen}).
The maximum power generation (black square) is given by Eq.~(\ref{Eq:Pmax}), with efficiency given by Eq.~(\ref{Eq:Eff-at-Pmax}). 
Each loop is formed (in the manner indicated by the arrows) by taking a given heat-engine and changing the resistance of the load upon it from zero up to infinity (i.e. increasing the bias from zero up to the stopping voltage).
The triangle marks that system's highest efficiency, while the square marks its highest power generation.
The details of these loops will depend on how the transmission function varies with bias (which will depend on the nature of the screening, etc.).  Loop 1 is for a system with a narrow transmission function
as in section~\ref{Sect:scatter-nonlin-Carnot}, which has a low power output, but is capable of 
achieving a high efficiency (close to Carnot efficiency). This system has a high efficiency at maximum power, which can be close to the Curzon-Ahlborn efficiency, if its parameters are tuned carefully \cite{elb09}. Loop 2 is a system with a transmission in the form of a Heaviside $\theta$-function, as in 
section~\ref{Sect:max-power-and-theta-funct}, 
its maximum efficiency is lower, but its maximum power is much higher.
With correct tuning it achieves the highest power generation of any system, as sketched here.
Only specific systems (those with transmission-functions which act as the correct type of band-pass filters) 
have loops which touch the envelope. 
All other systems will have power-efficiency curves significantly below the envelope, such as the dashed loop.
}
\end{figure}
%========================================

A given system has a given transmission function ${\cal T}_{ LR}(E)$, and as a result it has 
a given curve of efficiency against power output as a function of bias, typically a loop as sketched 
in Fig.~\ref{fig:eta_power},
with a given maximum power.
The phenomenological theory used for Fig.~\ref{fig:eta_power} gives this curve for linear response, but is unable to 
capture the maximum power of a given system, instead the horizontal axis in Fig.~\ref{fig:eta_power} gives
the system's normalized power; that is to say the power as a fraction of that system's unknown maximum power.
This hides the fact that the systems with the highest efficiencies typically also have the lowest 
maximum powers.  While this observation of a competition between efficiency and power is fairly common, to-date only scattering theory has explicitly shown that such a competition is unavoidable \cite{whitney-prl2014,whitney2015,BS2015}.
% For an example, see the ``loops'' sketched in Fig.~\ref{Fig:Eff-max-vs-Pgen}.

Refs.~\cite{whitney-prl2014,whitney2015} used an optimization method similar to 
 that which we presented in section~\ref{Sect:max-ZT-phonons} to find the envelope defined by these
 ``loops'' for all conceivable systems described by scattering theory.  That is to say that these works found the boundary which separates the region of efficiencies and powers
 achievable by systems that are described by scattering theory, and the region that no system can achieve
 whatever its transmission function.
They thus showed that one cannot get close to Carnot efficiency unless 
the power generation is much less than $P_{\rm gen}^{\rm max}$.  As we increase the desired power generation towards $P_{\rm gen}^{\rm max}$, the maximum possible efficiency decays monotonically.
The system which achieves the maximal efficiency for a given $P_{\rm gen}$ is one which has a transmission in the form of a band pass filter;
it lets though all particles in a window between $E_0$ and $E_1$, and blocks all other particles.
In general, $E_0$ and $E_1$ are given by an ugly transcendental equation, so there is no
 closed form algebraic expression for this maximal upper bound at arbitrary $P_{\rm gen}$.
 However, one can observe that larger $P_{\rm gen}$ requires a
 wider the band-pass filter (i.e. the greater the difference between $E_1$ and $E_0$), 
 up to the point where $E_1$ goes to infinity for $P_{\rm gen}\to P_{\rm gen}^{\rm max}$.
 
 In the limit $P_{\rm gen}\big/ P_{\rm gen}^{\rm max}\ll 1$, one can get an algebraic expression for the upper bound on heat-engine efficiency as a function of $P_{\rm gen}$; it is 
\begin{eqnarray}
\eta_{\rm eng} \big(P_{\rm gen}\big) =  \eta_{\rm eng}^{\rm Carnot} 
\left(1- 0.478
\sqrt{  {T_R \over T_L} \ {P_{\rm gen} \over P_{\rm gen}^{\rm max}}} \ + \ {\cal O}\left[P_{\rm gen} \big/ P_{\rm gen}^{\rm max}\right]  
\right) . \quad
\label{Eq:eta-eng-small-Pgen}
\end{eqnarray}
In the limit of maximum power generation,  $P_{\rm gen}=P_{\rm gen}^{\rm max}$, 
the upper bound on efficiency is 
\begin{eqnarray}
\eta_{\rm eng} (P_{\rm gen}^{\rm max}) 
&=& {\eta_{\rm eng}^{\rm Carnot}\over 1+0.936 (1+T_R/T_L) }.
\label{Eq:Eff-at-Pmax}
\end{eqnarray}
One might find it surprising that the efficiency at $P_{\rm gen}^{\rm max}$ 
is not vanishingly small, indeed it is more than one third of $\eta_{\rm eng}^{\rm Carnot}$.   
However,  this is less surprising when one recalls that it is the upper-bound on heat-flow that is at the origin of this effect.  As the power generated equals the heat-flow multiplied by the efficiency, the  efficiency must be  reasonably large when one achieves the maximum power output $P_{\rm gen}^{\rm max}$.

Refs.~\cite{whitney-prl2014,whitney2015} calculated similar expressions for the upper bound on refrigerator efficiency as a function of cooling power.   As pointed out in Eq.~(\ref{Eq:Jmax}), the maximum cooling power is 
$\half J^{\rm max}_{ L}$, and is achieved by a system with a transmission ${\cal T}_{ LR}= N\theta(E-\mu_{ L})$, where $\theta(E)$ is a  Heaviside function, and corresponds to blocking all particles with energies less than
Reservoir L's electrochemical potential $\mu_{ L}$ and letting through all those with energies above $\mu_{ L}$.
The refrigerator only gets close to Carnot efficiency for cooling powers much less than $\half J^{\rm max}_{ L}$.
As one increases the desired cooling power towards $\half J^{\rm max}_{ L}$, the maximum possible efficiency decays monotonically.

There are various differences between the expressions for the refrigerator and the heat-engine, but the basic picture remains the same.  The system that achieves maximum efficiency for given cooling power is one whose transmission takes the form of a band-pass filter, only letting through electrons with energies between $E_0$ and $E_1$ (although the form of $E_0$ and $E_1$ are inversed with respect to those of a heat-engine).
Larger cooling power $J_{ L}$ requires a wider band-pass filter, with the upper bound on the band-pass filter going to infinity if we wish to achieve $J_{ L}\to \half J^{\rm max}_{ L}$.
At lower cooling powers the upper bound on refrigerator efficiency
is given by
\begin{eqnarray}
\eta_{\rm fri}(J_{ L}) = \eta_{\rm fri}^{\rm Carnot} 
\left(1- 1.09
\sqrt{
\,{T_{ R} \over T_R-T_{ L}}\ {J_{ L} \over J_{ L}^{\rm max}} }\ + \ {\cal O}\left[ J_{ L} \big/ J_{ L}^{\rm max}\right]  \right),
\nonumber \\
\label{Eq:eta-fri-smallJ}
\end{eqnarray}

The biggest difference between the refrigerator and the heat engine is that the refrigerator's efficiency 
(coefficient of performance) vanishes 
at maximum cooling power.  In other words, one must supply an infinite amount of electrical power to achieve 
a cooling power equal to  $\half J^{\rm max}_{ L}$.  Although, one can get exponentially close to $\half J^{\rm max}_{ L}$ with a finite supply of electrical power.

Ref.~\cite{whitney2016} considered three-terminal machines modelled by scattering theory, and found that they had the same upper bounds on efficiency at given power output as those discussed above for two-terminal heat-engines
and refrigerators.  In that work the basic geometry of three-terminal heat-engines was that considered elsewhere in this review; one reservoir is at the hot temperature and supplies heat current to the system, while the power is generated by a current between the other reservoirs which are both at the cold temperature.  However, its proofs only apply to those three-terminal machines which can be described by scattering theory.

 \subsection{Maximizing efficiency when phonons also carry heat}
 \label{Sect:whitney2015}

%%%%%%%%%%%%%%%%%%%%%%%%%%%%
\begin{figure}
\centerline{\includegraphics[width=0.3\columnwidth]{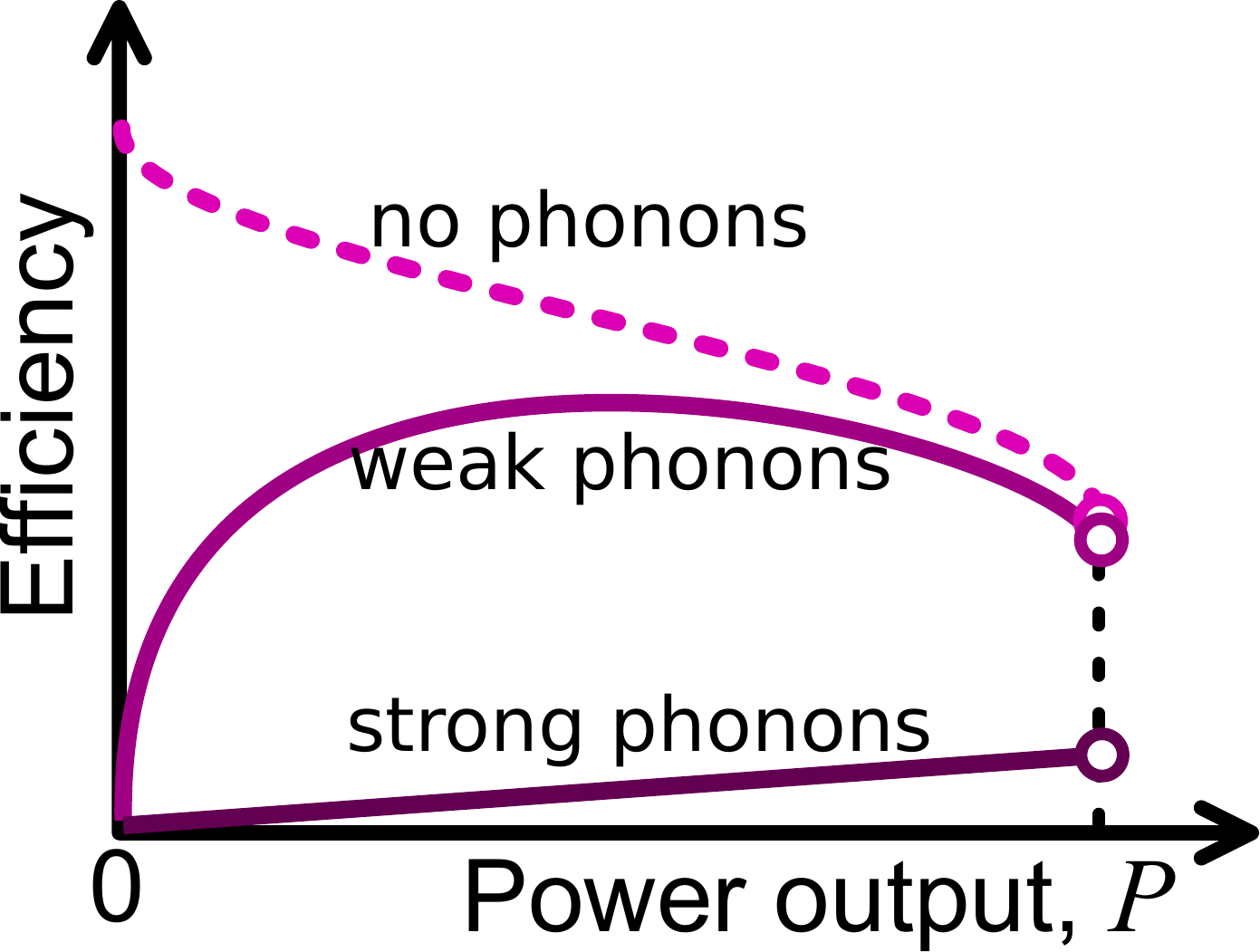}}
\caption{\label{Fig:eff-vs-power-phonons} 
Sketch of the maximum heat-engine efficiency as a function of the power it generates, in the presence of a phonon or photon heat flow from hot to cold, as sketched in Fig.~\ref{Fig:phonons-intro}.  More detailed curves can be seen in Ref.~\cite{whitney2015}. 
}
\end{figure}
%%%%%%%%%%%%%%%%%%%%%%%%%%

Here we consider phonon or photon heat flow in parallel with the electronic heat flow, as described in 
section~\ref{Sect:intro-phonons}, and ask what effect it has on the maximum efficiency at given power output.
In what follows we will refer to phonons, but the arguments we make could equally apply to photons.

For a heat-engine, if the heat flow out of the hot (left) reservoir in the absence of phonons 
is $J_{ L}$, then the heat flow in the presence of phonons will be  $J_{ L}+J_{\rm ph}$.
We will assume that the heat carried by the phonons, $J_{\rm ph}$, is dependent on the properties of the insulation between the hot reservoir L and the cold environment (assumed to be at the same temperature as reservoir R)
and on the temperatures  $T_{ L}$ and $T_{ R}$.
However, we will assume it is independent of the details of the thermoelectric systems (their transmission function, etc.) and of the bias across them.
In this case, the heat-engine's efficiency when generating power $P_{\rm gen}$ in the presence of the phonons is
\begin{eqnarray}
\eta^{\rm e+ph}_{\rm eng}(P_{\rm gen})\ =\ {P_{\rm gen} \over J_{L}(P_{\rm gen})+J_{\rm ph}} \ =\  {1 \over \eta_{\rm eng}^{-1} (P_{\rm gen}) +J_{\rm ph} / P_{\rm gen} },  
\label{Eq:eng-e+ph}
\end{eqnarray}
where $\eta_{\rm eng} (P_{\rm gen})$ is the heat-engine's efficiency 
in the absence of phonons (i.e.~for $J_{\rm ph}=0$).   
Given the maximum efficiency at given power in the absence of phonons, discussed 
in section~\ref{Sect:max-eta-given-P},
we can use this result to find the maximum efficiency for a given phonon heat flow, 
$J_{\rm ph}$.
An example of this is sketched in Fig.~\ref{Fig:eff-vs-power-phonons}.

The sketch in Fig.~\ref{Fig:eff-vs-power-phonons}, makes a number of points.
Firstly, Carnot efficiency is never possible for finite $J_{\rm ph}$.
Secondly,  phonons have a huge effect on the efficiency at small power output;
the efficiency vanishes at zero power output for any finite $J_{\rm ph}$,
with
\begin{eqnarray}
\eta^{\rm e+ph}_{\rm eng}(P_{\rm gen})=P_{\rm gen}\big/J_{\rm ph} \ \ \  \hbox{ for } \ P_{\rm gen} \ll J_{\rm ph}.
\label{Eq:eta-with-phonons-smallP}
\end{eqnarray}
For weak phonon heat flows, $J_{\rm ph} \ll P_{\rm gen}^{\rm max}$, the presence of the phonons has little 
effect on the efficiency near the maximum power output.
For strong phonon flow, where $J_{\rm ph} \gg  P_{\rm gen}^{\rm max}$, 
Eq.~(\ref{Eq:eta-with-phonons-smallP}) applies at all powers up to the maximum, $P_{\rm gen}^{\rm max}$.
Then, the efficiency is maximal when the power is maximal, where maximal power is the quantum bound given in Eq.~(\ref{Eq:Pmax}).
Section~\ref{Sect:max-eta-given-P} explained that this maximum power occurs for a system whose transmission is 
a Heaviside step function. Hence this result coincides with that 
in section~\ref{Sect:max-ZT-phonons} for strong phonon flows in the linear-response regime.

Now let us turn to the case of a refrigerator.
Whenever one is trying to refrigerate the cold (left) reservoir,
the presence of phonons carries a back flow of heat $J_{\rm ph}$ from hot to cold,
Hence to extract heat from reservoir L at rate $J$,
the refrigerator must actually extract heat at a rate $J_L=J+J_{\rm ph}$.
Here, for clarity, we take $J_{\rm ph}$ to be positive when $T_L< T_R$
(so it has the opposite sign from Eq.~(\ref{Eq:J_h,ph})).
\green{
The refrigerator's efficiency (coefficient of performance) 
is the heat current extracted, $J$, divided by the electrical power $P_{\rm abs}(J+J_{\rm ph})$ that is 
required to extract heat at the rate  $(J+J_{\rm ph})$.
Given that in the absence of phonons  $\eta_{\rm fri}(J)= J/P_{\rm abs}(J)$, we can write 
$P_{\rm abs} = (J+J_{\rm ph})\big/ \eta_{\rm fri}(J+J_{\rm ph})$.  Then  the efficiency (coefficient of performance) in the presence of phonons is }
\begin{eqnarray}
\eta_{\rm fri}^{\rm e+ph}(J) &=&  {J \, \eta_{\rm fri}(J+J_{\rm ph}) \over J+J_{\rm ph}} ,
\label{Eq:fri-e+ph}
\end{eqnarray}
where $\eta_{\rm fri} (J)$ is the refrigerator efficiency in the absence of phonons.
This means that once we have the maximum efficiency for given cooling power in the absence of phonons, as discussed in section~\ref{Sect:max-eta-given-P}, we can easily find the maximum efficiency at given cooling power in the presence of the phonon heat flow $J_{\rm ph}$.  The result is much the same as for a heat-engine, with one notable exception;
Eq.~(\ref{Eq:fri-e+ph}) means that the phonons reduce the maximum cooling power, 
so $J$ must now obey
\begin{eqnarray}
J&\leq& \half J_L^{\rm max} -J_{\rm ph},
\label{Eq:Jqb-phonons}
\end{eqnarray}
with $J_L^{\rm max}$ given in Eq.~(\ref{Eq:Jmax}).
Thus, the upper bound on cooling power reduces as $J_{\rm ph}$ increases.

The above bound has a direct consequence on the lowest temperature that can be achieved by the refrigerator,
because the refrigerator with the highest cooling power will be the one that achieves the lowest temperature.
To see this let us take one with the maximum cooling power, $\half J_L^{\rm max}$, which corresponds to a system with a Heaviside step transmission function, such as the point-contact discussed in section~\ref{Sect:scatter-pointcont}, see also Ref.~\cite{whitney2013-catastrophe}.
If reservoir L (the reservoir one wishes to refrigerate) is at the ambient temperature, $T_{ R}$,
when one starts the refrigeration, then
initially one has $J_{\rm ph}=0$ so heat is extracted at a rate equal to  $\half J_L^{\rm max}$.  However, as reservoir $L$ is cooled down through this lose of heat (reducing $T_L$),
$J_{\rm ph}$ grows and $J_L^{\rm max}$ shrinks, thus the heat extraction rate must go down.  
Well before $T_{ L}$ reaches zero, one arrives at the situation where $J_{\rm ph} =  \half J_L^{\rm max}$,
and any further cooling of reservoir $L$ is impossible.  Thus, if one has the $T_L$ dependence of $J_{\rm ph}$ for a given system,  the lowest temperature  
that reservoir $L$ can be refrigerated to, is given by the solution of the equation $J_{\rm ph} =  \half J_L^{\rm max}$.

We also note that, as with the heat-engine, phonons have a huge effect on the refrigerator efficiency at small cooling power.   For $J_{\rm ph} <  \half J_L^{\rm max}$, one has
\begin{eqnarray}
\eta^{\rm e+ph}_{\rm fri}(J)=J \ {\eta_{\rm fri}(J_{\rm ph}) \over J_{\rm ph}}\ \ \ \ \hbox{ for } \ J \ll J_{\rm ph},
\end{eqnarray}
which means that the efficiency  vanishes for small cooling power whenever phonons are present. 

\subsection{Lower limit on entropy production at given power output}
  
 Consider a two-terminal heat-engine that generates a power $P_{\rm gen}$ at an efficiency $\eta_{\rm eng}(P_{\rm gen})$.  This tells us that the heat flow out of the left (hot) reservoir is 
 $J_{ L}= P_{\rm gen}\big/\eta_{\rm eng}(P_{\rm gen})$, 
 while  the heat flow out of the right (cold) reservoir is 
 $J_{ R}= P_{\rm gen} - J_{ L} = P_{\rm gen}\left( 1- 1\big/\eta_{\rm eng}(P_{\rm gen})\right)$.
 Inserting these equations into 
 Eq.~(\ref{Eq:dotS-two-term}), we see that the rate of entropy production at given power generation 
 (for given $T_{ L}$ and $T_{ R}$) 
 is entirely determined by the efficiency at that power generation, by the relation
 \begin{eqnarray}
 \dot{\mathscr{S}}(P_{\rm gen}) &=& {P_{\rm gen} \over T_{ R}} \left({ \eta_{\rm eng}^{\rm Carnot} \over \eta_{\rm eng}(P_{\rm gen})} -1 \right) \,.
 \label{Eq:dotS-for-given-Pgen}
 \end{eqnarray}
 Thus entropy is produced for any efficiency less than Carnot efficiency.  One should not be confused by the factor 
 of $P_{\rm gen}$, it does not mean that any system which generates zero power will generate zero entropy.
 A system which generates zero power, may easily have a finite heat flow $J_{ L}$, in which case $\eta_{\rm eng}(P_{\rm gen}) =0$, then the above equation indicates that its entropy production rate will be finite.
 
 The form of Eqs.~(\ref{Eq:dotS-for-given-Pgen}) means that an upper bound on $\eta_{\rm eng}(P_{\rm gen})$
 immediately implies a lower bound on the rate of entropy production.  If we concentrate on the regime small
 power generation given by Eq.~(\ref{Eq:eta-eng-small-Pgen}), we find that the entropy production of a heat-engine
 generating a power $P_{\rm gen}$ must obey
 \begin{eqnarray}
  \dot{\mathscr{S}}(P_{\rm gen}) &\geq& 
  {0.478 P_{\rm gen}^{\rm max} \over\sqrt{ T_{ R} T_{ L}}} \  \left(\left( {P_{\rm gen} \over P_{\rm gen}^{\rm max}} \right)^{3/2} \ + \ {\cal O}\left[\left( {P_{\rm gen} \over P_{\rm gen}^{\rm max}} \right)^2 \right]  \ \right). \quad
 \end{eqnarray}
Taking this result together with Eq.~(\ref{Eq:max-dotS-no-work}), we see that quantum mechanics imposes both a lower bound and an upper bound on the rate of entropy production for a heat-engine.
 
 If we now consider a refrigerator which extracts heat $J_{ L}$ from the left (cold) reservoir with an efficiency (coefficient of performance) $\eta_{\rm fri}(J_{ L})$, we see that the heat flow out of the right (hot) reservoir  is $J_{ R}= - J_{ L}-P_{\rm abs}$, where the power absorbed by the refrigerator $P_{\rm abs}=J_{ L}\big/ P_{\rm abs}$.  Inserting these equations into 
 Eq.~(\ref{Eq:dotS-two-term}), we find that  the rate of entropy production of such a refrigerator is
 \begin{eqnarray}
 \dot{\mathscr{S}}(J_{ L}) &=& {J_{ L} \over T_{ R}} \ \left({1 \over \eta_{\rm eng}(J_{ L})} -{1 \over \eta_{\rm fri}^{\rm Carnot} } \right) \,.
 \label{Eq:dotS-for-given-JL}
 \end{eqnarray}
 As mentioned in section~\ref{Sect:max-eta-given-P}, 
 there is an upper-bound on refrigerator efficiency at given cooling power,
 much like the bound on heat-engine efficiency sketched in Fig.~\ref{Fig:Eff-max-vs-Pgen}.
Combining this efficiency bound with Eq.~(\ref{Eq:dotS-for-given-JL}) directly implies a lower bound on the entropy production of a refrigerator.
 For small cooling powers, we can insert Eq.~(\ref{Eq:eta-fri-smallJ}) into Eq.~(\ref{Eq:dotS-for-given-JL})
 to find that the entropy production of a refrigerator with cooling power $J_{ L}$ must obey
\begin{eqnarray}
\dot S \big(J_L\big) \  \geq \    
1.09 
{J^{\rm max}_{ L} \over T_{ L}}  \left( \sqrt{1-{T_{ L}\over T_{ R}}}  
\left({J_{ L} \over J_{ L}^{\rm max}} \right)^{3/2}\, +
 \ {\cal O}\left[\left( {J_{L} \over J_{L}^{\rm max}} \right)^2 \right]  \ \right).
 \nonumber \\
\label{Eq:dotS-fri-smallJ}
\end{eqnarray}
  
 It is easy to show that Eq.~(\ref{Eq:dotS-for-given-Pgen}) 
applies to the three-terminal heat-engines, 
 where one reservoir is at the hot temperature and two are at the same cold temperature, 
 with the power being generated between the two cold reservoirs.
Similarly  Eq.~(\ref{Eq:dotS-for-given-JL}) applies to three-terminal refrigerators,
where one extracts heat from a cold reservoir by driving an electrical current between two reservoirs at the same ambient temperature. 
 Since Ref.~\cite{whitney2016} has shown that such systems have the same efficiency bounds as two terminal systems, all results in this section apply equally to such three-terminal systems.

%% file: kubo.tex
%\section{Quantum theories II: Green-Kubo approach}
\section{Aspects of thermoelectricity in interacting systems}

\label{sec:interacting}

Strongly interacting systems 
are of great interest, since it appears that interactions are another avenue to large thermoelectric effects. 
Experimental results  on some strongly correlated materials
such as sodium cobalt  oxides revealed unusually large thermopower
values \cite{Terasaki1997,Wang2003}, in part attributed to strong
electron-electron interactions \cite{Peterson2007}.
A fundamental motivation for the study of 
interacting systems will be discussed in Sec.~\ref{sec:phaset}:
an analogy between a classical heat engine and a thermoelectric material
suggests, for strongly interacting systems, the possibility of large values 
of the thermoelectric figure of merit $ZT$ near electronic phase 
transitions \cite{Vining1997}.
In general, very little is known about the thermoelectric
properties of interacting systems: analytical results
are rare and numerical simulations challenging.
However, on the basis of the Green-Kubo formula, we will discuss 
a thermodynamic argument suggesting that the Carnot efficiency
is achieved in the thermodynamic limit for non-integrable 
momentum-conserving systems.

\subsection{Thermoelectricity and electronic phase transitions}
\label{sec:phaset}

A reasoning by Vining \cite{Vining1997} suggests that large values
of $ZT$ can be expected near electronic phase transitions.
First of all, we consider the thermal conductance at zero voltage:
\begin{equation}
K'\equiv \left( 
\frac{J_{h}}{\Delta T}
\right)_{\Delta V=0}=\frac{L_{hh}}{T^2}=K+GS\Pi. 
\end{equation} 
The thermoelectric figure of merit can then be written as
\begin{equation}
ZT=\gamma_K-1,
\quad \gamma_K\equiv \frac{K'}{K};
\end{equation}
obviously $ZT$ diverges if the ratio $\gamma_K$ diverges.

We now focus on the thermodynamic properties of the working fluid
itself rather than on transport. 
Consider an open system characterized by the number $N$
of particles, the chemical potential $\mu$ and the temperature $T$.
We have
\begin{equation}
dN=
\left. \frac{\partial N}{\partial \mu}\right|_T d\mu+
\left. \frac{\partial N}{\partial T}\right|_\mu dT,
\quad 
d \mathscr{S} =-\frac{\mu}{T}\,dN+ \frac{dU}{T}=
-\frac{\mu}{T}\,
\left(\left. \frac{\partial N}{\partial \mu}\right|_T d\mu+
\left. \frac{\partial N}{\partial T}\right|_\mu dT\right)
+\frac{1}{T}\,\left(
\left. \frac{\partial U}{\partial N}\right|_T dN+
\left. \frac{\partial U}{\partial T}\right|_N dT\right),
\end{equation}
where $U$ is the internal energy of the system. 
These equations can be written in a form similar to 
the coupled transport equations (\ref{eq:coupledlinear}):
\begin{eqnarray}
\left\{
\begin{array}{l}
dN=C_{NN} d\mu + C_{N\mathscr{S}} dT,
\\
\\
d\mathscr{S}=C_{\mathscr{S}N} d\mu + C_{\mathscr{S}\mathscr{S}} dT,
\end{array}
\right.
\label{eq:capacitymatrix}
\end{eqnarray}
where the \emph{capacity matrix} ${\bm C}$ has elements
\begin{equation}
C_{NN}=\left. \frac{\partial N}{\partial \mu}\right|_T,
\quad
C_{N\mathscr{S}}=\left. \frac{\partial N}{\partial T}\right|_\mu,
\quad
C_{\mathscr{S}N}=\frac{1}{T}
\left. \frac{\partial N}{\partial \mu}\right|_T
\left( \left.\frac{\partial U}{\partial N}\right|_T-\mu\right),
\quad
C_{\mathscr{S}\mathscr{S}}=\frac{1}{T}\,\left[
\left. \frac{\partial U}{\partial T}\right|_N+
\left. \frac{\partial N}{\partial T}\right|_\mu
\left( \left.\frac{\partial U}{\partial N}\right|_T-\mu\right)\right].
\end{equation}
Note that $C_{N\mathscr{S}}=C_{\mathscr{S}N}$ due to a Maxwell-type relation
\begin{equation}
\left(\frac{\partial N}{\partial T}\right)_\mu=
\left(\frac{\partial \mathscr{S}}{\partial \mu}\right)_T.
\end{equation}
Moreover, 
\begin{equation}
C_{\mathscr{S}\mathscr{S}}=
\left(\frac{\partial \mathscr{S}}{\partial T}\right)_\mu\equiv C_\mu
\end{equation}
is the entropy capacity at constant $\mu$. Finally, the entropy
capacity at constant $N$ is
\begin{equation}
C_{N}\equiv\left(\frac{\partial \mathscr{S}}{\partial T}\right)_N
=\frac{\det{\bm C}}{C_{NN}},
\end{equation}
where the last equality is derived after setting $dN=0$ in 
(\ref{eq:capacitymatrix}).

We now consider a thermodynamic \emph{cycle} consisting of 
two constant chemical potential strokes $d\mu$ apart and 
two constant particle number strokes $dN$ apart. 
The infinitesimal work performed by this cyclic process is
$-d\mu dN$ and we can compare it with the work 
$d\mathscr{S} dT$ performed by a Carnot cycle 
consisting of two isothermal strokes $dT$ apart and two adiabatic
strokes $d\mathscr{S}$ apart. The ratio between the 
heat to work conversion efficiencies
of the above two processes is therefore given by 
\begin{equation}
\frac{\eta}{\eta_C}=\frac{-d\mu dN}{d\mathscr{S} dT}.
\end{equation} 
As the Carnot efficiency for a cycle operating between 
temperatures $T$ and $T+dT$ is $\eta_C=dT/T$, we obtain
\begin{equation}
\eta=\frac{-d\mu dN}{Td\mathscr{S}}
=\frac{-d\mu (C_{NN} d\mu + C_{N\mathscr{S}} dT)}{T
(C_{\mathscr{S}N} d\mu + C_{\mathscr{S}\mathscr{S}} dT)}.
\end{equation}
This formula is analogous to Eq.~(\ref{eq:efficiency})
for the efficiency of thermoelectric transport. 
Similarly to Sec.~\ref{sec:ZT}, we can show that the maximum
of $\eta$ over $d\mu$, for a fixed $dT$, is given by
\begin{equation}
\eta_{\rm max}=
\eta_C\,
\frac{\sqrt{Z_{\rm th} T+1}-1}{\sqrt{Z_{\rm th}T+1}+1},
\label{eq:ZTth}
\end{equation}
where the \emph{thermodynamic figure of merit}
\begin{equation}
Z_{\rm th} T=\frac{C_{N\mathscr{S}}^2}{\det {\bm C}}= 
\gamma_{\mu N}-1, \quad \gamma_{\mu N}\equiv \frac{C_\mu}{C_N}.
\end{equation}
We point out that $Z_{\rm th} T$ is purely determined by the
properties of the working fluid, without referring to 
thermoelectric transport. 
Consequently, it does not include any contribution
from phonons, that instead affect the thermoelectric figure of merit $ZT$.

As a final step, we use the mapping $\mu\rightarrow -p$ and $N\to V$,
with $p$ and $V$ pressure and volume of a classical gas. We then 
consider the infinitesimal work $dpdV$ performed by a cycle consisting
of two isobaric strokes $dP$ apart and two isochoric strokes
$dV$ apart and compare it again with the work 
$d\mathscr{S} dT$ performed by a Carnot cycle.   
By using the same steps as above for the $\mu-N$ system,
we find that the maximum of the heat to work conversion efficiency
$\eta$ over $dp$, for a fixed $dT$, is given by
Eq.~(\ref{eq:ZTth}). 
The thermodynamic figure of merit for the $p-V$ systems reads
\begin{equation}
Z_{\rm th} T=
\gamma_{pV}-1, \quad \gamma_{pV}\equiv \frac{C_p}{C_V},
\end{equation}
where
\begin{equation}
C_p\equiv T \left(\frac{\partial \mathscr{S}}{\partial T}\right)_p,
\quad
C_V\equiv T \left(\frac{\partial \mathscr{S}}{\partial T}\right)_V
\end{equation}
are the heat capacity at constant pressure and volume, respectively. 
For a classical ideal (noninteracting) 
gas, $1<\gamma_{pV}\le \frac{5}{3}$, with the 
upper bound achieved for monatomic gases. Hence, $Z_{\rm th} T \le  
\frac{2}{3}$. 
On the other hand, the ratio $\gamma_{pV}$ (and $Z_{\rm th}$) can diverge for 
condensable gases, at the critical temperature $T_c$ between the gas 
phase and the two-phase region (gas-liquid coexistence). 
The analogy with a classical gas suggests the possibility of large values
of $Z_{\rm th}$ close to electronic phase transitions, strongly improving
the thermoelectric properties of the working fluid with respect to 
noninteracting systems. Indeed, it has been recently 
demonstrated \cite{Ouerdane2015} that $Z_{\rm th} T$ diverges
when approaching from the normal phase the critical point for
the transition to the superconducting phase.

It is worth mentioning here that the impact of a phase transition on the 
efficiency
of a cyclic quantum engine performing an Otto cycle was recently 
investigated
by Campisi and Fazio\cite{campisifazio}.
They considered interacting systems of size $N$, with the cycle
operated at the verge of a second-order phase transition. Their analysis 
was based on
finite-size scaling theory and their  key ingredient is the divergence of 
the specific heat with the system size at the phase transition.
They showed that, provided the critical exponents of the transition
fulfill a suitable condition, then at the thermodynamic limit 
$N\to\infty$ one can approach
the Carnot efficiency, $(\eta-\eta_C)\sim N^{-a}\to 0$ (with $a>0$), 
while keeping the ``power per resource'' fixed, namely the power $P\sim N$.
\green{It should be stressed that this means that the engine 
cannot achieve Carnot efficiency at finite power for any finite $N$, but it can do so in the limit $N\to \infty$.}
\red{
A similar result was obtained by 
Allahverdyan et al.~\cite{Allahverdyan13} when considering a generalized
Carnot cycle (i.e., not restricted to quasi-static processes), 
the working substance in contact with the thermal
baths being a quantum system of size $N$. In that paper, 
it was shown that it is possible to obtain $\eta\to\eta_C$ for $N\to\infty$, at finite
output power.
}

\subsection{Green-Kubo formula}
\label{sec:Kubo}

While the Landauer-B\"uttiker approach cannot be applied to interacting 
systems, the linear response regime can be numerically investigated 
in equilibrium simulations by using the Green-Kubo formula.
Such a formula is rooted in the fluctuation-dissipation theorem,
in that it relates the equilibrium noise (i.e. fluctuations) to the
linear response transport coefficients (i.e. dissipation). 
Indeed, the Green-Kubo formula expresses the Onsager kinetic coefficients
of Eq.~(\ref{eq:coupledlocal}) in terms of \emph{dynamic correlation functions} 
of the corresponding
current operators, calculated at thermodynamic equilibrium (see for
instance \cite{kubo,mahan}):
%\begin{eqnarray}
\begin{equation}
%L_{ab} &=& \lim_{\omega\to 0} {\rm Re} [L_{ab} (\omega)], 
\lambda_{ab} = \lim_{\omega\to 0} {\rm Re} [\lambda_{ab} (\omega)], 
%\\
\quad
%L_{ab}(\omega) &=& \lim_{\epsilon\to 0}
\lambda_{ab}(\omega) = \lim_{\epsilon\to 0}
\int_0^\infty\!\!\!\!dt e^{-i(\omega\!-\!i\epsilon)t}\lim_{\Omega\to\infty}\frac{1}{\Omega}
\int_0^\beta\!\!\!\!d\tau\langle \hat{J}_a \hat{J}_b (t+i\tau)\rangle, 
%\label{eq:GK}
\label{eq:kubo}
\end{equation}
%\end{eqnarray}
where $\beta=1/k_B T$,
% ($k_B$ it the Boltzmann constant),
$\langle \; \cdot \;\rangle = \left\{{\rm tr}[(\;\cdot\;) \exp(-\beta H)]\right\}/{\rm tr} [\exp(-\beta H)] $
denotes the thermodynamic expectation value at temperature $T$, 
$\Omega$ is the system's volume, and
the currents are $J_a=\langle {\hat J}_a \rangle$, with $\hat{J}_a$ 
being the total current operator.
Note that in extended systems, the operator 
${\hat J}_a=\int_\Omega d{\bm r} {\hat j}_a({\bm r})$ is an extensive quantity,
where ${\hat j}_a({\bm r})$ is the current density operator, 
satisfying the continuity equation
\begin{equation}
\frac{d\hat{\rho}_a({\bm r},t)}{d t} = \frac{i}{\hbar}\,[H,\hat{\rho}_a] = 
-\nabla \cdot {\hat{j}}_a({\bm r},t).
\label{eq:continuity}
\end{equation}
Here $\hat{\rho}_a$ is the density of the corresponding conserved quantity, 
that is, electric charge for the electric current and energy for 
the energy current.  
Eq.~(\ref{eq:continuity}) can be equally well written in classical 
mechanics, provided the commutator is substituted by the Poisson
bracket multiplied by the factor $i\hbar$. 
It can be shown that 
the real part of $\lambda_{ab}(\omega)$ can be decomposed into a
$\delta$-function at zero frequency defining a generalized \emph{Drude weight}
$D_{ab}$ (for $a=b$ this is the conventional Drude weight) 
and a regular part $\lambda_{ab}^{\rm reg}(\omega)$:
\begin{equation}
{\rm Re} \lambda_{ab}(\omega)=
2\pi D_{ab}\delta(\omega)+\lambda_{ab}^{\rm reg}(\omega).
\label{eq:Lreg}
\end{equation}
The matrix of Drude weights can be 
%within linear response 
also expressed in terms of time-averaged current-current correlations directly:
\begin{equation}
D_{ab}=\lim_{\bar{t}\to\infty}\frac{1}{{\bar{t}}}\int_0^{\bar{t}} 
dt \lim_{\Omega\to\infty}\frac{1}{\Omega}\int_0^\beta d\tau\langle \hat{J}_a (0) \hat{J}_b (t+i\tau)\rangle.
\label{drudematrix}
\end{equation}
It has  been shown that non-zero  Drude  weights,  ${D}_{ab}\ne  0$,
are a signature of ballistic
transport~\cite{Zotos1997,Zotos2004,Garst2001,H-M2005}, namely in the
thermodynamic limit the kinetic  coefficients $\lambda_{ab}$ 
diverge linearly with the system size. 

The linear response Green-Kubo formalism has been used to
investigate the thermoelectric properties of
one-dimensional integrable and non-integrable
strongly correlated quantum lattice models, see for instance 
\cite{chaikin1976,furukawa2005,prelovsek2005,Peterson2007,shastry2009,shastry2013}.
In spite of the generality and usefulness of the Green-Kubo formalism, 
there are a few significant limitations. 
First of all, it is a linear response theory, 
while many problems in nanoelectronics require a 
framework that can handle far from equilibrium quantum transport.
Moreover, the Green-Kubo formula is derived in the thermodynamic
limit and therefore its use for small system sizes is not 
well justified. Finally, the assumption of local thermal 
equilibrium is crucial \footnote{See, however, Ref.~\cite{Dhar2009}
which discussed a Green-Kubo formula for heat conductance
(rather than conductivity) in \emph{finite open} systems.}.
The nonequilibrium Green's function formalism (also referred to as
the Keldysh formalism) is often used 
instead of the Green-Kubo formula to analyze quantum transport in small 
systems \cite{datta}. 

\subsection{Conservation laws and thermoelectric transport}
\label{sec:cmotion}

The   way   in   which    the   dynamic   correlation   functions   in
Eq.~(\ref{eq:kubo})  decay  determines   the  ballistic,  anomalous  or
diffusive character of the heat and charge transport.  
It has been understood  that  this  decay  is directly  related  to  the
existence  of conserved dynamical quantities
\cite{Zotos1997,Zotos2004}. For quantum spin chains and 
under suitable conditions,
it  has been  proved that systems
possessing conservation  laws  exhibit  ballistic transport  at
finite temperature \cite{Ilievski2013}.

The following argument \cite{Benenti2013} highlights the role 
that conserved quantities play in the thermoelectric efficiency.
The  decay of time  correlations for the currents
can  be related  to the  existence of
conserved  quantities  by  using  \emph{Suzuki's formula}
\cite{Suzuki1971}, which  generalizes an inequality  proposed  by  Mazur
\cite{Mazur1969}.
Consider  a   system  of   size  $\Lambda$ along the direction of
the currents (we denote its volume as $\Omega(\Lambda)$) and 
Hamiltonian $H$, 
with a set of $M$  \emph{relevant conserved  quantities} ${Q}_m$,
$m=1,\ldots  ,M$, namely  the commutators  $[{H},Q_m]=0$.
A constant of motion $Q_m$ is by definition relevant if
it is not orthogonal to the currents under consideration,
in our case $\langle \hat{J}_e Q_m \rangle \ne 0$ and
$\langle \hat{J}_u Q_m \rangle \ne 0$.
%where $\langle\;\cdot\;\rangle$  denotes the  equilibrium
%thermodynamic  average.
It is assumed that the $M$ constants of motion are orthogonal,   i.e.,
$\langle   Q_m   Q_n\rangle   =   \langle   Q_n^2   \rangle
\delta_{mn}$ (this is always possible via a Gram-Schmidt procedure).
%Furthermore, let  us  assume  that the constants of motion $Q_m$  are
%linearly   extensive,  meaning   that $\langle   Q_m^2\rangle  \propto
%\Omega$.
Furthermore, we assume that the set $\{Q_m\}$ exhausts
all relevant
extensive
conserved quantities.
(in the thermodynamic limit
$\Omega\to\infty$).
Then using Suzuki's formula,
we can express the \emph{finite-size Drude weights}\footnote{Note that 
hereafter we shall use the simple thermal average correlator 
$\langle \hat{J}_a(0) \hat{J}_b(t) \rangle$ rather than 
the Kubo-Mori inner product  
$\int_0^\beta d\tau\langle \hat{J}_a (0) \hat{J}_b (t+i\tau)\rangle$;
see \cite{Ilievski2013} for a discussion of the assumptions needed
to justify the use of the simple thermal-averaged expression.}
\begin{equation}
d_{ab}(\Lambda)\equiv  \frac{1}{2\Omega(\Lambda)}
\lim_{\bar{t}\to\infty}\frac{1}{\bar{t}}
\int_0^{\bar{t}} dt \langle \hat{J}_a(0) \hat{J}_b(t) \rangle
\end{equation}
in terms of the relevant conserved quantities:
\begin{equation}
d_{ab}(\Lambda)=\frac{1}{2\Omega(\Lambda)}
\sum_{m=1}^M \frac{\langle \hat{J}_aQ_m \rangle \langle \hat{J}_bQ_m
  \rangle}{\langle Q_m^2 \rangle}.
\label{eq:finitesizedrude}
\end{equation}
On the other hand, the thermodynamic Drude weights can also be expressed
in terms of time-averaged
current-current correlations as
\begin{equation}\label{eq:Drude}
{D}_{ab}=\lim_{\bar{t}\to\infty}\lim_{\Lambda\to \infty}
\frac{1}{2\Omega({\Lambda}) \bar{t}}
\int_0^{\bar{t}} dt \langle \hat{J}_a(0) \hat{J}_b(t) \rangle .
\end{equation}
If the thermodynamic limit $\Lambda\to\infty$ commutes with the long-time
limit $\bar{t}\to\infty$, then the thermodynamic Drude weights ${D}_{ab}$
can be obtained as
\begin{equation}
{D}_{ab}=\lim_{\Lambda\to\infty} d_{ab}(\Lambda)\ .
\label{eq:drudeinfty}
\end{equation}
Moreover, if the limit does not vanish we can conclude that the presence
of relevant conservation laws yields non-zero generalized Drude weights,
which in turn imply that transport is ballistic, $\lambda_{ab}\sim \Lambda$.
As a consequence, the electrical conductivity is ballistic,
$\sigma\sim \lambda_{ee}
\sim \Lambda$, while the thermopower is asymptotically size-independent,
$S\sim \lambda_{eh}/\lambda_{ee}\sim \Lambda^0$.

We can see from Suzuki's formula that for systems with a single relevant
constant of motion ($M=1$), the ballistic contribution to $\det {\bm\lambda}$
vanishes, since it is proportional to ${D}_{ee}{D}_{hh}-
{D}_{eh}^2$, which is zero from Eqs. (\ref{eq:finitesizedrude})
and (\ref{eq:drudeinfty}).
Hence, $\det {\bm \lambda}$ grows slower than $\Lambda^2$,
and therefore the thermal conductivity $\kappa\sim \det{{\bm \lambda}}/
L_{ee}$ grows sub-ballistically, $\kappa\sim \Lambda^\alpha$, with $\alpha<1$.
Since $\sigma\sim\Lambda$ and $S\sim\Lambda^0$, we can conclude
that $ZT\sim \Lambda^{1-\alpha}$ \cite{Benenti2013}.
Hence $ZT$ diverges in the thermodynamic limit $\Lambda\to\infty$. This general
theoretical argument applies for instance to systems where momentum is the
only relevant conserved quantity.

Note that these 
conclusions for the thermal conductance and the figure of merit
do not hold when $M>1$, as it is typical for completely integrable systems.  
In that case we have, in general, 
$D_{ee}D_{hh}-D_{eh}^2\ne 0$, so that thermal conductance 
is ballistic and therefore $ZT$ is size-independent.

The above reasoning is not limited to quantum systems and has no dimensional 
restrictions; it has been illustrated by means of a diatomic 
chain of hard-point colliding particles \cite{Benenti2013}
(see details on the numerical simulation of classical 
reservoirs in \green{Appendix}~\ref{sec:numerics_c}),
where the divergence of the figure of merit with the system 
size cannot be explained in terms of the energy
filtering mechanism \cite{Saito2010}, 
in a two-dimensional 
system connected to reservoirs \cite{Benenti2014}, 
with the dynamics simulated by
the multiparticle collision dynamics method \cite{Kapral1999}
and in a one-dimensional gas of particles with nearest-neighbor Coulomb 
interaction,
modeling a screened Coulomb interaction between electrons \cite{Chen2015}. 
In all these (classical) models collisions are elastic 
and the component of momentum along the direction of 
the charge and heat flows is the only relevant constant of motion. 
We point out that it is a priori not excluded that
there exist models
where the long-time limit and the
thermodynamical limit do not commute when computing
the Drude weights.
However, numerical evidence shows that
for the models so far considered these two limits commute
\cite{Benenti2013,Benenti2014,Chen2015}.

Finally, we note that divergence of $ZT$ has been also predicted,
on different theoretical considerations, for an ideal homogeneous quantum wire 
with weak electron-electron interactions, 
in the limit of infinite wire length \cite{matveev}.

%% file: master.tex
\section{\green{Rate equations for quantum systems}}
\label{Sect:Qu-Master-Eqn}

In this section we consider an arbitrary quantum system coupled to multiple 
reservoirs of electrons, photons or phonons.
We take the system's Hamiltonian  (in the absence of the coupling to the 
reservoirs) to be
$\hat{\cal H}_{\rm s}$, and assume it is time-independent. 
In particular, we will allow for the possibility for strong interactions between 
electrons in the quantum system, in which case 
the scattering theory presented in chapter~\ref{Sect:scattering-theory} is not applicable.
If the coupling between the quantum system and the reservoirs is weak, then
we can model this situation with a quantum master equation.
The derivation requires that the system-reservoir coupling 
is weak enough that it has a very small effect on the system on the 
scale of the memory time associated with that reservoir.
Physically, the memory time is the time-scale on which a mode in the reservoir 
which was excited by a transition in the system will have an effect on the 
dynamics of the system.
For a reservoir of free-electrons, this memory time is of order $h/(\kB T)$,
so we require that the coupling to each reservoir is much less than the 
temperature of that reservoir.  For a reservoir of bosons (either photons or 
phonons), the memory time depends on both the bosonic spectrum and the 
temperature, but again the memory time decays with increasing temperature.
If the reservoir's total effect on the system is small during a memory time, we 
can treat the coupling to each reservoir mode using Fermi's golden rule, 
and neglect the possibility that the system interacts with two environment modes 
at the same time. 
Another way to say this is to say the system is in the regime of sequential tunnelling \cite{Schoeller97}. 
Then the evolution of the system's density matrix, $\rho(t)$,
in the basis of eigenstates of the system's Hamiltonian,
is given by a quantum master equation of a Markovian form.
In some cases, it has been shown that this master equation can be cast in a Lindblad form
\cite{Davies74,Davies76,Dumcke1985,lindblad,book:open-quantum,Alicki2006,Whitney2008,Rivas2010}.

In this review, we will restrict our analysis to systems for which this Markovian master equation is
particularly simple, because we will assume that quantum 
coherent superpositions 
do not play a role.  That is to say that we assume the off-diagonal 
elements of the system's density matrix are negligible at all times (where 
we take that density matrix to be written in the system's energy eigenbasis).  
The conditions under which this is a reasonable assumption are 
a little subtle and we postpone the discussion of them until in 
section~\ref{Sect:master-coherences}, although we note now that they can be 
safely neglected in all the machines considered in 
chapter~\ref{Sect:master-examples}.  
Upon neglecting the coherent superpositions, 
the quantum master equation reduces to a rate equation for the probability 
of occupying a given system state.
One can consider the case of superconducting \cite{Sothmann-Futterer-Governale-Konig2010} 
or ferromagnetic \cite{Wysokinski2012} reservoirs, but this can be done. 

This rate equation is sufficiently simple that it can be explained and used 
without a 
detailed understanding of its origin.  Thus, we present the rate
equation first, 
and only afterwards do we present the connection with the microscopic 
Hamiltonian 
for the quantum system and the reservoirs. 
The two ingredients that one needs to construct any rate equation are 
(i) the  states  and (ii) the transitions.   We start by defining the states. 
In the rate equation that we consider, the states are the 
many-body eigenstates of the quantum system when isolated from the reservoirs.
As an example, consider the Hamiltonian discussed in Table~\ref{Table:example-H}, it has four many-body states, 
$|0 \rangle$, $|1 \rangle$, $|2 \rangle$ and $|\rmd \rangle$, which we would use as the states for the rate equation.

%%%%%%%%%%%%%%%%%%%%%%%%%%%%%%%%%%%%%%
\begin{table}[b]
\centerline{\begin{tabular}{|c|c c|c c|} 
\hline
\!Many-body\! & \multicolumn{2}{|c|}{ 
Electronic states} & Electron- & Energy, \\
states, $|a\rangle$& 1 & 2 & number, $N_a$ & $E_a$ \\
\hline
$|0 \rangle$ & empty & empty & 0 & 0 \\
$|1 \rangle$ & full & empty & 1 & $\eps_1$ \\
$|2 \rangle$ & empty & full & 1 & $\eps_2$ \\
$|\rmd \rangle$ & full & full & 2 & $\eps_1\!+\!\eps_2\!+\!U\!$ \\
\hline
\end{tabular}}
\caption{\label{Table:example-H}
As an example, consider a system with two possible fermionic states,
such that its Hamiltonian is
$\hat{\cal H}^{\rm example}_{\rm s} 
= \epsilon_1 \, \hat{n}_1+\epsilon_1 \, \hat{n}_2+  U\,\hat{n}_1\hat{n}_2\,$,
with the number operator $n_i= \hat{d}^\dagger_i \hat{d}_i$, where 
$\hat{d}^\dagger_i$ and $\hat{d}_i$ are fermionic creation and annihilation 
operators for the state $i$.
The $U$-term is due to  Coulomb repulsion between electrons, it 
means that the energy for occupying both states is more than just the sum of 
occupying each
state individually. We list the four many-body eigenstates of this Hamiltonian, 
labelling them 
$|0 \rangle$, $|1 \rangle$, $|2 \rangle$, $|\rmd \rangle$ (where d stands for 
``double-occupancy''). 
}
\end{table}
%%%%%%%%%%%%%%%%%%%%%%%%%%%%%%%%%%%%%%

Now let us imagine that there are reservoirs coupled to the system which can 
induces changes in the system's state.  
The system will exchange electrons with electronic reservoirs,
changing the both system's state and its charge. 
It will exchange photons or phonons with the relevant reservoirs, 
changing the system's state without changing its charge.
Let us define the rate of each transition from system state $a$
to system state $b$ due to the coupling to reservoir $i$ as $\Gamma^{
(i)}_{ba}$. We then define the total rate of transition from state $a$ to 
$b$ as $\Gamma_{ba}$, so it is  the following sum over all reservoirs $i$,
\begin{align}
\Gamma_{ba} = \sum_i \Gamma^{ (i)}_{ba} \ .
\label{Eq:Gamma_ba}
\end{align}
Then the probability $P_b (t)$ that one finds the system in state $|b\rangle$ 
at time $t$
is given by the rate equation (or classical master equation)
\begin{align}
{\rmd \over \rmd t} P_b (t) 
= \sum_a \Big( \Gamma_{ba} \, P_a (t) \ -\ \Gamma_{ab} \, P_b (t) \Big)  \, ,
\label{Eq:master-eqn}
\end{align}
where the sum is over all system states (formally the sum is for $a \neq b$, 
but we do not need
to specify this because the term with $a=b$ is zero).
The first term in the sum is the rate at which probability arrives to the state 
$b$, while the
second term is the rate at which it leaves state $b$.
One can derive the rates $\Gamma_{ba}^{(i)}$ from the microscopic 
Hamiltonian,
as we do in section~\ref{Sect:transition-rates}, 
or one can treat these rates as phenomenological constants.
However, if one treats them phenomenologically, 
one must still ensure that these rates obey
a relation known as {\it  local-detailed balance}  \cite{Katz-Lebowitz-Spohn83},
sometimes also called a micro-reversibility relation \cite{Crooks1999} 
\green{
for a system with a non-degenerate Hamiltonian \cite{Dumcke1985},
}
\begin{align}
\Gamma^{(i)}_{ab}  
&= \Gamma^{(i)}_{ba} \,\exp\left[ -\DS^{(i)}_{ba}  \, \Big/\,\kB 
\right]\, ,
\label{Eq:rate-for-reverse-final}
\end{align}
where $\DS^{(i)}_{ba}$ 
is the change in entropy in reservoir $i$
when it induces a system transition from $a$ to $b$.
This entropy change is given by the Clausius relation
\begin{align}
\DS^{(i)}_{ba} ={\Delta Q^{(i)}_{ba} \over T_i} 
= {E_a-E_b -(N_a-N_b)\mu_i \over T_i}, 
\label{Eq:DeltaS_ba}
\end{align}
where $E_a$ and $N_a$ are the energy and electron-number for system state $a$
(see Table~\ref{Table:example-H} for examples of $E_a$ and $N_a$),
and $\mu _i$ is the electrochemical potential of reservoir $i$.
Here $\Delta Q^{(i)}_{ba}$ is the change in heat in reservoir $i$ 
associated with the transition $a \to b$, the reason it can be written as 
$E_a-E_b -(N_a-N_b)\mu_i$ will be discussed in 
section~\ref{Sect:transition-rates}, 
where Eq.~(\ref{Eq:rate-for-reverse-final}) will be derived.

The physical consequence of Eq.~(\ref{Eq:rate-for-reverse-final}) is that 
a transition that increases the entropy of the reservoir has a higher rate than the reverse 
process (which reduces the entropy of the reservoir).   
Eq.~(\ref{Eq:rate-for-reverse-final})
will be the crucial ingredient in showing that master equations obey the
laws of thermodynamics.

\green{We note that some works recast Eq.~(\ref{Eq:master-eqn}) as 
the matrix equation 
\begin{eqnarray}
{\rmd \over \rmd t} {\bf P} (t) 
\ =\ \bm{\Gamma}\ {\bf P}(t),
\label{Eq:master-eqn-matrix-form}
\end{eqnarray} 
where  ${\bf P} (t)$ is a column vector whose 
elements are $P_a(t)$.
The matrix $\bm{\Gamma}$ has off-diagonal elements given by $\Gamma_{ba}$, 
while its diagonal elements are defined to be $\Gamma_{bb} = -\sum_{a\neq b} 
\Gamma_{ab}$;
this means each column of the matrix  $\bm{\Gamma}$ sums to zero.
In all that follows in this chapter, it will be more convenient to work with 
Eq.~(\ref{Eq:master-eqn}) than Eq.~(\ref{Eq:master-eqn-matrix-form}).}

It is always helpful to visualize this rate equation as a network,
where each state is a vertex and each transition is a bond.
Each bond is labelled by the reservoir that induces the transition.
Examples of such networks are sketched in the insets of 
Figs.~\ref{Fig:2-term-sys},
\ref{Fig:3-term-sys1} and \ref{Fig:3-term-sys2}.
If multiple reservoirs can induce a transition between
two states, then we draw multiple bonds between those states (one for each 
reservoir);
see for example the two bond between states 0 and 1 in inset (a) of 
Fig.~\ref{Fig:2-term-sys},
one for reservoir L and one for reservoir R.
Similarly, if a given reservoir induces \green{multiple} transitions then 
there will be multiple bonds associated with that reservoir; for example there 
are four bond associated with 
reservoir L in inset (b) of Fig.~\ref{Fig:2-term-sys} (and four associated with 
reservoir R).

Just as there is a probability for each state in the network, 
we can define a probability current for each bond in the network.
Many observables, particularly particle and energy currents into the system 
from the reservoirs,
are given naturally in terms of these probability currents.
We define ${\cal I}_{ba}^{(i)}(t)$ as the probability current for the transition from 
state $a$ to state $b$ at time $t$ due to reservoir $i$, and we take it to be the 
probability flow \green{from $a$ to $b$ minus the probability flow from $b$ to $a$.} 
Thus
\begin{align}
{\cal I}_{ba}^{(i)}(t) \ =\  -{\cal I}_{ab}^{(i)}(t) \ =\  
\Gamma_{ba}^{(i)} \, P_a (t) - \Gamma_{ab}^{(i)} \, P_b (t) \ .
\label{Eq:probability-currents}
\end{align}
Then, the rate equation reads 
${\rmd \over \rmd t} P_b (t) 
= \sum_i \sum_a {\cal I}_{ba}^{(i)}(t)$.
This equation provides no more information than Eq.~(\ref{Eq:master-eqn}), but 
can be a convenient way of thinking
of certain aspects of the system's physics, particularly in the steady state.

\subsection{Steady state solution of the rate equation}

For a time-independent system Hamiltonian with time-independent couplings to 
the reservoirs,
there is a steady-state solution of the rate equation in 
Eq.~(\ref{Eq:master-eqn}), which 
corresponds to 
\begin{align}
{\rmd \over \rmd t} P_b (t) =0 \qquad \hbox{ for all $b$}.
\nonumber
\end{align}
Defining the solution to this equation as $P_a^{\rm steady}$, it must obey
\begin{align}
0 =  \sum_a  \Big( \Gamma_{ba} \, P_a^{\rm steady} \ -\ \Gamma_{ab} \, P_b^{\rm 
steady} \Big) 
\qquad \hbox{ for all $b$}. 
\label{Eq:steady-state-master-eqn}
\end{align}
This forms a set of simultaneous equations which can be solved to find the 
steady-state occupation probability for each state, $P_a^{\rm steady}$.

If the reservoir couplings induce transitions between all eigenstates, there is 
likely to be only one steady-state, although one should verify this for the 
system in question. 
When the steady-state is unique, then any system state will eventually relax to 
the state $P_b^{\rm steady}$, typically at a rate of 
order the slowest of the decay rates, $\{\Gamma_{ab}\}$.
We assume we are only interested in the response of the system on time-scale 
very much longer than this relaxation time, so the physics is entirely 
dominated by the
steady-state.

If we recast this in terms of  probability currents defined in 
Eq.~(\ref{Eq:probability-currents}), then the \green{steady-state probability currents are}
\begin{align}
{\cal I}_{ba}^{(i) \rm steady}(t) \ =\  -{\cal I}_{ab}^{(i) \rm 
steady}(t) \ =\  
\Gamma_{ba}^{(i)} \, P_a^{\rm steady} - \Gamma_{ab}^{(i)} \, P_b^{\rm 
steady} \ .
\label{Eq:probability-currents-steady}
\end{align}
These probability currents then obey a {\it Kirchhoff's law};
in other words the sum of probability currents into (or out of) vertex $a$ sum 
to zero, 
\begin{align}
0 =  \sum_i \sum_b  {\cal I}_{ba}^{(i)\,\rm steady} 
= \sum_i \sum_b {\cal I}_{ab}^{(i)\,\rm steady}  \qquad \hbox{ for 
all $a$}. 
\label{Eq:Kirchhoff-for-prob-currents}
\end{align}
Section~\ref{Sect:single-loop} will show that when the network of system states 
is simple enough, one can get useful information about the system's properties 
from this Kirchhoff law,
without needing to solve Eq.~(\ref{Eq:steady-state-master-eqn}).  
However, to get full information about any system, 
solving Eq.~(\ref{Eq:steady-state-master-eqn}) is unavoidable.

\subsection{Currents into the system, and power output}
\label{Sect:currents}

To understand the steady-state properties of the machines that interest us, 
we need the currents of particles and energy 
into the system from the various reservoirs.
The particle (electron) current into the system from reservoir $i$ is 
given by the 
probability currents associated with transitions involving reservoir $i$.
If that transition involves the system changing from a state $a$ in which the 
system contains $N_a$ electrons,
to a state $b$ in which the system contains $N_b$ electrons, then it is because 
$\left(N_b -N_a \right)$ electrons 
have flowed from reservoir $i$ into the system.  This transition occurs 
with a rate given by the 
probability current ${\cal I}_{ba}^{(i)}(t)$ to go from $a$ to $b$ due to 
the coupling to
reservoir $i$ given by Eq.~(\ref{Eq:probability-currents}).  
The particle current into the system from reservoir $i$,
is given by summing over all transitions involving $i$.
Hence, the particle current into the system from reservoir $i$ is 
\begin{align}
\Jparticlei(t) = \half \sum_{ab} \left(N_b -N_a \right) {\cal 
I}_{ba}^{(i)}(t),
\label{Eq:I^N}
\end{align} 
where the factor of $\half$ is due to the fact that the sum over $a$ and $b$ 
counts each transition twice.
By analogy, the energy current out of reservoir $i$ 
into the system is
\begin{align}
\Jenergyi(t) = \half\sum_{ab} \left(E_b -E_a \right)  {\cal 
I}_{ba}^{(i)}(t).
\label{Eq:I^E}
\end{align} 
The steady-state particle and energy currents are given by taking  ${\cal 
I}_{ba}^{(i)}(t)= {\cal I}_{ba}^{(i) \rm steady}$ given by 
\green{Eq.~(\ref{Eq:probability-currents-steady})}.
From the above two currents, 
we get the electrical current, $\Jelectrici$, and heat current, $\Jheati$,
flowing out of reservoir $i$ into the quantum system:
\begin{align}
\Jelectrici &= e \Jparticlei, 
\\
\Jheati &=  \Jenergyi - \mu_i \Jparticlei, 
\label{Eq:J}
 \end{align}
where $e$ is the electronic charge (so $e$ is negative).  
Physically, the heat current $\Jheati$ is just the energy current measured 
from the reservoir's electrochemical potential.
This definition make sense from a microscopic point of view; electrons above a 
reservoir's electrochemical potential
reduce the heat in that reservoir when they escape (making the Fermi 
distribution infinitesimally narrower), 
but electrons below the electrochemical potential increase the heat in that reservoir 
when they escape (making the Fermi distribution infinitesimally broader).  

If reservoir $i$ is a reservoir of non-interacting bosons, such as photons 
or phonons, 
the formulas for particle and energy flow are the same. 
However, since photons and phonons are uncharged, they do not carry any 
electrical current. 

The rate equation should be constructed such that each transition conserves both 
energy and electron number
(although there is no requirement that it conserves the number of photons or 
phonons).  The energy and electron number in the quantum system become constant 
once the system reaches its steady-state, so there can be no net flow of 
electrons or energy into the system,
thus the energy currents and particle currents obey
\begin{align}
\Jenergy^{\rm (sum)} &\equiv \sum_i  \Jenergyi =0\,, 
\label{Eq:energy-conservation}
\\
\Jparticle^{\rm (sum)}  &\equiv \sum_i \Jelectrici =0\, ,
\end{align}
where the sums are over all reservoirs.
To see explicitly that the steady-state obeys the first of these equations,
 we note that \green{
 Eqs.~(\ref{Eq:probability-currents-steady},\ref {Eq:I^E}) }
 mean that $\Jenergy^{\rm (sum)} = -\half \sum_{ab} \left(E_b -E_a \right) 
\left(\Gamma_{ba}  P_a^{\rm steady} - \Gamma_{ab}  P_b^{\rm steady} \right)$.
If we now exchange dummy variables $a \leftrightarrow b$ in the term  containing 
$E_a$, we get
 \begin{align}
\Jenergy^{\rm (sum)} = -\sum_b E_b  \sum_{a}  
\left(\Gamma_{ba}  P_a^{\rm steady} - \Gamma_{ab}  P_b^{\rm steady} \right).
\end{align}  
Then, Eq.~(\ref{Eq:steady-state-master-eqn}) immediately gives 
$\Jenergy^{\rm (sum)} =0$ as required.  The proof that $\Jparticle^{\rm (sum)}=0$ is the same, 
except that one replaces $E_b,E_a$ with $N_b,N_a$.

Now we turn to calculating the electrical power that the system generates in a given reservoir. 
Injecting a particle into reservoir 
$i$ requires a work equal to the \green{reservoir's} electrochemical potential, $\mu_i$.  Just 
as moving a classical particle up a hill requires a work equal to the potential 
energy.  This electric power that the system generates could be stored in the form of electric work (taking the reservoir to be one plate in a capacitor), or it could be immediately
converted into another form of work.
In the latter case, an example would be an ideal electric motor connected between the reservoir and ground
(a reservoir with electrochemical potential equal to zero).
Such a motor turns electrical work into mechanical work without losses, but from the point of view of the electric circuit, it is just a load.
The power that the system generates and sends into the load connected to reservoir $i$ is 
\begin{align}
P_{\rm gen}^{(i)}=-\mu_i J_{\rho,i}, 
\label{Eq:P_gen^i}
\end{align} 
where $J_{\rho,i}$ is the particle current into the system from reservoir $i$ 
(i.e. it is the number of electrons that flow from reservoir $i$ to the system per unit time).
The negative sign is because of our convention for currents. 
This convention means that if $\mu_i$ is larger than all other electrochemical potentials, 
then the current $\Jelectrici$ into the system from reservoir $i$ should be negative 
if we want to do work by moving charge from the reservoirs with lower electrochemical potentials 
to reservoir $i$ with its higher electrochemical potential.
The power generated in \green{reservoir} $i$ can be cast in the familiar form of 
voltage $\times$ electrical current, by noting that $\mu_i=eV_i$ so
\begin{align}
P_{\rm gen}^{(i)}  = -V_i \Jelectrici\, .
\label{Eq:P_gen=VI}
\end{align}
Note, that the power generated in reservoir $i$ depends on the definition of ground (the energy from which all electrochemical potentials are measured).  
However, since $\sum_i \Jparticlei=0$, the total power generated (summed over all reservoirs) 
is independent of any overall shift of the electrochemical potential with respect to ground, as one 
expects.

Similarly, we can clearly see the {\it Joule heating} effect, if we consider the case where the reservoirs are all at the same temperatures (so the system cannot perform any thermoelectric power generation), but reservoir $i$ is maintained at a electrochemical potential $\mu_i=eV_i$ by a power supply.
Then, the system will act as a resistance (usually a non-linear resistance), which absorbs a power 
$P_{\rm abs}^{(i)} = -P_{\rm gen}^{(i)}$
from the power supply coupled to reservoir $i$.  Eq.~(\ref{Eq:heat+power}) will tells us that 
the electrical power absorbed by the system is radiated into the reservoirs as heat. This is Joule heating,
with the usual form of voltage $\times$ electrical current in Eq.~(\ref{Eq:P_gen=VI}).
The negative signs ensure that the power absorbed, $P_{\rm abs}^{(i)}$, is positive, when currents flow from regions of high electrochemical potential to regions of lower electrochemical potential.

\subsection{From the microscopic Hamiltonian to the rate equation}
\label{Sect:transition-rates}

\green{
Here we discuss the derivation of the above rate equation from a microscopic Hamiltonian 
for the system and the reservoirs.  The derivation assumes that the coupling between the system and the reservoirs is weak enough that one can apply a Fermi golden rule approximation.  
The golden-rule treatment of such system-reservoir problems, in which one assumes the system-reservoir coupling is a perturbation that can be treated to lowest order, has a long history.  Depending on the community and context it is known as the Redfield \cite{redfield} or Bloch-Redfield \cite{Bloch57} approximation, 
the sequential tunnelling approximation in transport theory (see e.g.~\cite{Schoeller97}), 
or the weak-coupling limit of the Nakajima-Zwanzig model \cite{Nakajima58,Zwanzig60}.  
A more rigorous treatment well known in the mathematical physics community is the weak-coupling limit of quantum-mechanical master equations in Refs.~\cite{Davies74,Davies76,Dumcke1985}.
For students looking to learn these techniques we recommend Ref.~\cite{Bellac-qm-book}, with Refs.~\cite{Atom-Photon-Interactions-book,Blum-book,book:open-quantum} being good alternatives.  
These works provide a good base from which to attack more technical reviews such as that of 
the sequential tunnelling approximation  in Ref.~\cite{Schoeller97}.
}

Consider a finite size quantum system with Hamiltonian $\hat{\cal H}_{\rm s}$ 
coupled to a number of reservoirs.
They may be reservoirs of non-interacting electrons, 
which can tunnel between the reservoirs and the quantum system.
Alternatively, they may be reservoirs of non-interacting photons or phonons, 
whose emission or absorption induce transitions within the quantum system.  
Then, the Hamiltonian for the system plus the reservoirs will be
\begin{align}
\hat{\cal H}_{\rm total} &= \hat{\cal H}_{\rm s} 
+\sum_{i \in {\rm el}}
\left( \hat{\cal V}_{\rm el}^{(i)} +\hat{\cal H}_{\rm el}^{(i)} 
\right)
+\sum_{i \in {\rm ph}} 
\left( \hat{\cal V}_{\rm ph}^{(i)} +\hat{\cal H}_{\rm ph}^{(i)} 
\right).
\label{Eq:H_total-1}
\end{align}
The first sum is over all reservoirs of non-interacting electrons. 
The second sum 
is over all reservoirs of non-interacting bosonic modes, which could be photons 
or phonons. \green{In all cases, we assume this total Hamiltonian is bounded from below.}

If reservoir $i$ consists of non-interacting electrons, then its 
Hamiltonian is
\begin{align}
 \hat{\cal H}_{\rm el}^{(i)}  &= \sum_\gamma \,E_{i;\gamma}\, 
\hat{c}_{i;\gamma}^\dagger \,\hat{c}_{i;\gamma},
\end{align}
where $E_{i;\gamma}$, $\hat{c}_{i;\gamma}^\dagger$ and $\hat{c}_{i;\gamma}$ are 
respectively the energy, the creation operator and the annihilation operator 
for the fermionic state $\gamma$ in reservoir $i$.  The coupling to such a reservoir 
induces transitions in the system which changes its charge state by one.
If we define $\hat{d}^\dagger_\alpha$ and $\hat{d}_\alpha$ as the creation and annihilation operators for
electron state $\alpha$ in the system, then
\begin{align}
 \hat{\cal V}_{\rm el}^{(i)} \ \ &=\  \sum_\gamma \left(\hat{V}_{{\rm el}; i}(E_\gamma)  \,\hat{c}_{i;\gamma}^\dagger  
 \ +\  {\hat V}_{{\rm el}; i}^\dagger  (E_\gamma)  \,\hat{c}_{i;\gamma} \right)\, , 
 \qquad 
 \hbox{ with } \  {\hat V}_{{\rm el}; i}(E_\gamma)\ = \ \sum_\alpha V^{{\rm el};i}_{\alpha}(E_\gamma)\  \hat{d}_\alpha \ ,
\label{Eq:V_el}
\end{align}
where the complex number $V^{{\rm el};i}_{\alpha}$ is the matrix element for the transition under consideration.
We assume that the mode $\gamma$ in reservoir $i$ is entirely determined by its energy $E_\gamma$.
If we wish to include internal degrees of freedom of the reservoir 
(coupling to different spin-states or  multiple modes of the reservoir), we 
treat it as multiple reservoirs, each with one degree of freedom.
For example, a reservoir of electrons with spin-up and spin-down can be treated 
as two reservoirs, one of spin-up electrons and the other with spin-down 
electrons.  In this manner it is easy to take into account different coupling 
to different spin-states or reservoir modes, spin-accumulations in the 
reservoirs (different electrochemical potentials for spin up and spin down), etc.

If reservoir $i$ consists of non-interacting photons or phonons (or some 
other chargeless bosonic excitation) it has a Hamiltonian
\begin{align}
 \hat{\cal H}_{\rm ph}^{(i)}  &= \sum_\gamma \,E_{i;\gamma}\, 
\hat{b}_{i;\gamma}^\dagger \,\hat{b}_{i;\gamma},
\end{align}
where $E_{i;\gamma}$, $\hat{b}_{i;\gamma}^\dagger$  and $\hat{b}_{i;\gamma}$ 
are respectively the energy, the creation operator and the annihilation 
operator for the bosonic 
state $\gamma$ in reservoir $i$.  The coupling to such a reservoir 
induces transitions in the system which move system electrons from state $\alpha$ to state $\beta$,
\begin{align}
 \hat{\cal V}_{\rm ph}^{(i)}\ \ &= \ \sum_{\gamma} \, \hat{V}_{{\rm ph};i} (E_\gamma)\, 
 \left(\hat{b}_{i;\gamma}^\dagger + \hat{b}_{i;\gamma}\right)\ , 
 \qquad \hbox{ with } \ 
  \hat{V}_{{\rm ph};i} (E_\gamma)\ = \ \sum_{\alpha\beta} V_{\beta\alpha}^{{\rm ph};i}(E_\gamma)
 \ \hat{d}_\beta^\dagger 
\hat{d}_{\alpha},
\label{Eq:V_ph}
\end{align}
where the complex number $V_{\alpha\beta}^{{\rm ph};i}(E_\gamma)$ is the matrix element for the transition being considered. The Hermiticity of the Hamiltonian requires that  $V_{\alpha\beta}^{{\rm ph};i}(E_\gamma)$
is the complex conjugate of $V_{\beta\alpha}^{{\rm ph};i}(E_\gamma)$.

We now use Fermi's golden-rule to calculate the transition rates for the above 
microscopic Hamiltonian. In this context the golden-rule is a perturbative treatment to lowest order in the system-reservoir coupling.
\green{As mentioned above, a good introduction is Ref.~\cite{Bellac-qm-book},
with more technical alternatives being Refs.~\cite{Atom-Photon-Interactions-book,Blum-book,book:open-quantum}.   
A powerful diagrammatic treatment of this type of problem is reviewed in  Ref.~\cite{Schoeller97},
where the golden-rule approximation is referred to as   the sequential tunnelling 
approximation.}

\green{We start be writing the system in terms of its many-body eigenbasis, an example of which is given in 
Table~\ref{Table:example-H}. This involves passing from second-quantization back to first quantization. This is the opposite direction from that taken in most textbooks (which go from first quantization to second quantization), so Appendix~\ref{Append:second-to-first-quantization} gives a a quick summary of the transformation in the opposite direction for a two-state system similar to that in Table~\ref{Table:example-H}. 
This transformation allows us to write the system in terms of a set of many-body eigenstates.  In this basis the
system dynamics in the absence of the coupling to the reservoirs are trivial, because ${\cal H}_{\rm s}$ is a diagonal-matrix, and so does not induce transitions between states.  This means the only transitions between
a many-body eigenstate $a$ and a many-body eigenstate $b$ are due to the coupling to a reservoir.
Then each such transition in the system is associated with either absorbing a particle from a reservoir
or emitting a particle into a reservoir.}

For transitions which involve an electron moving from the reservoir to the 
system, we know that 
matrix elements containing ${\cal V}_{\rm el;i}$ are only non-zero when
the many-body system states $|a\rangle$ and $|b\rangle$ differ by one unit of charge. 
Energy conservation tells us that the rate of transition from state $a$ to 
state $b$ 
depends on the density of electrons at energy $\Omega_{ba}$ in the reservoir, 
where we define
\begin{align}
\Omega_{ba}=E_b-E_a \, ,
\end{align}
where $E_a$ is the energy of the system state $a$.
The density of electrons at energy $\Omega_{ba}$ in reservoir $i$ is  
$\nu_{i}(\Omega_{ba}) f_i(\Omega_{ba})$,
where
$\nu_{i}(E)$ is the density of reservoir states at energy $E$,
and  $f_i(E)$  is the Fermi factor for the reservoir.  Here 
\begin{align}
f_i(E) &= 1\big/\big(1+\exp[(E-\mu_i)/\kB T_i]\big),
\end{align}
with $\mu_i$ being the reservoir's electrochemical potential, and 
$T_i$ being its temperature.
For a transition from system state $a$ to system state $b$, which is 
achieved by the
system absorbing an electron from the reservoir $i$,
the golden-rule transition rate is 
\begin{align}
\Gamma^{({\rm el};i+)}_{ba} = {1\over h} \nu_{i}(\Omega_{ba}) \ 
f_i(\Omega_{ba}) \,
\ \Big| \bigbra{b} \hat{V}_{{\rm el};i}(\Omega_{ba}) \bigket{a}  
\Big|^2, 
\label{Eq:Gamma_el_ba-plus}
\end{align}
where the superscript ``$+$'' indicates that the transition from $a$ to $b$ adds an electron to the system,
and  $\hat{V}_{{\rm el};i}(E)$ is given in 
Eq.~(\ref{Eq:V_el}).  

For transitions which involve an electron moving from the system to the 
reservoir,
energy conservation tells us that the rate of transition from state $a$ to 
state $b$ 
depends on the density of empty electron states at energy $-\Omega_{ba}$,
this density is hence given by $\nu_{i}(-\Omega_{ba}) 
[1-f_i(-\Omega_{ba})]=\nu_{i}(-\Omega_{ba}) f_i(\Omega_{ba})$.
Thus for a transition from system state $a$ to system state $b$, which is 
achieved by the
system emitting an electron into reservoir $i$,
the golden-rule transition rate is 
\begin{align}
\Gamma^{({\rm el};i-)}_{ba} =  {1\over h} \nu_{i}(-\Omega_{ba}) \ 
f_i(\Omega_{ba}) \,
\ \Big|\bigbra{b} \hat{V}_{{\rm el};i}^\dagger (-\Omega_{ba})\bigket{a} 
\Big|^2 ,
\label{Eq:Gamma_el_ba-minus}
\end{align}
where the superscript ``$-$'' indicates that the system loses an electron 
during the transition from $a$ to $b$.

The structure is similar for bosonic excitations (phonons or photons) as it was 
for electrons, 
except that the fermion functions are replaced by bosonic functions, 
\begin{align}
n_i (E)=1\big/\big(\exp[E/\kB T_i ] -1\big), 
\end{align}
with $E>0$.
A transition from system state $a$ to system state $b$ with $E_b > E_a$, 
involves the system absorbing a bosonic excitation 
(photon or phonon) with energy $\Omega_{ba}=E_b-E_a$ from the reservoir.  
The golden-rule rate for this transition is 
\begin{align}
\Gamma^{\rm (ph;i +)}_{ba} = {1\over h} \nu_{i }(\Omega_{ba}) \ n_i 
(\Omega_{ba}) \,
\ \Big|\bigbra{b}\hat{V}_{{\rm ph};i}(\Omega_{ba})\bigket{a} 
\Big|^2 ,
\label{Eq:Gamma_ph_ba-plus}
\end{align}
where $\hat{V}_{{\rm ph};i}(E)$ is the operator in Eq.~(\ref{Eq:V_ph}).  
We use the superscript ``$+$'' to indicate that the system has gained energy 
during the transition.
One could say that the system has gained ``one bosonic excitation'' in analogy 
with what we said for electrons, however we avoid this language because the 
number of bosonic excitations in the system is not well-defined (since bosons 
such as photons and phonons need not be conserved by the system Hamiltonian).

Similarly, a system transition with $E_b < E_a$ (so $\Omega_{ba}<0$) without a 
change in electron number in the system, 
involves the system emitting a photon or phonon with energy $-\Omega_{ba}$ into 
the reservoir.
The golden-rule rate for this transition is 
\begin{align}
\Gamma^{\rm (ph;i -)}_{ba} = - {1\over h} \nu_{i }(-\Omega_{ba}) \ n_i 
(\Omega_{ba}) 
\ \Big|\bigbra{b} \hat{V}_{{\rm ph};i}(-\Omega_{ba}) \bigket{a} 
\Big|^2, 
\label{Eq:Gamma_ph_ba-minus}
\end{align}
where we have used the fact that $1+n_i (-E) = -n_i (E)$. 
The superscript ``$-$'' indicates that the system has lost energy in the 
transition.

%-----------------------------------
\subsection{Neglecting coherent superpositions in the rate equation}
\label{Sect:master-coherences}

\green{
In our rate equation analysis we have taken the system's density matrix in the basis defined   by 
the many-body eigenbasis of the system Hamiltonian ${\cal H}_{\rm s}$, and then we have neglected the off-diagonal elements of this matrix.}
This neglects quantum coherent superpositions, which is why the master equation reduces to a classical rate equation.  We can then directly apply results from the thermodynamics of 
stochastic processes.
The conditions under which the quantum coherent superpositions
(off-diagonal elements) can be neglected are as follows.
Firstly, one must start with a state which contains no coherent superpositions,
and secondly the interaction with the reservoirs should not generate any 
coherent superpositions.
Let us now discuss each of the conditions in more detail.

The first condition is that the system's initial state should contain no 
quantum coherent superpositions,
by which we mean that we start at time $t_0$ with a density matrix which is a 
product  of density matrices for the system and for each reservoir. 
Each reservoir's density matrix is assumed to start in a thermal state of its 
Hamiltonian (neglecting the coupling to the system) at its own temperature.
However, we also require that the system's density-matrix is diagonal in the energy 
eigenbasis of its Hamiltonian, $\hat{\cal H}_{\rm s}$ (recall that this is the 
Hamiltonian of the system if one neglects the coupling to the reservoirs).  Then 
the initial state of the system is 
\begin{align}
\rho^{\rm (s)}_{ab}(t_0) = P_a(t_0)\ \delta_{ab}\ ,
\label{Eq:rho-sys-diag}
\end{align} 
where $\delta_{ab}$ is a Kronecker delta-function, 
and  $P_a (t_0)$ is the probability that the system is in eigenstate $a$ of  
$\hat{\cal H}_{\rm s}$.    
Since such states have no quantum coherent superpositions, they would be 
time-independent 
in the absence of coupling to the reservoirs.  In contrast, any state 
containing superpositions would undergo coherent oscillations (in the absence 
of reservoirs coupling) at a frequency 
given by the energy difference between the states in the superposition.
A natural  initial system state is a thermal state at some 
temperature $T$, in which $P_a(t_0)= \exp[-E_a/(\kB T)]\big/Z$ 
for an eigenstate with energy $E_a$,
where the partition function $Z= \sum_a \exp[-E_a/(\kB T)]$ with the sum being 
over all eigenstates of 
$\hat{\cal H}_{\rm s}$.   However, we emphasize that any initial 
state that obeys Eq.~(\ref{Eq:rho-sys-diag}) is acceptable.

The second condition is that coupling to the reservoirs does not generate any 
coherent superpositions
in the system.   
This condition depends on the nature of the coupling between 
the system and the reservoirs.  If the coupling turns one system eigenstate 
into another system 
eigenstate, as is the case for all examples in
section~\ref{Sect:master-examples}, 
then it does not create a coherent superposition in the system.
A counter-example of a system which does create coherent superpositions is 
considered in Ref.~\cite{Levy-Kosloff2014}, 
which points out that one has to be careful in the treatment of the system if one 
wants to
get dynamics that obey the second law of thermodynamics.
Treating such superposition-generating cases is beyond the scope of this review, however it is instructive 
to take a moment to understand why 
coherent superpositions are generated 
(although the following explanation will be more clear after having read some of the examples in section~\ref{Sect:master-examples}).
Consider a system with two possible electronic states, 
as in Fig.~\ref{Fig:3-term-sys1} discussed in section~\ref{Sect:3-term-sys1}, 
but in which the 
system Hamiltonian, ${\cal H}_{\rm s}$, contains a direct tunnel coupling between states 1 and 2.
\green{Such a system Hamiltonian is discussed in Appendix~\ref{Append:second-to-first-quantization}.}
The many-body eigenstates with $N_a=0$ or $N_a=2$ 
are respectively $|0\rangle$ and $|\rmd\rangle$, as defined in 
Table~\ref{Table:example-H}. 
However, the tunnel coupling between state 1 and 2 means that the many-body 
eigenstates of ${\cal H}_{\rm s}$ with $N_a=1$ are superpositions of $|1\rangle$ and $|2\rangle$.  The 
first of these eigenstates, 
$|+\rangle$, is a bonding state, and contains a superposition which is a sum of 
$|1\rangle$ and $|2\rangle$.
The second of these, $|-\rangle$, is an anti-bonding state, and contains a 
superposition which is a difference of $|1\rangle$ and $|2\rangle$.
These states have energies $E_+$ and  $E_-$, with the bonding state energy $E_+$
being less than the anti-bonding state energy $E_-$.
Now we assume, as in section~\ref{Sect:3-term-sys1}, that reservoir L is tunnel 
coupled to system state 1 but not to system state 2. 
The coupling to reservoir L can 
\green{
be in two regimes;
}
\begin{itemize}
\item[(i)]
If the energy difference $(E_--E_+)$ is much larger than the coupling between state $|1\rangle$ and reservoir L,
we can assume that the reservoir mode with energy $E_+$ will couple to the system state $|+\rangle$
and the reservoir mode with  energy $E_-$ will couple to the system state $|-\rangle$.
Since there is no coherence between reservoir modes with energies $E_+$ and $E_-$, this will
not generate any coherence in the system, and then the rate equation analysis we consider here should apply.
In this case one has to be sure to take the states in the 
\green{rate equation as the system's} 
many-body eigenstates, i.e.~ $|0\rangle$,   $|+\rangle$,   $|-\rangle$,  and  $|d\rangle$.

\item[(ii)]
If the energy difference $(E_--E_+)$ is of the order of the coupling between state $|1\rangle$ and reservoir L,
we can assume that a reservoir mode which couples to  $|+\rangle$ will also couple to  $|-\rangle$.
Consider the system to be  in the state $|0\rangle$, when an electron from that mode of reservoir L tunnels into it.
Then the system will arrive at a state which is a coherent superposition of $|+\rangle$ and  $|-\rangle$.
Thus, the coupling to the reservoir will generate a coherent superposition of many-body eigenstates within the system.  
This means that even if one starts with 
the system's density matrix in a diagonal state, the coupling to the reservoir 
will generate off-diagonal terms. 
In this case, the theory presented in this review is insufficient, and one must 
treat the evolution of the full
density matrix \cite{Levy-Kosloff2014}, rather than just its diagonal elements.
\end{itemize}
\green{
Thus the rate equations discussed in this chapter (which neglect coherences) apply to problems of the type in regime (i) but not in regime (ii).  More generally,  the rate equations method requires that the 
coupling to reservoirs is smaller than {\it any} energy scale in the quantum system.  This requires
that the quantum system has {\it no degeneracies} between its many-body eigenstates.
}

While this review concentrates on steady-states of systems with 
time-independent Hamiltonians, the rate equation
technique discussed here applies to any time-dependent problem in which 
coherent superpositions of system states are absent.
The rate equation technique applies to arbitrary time-dependence of the system-reservoir 
couplings and system's energy-levels, so long as they obey the two conditions 
discussed above at all times, along with a third condition.
\green{This third condition is that the time-dependence of ${\cal H}_{\rm s}$ does not generate superpositions
of system eigenstates (this can be checked in the absence of the coupling to the reservoirs).  Two examples of time-dependences which do not generate superpositions
of system eigenstates are (i) arbitrary time-dependences of the eigenenergies of ${\cal H}_{\rm s}$, but with unchanging eigenstates (so $\hat {\cal H}_{\rm s}(t_1)$ commutes with $\hat 
{\cal H}_{\rm s}(t_2)$ for all times $t_1 $ and $t_2$ during the evolution),
or (ii) adiabatically slow evolution of ${\cal H}_{\rm s}$.}

\green{
We mention systems which do not  satisfy all the above conditions in section~\ref{Sect:beyond-master}.}

%-----------------------------------
\subsection{Local-detailed balance for transition rates}
\label{Sect:local-detailed}

Here we show that the golden-rule transition rates, discussed in the previous 
section, satisfy the relation in Eq.~(\ref{Eq:rate-for-reverse-final}). 
This relation is called ``local detailed balance'' 
\cite{Katz-Lebowitz-Spohn83,Dumcke1985}, 
and the word ``local'' is crucial, because it is different from the usual detailed balance condition.
``Detailed balance'' is a property of the occupation probabilities of system 
states {\it at equilibrium} (see section~\ref{Sect:master-zeroth-law}).
''Local detailed balance'' in Eq.~(\ref{Eq:rate-for-reverse-final}) is a statement about transition rates, 
which is true because each reservoir is in its own {\it local} equilibrium,
irrespective of whether the system \green{ (in which the reservoir induces transitions)} 
is in equilibrium or not.

For a transition induced by the coupling to electron reservoir $i$,
if the transition from system state $a$ to system state $b$ involves the system 
absorbing an electron, then the transition from $b$ to $a$ must involve the 
system emitting an electron.
The former transition is given by Eq.~(\ref{Eq:Gamma_el_ba-plus}),
while the latter is given by Eq.~(\ref{Eq:Gamma_el_ba-minus}) with $a 
\leftrightarrow b$. 
Then since $f_i(-E) =f_i(E) \exp\left[(E-\mu_i)\big/(\kB 
T_i)\right] $
and $\Omega_{ab}=-\Omega_{ba}$,
we find 
\begin{align}
\Gamma^{\rm (el;i-)}_{ab}  
&= \Gamma^{\rm (el;i+)}_{ba}  \ 
\exp\left[ {\Omega_{ba} - \mu_i \over \kB T_i}\right] \, .
\label{Eq:rate-for-reverse-electron-plus}
\end{align}
Now we note that the change in heat in reservoir $i$, 
when it emits or absorbs an electron, is equal to 
the change in that reservoir's 
energy measured from its electrochemical potential.
If the system state changes from $a$ to $b$ because it absorbs an electron from
reservoir $i$
(so the electron leaving the reservoir has energy $\Omega_{ba}$), 
then the change in heat in reservoir $i$ is  
\begin{align}
\Delta Q_{ba}^{(i+)} = - \big( \Omega_{ba} -\mu_i \big)\ .
\end{align}
Alternatively, if the system state changes from $a$ to $b$ because it emits an 
electron into reservoir $i$ 
(so the electron enters the reservoir with energy $-\Omega_{ba}$), 
then the change in heat in reservoir $i$ is  
\begin{align}
\Delta Q_{ba}^{(i-)} = - \Omega_{ba} -\mu_i \ .
\end{align}
We then use the Clausius relation to
define the change in reservoir $i$'s entropy, $\DS^{(i)}_{ba}$, 
when it changes the system state from $a$ to $b$, 
as the above change in heat divided by the reservoir's temperature.
Substituting this definition of change 
into Eq.~(\ref{Eq:rate-for-reverse-electron-plus}), one can see that one gets 
the {\it local-detailed balance} given in 
Eq.~(\ref{Eq:rate-for-reverse-final}), 
irrespective of whether the system absorbs an electron ($N_b-N_a=1$) or emits 
an electron
($N_b-N_a=-1$).  
Note that we do not need to label the transitions with ``$+$'' or ``$-$'' for 
absorption or emission, 
because this is completely determined by the states $a$ and $b$
(or more precisely their electron-number, $N_a$ and $N_b$).

Now we turn to considering 
transitions from state $a$ to $b$ which involving the system absorbing or 
emitting a bosonic excitation (photon or phonon).  We can make a similar 
argument as 
that for electrons above.
We take Eq.~(\ref{Eq:Gamma_ph_ba-minus}) with $a \leftrightarrow b$, and 
comparing it to Eq.~(\ref{Eq:Gamma_ph_ba-plus})
Since $n_i (-E) =-n_i (E) \exp\left[E\big/(\kB T_i )\right] $
and $\Omega_{ab}=-\Omega_{ba}$, we recover 
Eq.~(\ref{Eq:rate-for-reverse-final}) for bosons.
For this, we define $\Delta Q^{(i)}_{ba}$  as the change in the heat in 
the bosonic reservoir,
which is equal to the change in energy of that reservoir (since there is no 
chemical potential for the photons or phonons), and thus equals $-\Omega_{ba}$.

%-----------------------------------
\subsection{Equilibrium and the zeroth law of thermodynamics}
\label{Sect:master-zeroth-law}

If all reservoirs are at the same temperature and same electrochemical potential as 
each other,
$T_i=T$ and $\mu_i=\mu$, then the reservoirs are in equilibrium with 
each other.
Any system coupled between them should also achieve equilibrium at the same 
temperature
and electrochemical potential. \green{We can make a dynamical definition of equilibrium, and thereby a dynamical 
formulation of the zeroth law of thermodynamics.
Dynamically, equilibrium} implies that the system is in the state of {\it detailed balance},
which means that the system state is such that the transitions $a \to b$ 
and $b \to a$ occurs at the same rate. 
Thus, if we define the occupation probability for state $a$ at equilibrium as 
$P^{\rm eq}_a$, then it must obey
\begin{align}
\Gamma^{(i)}_{ab} P^{\rm eq}_b = \Gamma^{(i)}_{ba} P^{\rm eq}_a 
\qquad \hbox{ for all }\, a,b, i. 
\label{Eq:detailed-balance}
\end{align}
Taking Eq.~(\ref{Eq:rate-for-reverse-final}) for $T_i=T$ and 
$\mu_i=\mu$, we have 
\begin{align}
{\Gamma^{(i)}_{ab} \, =\, \Gamma^{(i)}_{ba}} \,
\exp\left[ -{\Omega_{ba} -(N_b-N_a)\mu \over \kB T}\right]\, ,
\end{align}
where the exponent is the same for all reservoirs.
Then Eq.~(\ref{Eq:detailed-balance}) corresponds to 
\begin{align}
P_a (t) \ =\ P^{\rm eq}_a \ \equiv \ {1 \over Z} \exp\left[-{E_a - N_a \mu 
\over \kB T}
\right]
\end{align}
with normalization $Z= \sum_a  \exp\left[-(E_a-N_a\mu)\big/(\kB T)\right]$, 
where $a$ is summed over all system states.  This is the state one would 
naively write down for a system in equilibrium at temperature $T$ 
and electrochemical potential $\mu$, and here we have shown that it is indeed the 
equilibrium state.
Since this state satisfies detailed balance, it is a steady-state where all 
electron currents and heat currents are zero.

The zeroth law of thermodynamics states that if system A is in equilibrium with a system B and with a system C, then systems B and C must also be in equilibrium with each other.  
Let us consider  two reservoirs (A and B) which we know to be in equilibrium with each other. 
\green{Let us take a dynamical definition of equilibrium, which says systems are in equilibrium
if there is no heat current or particle current between them when they are coupled to each other.}
Then, if a small quantum system C is in equilibrium with reservoir A (as modelled by a rate equation), 
the above rate equation analysis is sufficient to prove that system C will also be in equilibrium with reservoir B.

\green{This dynamical formulation of the zeroth law based on the dynamical definition of equilibrium 
assumes that the coupling between systems can support independent particle and heat currents.}
This presents a minor problem for simple quantum systems modelled by the rate equation, since some of the most interesting ones do not allow independent particle and heat currents. In other words, the value of the particle current completely determines the energy current, which is often referred to as {\it tight coupling} between these two types of current.  An explicit example of such a system is that in Fig.~\ref{Fig:3-term-sys2}
in the situation where the transitions indicated by the dashed lines in the inset are absent.
Then each electron leaving reservoir L carries an energy of exactly $\eps_1$ out of that reservoir
irrespective of the biases and temperatures of the different reservoirs,
so the energy current, $\JenergyL$, is not independent of the particle current, $\JparticleL$, because 
$\JenergyL= \eps_1\JparticleL$.
Under such circumstances a system can satisfy Eq.~(\ref{Eq:detailed-balance})
even when it is coupled to multiple reservoirs at different temperatures and chemical potential,
so long as there is a specific relationship between system parameters, reservoir temperatures
and reservoir biases.  Such situations typically correspond to situations under which the quantum system acts as a Carnot efficient machine (generating no entropy).  
This situation is discussed in detail in section~\ref{Sect:rule-Carnot}.
However, this poses a problem for the definition of equilibrium for such systems.
A resolution of this problem is to define equilibrium between a reservoir and a quantum system by saying that there is no particle or heat flow between them when they are coupled to each other, even when all the system parameters (energy gaps, etc) are varied a little.  This works because the systems which satisfy Eq.~(\ref{Eq:detailed-balance}) when coupled to multiple reservoirs not in equilibrium with each other, 
only do so for specific values of their parameters.  If we change those system parameters a little, 
one will observe a violation of Eq.~(\ref{Eq:detailed-balance}) which will result in  particle and heat currents,
unless the system is truly in equilibrium with all the reservoirs it is in contact with.

%-----------------------------------
\subsection{First law of thermodynamics}

The first law of thermodynamics in the steady-state follows from 
Eq.~(\ref{Eq:energy-conservation}), 
which is a direct consequence of the fact that the rate equation conserves 
energy.
Combining it with  Eqs.~(\ref{Eq:J}) and (\ref{Eq:P_gen^i}), we get 
\begin{align}
\sum_i \Jheati  \ =\ \sum_i P_{\rm gen}^{(i)} \ ,
\label{Eq:heat+power}
\end{align}
where $i$ is summed over all reservoirs.
The left hand side is the total heat-current into \green{the system from} the reservoirs, and 
the right hand side is the total power generated by the system.
Thus Eq.~(\ref{Eq:heat+power}) corresponds to the first law of thermodynamics, 
since it says that 
the rate of work production (electrical power) equals the rate of heat 
absorption (total heat current).
Note that the equality between power generated and heat absorbed only holds when we sum over all reservoirs,
in general it does not hold at the level of any given
reservoir.

\subsection{Second law of thermodynamics}
\label{Sect:2ndlaw}

Here we present a proof, taken from  Ref.~\cite{review-vandenBroeck}, 
that the rate equation for 
any system of discrete states fulfills the second law of thermodynamics.
This proof is similar in style to the much older proofs by Spohn \cite{Spohn78} and  Alicki \cite{alicki79}
for more complicated quantum master equations (which include coherence), see also Ref.~\cite{Peres-book-2nd-law}.
It is convenient not to take the steady-state limit until the end of 
the derivation, so we assume that the system has dynamics (which implies that the probabilities 
$P_a(t)$ are time-dependent).
We start by noting that the rate of change of entropy in reservoir $i$ can 
be
written as 
\begin{align}
{\rmd \over \rmd t} \mathscr{S}^{(i)} (t)
\ =\ - {\Jheati(t) \over T_i} 
\ =\ \green{1 \over 2} \sum_{ab}
{\cal I}_{ba}^{(i)}(t) \ \DS^{(i)}_{ba}, 
\label{Eq:dotS_i}
\end{align}
where $\DS^{(i)}_{ba}$ and ${\cal I}^{(i)}_{ba}$ are given by 
Eqs.~(\ref{Eq:DeltaS_ba},\ref{Eq:probability-currents}).
\green{The factor of a half comes from the fact that the sum over all $a$ and $b$ counts all transitions twice.}
We recall that $\DS^{(i)}_{ba}$ is the entropy change of reservoir $i$ 
when that reservoir induces a system transition $a \to b$, 
while ${\cal I}^{(i)}_{ba}$
is the probability current associated with this transition at time $t$.
Since the system state is typically non-thermal,
we cannot use Clausius' definition to calculate its rate of change of entropy. 
Instead, we use the Shannon entropy,
\begin{align}
\mathscr{S}_{\rm sys} = -\kB \sum_b P_b(t) \ln[P_b(t)].
\label{eq:Shannon}
\end{align}
Its time-derivative is simplified by the probability conservation condition 
$\sum_b {\rmd \over \rmd t} P_b(t) =0$,
we then use Eq.~(\ref{Eq:probability-currents}) to write
\begin{align}
{\rmd \over \rmd t} \mathscr{S}_{\rm sys}(t) 
% \ =\ - \kB \sum_b  {\rmd P_b(t)\over \rmd t} \,\ln [P_b(t)] 
\ =\ -\kB \sum_{abi} \, {\cal I}_{ba}^{(i)}(t) \,\ln 
\big[P_b(t)\big] \ .
\label{Eq:dotS_sys}
\end{align}
To proceed with the proof, we write ${\rmd \over \rmd t} \mathscr{S}_{\rm sys}$ as two copies terms of 
the 
right hand side of Eq.~(\ref{Eq:dotS_sys}) each divided by two,
and then interchange the dummy-indices $a\leftrightarrow b$ in \green{one of the} term.
Then since ${\cal I}_{ba}^{(i)}(t)=-{\cal I}_{ab}^{(i)}(t)$, we can 
write
\begin{align}
{\rmd \over \rmd t} \mathscr{S}_{\rm sys}(t) 
\ =\  \green{{1 \over 2}} \,\kB \sum_{abi} {\cal I}_{ba}^{(i)}(t)
\ \Big(\ln \big[P_a(t)\big]-\ln\big[P_b(t)\big] \Big). 
\label{Eq:dotS_sys-modified}
\end{align}
The total entropy of the system and the reservoirs at time $t$ is
\begin{align}
\mathscr{S}(t) = \mathscr{S}_{\rm sys}(t) +  \sum_i \mathscr{S}^{(i)}(t). 
\end{align}
Given Eqs.~(\ref{Eq:dotS_i}) and (\ref{Eq:dotS_sys-modified})
we conclude that the total entropy obeys
\begin{align}
{\rmd \over \rmd t} \mathscr{S}(t) 
\ =\ {\rmd \over \rmd t} \mathscr{S}_{\rm sys}(t)  + \sum_i  {\rmd \over \rmd t} 
\mathscr{S}^{(i)}(t) 
\ =\ \green{{1 \over 2}}\,\kB  \sum_{abi} {\cal I}_{ba}^{(i)}(t)
\ \Big(\ln \big[P_a(t)\big]-\ln\big[P_b(t)\big]+ {\DS^{(i)}_{ba}\big/ 
\kB} \Big). 
\end{align}
Now let us write  ${\cal I}_{ba}^{(i)}(t)$ in terms of rates, as in 
Eq.~(\ref{Eq:probability-currents}), and
use Eq.~(\ref{Eq:rate-for-reverse-final}) 
to write $\Gamma^{(i)}_{ab}$ in terms of $\Gamma^{(i)}_{ba}$.  
Then writing 
$\ln\big[P_b(t)\big]-\DS^{(i)}_{ab}\big/\kB = 
\ln\big[P_b(t)\e^{-\DS^{(i)}_{ba}/\kB}\big]$,
we get
\begin{align}
{\rmd \over \rmd t} \mathscr{S}(t) 
&= \green{{1 \over 2}}\,\kB\sum_{abi}   \Gamma^{(i)}_{ba} \ \Big( \, P_a(t)  
- P_b(t) \exp\big[{-\DS^{(i)}_{ba}/\kB}\big] \Big) 
\ \Bigg(\ln \big[P_a(t)\big]
- \ln\left[P_b(t) \exp\big[{-\DS^{(i)}_{ba}/\kB}\big]  \right] \Bigg). 
\label{Eq:dotS_total}
\end{align}
To arrive at the second-law, we must prove that this quantity cannot be negative.
To do so, we note that the only non-zero contributions to the sum are those 
with  $a\neq b$
(since $\DS^{(i)}_{bb}=0$), and that $\Gamma^{(i)}_{ba} \geq 0$
for all such  contributions. Next, we note that each term in the sum takes the 
form 
$(x-y)\big(\ln[x]-\ln[y]\big)$.  Since $\ln[x]$ is a monotonically increasing 
function of $x$, we have $(x-y)\big(\ln[x]-\ln[y]\big)\geq0$ for all $x,y$.
Thus we can conclude that none of the terms in the sum over $i$, $a$ and 
$b$
in Eq.~(\ref{Eq:dotS_total}) are negative.
Thus we have proven that any such rate equation will obey the
second-law of thermodynamics, 
in the form 
\begin{align}
{\rmd \over \rmd t}  \mathscr{S}(t) \ \geq \ 0
\label{Eq:2ndlaw}
\end{align}
We did not take the steady-state limit to get this result, so it applies 
even when the system state is time-dependent, for an arbitrary initial system 
state.  
In the steady-state limit, we have ${\rmd \over \rmd t} \mathscr{S}_{\rm sys}=0$, because
${\rmd \over \rmd t} P_b(t)=0$ for all $b$. However, assuming this at the 
beginning 
of the derivation does not simplify the 
proof of the second law.

One should not forget that the result in Eq.~(\ref{Eq:2ndlaw}) 
is for the entropy production averaged over a large number of transitions. It 
is thus only directly applicable to a given system in a situation where 
fluctuations about this average are small enough to be neglected.
This is typically the case for system responses on time-scales much longer than 
those for a transition in the system.
Since transitions are uncorrelated, we can apply central limit theorem, then 
the average 
entropy production calculated above scales like the number of transitions 
(which grows linearly in time), while fluctuations scale like the square-root of 
the number of transitions.  
Thus for long enough times, the fluctuations will become much less than the 
average, 
at which point we can neglect the fluctuations, and the second-law becomes a 
true ``law''
(applicable to any system under any conditions).  
However, on any shorter time-scale the second law is a universal statement that 
applies {\it only} to the average entropy production. 
Much more useful at such short times are
certain universal results known as {\it fluctuation theorems}, since these include fluctuations about the average, 
see sections~\ref{Sect:traj}.

\subsection{Efficiency of ``single-loop'' machines}
\label{Sect:single-loop}

\begin{figure}
\centerline{\includegraphics[width=0.55\textwidth]{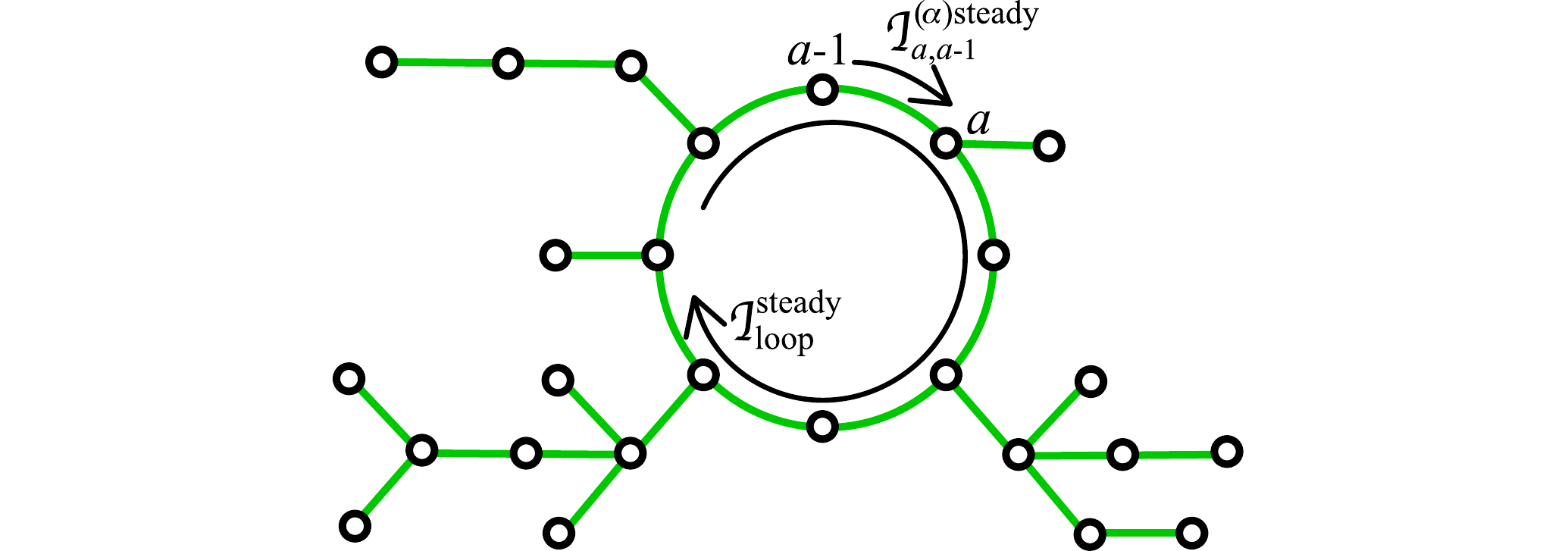}}
\caption{\label{Fig:single-loop}
The network of system states (vertices) and the transitions (bonds) between 
them, 
for a system in which the network has a {\it single loop}.  
The probability current from state $a-1$ to state $a$ in the loop is 
${\cal I}^{(i) {\rm steady}}_{a,a-1}$,
 where $i$ is the reservoir associated with the transition $a-1 \to a$.
 Concrete examples would be that in inset (a) of Fig.~\ref{Fig:2-term-sys}, or 
those in the the insets of 
 Figs.~\ref{Fig:3-term-sys1} and \ref{Fig:3-term-sys2} in cases where the 
dashed bonds can be neglected.
}
\end{figure}

Here we restrict our interest to the simplest machines, those whose rate 
equations 
correspond to a network that contains a single loop.
Concrete examples would be that in inset (a) of Fig.~\ref{Fig:2-term-sys}, or 
those in the the insets of 
 Figs.~\ref{Fig:3-term-sys1} and \ref{Fig:3-term-sys2} in cases where the 
dashed bonds can be neglected.
Fig.~\ref{Fig:single-loop} shows a more complicated network, which contains 
multiple side-branches but  
still only a single loop.
We will find the steady-state efficiency of such a machine using the 
Kirchhoff's law in 
Eq.~(\ref{Eq:Kirchhoff-for-prob-currents}), without needing to solve the set of 
simultaneous equations for the steady-state occupation probabilities given by
Eqs.~(\ref{Eq:steady-state-master-eqn}).
The logic followed in this section is inspired by Ref.~\cite{Einax-Nitzan2016,Einax-Nitzan2016b}.

The first thing to note is that Kirchhoff's law, 
Eq.~(\ref{Eq:Kirchhoff-for-prob-currents}), means there can be no steady-state 
probability current in the side branches on the network,
so ${\cal I}_{ba}^{(i) \rm steady}=0$ for $a$ and $b$ anywhere except in 
the loop.  
This can be proven by starting at the ends of each branch, where the the 
probability current is obviously zero, and then using Kirchhoff's law to see 
that bonds one step nearer to the loop
have zero probability current, and so forth, until one has addressed all bonds 
in each side-branch.
Then the only non-zero probability currents are on bonds in the loop,
for which Kirchhoff's law implies that the probability current on every bond in 
the loop is the same,
${\cal I}_{a,a-1}^{\rm (i)steady}= {\cal I}_{\rm loop}^{\rm steady}$ where 
$a-1$ and $a$ label any two neighbouring states (vertices) on the loop, and 
$i$ is the reservoir which is associated with the transition (bond)  from 
$a-1$ to $a$.

The steady-state currents of particles and energy
that enable the machine to convert a heat flow into power (or power into a heat 
flow), are proportional to the probability currents.  Thus, only the 
transitions in the loop are relevant, and they all have the same probability 
current,   ${\cal I}_{\rm loop}^{\rm steady}$.
From Eqs.~(\ref{Eq:I^N}-\ref{Eq:J}), we have the 
particle and energy currents  into the system from reservoir $i$ as
\begin{align}
\Jparticlei = {\cal I}_{\rm loop}^{\rm steady} \times 
\sum_{a \,\in \,\{i\}}  \Big(N_{a}-N_{a-1}\Big)  \, , 
\qquad \qquad
\Jenergyi = {\cal I}_{\rm loop}^{\rm steady} \times 
\sum_{a \,\in \,\{i\}}  \Big(E_{a}-E_{a-1}\Big)  \, .
\label{Eq:I^N+I^E-in-terms-of-I_loop}
\end{align}
Here ``$a \in \{i\}$'' indicates that the sum over all $a$ for 
which the transition $(a-1) \to a$ is associated with reservoir $i$,
while  $N_a$ and $E_a$ are the electron number and energy of system state $a$.
Thus, the heat current into the system from reservoir $i$ is
\begin{align}
\Jheati = {\cal I}_{\rm loop}^{\rm steady} \times 
\Delta Q_{\rm loop}^{(i)},
\label{Eq:J-in-terms-of-I_loop}
\end{align}
where $\Delta Q_{\rm loop}^{(i)}$ is the heat that enters the system from 
reservoir
$i$ in the transitions that form the loop, so 
\begin{align}
\Delta Q_{\rm loop}^{(i)} \equiv \sum_{a \,\in \,\{i\}}  
\Big(E_{a}-E_{a-1}- \mu_i \big(N_a-N_{a-1} \big)\Big).  \ 
\end{align}
The power generated in the electronic reservoir $i$ is 
\begin{align}
P_{\rm gen}^{(i)} =   {\cal I}_{\rm loop}^{\rm steady} \times 
\Delta W_{\rm loop}^{(i)},
\end{align}
where $\Delta W_{\rm loop}^{(i)}$ is the work done by the system on 
reservoir
$i$ in the transitions that form the loop, so 
\begin{align}
\Delta W_{\rm loop}^{(i)} \equiv - \sum_{a\,\in \,\{i\}}   \mu_i \big(N_a-N_{a-1} \big).  \ 
\end{align}
\green{
Next let us define $P_{\rm gen}$ as the sum of the power generated in all reservoirs,
and $J_{\rm heat}$ as the total heat absorbed from all reservoirs. 
Then 
\begin{align}
P_{\rm gen} &=  {\cal I}_{\rm loop}^{\rm steady} \times \Delta W^{\rm (gen)}_{\rm loop}
\nonumber 
\\
J_{\rm heat} &=  {\cal I}_{\rm loop}^{\rm steady} \times \Delta Q^{\rm (heat)}_{\rm loop}
\nonumber
\end{align}
where $\Delta W^{\rm (gen)}_{\rm loop}$ is the sum of  $\Delta W_{\rm 
loop}^{(i)}$ over all electronic reservoirs $i$ in which the 
electrical power is generated, and
 $\Delta Q^{\rm (heat)}_{\rm loop}$ is the sum of  $\Delta Q_{\rm 
loop}^{(i)}$ over all reservoirs $i$ which act as heat sources.
Since the heat-engine efficiency is $\eta_{\rm eng} = {P_{\rm gen} \big/ J_{\rm heat}}$, 
this efficiency is independent of the probability current ${\cal I}_{\rm loop}^{\rm steady}$ 
in a single-loop steady-state machine; it is simply}
\begin{align}
\eta_{\rm eng}  \ = \ 
{\Delta W^{\rm (gen)}_{\rm loop} \over  \Delta Q^{\rm (heat)}_{\rm loop} }.
\label{Eq:single-loop-eng}
\end{align}
Similarly, the coefficient of performance, $\eta_{\rm fri} = { J_{\rm cold}\big/ P_{\rm abs} }$, 
of a single-loop steady-state 
refrigerator is 
\begin{align}
\eta_{\rm fri} \ = \  {\Delta 
Q^{\rm (cold)}_{\rm loop}
\over  \Delta W^{\rm (abs)}_{\rm loop} }.
\label{Eq:single-loop-fridge}
\end{align}
Here, $\Delta W^{\rm (abs)}_{\rm loop}$ is the sum of  $-\Delta W_{\rm 
loop}^{(i)}$ over all electronic reservoirs $i$ which supply the 
electrical power absorbed by the machine, 
while  $\Delta Q^{\rm (cold)}_{\rm loop}$ is the sum of  $\Delta Q_{\rm 
loop}^{(i)}$ over all reservoirs $i$ being refrigerated.

This shows that one can find the efficiencies without solving the steady-state 
equation,
because they are given by ratios in which ${\cal I}_{\rm loop}^{\rm steady}$ 
cancels out.
As such, they only depend on the energy, $E_a$,  and particle number, $N_a$, for 
each state in the loop, as shown in 
Eqs.~(\ref{Eq:single-loop-eng},\ref{Eq:single-loop-fridge}).
In contrast, we cannot find other quantities, such as the power generated, 
without 
knowing ${\cal I}_{\rm loop}^{\rm steady}$. 
The only way to find ${\cal I}_{\rm loop}^{\rm steady}$ is 
to solve the full steady-state problem, given by the simultaneous equations in 
Eq.~(\ref{Eq:steady-state-master-eqn}),
and then use Eq.~(\ref{Eq:probability-currents}) to find the probability 
current at some 
point in the loop. 
We also note that if the machine's network contains multiple loops, then  
the probability currents do not drop out of the efficiencies, 
so one cannot find the efficiency without finding the steady-state solution 
of the rate equations.

%========================================
%========================================

 \subsection{Stochastic thermodynamics for rate equations}
\label{Sect:traj}

Most of the time, the simplest way to get quantitative results about the 
steady-state is to solve the 
steady-state equations, see 
sections~\ref{Sect:2-term-sys}-\ref{Sect:3-term-sys2} for specific examples.
However, one can also think in terms of stochastic trajectories that explore the space of system 
states with time.
The price to be paid is that the number of such trajectories increases 
exponentially with time.
Despite this, certain useful results can arrived at from considering these 
trajectories.  The study of such trajectories is known as {\it stochastic thermodynamics} \cite{seifert,review-Broeck-Esposito}.
Here, we apply ideas of stochastic thermodynamics to the rate equations introduced above, 
and show how they can be used to derive fluctuation theorems,
and a simple rule for the achievability of Carnot efficiency.
Section~\ref{sec:machines} briefly discusses stochastic thermodynamics in contexts other than the rate equations considered here.

Entropy production is a probabilistic process, and 
the second law of thermodynamics is only a statement about  
{\it average} entropy production. 
In most macroscopic situations, the statistical fluctuations about this average 
are 
extremely small, and can be neglected.  However, the fluctuations may be 
significant in nanoscale system, and deserve closer study. 
Suppose that we are able to resolve 
individual transitions in the system, then we would be able to follow the 
entropy change of the system and reservoirs transition by transition.
On the scale of a few transitions, we expect to see significant violations of 
the second law.
While some aspects of these violations are system specific, there are certain 
{\it universal}
results for these violations known as {\it fluctuation theorems}.
Deriving and understanding the meaning of such fluctuation theorems in quantum systems
is crucial to understand the quantum thermodynamics of such systems, 
for a review see \cite{Campisi-review}.

%--------------------------------------------------------
\subsubsection{Stochastic trajectories}

Let us now define a trajectory of the system dynamics, $\zeta$, as a given 
series of 
$n$ transitions; the system starts in a state $ a_0$ at time $t_0$, 
followed by a transition to state $a_1$ at time $t_1$ due to the coupling to 
reservoir $i_1$, 
followed  by a transition to state $a_2$ at time $t_2$ due to the coupling to 
reservoir $i_2$,
and so forth, until the system makes a transition to its final state $a_n$ at 
time $t_n$, and remains in this state until time $t$.  Obviously, we take $t_0 
< t_1 <t_2< \cdots < t_n < t$.
Let us denote this trajectory as
\begin{align}
\zeta \ \equiv 
\begin{array}{c} \phantom{i_0} \\ | \\ t_0 \end{array} 
\hskip 1mm \raisebox{6pt}{$a_0$} 
\hskip -7.3mm {\xrightarrow{\hspace*{10mm}} }
\hskip -2.5mm \begin{array}{c} i_1 \\ {\bm |} \\ t_1 \end{array}  
\hskip 1mm \raisebox{6pt}{$a_1$} 
\hskip -7.5mm {\xrightarrow{\hspace*{10mm}} }
\hskip -2.5mm \begin{array}{c} i_2 \\  {\bm |} \\ t_2 \end{array}  
\hskip 1mm \raisebox{6pt}{$a_2$} 
\hskip -7.5mm {\xrightarrow{\hspace*{10mm}} }
\  \cdots \ 
\hskip 2.5mm \raisebox{6pt}{$a_{n-1}$} 
\hskip -8.5mm {\xrightarrow{\hspace*{10mm}} }
\hskip -2.5mm \begin{array}{c} i_n \\  {\bm |} \\ t_n \end{array} 
\hskip 1mm \raisebox{6pt}{$a_n$} 
\hskip -7.5mm {\xrightarrow{\hspace*{10mm}} }
\hskip -2.5mm \begin{array}{c} \phantom{i_0} \\ | \\ t \end{array}. 
\label{Eq:def-trajectory-zeta}
\end{align}
We will compare this trajectory with its {\it time-reverse} which we call 
$\bar{\zeta}$,
evolving in a system with {\it time-reversed parameters}.
The time-reversing of the trajectory means
\begin{align}
\bar \zeta \ \equiv 
\hskip -2mm \begin{array}{c} \phantom{i_0} \\ | \\ t_0 \end{array} 
\hskip 1mm \raisebox{6pt}{$a_n$} 
\hskip -7.3mm {\xrightarrow{\hspace*{10mm}} }
\hskip -2.5mm \begin{array}{c} i_n \\ {\bm |} \\ {\bar t}_n \end{array}  
\hskip 0mm \raisebox{6pt}{$a_{n-1}$} 
\hskip -9mm {\xrightarrow{\hspace*{11mm}} }
\hskip -3.5mm \begin{array}{c} i_{n-1} \\  {\bm |} \\  {\bar t}_{n-1}  
\end{array}  
\hskip -.5mm \raisebox{6pt}{$a_{n-2}$} 
\hskip -10mm {\xrightarrow{\hspace*{11mm}} }
\  \cdots \ 
\hskip 4mm \raisebox{6pt}{$a_1$} 
\hskip -6.5mm {\xrightarrow{\hspace*{8mm}} }
\hskip -2.5mm \begin{array}{c} i_1 \\  {\bm |} \\ {\bar t}_1 \end{array} 
\hskip 1mm \raisebox{6pt}{$a_0$} 
\hskip -7.5mm {\xrightarrow{\hspace*{8mm}} }
\hskip -2.5mm \begin{array}{c} \phantom{i_0} \\ | \\ t \end{array}, 
\label{Eq:def-trajectory-barzeta}
\end{align}
where the time ${\bar t}_k \equiv t_0+t-t_k$, so 
$t_0 <{\bar t}_n < \cdots < {\bar t}_2< {\bar t}_{1} < t$.
When we say that the system has {\it time-reversed parameters} \footnote{We assume the time-dependences do not violate the conditions which allow the use of the rate equation discussed here,  see the last paragraph of section~\ref{Sect:master-coherences}.}, we mean that we consider the evolution in a 
different system whose parameters (transition rates, system Hamiltonian, etc.) 
are related to those of the original system by
\begin{align}
\overline{\Gamma}_{ba}^{\rm \,(i)} (\tau) = \Gamma_{ba}^{\rm(i)} 
(t_0+t-\tau),
\qquad
\overline{\cal H}_{\rm s}(\tau)= \hat{\cal H}_{\rm s}(t_0+t-\tau).
\label{Eq:time-reversed-rates}
\end{align}
If one only considers time-independent ${\cal H}_{\rm s}$ and time-independent rates, as we do in most of this review, 
then the time-reversing of these parameters can be forgotten,  and $\zeta$ and 
$\bar\zeta$ are time-reversed trajectories 
in the same system.  However, we think it is important to keep the 
time-dependence in the rates in derivation, 
so one can see which of the results we get for time-independent couplings 
cannot be trivially extended to time-dependent couplings.

The Markovian nature of the rate equation governing this dynamics 
means that the probability of following such paths is the product of the 
probabilities for each transitions.
For a single transition, the probability for the system to remain in state 
$a_m$ from time $t_m$ to time $t_{m+1}$ and then make a transition to state 
$a_{m+1}$ at time $t_{m+1}$ through an interaction with reservoir 
$i_{m+1}$ is 
$\Gamma^{(i_{m+1})}_{a_{m+1} a_m}(t_{m+1}) 
\exp\left[-\int_{t_m}^{t_{m+1}}\Gamma_{a_m}(\tau) \rmd \tau \right]$,
where we define $\Gamma_{a}(t)\equiv \sum_{b,i} \Gamma^{(i)}_{ba}(t)$ 
as the total rate of leaving state $a$ at time $t$ (so the sum is over 
transitions to any state $b$ due to the coupling to any reservoir $i$).
Thus the probability of trajectory $\zeta$ is
\begin{align}
P(\zeta) =  \exp\left[{\textstyle -\int_{t_n}^{t}\Gamma_{a_n}(\tau) \rmd \tau} 
\right] \ \prod_{m=0}^{n-1}\, 
\Gamma^{(i_{m+1})}_{a_{m+1} a_m} (t_{m+1})
\, \exp\left[{\textstyle -\int_{t_m}^{t_{m+1}}\Gamma_{a_m}(\tau) \rmd \tau 
}\right].
\label{Eq:prob-traj-zeta}
\end{align}
The equivalent expression for the probability of the time-reversed path in the 
time-reversed system
is 
\begin{align}
\overline{P}\left(\bar{\zeta}\right) =  \exp\left[{\textstyle -\int_{{\bar 
t}_1}^{t}\overline\Gamma_{a_0}(\tau) \rmd \tau} \right] \ \prod_{m=1}^{n}\, 
\overline\Gamma^{\,(i_m)}_{a_{m-1} a_m} \left({\bar t}_m\right)
\, \exp\left[{\textstyle -\int_{{\bar t}_{m+1}}^{{\bar t}_{m} 
}\overline\Gamma_{a_m}(\tau) \rmd \tau }\right],
\label{Eq:prob-traj-barzeta}
\end{align}
where we recall that ${\bar t}_m$ is defined below 
Eq.~(\ref{Eq:def-trajectory-barzeta}), and for compactness we define
${\bar t}_{n+1}\equiv t_0$.
We now replace all time-reversed system rates using 
Eq.~(\ref{Eq:time-reversed-rates}), 
noting that
\begin{align}
\int_{{\bar t}_{m+1}}^{{\bar t}_{m}}\overline\Gamma_{a_m}(\tau) \rmd \tau 
= \int_{t_{m}}^{t_{m+1}} \Gamma_{a_m}(\tau) \rmd \tau , 
\end{align}
one sees that
the exponents in $\overline{P}\left(\bar{\zeta}\right)$ and $P(\zeta)$ are the same.
 As a result,
\begin{align}
\overline{P}\left({\bar{\zeta}}\right) \ =\ P(\zeta)  \ \times \  
{\displaystyle {\prod_{m=1}^{n}\, \Gamma^{(i_m)}_{a_{m-1} a_{m}}(t_m)} \over
{\phantom{-}\displaystyle \prod_{m=0}^{n-1}\, \Gamma^{(i_{m+1})}_{a_{m+1} a_{m}}(t_{m+1}) }
\phantom{-}}\ ,
\end{align}
at which point one can use Eq.~(\ref{Eq:rate-for-reverse-final}) to arrive 
directly at
\begin{align}
\overline{P}\left({\bar{\zeta}}\right) = P(\zeta) \ \times \  \exp\left[- {1 \over 
\kB}\Delta \mathscr{S}_{\rm res}(\zeta) \,\right],
\label{Eq:trajectory-fluct-rel}
\end{align}
where $\Delta\mathscr{S}_{\rm res}(\zeta)=\sum_{k=1}^{n} \Delta 
\mathscr{S}^{(i_k)}_{a_ka_{k-1}}$ is the total change in entropy in the reservoirs 
during trajectory $\zeta$.
We recall that the left hand side of Eq.~(\ref{Eq:trajectory-fluct-rel}) is the 
probability to follow the time-reversed trajectory in the system with 
time-reversed reservoir couplings, defined by Eq.~(\ref{Eq:time-reversed-rates}). 

Eq.~(\ref{Eq:trajectory-fluct-rel}) is a crucial relation, 
which we will use to derive a simple rule for achieving Carnot 
efficiency (see section~\ref{Sect:rule-Carnot}), and to derive some fluctuation theorems
(see sections~\ref{Sect:fluct-theorem-steady-state}, \ref{Sect:Crooks} and \ref{Sect:Seifert}).
In general, Eq.~(\ref{Eq:trajectory-fluct-rel}) is a relation between 
trajectories in two {\it different} systems; 
\green{one with its parameters time-reversed with respect to the other one} 
(we placed a bar over $P$ on the left hand side to recall this).
However, in the case of systems where \green{${\cal H}_{\rm s}$ and} the reservoir couplings are 
time-independent, 
then the relation becomes one between a
trajectory and its time-reverse in the {\it same} system (so one can drop the 
bar over $P$).

%--------------------------------------------------------------------------
\subsubsection{Fluctuation theorem for the steady-state}
\label{Sect:fluct-theorem-steady-state}

One should not forget that the steady-state is a state in which  the {\it 
average} occupation probabilities for each system state
do not vary in time.  However, this does not mean there are not time-dependent 
fluctuations about this average.  Such fluctuations will typically dominate on 
short time-scales,
while becoming irrelevant on long enough time-scales.  Thus 
to observe such fluctuations, one has to design the system so one can see the 
short time dynamics
of the system (ideally on the time-scale of individual transitions).

The objective of this section is to derive
the steady-state fluctuation relation of Evans and Searles 
\cite{Evans-Searles1994},
which says that the probability $P(-\Delta\mathscr{S},t)$ that the system undergoes a 
fluctuation that produces 
entropy $-\Delta\mathscr{S}$ in a time $t$ (i.e.\ it reduces the total entropy of system 
and reservoirs) is
\begin{align}
P(-\Delta\mathscr{S},t) =P(\Delta\mathscr{S},t) \exp\big[-\Delta\mathscr{S}\big/ \kB\big]\ .
\label{Eq:detailled-fluctuation-theorem}
\end{align}
Thus, the entropy can be reduced, but it is always more likely to be produced.
The intriguing thing is that while the distribution of entropy production, 
$P(\Delta\mathscr{S},t)$, 
is system specific for $\Delta\mathscr{S} >0$, the relation 
in Eq.~(\ref{Eq:detailled-fluctuation-theorem}) 
between $P(-\Delta\mathscr{S},t)$ and $P(\Delta\mathscr{S},t)$ is universally true for the 
steady-state response 
of any system.

Here, we reproduce the proof of the fluctuation theorem in 
Eq.~(\ref{Eq:detailled-fluctuation-theorem}) 
in Ref.~\cite{Seifert-PRL2005,review-Seifert2007,seifert,review-Broeck-Esposito}, which is based on the 
relation for individual trajectories in Eq.~(\ref{Eq:trajectory-fluct-rel}).
However, the first step is to assign an entropy to the system's initial and 
final state, 
even though these states are typically non-thermal distributions, and are not 
defined 
by a temperature. For this, Seifert \cite{Seifert-PRL2005,review-Seifert2007} 
argued that the entropy $\mathscr{S}^{\rm sys}_a$ that one should assign to system state 
$a$ is
\begin{align}
\mathscr{S}^{\rm sys}_a =-\kB \ln[P_{a}],
\label{Eq:S_a^sys}
\end{align} 
where $P_{a}$ is the occupation probability for state $a$. 
This choice can be motivated by noting that if one sums over 
all initial states,
the entropy of the system would be $\mathscr{S}^{\rm sys} = \sum_a P_a \mathscr{S}^{\rm sys}_a = 
-\kB \sum_a P_a \ln[P_a]$ 
which corresponds to the Shannon entropy.  We refer the reader to 
Refs.~\cite{review-Seifert2007} for an explanation of the other reasons for 
choosing 
Eq.~(\ref{Eq:S_a^sys}) as the entropy associated with a given system state.   
Given Eq.~(\ref{Eq:S_a^sys}), we see that the total change in entropy of the 
system and environment, 
$\Delta\mathscr{S}(\zeta)$, associated with trajectory $\zeta$ from $a_0$ at 
time $t_0$ to $a$ 
at time $t$ is 
\begin{align}
\Delta\mathscr{S}(\zeta) \ = \   
\Delta\mathscr{S}_{\rm res}(\zeta) + \mathscr{S}^{\rm sys}_{a} (t) - \mathscr{S}^{\rm sys}_{a_0} (t_0) 
\ = \
\Delta\mathscr{S}_{\rm res}(\zeta) + \kB \, \ln\left[P_{a_0}(t_0) \,\big/\, P_a(t)\right] 
,
\label{Eq:DeltaS_tot}
\end{align}
where $\Delta\mathscr{S}_{\rm res}(\zeta)$ is defined below 
Eq.~(\ref{Eq:trajectory-fluct-rel}).
This means that
\begin{align}
 {P_{a_0}(t_0)  \over P_{a}(t)}\exp\left[-{\Delta\mathscr{S}(\zeta)\over \kB} 
\right]  \ &= \  \exp\left[-{\Delta\mathscr{S}_{\rm res}(\zeta)\over \kB} \right]  ,
\label{Eq:exp-DeltaS_tot}
\end{align}
which we combine with Eq.~(\ref{Eq:trajectory-fluct-rel}) to get
\begin{align}
P\left(\zeta\right) \ P_{a_0}(t_0) \ &= \ {\overline{P}}\left(\bar\zeta\right)\ 
P_{a}(t) \ 
\exp\left[-\Delta\mathscr{S}(\zeta)\big/ \kB \right]\ , 
\label{Eq:P-zeta-versus-P-barzeta}
\end{align}
where we use the fact that $\Delta\mathscr{S}(\bar\zeta)=-\Delta\mathscr{S}(\zeta)$.

The probability that the system produces an entropy $-\Delta\mathscr{S}$ during the time 
from $t_0$ to $t$
can be written as the following sum over trajectories:
\begin{align}
P(\Delta\mathscr{S},t) = \sum_{a,a_0}\  \sum_{\zeta \in \{a_0,t_0\to a, t\}} 
\delta \left[\Delta\mathscr{S}- \Delta\mathscr{S}_{\scriptstyle{\rm  tot}}(\zeta) \right] \ 
P(\zeta) \ P_{a_0}(t_0),   
\label{Eq:P-DeltaS-as-traj-sum}
\end{align}
where the sum is over all trajectories from $a_0$ at time $t_0$ to $a$ at time 
$t$, but the Dirac $\delta$-function picks out only those trajectories 
which generate a total entropy equal to $\Delta\mathscr{S}$.
Substituting Eq.~(\ref{Eq:P-zeta-versus-P-barzeta}) into the 
right hand side of 
Eq.~(\ref{Eq:P-DeltaS-as-traj-sum}), and then making the substitution 
$\Delta\mathscr{S}(\zeta)=-\Delta\mathscr{S}(\bar\zeta)$ in the $\delta$-function, 
one gets
\begin{align}
P(\Delta\mathscr{S},t) =   \e^{\Delta\mathscr{S}/\kB} \sum_{a,a_0}\  \sum_{\zeta \in \{a_0,t_0\to 
a, t\}} 
\delta \left[\Delta\mathscr{S}+ \Delta\mathscr{S}_{\scriptstyle{\rm  tot}}(\bar\zeta) \right] \, 
{\overline{P}}\left(\bar\zeta\right)  P_{a}(t),   
\label{Eq:P(DeltaS)-to-barP}
\end{align}
where we have used the presence of the $\delta$-function to replace 
$\Delta\mathscr{S}(\bar\zeta)$ by $-\Delta\mathscr{S}$ in the exponent, and then noted that it 
simply forms a constant prefactor on the sums.  
The fact we are considering time-independent reservoir coupling means that we 
can drop the bar over $P$.
Next  we replace $\bar\zeta$ by $\zeta$, noting that the sum now runs over all 
paths $\zeta$ from 
$a$ at time $t_0$ to $a_0$ at time $t$.  Then Eq.~(\ref{Eq:P(DeltaS)-to-barP}) becomes
\begin{align}
P(\Delta\mathscr{S},t) \e^{-\Delta\mathscr{S}/\kB} = \! \sum_{a,a_0}\  \sum_{\zeta \in \{a,t_0\to 
a_0, t\}} 
\delta \left[\Delta\mathscr{S}+ \Delta\mathscr{S}_{\scriptstyle{\rm  tot}}(\zeta) \right] \, 
P\left(\zeta\right)  P_{a}(t)   \ .
\label{Eq:P(DeltaS)-to-P}
\end{align}
Finally, since we are in the steady-state $P_{a}(t)=P_{a}(t_0)=P_a^{\rm 
steady}$,
we see by comparison with Eq.~(\ref{Eq:P-DeltaS-as-traj-sum}) that the right 
hand side is simply $P(-\Delta\mathscr{S},t)$
% (with dummy variables $a_0$ and $a$ interchanged).
Thus, we have used the fact the system is in a steady-state to prove the 
fluctuation relation 
in Eq.~(\ref{Eq:detailled-fluctuation-theorem}).

%-------------------------------------
\subsubsection{A rule for achieving Carnot efficiency}
\label{Sect:rule-Carnot}

We can use the trajectories introduced in section~\ref{Sect:traj} to derive a 
simple rule 
for achieving Carnot efficiency.    The rule enables one to tell if a given 
machine can be 
Carnot efficient or not, without having to solve the steady-state rate 
equation. 
Carnot efficiency is only achievable if the machine produces no entropy on 
average,
$\langle \Delta\mathscr{S}(t;t_0) \rangle=0$.  
Our objective is to find out what this means in terms of trajectories.
One can write the average entropy production in terms of trajectories
as
\begin{align}
\langle \Delta\mathscr{S}(t;t_0) \rangle 
= \sum_{a_0,a}\ \sum_{\zeta \in \{a_0,t_0 \to a,t \} } \Delta\mathscr{S}(\zeta) 
\,P(\zeta) \, P_{a_0}(t_0) .
\end{align}
Let us now write $\langle \Delta\mathscr{S}(t;t_0)\rangle$ 
as  two copies of this sum each divided by two.
Since we sum over all trajectories and over all $a_0,a$, we can replace all 
trajectories by their time-reverse 
while interchanging $a_0$ and $a$ without changing the result of this sum.
Upon doing this we have 
\begin{align}
\langle \Delta\mathscr{S}(t;t_0) \rangle =
\sum_{a_0,a}\  \sum_{\zeta \in \{a_0,t_0 \to a,t \} } {\Delta\mathscr{S}(\zeta) 
\over 2} 
\left( P(\zeta) \, P_{a_0}(t_0) - P(\bar\zeta) \, P_{a}(t_0) \right)\, ,
\end{align}
where we used the fact that $\Delta\mathscr{S}(\bar\zeta) =-\Delta\mathscr{S}(\zeta) $.
Now since we are considering a system with time-independent couplings in the 
steady-state, we can 
use  $P_a(t_0)=P_a(t)=P_a^{\rm steady}$ to 
replace $P_a(t_0)$ by $P_a(t)$ and 
substitute in Eq.~(\ref{Eq:P-zeta-versus-P-barzeta}) (we drop the bar over $P$ in  Eq.~(\ref{Eq:P-zeta-versus-P-barzeta}) because we are considering a time-independent situation).
Then we have
\begin{align}
\langle \Delta\mathscr{S}(t;t_0) \rangle =& \half 
\sum_{a_0,a}\ \sum_{\zeta \in \{a_0,t_0 \to a,t \} } \Delta\mathscr{S}(\zeta) 
\Big(1 - \exp\left[- \Delta\mathscr{S}(\zeta)\big/ \kB \right]  \Big) 
 \ P(\zeta) \ P_{a_0}^{\rm steady}\ .
\end{align}
The term containing $ \Delta\mathscr{S}(\zeta)$ takes the form $x(1-\e^{-x})$, 
and this is greater than or equal to zero for all $x$.
Since all other factors are probabilities (and so not negative),
we see that $\langle \Delta\mathscr{S}(t;t_0) \rangle$ is never negative.  This 
constitutes another proof that the rate equation obeys the second-law 
of thermodynamics.
However, it also gives us more information;
The only way to arrive at $\langle \Delta\mathscr{S}(t;t_0) \rangle=0$, is for 
every term in the sum to be zero.
Thus to achieve Carnot efficiency each trajectory $\zeta$ that the system could 
follow 
must generate zero entropy, $\Delta\mathscr{S}(\zeta)=0$.

If Carnot efficiency requires that no trajectory generates entropy, it is 
obviously necessary (but not sufficient) that no
{\it closed} trajectory $\zeta_{\rm closed}$ generates entropy
(a closed trajectory being one which starts and ends at the same system state, 
$a=a_0$).
We see from Eq.~(\ref{Eq:DeltaS_tot}) that
$\Delta\mathscr{S}(\zeta_{\rm closed})=\Delta\mathscr{S}_{\rm res}(\zeta_{\rm closed})$ for such a trajectory, 
which makes the entropy that it generates independent of the
occupation probabilities of the steady-state.  
Thus, without solving the steady-state equation, \green{our objective is} to find 
the conditions under which the  entropy generated in the reservoirs around all 
closed trajectories is zero 
(or to show that no such conditions exist). 
Note that the closed trajectory may involve entropy flow from one reservoir to 
another, 
but the sum of the entropy change over all reservoir for the closed trajectory 
must be zero.
While the closed trajectories for long time response (relevant  to the 
steady-state) are very long, they can be 
broken into many {\it primitive closed trajectories}.
By ``primitive closed trajectories'', we simply mean a finite set of closed 
trajectories
from which all other closed trajectories can be constructed (it is largely a 
matter of convenience how one chooses this set).  
For systems with a relatively small number of states, 
there are relatively few such primitive trajectories, and they are fairly 
short. 
If the entropy generated around these closed primitive trajectories is zero, 
then
the entropy generated for all closed trajectories is zero.  In addition, 
self-retracing closed 
trajectories never generates any entropy in the reservoirs, so one can focus ones 
attention on 
those which  do not self-trace.

Requiring that every closed trajectory must produce zero entropy in the 
reservoirs is 
obviously a necessary condition for the system to be Carnot efficient.  
However, we will now argue that it is also a sufficient condition.  The fact 
that closed trajectories 
generate no entropy in the reservoirs, means that \green{every} open trajectory from 
$a_0$ to $a$ generates the same entropy in the reservoirs.  Thus one can always 
choose $P_a$ with respect to $P_{a_0}$ such that the entropy change in the 
system $\mathscr{S}^{\rm sys}_a-\mathscr{S}^{\rm sys}_{a_0}$
is equal and opposite to that entropy change in the reservoirs.  Doing this for 
all $a$ gives a unique value of $P_a$ for each $a$ which satisfies the 
condition that no open trajectory produces any entropy.
\green{However, it is not guaranteed that this recipe for choosing $P_a$ is the steady-state solution of the
rate equation.  If is not the steady-state solution, it will not be consistent with the assumptions 
made to get this far, and the recipe will not be valid.  To show that this is not a problem and that 
the recipe does give a solution that coincides with the steady-state one, it is sufficient
to focus on} 
trajectories associated with a single transition from state $a$ 
to state $b$
due to a single interaction with reservoir $i$,
the condition that the total entropy does not change for such a 
single-transition trajectory is 
$0 = \Delta\mathscr{S}_{ba}^{(i)}/\kB - (\ln P_b -\ln P_a )$ for all $a,b,i$. 
Given
Eq.~(\ref{Eq:rate-for-reverse-final}), this means that 
\begin{align}
\Gamma_{ba}^{(i)} P_a = \Gamma_{ab}^{(i)} P_b\qquad  \hbox{ for all } 
a,b,i.
\label{Eq:master-detailed-balance-for-Carnot}
\end{align} 
This is reminiscent of the detailed balance relation discussed in 
section~\ref{Sect:master-zeroth-law},
and it is trivial to see that it satisfies the condition for a steady-state, 
Eq.~(\ref{Eq:steady-state-master-eqn}). 
\green{
This is sufficient to see that the probabilities, $P_a$, given by the above recipe do coincide with the steady-state of such a system in which no closed trajectory generates any entropy.
This, in turn, means that open trajectories in such a system (in its steady-state) generate no entropy.
}  

Hence, requiring every closed trajectory 
to produce zero entropy is both a necessary and a sufficient condition for the 
system to have Carnot efficiency.
These results will be useful enough that we call it a ``rule''. 
\begin{itemize}
\item[] {\bf Rule for achieving Carnot efficiency:} The requirement to achieve 
Carnot efficiency in the steady-state, is that {\it all} closed trajectories in 
the system's 
state-space must generate zero entropy in the reservoirs. 
For this, it is sufficient to verify the absence of entropy generation for 
every primitive closed trajectory which is not self-retracing.   
\end{itemize}
What is nice about this rule, is that one does not have to solve the 
steady-state rate equation
to see if a system is Carnot efficient or not, one just has to inspect the 
primitive closed trajectories.  
Section~\ref{Sect:master-examples} shows how this rule can easily be applied to
a variety of concrete systems. 

For the machines with two or three reservoirs that we know of (see section~\ref{Sect:master-examples}), 
the ones that achieve Carnot efficiency obey a {\it tight coupling} condition.
Unlike in linear response, it is not clear if this is a necessary requirement, or simply a convenient manner to
achieve a system that easily satisfies the above rule.
A system obeys the tight coupling condition if every electron entering or leaving it from a given reservoir 
carries exactly the same amount of energy.  This implies that the ratio of energy current to particle current,
$\Jenergyi/\Jparticlei$, is a constant determined by system properties, independent of all reservoir biases and temperatures.
An explicit example of such a system is that in Fig.~\ref{Fig:3-term-sys2}
in the situation where the transitions indicated by the dashed lines in the inset are absent.
Then each electron leaving reservoir L carries an energy of exactly $\eps_1$ out of the reservoir,
so $\JenergyL= \eps_1\JparticleL$.  If, in contrast, we allow the transitions marked by the dashed lines in the inset of Fig.~\ref{Fig:3-term-sys2}, then the system does not obey the tight coupling condition, for example 
an electron leaving reservoir L can carry energy $\eps_1$ or $\eps_2$. Then the ratio of energy current to particle current will depend on transition rates, which in turn depend on reservoir biases and temperatures. 
At a hand-waving level, one can see why tight coupling makes it easier to satisfy the above rule for Carnot efficiency.
If one does not have tight coupling, it is because at least one reservoir couples to two system transitions
with different energies.  In this case, it is likely that there are at least two loops in the system,
in which case it is harder to tune all parameters to ensure that no loop generates any entropy.

It is important to note that Eq.~(\ref{Eq:master-detailed-balance-for-Carnot}) implies that there are no currents flowing in the system, cf.\ section~\ref{Sect:currents}.  Thus the machine is Carnot efficient, but produces no power.  However, if we make a small change in the parameters (typically changing the electrochemical potential of a reservoir), we can get a machine which generates  a small (but finite) amount of power
at an efficiency which is only very slightly less than that of Carnot.

Finally, we note that the above arguments mean that a Carnot efficient machine exhibits no fluctuations in its 
entropy production.  At no moment does it have a fluctuation which increases or reduces entropy.

\subsubsection{Crooks' fluctuation theorem}
\label{Sect:Crooks}

Having derived the steady-state fluctuation theorem 
in section~\ref{Sect:fluct-theorem-steady-state}, we note that we can get 
Crooks' fluctuation theorem \cite{Crooks1999} from an almost identical derivation.
The difference is in the choice of system and protocol.
We assume the set-up has time-dependent parameters,
so we cannot drop the bar over $P$ on the right hand side of Eq.~(\ref{Eq:P(DeltaS)-to-barP}).
Although, we assume that the time-dependence does not violate the conditions in the last paragraph of 
section~\ref{Sect:master-coherences}, which allow the use of the rate equation discussed here.
This means the system is not in a steady state, none the less  we assume that the initial and final state of the system are the same, so $P_a(t)=P_a(t_0)$ for all $a$.
Crooks \cite{Crooks1999} pointed out that this rather restrictive assumption is natural in certain non-steady-state situations.
For example, suppose we start with the system in a thermal state of ${\cal H}_{\rm sys}(t_0)$ at the temperature equal to that of reservoir $i$.  We can then manipulate the system as we wish, changing  ${\cal H}_{\rm sys}$, turning on and off couplings to different reservoirs, etc., up until some time $t'$.  We then take the system's Hamiltonian back to its value at $t_0$, 
and decouple the system from all reservoirs except reservoir $i$.
 Whatever the state of the system at time $t'$, it will relax towards a state identical to its state at $t_0$.  
 If the time $t-t'$ is large enough, the
 state of the system at time $t$ will be practically indistinguishable from its state at $t_0$, and we will have
 $P_a(t)=P_a(t_0)$ for all $a$.

 Armed with this information, we see that the derivation in section~\ref{Sect:fluct-theorem-steady-state} applies to 
 the evolution from time $t_0$ to time $t$ for any time-dependent parameters, so long as the system state obeys 
 $P_a(t)=P_a(t_0)$ for all $a$.  The only difference in the derivation is that we cannot drop the bar over $P(\zeta)$ on the right hand side when we go from Eq.~(\ref{Eq:P(DeltaS)-to-barP}) to Eq.~(\ref{Eq:P(DeltaS)-to-P}).
 Thus instead, when we replace  $P_a(t)$ by $P_a(t_0)$, we get
\begin{align}
P(\Delta\mathscr{S},t) \e^{-\Delta\mathscr{S}/\kB} = \! \sum_{a,a_0}\  \sum_{\zeta \in \{a,t_0\to 
a_0, t\}} 
\delta\left[\Delta\mathscr{S}+\Delta\mathscr{S}_{\scriptstyle{\rm  tot}}(\zeta)\right]
\overline{P}\left(\bar{\zeta}\right)  P_{a}(t_0)  . 
\label{Eq:P(DeltaS)-to-P-Crooks}
\end{align}
Comparing this equation with Eq.~(\ref{Eq:P-DeltaS-as-traj-sum}), we see that the right-hand side is the probability for the system undergoes a fluctuation which produces entropy $-\Delta\mathscr{S}$ between time $t_0$ and time $t$, if the system is evolving under the time-reversed parameters compared to the original system, see Eq.~(\ref{Eq:time-reversed-rates}) .
Thus, we arrive at Crooks' fluctuation theorem,
\begin{align}
\overline{P}(-\Delta\mathscr{S},t) =P(\Delta\mathscr{S},t) \exp\big[-\Delta\mathscr{S}\big/ \kB\big]\ .
\label{Eq:Crooks}
\end{align}
This relations differs from the steady-state one, Eq.~(\ref{Eq:detailled-fluctuation-theorem}), by the bar over $P$,
which indicates that the equality relates the probability of the change of entropy $\Delta\mathscr{S}$ in a time-dependent problem, 
and the probability of the opposite change of entropy, $-\Delta\mathscr{S}$, in a system with {\it time-reversed parameters}.

%------------------------------------------------------
\subsubsection{Non-equilibrium partition identity}
\label{Sect:Seifert}

The above analysis of the trajectories gives us 
all the ingredients necessary to derive a different fluctuation theorem 
known as the 
{\it non-equilibrium partition identity} 
\cite{Yamada-Kawasaki1967,Carberry-JChemPhys-2004}.
This theorem is less powerful than the steady-state and Crooks' fluctuation  theorems,
but it is more general.  It is applicable to any time-dependent problem, with any resulting
time-dependence of the system state, so long as the conditions are fulfilled which allow one to use
the rate equation discussed here (see the last paragraph of section~\ref{Sect:master-coherences}).
The theorem states that 
\begin{align}
\left\langle \e^{-\Delta\mathscr{S}/\kB} \right\rangle \ = \ 1 , 
\label{Eq:Integral-fluct-theorem}
\end{align}
so on average $\e^{-\Delta\mathscr{S}/\kB}$ is unity.  
This is an {\it integral fluctuation theorem}, meaning 
it is a statement about the whole probability distribution.
This is in contrast with the steady-state fluctuation relation, 
Eq.~(\ref{Eq:detailled-fluctuation-theorem}), which is a 
relation between probabilities to produce specific entropies.
As a result, Eq.~(\ref{Eq:Integral-fluct-theorem}) contains much 
less information than the steady-state fluctuation relation;
Eq.~(\ref{Eq:detailled-fluctuation-theorem}) directly implies 
Eq.~(\ref{Eq:Integral-fluct-theorem}) --- integrating the former over all 
$\Delta\mathscr{S}$ gives the latter  ---
but Eq.~(\ref{Eq:Integral-fluct-theorem}) does not imply
 Eq.~(\ref{Eq:detailled-fluctuation-theorem}) \cite{Carberry-JChemPhys-2004}.
However, we will follow \cite{Seifert-PRL2005,review-Seifert2007}, and show 
that 
the non-equilibrium partition identity in Eq.~(\ref{Eq:Integral-fluct-theorem}) 
is valid for any rate equation, even when the system is not in the 
steady-state or when the problem has time-dependent parameters.

The proof is carried out by considering the following sum over trajectories,
\begin{align}
\left\langle \e^{-\Delta\mathscr{S}/\kB} \right\rangle \ &= \  
\sum_{a_0,a_n} \sum_{\zeta \in  \{a_0,t_0 \to a,t\}}    P(\zeta) \, P_{a_0} 
(t_0) 
\ \e^{-\Delta\mathscr{S}(\zeta)/ \kB}\,,
\label{Eq:exponential-entropy-as-traj-sum}
\end{align}
where $\Delta\mathscr{S}(\zeta)$ is defined as in 
section~\ref{Sect:fluct-theorem-steady-state}.
Now substituting in Eq.~(\ref{Eq:P-zeta-versus-P-barzeta}), 
and noting that $\Delta\mathscr{S}(\bar\zeta)=-\Delta\mathscr{S}(\zeta)$, 
we see that
\begin{align}
\left\langle  \e^{-\Delta\mathscr{S}/\kB} \right\rangle \ &= \  
\sum_{a_0,a} \sum_{\bar\zeta \in  \{a,t_0 \to a_0,t\}}  
{\overline{P}}\left({\bar{\zeta}}\right) \,P_{a}(t) \, ,
\label{Eq:derivation-Integral-fluct-theorum}
\end{align}
where we have used the fact that a sum over $\zeta \in \{a_0,t_0 \to a,t\}$ is 
the same as a sum over
$\bar\zeta \in \{a,t_0 \to a_0,t\}$.
Now we note that this is a sum over all paths from $a$ to $a_0$ in the 
time-reversed system,
which is the system whose time-dependent parameters are given by 
Eq.~(\ref{Eq:time-reversed-rates}).  However, irrespective of what the 
time-dependence of the coupling is in the original system, the time-reversed 
system could in principle exist, and thus must respect probability 
conservation.  Probability conservation means that the sum over all paths $\bar\zeta$ from $a$ to $a_0$ also summed over all final states $a_0$ must be one at all times:
\begin{align}
\sum_{a_0} \sum_{\bar\zeta \in  \{a,t_0 \to a_0,t\}}  {\overline P}\left({\bar{\zeta}}\right) = 1\,.
\end{align} 
Substituting this into Eq.~(\ref{Eq:derivation-Integral-fluct-theorum}) 
and summing over $a$ leads immediately to the 
{\it non-equilibrium partition identity} in 
Eq.~(\ref{Eq:Integral-fluct-theorem}).

\green{
If one writes the non-equilibrium partition identity in Eq.~(\ref{Eq:Integral-fluct-theorem})}
as $\left\langle 1-\exp\big[- \Delta\mathscr{S}\big/\kB\big]  
\right\rangle =0$ 
and then notes that 
$x \geq 1-\e^{-x}$ for all $x$, one immediately sees that this identity implies 
$ \left\langle \Delta\mathscr{S} \right\rangle \geq 0$.  Thus the rate 
equation obeys second law of thermodynamics
regardless of the time-dependence of the problem (although we recall that this rate equation only
applies for time-dependences with fulfill the conditions in the last paragraph of section~\ref{Sect:master-coherences}). 
However, 
\green{Eq.~(\ref{Eq:Integral-fluct-theorem})}
gives us more information than the second law, because it is an identity, when the second law is only an 
inequality.

\subsubsection{Fluctuations exhibiting Carnot efficiency are the least likely}

We close this section by  mentioning the intriguing work \cite{unlikely-Carnot} which showed that machines described by stochastic thermodynamics are less likely to have a ``Carnot efficient'' fluctuation than any other fluctuation.
To understand what this means consider a machine operating in the steady-state whose average efficiency is less than that of Carnot.  \green{Its entropy production will fluctuate as described by the fluctuation relation in section~\ref{Sect:fluct-theorem-steady-state}, thus there are fluctuations in its efficiency.
When the entropy production is negative for a short time, due to such a fluctuation, the efficiency will  exceed Carnot efficiency during that time.}
Ref.~\cite{unlikely-Carnot} considered the rate at which such efficiency fluctuations decay in the long time limit.
The decay is only zero at the average efficiency, which guarantees that the efficiency measured over long enough times is always the average efficiency (since fluctuations average out over long times).
All fluctuations decay exponentially with time, and broadly speaking the rate of decay is larger for large fluctuation, as one might guess.  However, the decay rate is {\it not} a monotonic function of the size of the fluctuation. Remarkably, the decay rate is maximal for a fluctuation which corresponds to the
Carnot efficiency, irrespective of whether this fluctuation is large (i.e.~when the average efficiency is much less than Carnot)
or small (i.e.~when the average efficiency is close to that of Carnot).
Thus in the long time limit, a fluctuation exhibiting the Carnot efficiency is exponentially less probable than 
any other fluctuation; this means it is less probable than a fluctuation exhibiting an efficiency larger than Carnot efficiency.

Ref.~\cite{unlikely-Carnot} shows that this observation holds in the limit of the average efficiency tending towards  Carnot efficiency, with the decay rate being zero at the average efficiency, and rapidly sweeping up to its maximum value  at the Carnot efficiency. 
The case of a machine with exactly the Carnot efficiency is a bit special, because such a machine exhibits no fluctuations at all, as mentioned at the 
end of section~\ref{Sect:rule-Carnot}.  Thus the rate of decay of such fluctuations is irrelevant simply because their magnitude is zero.

\green{
Another recent work \cite{Jiang-Agarwalla-Segal} indicates that the suppression of those fluctuations which give  Carnot efficiency is a consequence of time-reversal symmetry.  That work considers three-terminal systems in which an external magnetic field breaks time-reversal symmetry.
They find that the decay rate is still maximal for a given fluctuation of efficiency, but that the efficiency in question can be large or smaller than the Carnot efficiency (depending on the value of the magnetic field and other system parameters).
}

\subsection{Beyond rate equations}
\label{Sect:beyond-master}

\green{
While we concentrate on time-independent situations in this review, for completeness we mention
that one has to be a bit more careful with the derivation for time-dependent problems. 
Broadly speaking, the approaches cited in section~\ref{Sect:transition-rates} work reasonably well when the time-dependence is slow on the scale of the system dynamics and on the scale of the reservoir memory times. In this adiabatic regime, one usually finds the same rate equations with time-dependent parameters, but care should be taken that non-adiabatic corrections to this approximation are indeed small.     
Most work on this situation has been for systems where coherences cannot be neglected \cite{Davies78}
(i.e.~when the assumptions in section~\ref{Sect:master-coherences} are not satisfied)
for which the situation is richer even for slow driving. 
Then reservoir induced decoherence can destroy interference effects in the system, and thereby completely change the final state of the system 
\cite{Gefen-Ben-Jacob-Caldeira87, Shimshoni-Gefen91,Grifoni-Hanggi98,Albash2012}.
}

\green{
There are numerous methods that go beyond simple rate equations, and so can capture features of time-independent and time-dependent systems that rate equations cannot.  Here we mention some of the more popular methods which work for various situations which do not satisfy the requirements in section \ref{Sect:master-coherences}.
}

\subsubsection{The Lindblad equation: a markovian master equation with coherences}

\green{
Here we mention more complicated master equations which included quantum coherences, and so go beyond those discussed elsewhere in this section.
The best understood of such equations are those for Markovian dynamics, such as the Lindblad equation \cite{lindblad,book:open-quantum,Alicki-semigroup}.   
These equations look a little like the rate equations in this section, 
however rather than give the rate of change of the occupation probability of the $n$ system states, they give the rate of change of the $n \times n$ system density matrix. As such, their structure is rather more complicated,
however they can be used to treat problems in which the evolution generates off-diagonal elements in the system's density matrix.  
There has been a great deal of work over many years on the type of master equation with coherence 
known as the Lindblad equation.
Dynamics under this equation is well reviewed in textbooks \cite{book:open-quantum,Alicki-semigroup}.  These systems have long since been show to obey the laws of thermodynamics; the various proofs and some associated controversies are nicely reviewed in 
Ref.~\cite{Kosloff-review}, which contains an extensive bibliography of the original works on this subject, such as
Refs~\cite{Davies74,Davies76,Spohn78,alicki79} and many more.
We have nothing to add to this review, although we hope that reading the proofs in this section (for systems without coherences) will provide a good preparation for the proofs for the more complicated Lindblad equation.
}

\subsubsection{\green{Quantum master equations from golden-rule: Bloch-Redfield or  Nakajima-Zwanzig}}

\green{
Quantum master equations that include coherences can derived from approximate methods based on a golden-rule treatment of such system-reservoir problems, in which one assumes the system-reservoir coupling is a perturbation that can be treated to lowest order.  
Depending on the community and context such equations are known as the Redfield \cite{redfield} or Bloch-Redfield \cite{Bloch57}, 
the sequential tunnelling approximation in transport theory (see e.g.~\cite{Schoeller97}), 
or the weak-coupling limit of the Nakajima-Zwanzig model \cite{Nakajima58,Zwanzig60}.  
A more rigorous treatment known in the mathematical physics community is the weak-coupling limit of quantum-mechanical master equations in Refs.~\cite{Davies74,Davies76,Dumcke1985}.
The rate equations that we presented  above are taken from these quantum master equations,
under the additional assumption that coherences are not important.
}

\green{
Once one includes the coherences, the perturbative quantum master equation
is believed to be a reasonable approximation whenever the memory time is significantly shorter than the
dissipative timescales (the typical timescale between interactions of the system with the reservoirs), 
even if the system's dynamics are rapid on the timescale of the memory time.
This belief is based on estimating the next order in perturbation theory, and finding it to be small in this case.
This master equation gives the Lindblad equation directly when one takes the memory time to zero 
(unlike in some other derivations of the Lindblad equation, no course-graining of the dynamics is necessary).
For finite memory times, it look similar to a Lindblad equation,
but its slightly different structure makes it hard to prove that it respect positivity
(i.e. that it never generates negative probabilities), see  \cite{Whitney2008} for a proof of positivity in a particular system.
There is not yet a consensus on whether it obeys the laws of thermodynamics 
in the regime where it does not coincide with the Lindblad equation, 
although it has recently been claimed that it does obey the second law 
\cite{whitney-fluct2016}.
}

\green{While the Bloch-Redfield approximation relies on weak coupling between the system and the reservoirs,
one can sometimes use a  well known trick to treat a simple system which is strongly coupled to its reservoirs.
The trick is to perform a polaron transformation on the total Hamiltonian of the system and its reservoirs, see ``Small polaron theory'' in chapter 7 of \cite{mahan} and Refs.~\cite{Zwerger83a,Zwerger83b,Silbey-Harris,Aslangul86,Dekker87,Aslangul88}. Under the right condition, this enables one to transform the problem into that of a system weakly coupled to reservoirs (although the transformation redefines exactly what one calls
the system and what one calls the reservoir), which can then be treated with the Bloch-Redfield or Lindblad approaches.  This polaron transformation followed by a weak-coupling approximation is a more rigorous and transparent version of the ``non-interacting blip'' approximation of  Ref.~\cite{Leggett-review}.
This method was used in Ref.~\cite{Gelbwaser2015} to treat a quantum heat engine which is strongly coupled to its reservoirs.       
}

\subsubsection{Non-equilibrium Green's functions and real-time transport theory}
\label{Sect:Keldysh}

\green{
Non-equilibrium Green's functions are a powerful method for modelling the transport properties of 
quantum systems,  see for example chapter 4 of Ref.~\cite{DiVentra-book}.
There is current progress in using this method to
calculate the properties of heat-to-work conversion, and prove the laws of thermodynamics, 
for far-from-equilibrium systems that cannot be modelled by either Landauer scattering theory or Lindblad master equations.
These systems are typically those which exhibit interactions (so scattering theory is inapplicable), and are not weakly coupled to the reservoirs (so their dynamics exhibit memory effects not captured by rate equations, Lindblad equations or Bloch-Redfield equations).
}

\green{
While there are numerous works which use non-equilibrium Keldysh versions of energy Green's functions to calculate heat engine or refrigeration efficiencies for various quantum systems,
we only know of a few works which pose the question of whether such systems obey the laws of thermodynamics 
when far from equilibrium \cite{Sanchez-2ndlaw2014,Esposito-2ndlaw,Esposito2015,Bruch-2016,Sanchez-2ndlaw-2016}. These works have found the heat and charge currents for certain systems and shown that 
they obey the laws of thermodynamics. 
Refs.~\cite{Sanchez-2ndlaw2014,Esposito-2ndlaw,Esposito2015,Bruch-2016}
do this for non-interacting systems (quadratic Hamiltonians), while Refs.~\cite{Sanchez-2ndlaw-2016} treats interacting systems with adiabatic driving.
One extremely recent work used similar methods for non-equilibrium propagators in time (rather than energy) on the Keldysh contour \cite{whitney-fluct2016}, it claims that one can use a method known as real-time transport theory \cite{Schoeller-Schon1994,Konig96,Konig97,Leijnse2008,Schoeller2009,Wegewijs2014,Sothmann2014,Schulenborg2016} 
to prove the 
second law of thermodynamics, and the fluctuation theorems in sections~\ref{Sect:fluct-theorem-steady-state}-\ref{Sect:Seifert}, for an arbitrary interacting quantum system with or without time-dependent driving.
All these works raise a number questions, and we feel it is much too soon to write a definitive review of these methods.
}

\section{Rate equations --  Examples}
\label{Sect:master-examples}

In this chapter, we use the rate equations introduced in chapter~\ref{Sect:Qu-Master-Eqn}
to model three examples of machines which carry out heat-to-work conversion.  
The machines are sketched in 
Figs.~\ref{Fig:2-term-sys}, \ref{Fig:3-term-sys1} and  \ref{Fig:3-term-sys2}. 
The examples are presented in order of increasing complexity, 
but the discussion of each example is self-contained
(thus there is 
some repetition from example to example).
In each case, we use the results in sections~\ref{Sect:single-loop} 
and \ref{Sect:rule-Carnot}
to get information about the efficiencies through a simple
inspection of the system-states and  transition
(without solving the steady-state rate equation).
In each case, we also present the solution of the steady-state rate 
equation, 
Eq.~(\ref{Eq:steady-state-master-eqn}), and the use of  
Eqs.~(\ref{Eq:I^N}-\ref{Eq:J}) 
to calculate all currents (heat, energy, particle, and charge) and the power 
generated.

The first machine (section~\ref{Sect:2-term-sys}) 
is a two-terminal device, which is the quantum equivalent of a thermoelectric.
It allows the flow of electrons between the
reservoirs, but any heat flow is accompanied by a charge flow,
and vice-versa, due to the energy selectivity of the quantum dot. 
As such, it exhibits strong Seebeck and Peltier effects.
To make a steady-state heat-engine one needs two such devices with opposite 
thermoelectric response in a thermocouple geometry; this means they are
coupled between three macroscopic electronic reservoirs with the central 
macroscopic 
electronic reservoir being hotter due to coupling to some sort of external heat 
source (see Fig.~\ref{Fig:thermocouple}a). 
The same device can be a steady-state refrigerator, cooling the central 
macroscopic reservoir, if one applies an electrical current though the 
thermoelectrics.

The other two machines (sections~\ref{Sect:3-term-sys1} and 
\ref{Sect:3-term-sys2}) are three-terminal machines
which act as the quantum equivalent of a thermocouple (see Fig.~\ref{Fig:thermocouple}b).  
Heat (but not charge) 
is injected from a hot-reservoir, and causes an electric current between two 
other reservoirs (L and R).  In one case the heat source is a bosonic bath 
(photons or phonons), while in the other case the heat source is an electronic 
bath which is capacitively coupled to the rest of the device.

\subsection{Thermoelectric dot between two electronic reservoirs}
\label{Sect:2-term-sys}

We follow Refs.~\cite{Tsaousidou2007,elb09,Kennes2013,Murphy-Mukerjee-Moore2008,Taylor-Segal2015,Szukiewicz2016},  
and consider a quantum dot between two electronic reservoirs. Let us assume 
that the dot's level-spacing is large enough that there is only one dot-state 
within a window of order temperature of the reservoirs' electrochemical potentials.
Then we can treat the dot as having only one level, at energy $\eps_1$, as 
sketched in Fig.~\ref{Fig:2-term-sys}. 
Further let us assume that the dot is in the Coulomb blockade regime, where 
the charging energy for double-occupation is $U$.
\green{
While we only consider a single dot-state here, the rate equation approach has also been used to study multi-level quantum dots with Coulomb blockade effects~\cite{erdman}.}

%---------------------------------------------------
\subsubsection{Solving the problem without spin or double-occupancy}
\label{Sect:spinless}

The simplest case is that in which we neglect spin and assume the charging 
energy for 
double-occupancy, $U$, is much bigger than all other energy scales 
(temperatures, biases, etc.)
Then we only have two system states 0 (dot-level empty) and 1 (dot-level singly 
occupied), with energies 
$E_0=0$ and $E_1=\eps_1$ respectively.
Then, the rate equation for the dot's dynamics is 
\begin{align}
{\rmd  \over \rmd t}
\left( \begin{array}{c} P_0(t) \\ P_1(t) \end{array}\right) \, = \,
\left( \begin{array}{cc}
-\Gamma_{10} & \Gamma_{01} \\ 
\Gamma_{10} & -\Gamma_{01} 
\end{array}\right)\, 
\left( \begin{array}{c} P_0(t) \\ P_1(t) \end{array}\right) \, ,
\end{align}
where $\Gamma_{ba}=\Gamma^{(\rm L)}_{ba}+\Gamma^{(\rm R)}_{ba}$.
and these rates obey Eq.~(\ref{Eq:rate-for-reverse-final}).
 \green{In many cases, it may be sufficient to treat these rates as phenomenological parameters,} 
 however if one wishes to relate them to the Hamiltonian of the system and reservoirs, 
as in section~\ref{Sect:transition-rates}, one has
\begin{align}
\Gamma^{(i)}_{10} =& {1 \over h} \nu_i \big(\eps_1)  \, 
 f_i\big(\eps_1\big) \, \big|V_i(\eps_1) \big|^2
\end{align}
where $V_i(\eps_1)$ is the tunnel-coupling of the system to reservoir $i$, for $i={\rm L,R}$.
From this Eq.~(\ref{Eq:rate-for-reverse-final}) gives $\Gamma^{\rm i}_{01}$, 
as discussed in section~\ref{Sect:local-detailed}.

%========================================
\begin{figure}
\centerline{\includegraphics[width=0.45\textwidth]{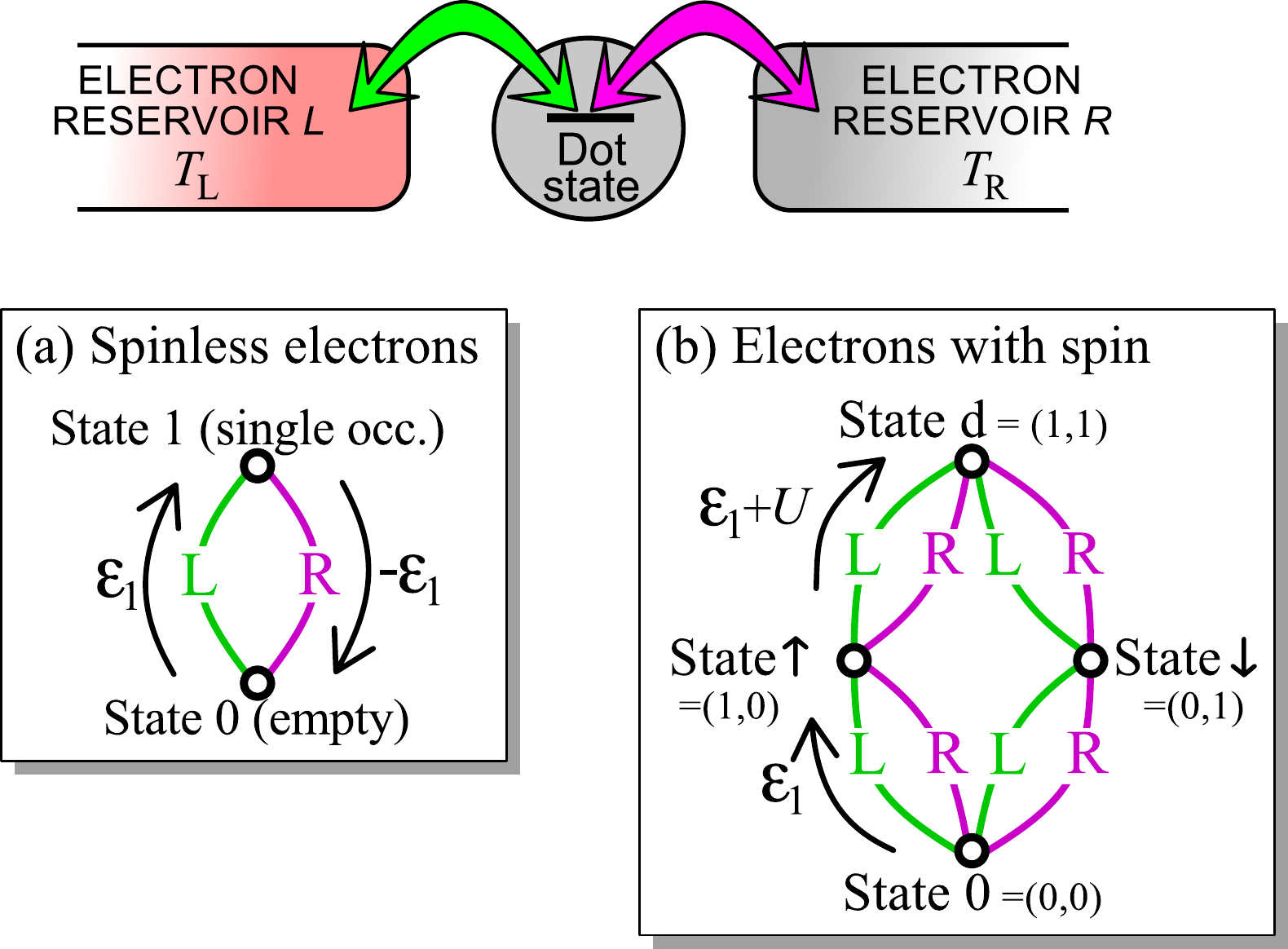}
}
\caption{\label{Fig:2-term-sys}
A single-level quantum dot in the Coulomb blockade regime, coupled to two 
reservoirs 
at different temperatures, $T_{\rm L}$ and $T_{\rm R}$. 
In inset (a) we show the two dot states that exist if we neglect the electron's 
spin, and assume the charging energy
is too high for the dot to ever be doubly-occupied, so the dot makes 
transitions between state 0 (empty) 
and state 1 (single-occupancy).
In inset (b) we include spin and double-occupancy, so the dot has four possible 
states
$(0,0)$, $(1,0)$, $(0,1)$ and $(1,1)$, where the first number and second number 
in the brackets are  the occupancy of the $\up$-state and $\dn$ state, 
respectively.
When reservoirs L and R are at different temperatures, the dot can act as a 
thermoelectric.
Each inset also indicated the energy the reservoir gives to the system during 
the transition marked by the arrow.
}
\end{figure}
%========================================

The particle current into the system from the reservoirs are 
\begin{align} 
\JparticleL = -\JparticleR = {\cal I}_{10}^{\rm (L) \,steady}
\end{align}
Without loss of generality, we define the zero of energy to coincide with
reservoir L's electrochemical potential, so that $\mu_{\rm L}=0$.
We define $\mu$ as the difference in electrochemical potential between the reservoirs, 
so $\mu=\mu_{\rm R}-\mu_{\rm L}$.
Then the heat currents out of reservoirs L and R are
\begin{align}
\JheatL \, &=\, \JenergyL  \, = \, \eps_1 \JparticleL 
,\qquad
\JheatR   \, = \, (\mu-\eps_1) \JparticleL ,
\label{Eq:master-two-term-simple-Js}
\end{align}
where we have used the fact that $\JenergyR =  -\JenergyL$ and 
$ \JparticleR= - \JparticleL$.
Note that these direct relationships between the heat currents and the particle currents are 
a consequence of the fact that the nature of the system means that every electron leaving a given reservoir carries the same amount of heat; for example every electron entering from reservoir L carries heat $\eps_1$.
This is thus an example of {\it tight coupling}, which section~\ref{Sect:rule-Carnot} mentions as a common 
pre-requisite for Carnot efficiency.  

The power generated is  
\begin{align}
P_{\rm gen} \, = \,  -\mu \JparticleR \, =\, \mu \JparticleL .
\label{Eq:P_gen-master}
\end{align}
Since we are in the steady-state, the system entropy does not change with time 
($\rmd \mathscr{S}_{\rm  sys}\big/\rmd t=0$), thus
the rate of total entropy production is 
\begin{align}
{\rmd \mathscr{S} \over \rmd t} 
\ =\ {\rmd \over \rmd t} \left(\mathscr{S}^{(\rm L)}_{\rm res} +\mathscr{S}^{(\rm R)}_{\rm res} 
\right)
\ = \  -{\JheatL \over T_{\rm L}} - {\JheatR \over T_{\rm R}} 
\ =\  {\JparticleL \over T_{\rm R}}\,\Big( \eps_1 \big(1-T_{\rm 
R}\big/T_{\rm L}\big) -\mu \Big) \ .
\label{Eq:dotS-2term}
 \end{align}

If we use the system as a heat engine, where reservoir L is the heat source
($T_{\rm L} > T_{\rm R}$) that induces the power generation,
then the efficiency is
\begin{align}
\eta_{\rm eng} \, \equiv\, P_{\rm gen} \big/ \JheatL \ =\ \mu\big/ \eps_1\,, 
\label{Eq:efficiency-engine-two-term-master}
\end{align}
where $\mu$ has the same sign as $\eps_1$ to ensure that $P_{\rm gen} >0$.
If, in contrast, we use the system as a refrigerator, where reservoir L is the 
one being cooled
($T_{\rm L} < T_{\rm R}$), and the system absorbing electrical power $P_{\rm abs}=-P_{\rm 
gen}$ to 
carry out the cooling, then the efficiency is   
\begin{align}
\eta_{\rm fri} \, \equiv\, \JheatL \big/ P_{\rm abs} \ =\ \eps_1\big/ (-\mu) \,,
\label{Eq:efficiency-fridge-two-term-master}
\end{align}
where $\mu$ has the opposite sign from $\eps_1$ to ensure $J_{\rm h,L} >0$.

Inset (a) of Fig.~\ref{Fig:2-term-sys} makes it clear that this system is a 
single-loop machine,
so it is not surprising that the efficiencies given above coincide with 
Eqs.~(\ref{Eq:single-loop-eng},\ref{Eq:single-loop-fridge}).  To see this one 
notes that
the heat flow $\JheatL$ is associated with the transition
\hbox{
$\raisebox{6pt}{${\scriptstyle 0}$} \hskip -2.5mm\xrightarrow{\hspace*{3mm}} 
\hskip -1.5mm
\raisebox{8pt}{${\scriptstyle {\rm L}}$} \hskip -1.1mm {\bm |} \hskip -1.5mm
\xrightarrow{\hspace*{3mm}} \hskip -2mm \raisebox{6pt}{${\scriptstyle 1}$}$} 
and its time-reserve.
Similarly, power is generation or absorption only 
occurs when an electron is injected into reservoir R,
which is only associated with 
\hbox{
$\raisebox{6pt}{${\scriptstyle 1}$} \hskip -2.5mm\xrightarrow{\hspace*{3mm}} 
\hskip -1.5mm
\raisebox{8pt}{${\scriptstyle {\rm R}}$} \hskip -1.1mm {\bm |} \hskip -2.3mm
\xrightarrow{\hspace*{3mm}} \hskip -2mm \raisebox{6pt}{${\scriptstyle 0}$}$} 
and its time-reserve.
Thus, there is only one term in the numerator and denominators of 
Eqs.~(\ref{Eq:single-loop-eng},\ref{Eq:single-loop-fridge}), and the results 
coincide with
Eqs.~(\ref{Eq:efficiency-engine-two-term-master},\ref{Eq:efficiency-fridge-two-term-master}).

One could immediately get information about the efficiency of this device, 
by inspecting the system states and transitions without solving the equation 
for the 
steady-state of the rate equation
(using the results in  sections~\ref{Sect:single-loop} 
and \ref{Sect:rule-Carnot}).  However, in this case the rate equation is 
simple enough that it
is as easy just to solve it.  Thus, we present the steady-state solution of the
rate equation first, and afterwards show that it fits with the results in 
sections~\ref{Sect:single-loop} and \ref{Sect:rule-Carnot}.

One can use $P_1(t)=1-P_0(t)$, 
to reduces the rate equation to
${\rmd \over \rmd t}P_0(t) = \Gamma_{01}-\left(\Gamma_{10}+\Gamma_{01}\right) 
P_0(t)$. 
The steady-state is given by $\rmd P_0(t) \big/\rmd t =0$, and so
\begin{align}
P_0^{\rm steady} = {\Gamma_{01} \over \Gamma_{10}+\Gamma_{01}}\, , 
\qquad 
P_1^{\rm steady} = {\Gamma_{10} \over \Gamma_{10}+\Gamma_{01}}\, .
\end{align}
From Eq.~(\ref{Eq:I^N}), we get the steady-state particle current 
$\JparticleL = \Gamma^{\rm (L)}_{10}P^{\rm steady}_0 
- \Gamma^{\rm (L)}_{01}P^{\rm steady}_1$. 
Substituting in the above results gives  
\begin{align}
\JparticleL \ =\ { \Gamma^{\rm (L)}_{10}  \Gamma^{\rm (R)}_{01} 
- \Gamma^{\rm (R)}_{10}  \Gamma^{\rm (L)}_{01} 
\over \Gamma^{\rm (L)}_{10} + \Gamma^{\rm (R)}_{10} 
+ \Gamma^{\rm (L)}_{01} + \Gamma^{\rm (R)}_{01} }
 \ =\ { \Gamma^{\rm (L)}_{10}  \Gamma^{\rm (R)}_{10} 
\left(\e^{-\DS^{\rm R}_{10}\big/\kB} - \e^{-\DS^{\rm L}_{10}\big/\kB} 
\right)
\over \Gamma^{\rm (L)}_{10} \left(1+ \e^{-\DS^{\rm L}_{10}\big/\kB} \right)
+ \Gamma^{\rm (R)}_{10}  \left(1+ \e^{-\DS^{\rm R}_{10}\big/\kB} \right) }\,,
\label{Eq:I^N_L-two-term}
\end{align}
where we get the right hand equality by using Eq.~(\ref{Eq:rate-for-reverse-final}).
All other currents are then given by Eq.~(\ref{Eq:master-two-term-simple-Js}).

The machine operates {\it reversibly} (in the thermodynamic sense) if the rate 
of entropy production, given by Eq.~(\ref{Eq:dotS-2term}), is zero.  
This is achieved when one chooses the chemical 
potential difference 
\begin{eqnarray}
\mu = \eps_1 \big(1-T_{\rm R}\big/T_{\rm L}\big) .
\label{Eq:mu-for-reversible}
\end{eqnarray}
In this case Eqs.~(\ref{Eq:efficiency-engine-two-term-master},
\ref{Eq:efficiency-fridge-two-term-master}) become the relevant Carnot 
efficiencies
$\eta_{\rm eng} =1-T_{\rm R}\big/T_{\rm L}$ and 
$\eta_{\rm fri}= \big(T_{\rm R}\big/T_{\rm L} -1\big)^{-1}$.
However, there is a price to pay to achieve this efficiency, the price is that 
the power output is zero.  This is because Eq.~(\ref{Eq:mu-for-reversible}) 
implies $\DS^{\rm L}_{10}=\DS^{\rm R}_{10}$, and 
Eq.~(\ref{Eq:I^N_L-two-term}) then means that the particle current 
$\JparticleL=0$.
The way to get a non-zero power output is to slightly reduce  $\mu$, so 
$\big(\rmd \mathscr{S}\big/ \rmd t\big)$ becomes slightly positive 
and a small but finite power is produced. Of course, now $\mu$ is slightly less 
than 
$\eps_1(1-T_0/T_{\rm H})$, and so the efficiency is slightly less than that of 
Carnot.

The above derivation is a full treatment of the problem, giving all currents of 
heat, charge, etc.  
However, if one only wants to answer the question of whether the system can 
achieve Carnot efficiency, it would be sufficient to use the rule in 
section~\ref{Sect:traj}.  In this case, the full derivation was so simple that this rule
is not really simpler than the full derivation.  
However, it is worth seeing  how the rule applies in this case, 
before applying it to more complicated situations.
The system dynamics explore all trajectories on the very simple network shown 
in the inset (a) of Fig.~\ref{Fig:2-term-sys}.  There are only two primitive 
closed trajectories
in this state space.  The first is
\begin{align}
\zeta_1\ = \ 
\hskip 2mm \raisebox{6pt}{$0$} 
\hskip -2mm {\xrightarrow{\hspace*{3mm}} } \hskip -2mm
\begin{array}{c} {\rm L} \\ {\bm |} \\ \phantom{L} \end{array}  
\hskip 2mm \raisebox{6pt}{$1$} 
\hskip -7mm {\xrightarrow{\hspace*{10mm}} } \hskip -2mm
\begin{array}{c} {\rm R} \\ {\bm |} \\ \phantom{L} \end{array}  \hskip -4mm 
\hskip 4mm \raisebox{6pt}{$0$} 
\hskip -5mm {\xrightarrow{\hspace*{3mm}} }  
\end{align} 
and the second is $\bar{\zeta}_1$, which is the time-reverse of $\zeta_1$.
For the transition 
$\raisebox{6pt}{${\scriptstyle 0}$} \hskip -2.5mm\xrightarrow{\hspace*{3mm}} 
\hskip -1.5mm
\raisebox{8pt}{${\scriptstyle {\rm L}}$} \hskip -1.1mm {\bm |} \hskip -1.5mm
\xrightarrow{\hspace*{3mm}} \hskip -2mm \raisebox{6pt}{${\scriptstyle 1}$}$
the entropy change in reservoir L is $-\eps_1/T_{\rm L}$,
while for the transition 
$\raisebox{6pt}{${\scriptstyle 1}$} \hskip -2.5mm\xrightarrow{\hspace*{3mm}} 
\hskip -1.5mm
\raisebox{8pt}{${\scriptstyle {\rm R}}$} \hskip -1.1mm {\bm |} \hskip -2.3mm
\xrightarrow{\hspace*{3mm}} \hskip -2mm \raisebox{6pt}{${\scriptstyle 0}$}$
the entropy change of reservoir R is $(\eps_1-\mu_{\rm R})/T_{\rm R}$.  
Thus, the sum of entropy changes in all reservoirs during the closed trajectory 
$\zeta_1$ is
\begin{align}
\Delta\mathscr{S}_{\rm res}(\zeta_1) \ =\ -\eps_1/T_{\rm L} + (\eps_1-\mu_{\rm R})/T_{\rm 
R}\,.
\end{align}
The sum of entropy changes in all reservoirs during $\bar{\zeta}_1$ is
$\Delta\mathscr{S}_{\rm res}(\bar{\zeta}_1)=-\Delta\mathscr{S}_{\rm res}(\zeta_1)$.
Section~\ref{Sect:traj}'s rule says that
Carnot efficiency is only achieved if $\Delta\mathscr{S}_{\rm res}(\zeta_1)=0$.
Without further algebra,
this gives the result that this system can achieve Carnot efficiency if it 
obeys 
Eq.~(\ref{Eq:mu-for-reversible}).

%------------------------------------------------------------------------
\subsection{Including spin and double-occupancy}

If we include spin and the possibility of double-occupancy in the problem 
described above, the physics becomes
more complicated.  We will show here how to treat this case, and reproduce the 
result of Ref.~\cite{Murphy-Mukerjee-Moore2008,Taylor-Segal2015}, that 
the system can achieve Carnot efficiency if the charging energy vanished 
($U=0$) or diverges ($U=\infty$), but not if $U$ is finite.  

In the case where we include spin, there are four states labelled by 
$(n_\up,n_\dn)$, where 
$n_\sigma=1$ if the dot's electron state with spin-$\sigma$ is full, and
$n_\sigma=0$ if the electron state with spin-$\sigma$ is empty.
To simplify the notation we refer to $(0,0)$ as state ``0'',
$(1,0)$ as state ``$\up$'',
$(0,1)$ as state ``$\dn$'',
and
$(1,1)$ as state ``d'' (for double occupation).
These states have energy $E_0=0$, $E_\up=E_\dn=\eps_1$ and $E_\rmd = 2\eps_1+U$, 
respectively,
where $U$ is the Coulomb charging energy that must be paid if one wishes to 
place two electrons on the dot. 
A similar model with ferromagnetic leads is treated in Ref.~\cite{Szukiewicz15},
but we will restrict ourselves to non-magnetic reservoirs.
Then the rate equation is
\begin{align}
{\rmd  \over \rmd t}
\left( \!\! \begin{array}{c} P_0 \\ P_\up  \\ P_\dn  \\  P_\rmd  \end{array} 
\!\! \right)  = 
\left(\!\! \begin{array}{cccc}
 -\Gamma_{\up0}-\Gamma_{\dn0} \!\! & \Gamma_{0\up} & \Gamma_{0\dn} & 0 \\ 
\Gamma_{\up0} &\!\! -\Gamma_{0\up}-\Gamma_{\rmd\up} \!\! &0 & \Gamma_{\up\rmd}\\
\Gamma_{\dn0} & 0& \!\! -\Gamma_{0\dn}-\Gamma_{d\dn} \!\! & \Gamma_{\dn\rmd}\\
0& \Gamma_{\rmd\up}  & \Gamma_{\rmd\dn} & 
\!\!-\Gamma_{\up\rmd}-\Gamma_{\dn\rmd} \\
\end{array} \!\! \right)
\, 
\left(\!\! \begin{array}{c} P_0 \\ P_\up \\ P_\dn  \\  P_\rmd 
\end{array}\!\!\right)\,, 
\end{align}
where $\Gamma_{ba}=\Gamma^{\rm (L)}_{ba}+\Gamma^{\rm (R)}_{ba}$.

Using section~\ref{Sect:transition-rates}, we have the rates involving an 
electron entering the dot are
\begin{subequations}
\label{Eq:rates-with-spin&double-occ}
\begin{align}
\Gamma^{\rm i}_{\up 0} \ =\ \Gamma^{\rm i}_{\dn 0} \ =& \ \ {1 \over h} \nu_i 
\big(\eps_1)  \, 
 f_i\big(\eps_1\big) \, \big|V_i(\eps_1) \big|^2
 \\
\Gamma^{\rm i}_{d\up} \ =\ \Gamma^{\rm i}_{d\dn} \ =& \ \ {1 \over h} \nu_i 
\big(\eps_1+U)  \, 
 f_i\big(\eps_1+U\big) \, \big|V_i(\eps_1+U) \big|^2
\end{align}
\end{subequations}
for $i={\rm L,R}$,  
The remaining rates, those involving an electron leaving the dot, are related 
to these via Eq.~(\ref{Eq:rate-for-reverse-final}), as discussed in 
section~\ref{Sect:local-detailed}.

Taking the steady-state limit in which the left hand side of the rate 
equation is zero,
and then using the fact that $P_\rmd = 1-P_0-P_\up-P_\dn$, we get three 
simultaneous equations, 
\begin{align}
(\Gamma_{\up0}+\Gamma_{\dn0})P^{\rm steady}_0 
-  \Gamma_{0\up} P^{\rm steady}_\up - \Gamma_{0\dn}P^{\rm steady}_\dn \ \ &=\ 0\ ,
\nonumber\\
(\Gamma_{\up\rmd}-\Gamma_{\up0}) P^{\rm steady}_0 
+(\Gamma_{0\up}+\Gamma_{\rmd\up}+\Gamma_{\up\rmd})P^{\rm steady}_\up 
+ \Gamma_{\up\rmd} P^{\rm steady}_\dn \ \ &=\ \Gamma_{\up\rmd} \ ,
\nonumber\\
(\Gamma_{\dn\rmd}-\Gamma_{\dn0}) P^{\rm steady}_0 
+ \Gamma_{\dn\rmd} P^{\rm 
steady}_\up+(\Gamma_{0\dn}+\Gamma_{\rmd\dn}+\Gamma_{\dn\rmd})P^{\rm steady}_\dn 
\ \ &= \ \Gamma_{\dn\rmd}  \ .
\nonumber\\
\label{Eq:simult-eqns-with-spin&double-occ}
\end{align}
It is not difficult to solve this set of simultaneous equations using the 
standard methods,
but it is tedious. The solutions are long algebraic expressions, which  are 
hard to simplify to something easily comprehensible. 
Thus we do not proceed further with this, if the reader wants
the solution, they can evaluate it themselves, or use computer algebra software 
(such as Wolfram's Mathematica) to do so \footnote{Our experience is that Mathematica gives the algebraic solution of the problem most easily if one does the following.  Solve Eq.~(\ref{Eq:simult-eqns-with-spin&double-occ}) for arbitrary rates first, and then substitute the rates of interest, given by Eqs.~(\ref{Eq:rates-with-spin&double-occ}), into the solution.  However, this may depend on the version of the software used.}.

Once one has this solution, one can use Eq.~(\ref{Eq:probability-currents}) 
to write down the probability current for each transitions, ${\cal 
I}_{ba}^{(i)}$. 
Then the particle and energy currents are given by
\begin{align}
\JparticleL \ =\  -\JparticleR \ =& \ \  
{\cal I}^{\rm (L)}_{\up 0}+{\cal I}^{\rm (L)}_{\dn 0}+{\cal I}^{\rm (L)}_{\rmd 
\up}+{\cal I}^{\rm (L)}_{\rmd\dn}\,,
\\
\JenergyL \ =\  -\JenergyR \ =& \ \  
\eps_1\left({\cal I}^{\rm (L)}_{\up 0}+{\cal I}^{\rm (L)}_{\dn 0}\right)
+(\eps_1+U) \left({\cal I}^{\rm (L)}_{\rmd \up}+{\cal I}^{\rm 
(L)}_{\rmd\dn}\right)\,,
\end{align}
while the heat currents are $\JheatL=\JenergyL$ and 
$\JheatR=\JenergyR-\mu \JparticleR 
= \mu \JparticleL -\JenergyL$.

If one wishes to operate the machine as a heat-engine, using reservoir L as the 
heat source 
($T_{\rm L} > T_{\rm R}$) one must choose $\mu$ such that the power generated 
is positive, $P_{\rm gen} > 0$, with $P_{\rm gen}$ 
given by Eq.~(\ref{Eq:P_gen-master}).
Then the heat-engine's efficiency is $\eta_{\rm eng}=P_{\rm gen}\big/ \JheatL$.
If one wishes to operate the machine as a refrigerator, 
extracting heat from a cold reservoir L ($T_{\rm L}< T_{\rm R})$, then one must 
choose
$\mu$ such that the power absorbed is positive, $P_{\rm abs} \equiv -P_{\rm 
gen} > 0$, 
with $P_{\rm gen}$ given by Eq.~(\ref{Eq:P_gen-master}). 
Then the refrigerator's efficiency or coefficient of performance is
$\eta_{\rm fri}=\JheatL\big/ P_{\rm abs}$.

Since the state-space network of the machine (inset (b) of 
Fig.~\ref{Fig:2-term-sys}) 
has multiple loops, we cannot use 
Eqs.~(\ref{Eq:single-loop-eng},\ref{Eq:single-loop-fridge}) 
to get the efficiencies without solving the steady-state rate equation. 
However, we can still use section~\ref{Sect:traj}'s rule to look at the 
conditions
for Carnot efficiency.  With this, we can reproduce an interesting result
in Ref.~\cite{Murphy-Mukerjee-Moore2008,Taylor-Segal2015}, which showed 
that one can have Carnot efficiency for $U=0$ or $U=\infty$, but {\it not} for 
$U$ between the two.  Refs.~\cite{Murphy-Mukerjee-Moore2008,Taylor-Segal2015} 
were for linear response (where having $ZT \to \infty$ is equivalent to Carnot efficiency); 
Ref.~\cite{Murphy-Mukerjee-Moore2008} considering weak coupling to 
the reservoirs, while Ref.~\cite{Taylor-Segal2015} considered arbitrarily strong coupling to 
the reservoirs. Here we will show that the same conclusion can be made beyond 
the linear-response regime, in the context of our rate 
equations for a system weakly coupled to the reservoirs.

Section~\ref{Sect:traj}'s rule requires that we calculate the entropy change 
around 
the primitive closed trajectories of the system space, which is sketched in 
inset (b) of Fig.~\ref{Fig:2-term-sys}.
There are eight such primitive trajectories which visit two states  without 
being self-retracing;
they are

\begin{align}
\zeta_1 = 
\hskip 2mm \raisebox{6pt}{$0$} 
\hskip -2mm {\xrightarrow{\hspace*{3mm}} } \hskip -2mm
\begin{array}{c} {\scriptstyle {\rm L}} \\ {\bm |} \\ \phantom{{\scriptstyle 
L}} \end{array}  
\hskip 2mm \raisebox{6pt}{$\up$} 
\hskip -6.5mm {\xrightarrow{\hspace*{8mm}} } \hskip -2mm
\begin{array}{c} {\scriptstyle {\rm R}} \\ {\bm |} \\ \phantom{{\scriptstyle 
L}} \end{array}  \hskip -4mm 
\hskip 4mm \raisebox{6pt}{$0$} 
\hskip -4.5mm {\xrightarrow{\hspace*{3mm}} }   
\,,\qquad \quad  &
\zeta_2 = 
\hskip 2mm \raisebox{6pt}{$0$} 
\hskip -2mm {\xrightarrow{\hspace*{3mm}} } \hskip -2mm
\begin{array}{c} {\scriptstyle {\rm L}} \\ {\bm |} \\ \phantom{{\scriptstyle 
L}} \end{array}  
\hskip 2mm \raisebox{6pt}{$\dn$} 
\hskip -6.5mm {\xrightarrow{\hspace*{8mm}} } \hskip -2mm
\begin{array}{c} {\scriptstyle {\rm R}} \\ {\bm |} \\ \phantom{{\scriptstyle 
L}} \end{array}  \hskip -4mm 
\hskip 4mm \raisebox{6pt}{$0$} 
\hskip -4.5mm {\xrightarrow{\hspace*{3mm}} }   
\,,\\
\zeta_3 = 
\hskip 2mm \raisebox{6pt}{$\up$} 
\hskip -2mm {\xrightarrow{\hspace*{3mm}} } \hskip -2mm
\begin{array}{c} {\scriptstyle {\rm L}} \\ {\bm |} \\ \phantom{{\scriptstyle 
L}} \end{array}  
\hskip 2mm \raisebox{6pt}{$\rmd$} 
\hskip -6.5mm {\xrightarrow{\hspace*{8mm}} } \hskip -2mm
\begin{array}{c} {\scriptstyle {\rm R}} \\ {\bm |} \\ \phantom{{\scriptstyle 
L}} \end{array}  \hskip -4mm 
\hskip 4mm \raisebox{6pt}{$\up$} 
\hskip -4.5mm {\xrightarrow{\hspace*{3mm}} }   
\,,\qquad \quad  &
\zeta_4 = 
\hskip 2mm \raisebox{6pt}{$\dn$} 
\hskip -2mm {\xrightarrow{\hspace*{3mm}} } \hskip -2mm
\begin{array}{c} {\scriptstyle {\rm L}} \\ {\bm |} \\ \phantom{{\scriptstyle 
L}} \end{array}  
\hskip 2mm \raisebox{6pt}{$\rmd$} 
\hskip -6.5mm {\xrightarrow{\hspace*{8mm}} } \hskip -2mm
\begin{array}{c} {\scriptstyle {\rm R}} \\ {\bm |} \\ \phantom{{\scriptstyle 
L}} \end{array}  \hskip -4mm 
\hskip 4mm \raisebox{6pt}{$\dn$} 
\hskip -4.5mm {\xrightarrow{\hspace*{3mm}} }  \,, 
\end{align}
plus four trajectories which are the time-reverse of these.
Since the system state has the same energy for $\up$ and $\dn$,
the entropy change associated with trajectories $\zeta_1$ and $\zeta_2$
is the same, as is that associated with trajectories $\zeta_3$ and $\zeta_4$.
Following the same logic as for trajectory $\zeta_1$ in 
section~\ref{Sect:spinless}, we have
\begin{align}
\Delta\mathscr{S}_{\rm res}(\zeta_1)=\Delta\mathscr{S}_{\rm res}(\zeta_2) \ &=\  
 -{\eps_1\over T_{\rm L}} + {\eps_1-\mu\over T_{\rm R}}\,,
 \label{Eq:DeltaS-zeta1}
\\
\Delta\mathscr{S}_{\rm res}(\zeta_3)=\Delta\mathscr{S}_{\rm res}(\zeta_4) \ &=\  
 -{\eps_1+U\over T_{\rm L}} + {\eps_1+U-\mu \over T_{\rm R}}\,.
 \label{Eq:DeltaS-zeta3}
\end{align}
There are two further primitive trajectories which visit four states without being 
self-retracing;
\begin{align}
\zeta_5\ =
\hskip 2mm \raisebox{6pt}{$0$} 
\hskip -2mm {\xrightarrow{\hspace*{3mm}} } \hskip -2mm
\begin{array}{c} {\scriptstyle {\rm R}} \\ {\bm |} \\ \phantom{{\scriptstyle 
L}} \end{array}  
\hskip 2mm \raisebox{6pt}{$\up$} 
\hskip -6.5mm {\xrightarrow{\hspace*{8mm}} } \hskip -2mm
\begin{array}{c} {\scriptstyle {\rm R}} \\ {\bm |} \\ \phantom{{\scriptstyle 
L}} \end{array} 
\hskip 2mm \raisebox{6pt}{$\rmd$} 
\hskip -6.5mm {\xrightarrow{\hspace*{8mm}} } \hskip -2mm
\begin{array}{c} {\scriptstyle {\rm R}} \\ {\bm |} \\ \phantom{{\scriptstyle 
L}} \end{array} 
\hskip 2mm \raisebox{6pt}{$\dn$} 
\hskip -6.5mm {\xrightarrow{\hspace*{8mm}} } \hskip -2mm
\begin{array}{c} {\scriptstyle {\rm R}} \\ {\bm |} \\ \phantom{{\scriptstyle 
L}} \end{array} 
\hskip 1mm \raisebox{6pt}{$0$} 
\hskip -5.5mm {\xrightarrow{\hspace*{3mm}} } 
\end{align}   
and its time-reverse.
It is trivial to see that $\Delta\mathscr{S}_{\rm res}(\zeta_5)=0$.
One might imagine that there are more primitive trajectories which visit four 
states without being self-retracing;
such as
\begin{align}
\hskip 2mm \raisebox{6pt}{$0$} 
\hskip -2mm {\xrightarrow{\hspace*{3mm}} } \hskip -2mm
\begin{array}{c} {\scriptstyle i_1} \\ {\bm |} \\ \phantom{{\scriptstyle 
L}} \end{array}  
\hskip 2mm \raisebox{6pt}{$\up$} 
\hskip -6.5mm {\xrightarrow{\hspace*{8mm}} } \hskip -2mm
\begin{array}{c} {\scriptstyle i_2} \\ {\bm |} \\ \phantom{{\scriptstyle 
L}} \end{array} 
\hskip 2mm \raisebox{6pt}{$\rmd$} 
\hskip -6.5mm {\xrightarrow{\hspace*{8mm}} } \hskip -2mm
\begin{array}{c} {\scriptstyle i_3} \\ {\bm |} \\ \phantom{{\scriptstyle 
L}} \end{array} 
\hskip 2mm \raisebox{6pt}{$\dn$} 
\hskip -6.5mm {\xrightarrow{\hspace*{8mm}} } \hskip -2mm
\begin{array}{c} {\scriptstyle i_4} \\ {\bm |} \\ \phantom{{\scriptstyle 
L}} \end{array} 
\hskip 1mm \raisebox{6pt}{$0$} 
\hskip -5.5mm {\xrightarrow{\hspace*{3mm}} }\,, 
\end{align}   
where each $i_i$ can be L or R (giving $2^4$ different trajectories).
However, each of them can be composed out of  $\zeta_5$ plus suitable 
combinations of 
$\zeta_1, \cdots, \zeta_4$.
For example
\begin{align}
\hskip 2mm \raisebox{6pt}{$0$} 
\hskip -2mm {\xrightarrow{\hspace*{3mm}} } \hskip -2mm
\begin{array}{c} {\scriptstyle {\rm R}} \\ {\bm |} \\ \phantom{{\scriptstyle 
L}} \end{array}  
\hskip 2mm \raisebox{6pt}{$\up$} 
\hskip -6.5mm {\xrightarrow{\hspace*{8mm}} } \hskip -2mm
\begin{array}{c} {\scriptstyle {\rm L}} \\ {\bm |} \\ \phantom{{\scriptstyle 
L}} \end{array} 
\hskip 2mm \raisebox{6pt}{$\rmd$} 
\hskip -6.5mm {\xrightarrow{\hspace*{8mm}} } \hskip -2mm
\begin{array}{c} {\scriptstyle {\rm R}} \\ {\bm |} \\ \phantom{{\scriptstyle 
L}} \end{array} 
\hskip 2mm \raisebox{6pt}{$\dn$} 
\hskip -6.5mm {\xrightarrow{\hspace*{8mm}} } \hskip -2mm
\begin{array}{c} {\scriptstyle {\rm R}} \\ {\bm |} \\ \phantom{{\scriptstyle 
L}} \end{array} 
\hskip 1mm \raisebox{6pt}{$0$} 
\hskip -5.5mm {\xrightarrow{\hspace*{3mm}} } 
\ \ \ = \ \zeta_{5}+\zeta_{3}\,.
\end{align} 
It is now trivial to see that  if $U=0$,   Eq.~(\ref{Eq:DeltaS-zeta1}) and 
Eq.~(\ref{Eq:DeltaS-zeta3}) are the same.
Then if we choose $\mu$ to satisfy Eq.~(\ref{Eq:mu-for-reversible}), we have 
zero entropy production of all closed trajectories.
Then section~\ref{Sect:traj}'s rule means that the system will be Carnot 
efficient.
However as soon as $U\neq 0$, \green{there is no choice of $\mu$ for which} 
both Eq.~(\ref{Eq:DeltaS-zeta1}) and Eq.~(\ref{Eq:DeltaS-zeta3}) are equal to zero 
for $T_{\rm L}\neq T_{\rm R}$.
Thus the system can never be Carnot efficient for finite $U$.

The exception to this is the limit $U\to\infty$, for which there is never 
enough energy for the system to be doubly-occupied.  Thus, $P_\rmd =0$ and 
$\Gamma_{\up\rmd}=\Gamma_{\rmd\up}=\Gamma_{\dn\rmd}=\Gamma_{\rmd\dn}=0$.
The primitive trajectories involving $\rmd$ drop out of the dynamics, so the 
only relevant primitive trajectories are $\zeta_1$, $\zeta_2$ and their 
time-reverses.
In this $U=\infty$ case (as in the $U=0$ case), we can choose $\mu$ to satisfy 
Eq.~(\ref{Eq:mu-for-reversible}), 
and have that no closed trajectory generating entropy, then 
section~\ref{Sect:traj}'s rule means that the system will be Carnot efficient.
Hence, we conclude that Carnot efficiency is achievable for $U=0$ or $U=\infty$ by having $\mu$ satisfy Eq.~(\ref{Eq:mu-for-reversible}), however Carnot efficiency is never possible for finite $U$.

%----------------------------------------
\subsection{Machine with one bosonic and two electronic reservoirs}
\label{Sect:3-term-sys1}

\begin{figure}
\centerline{\includegraphics[width=0.75\textwidth]{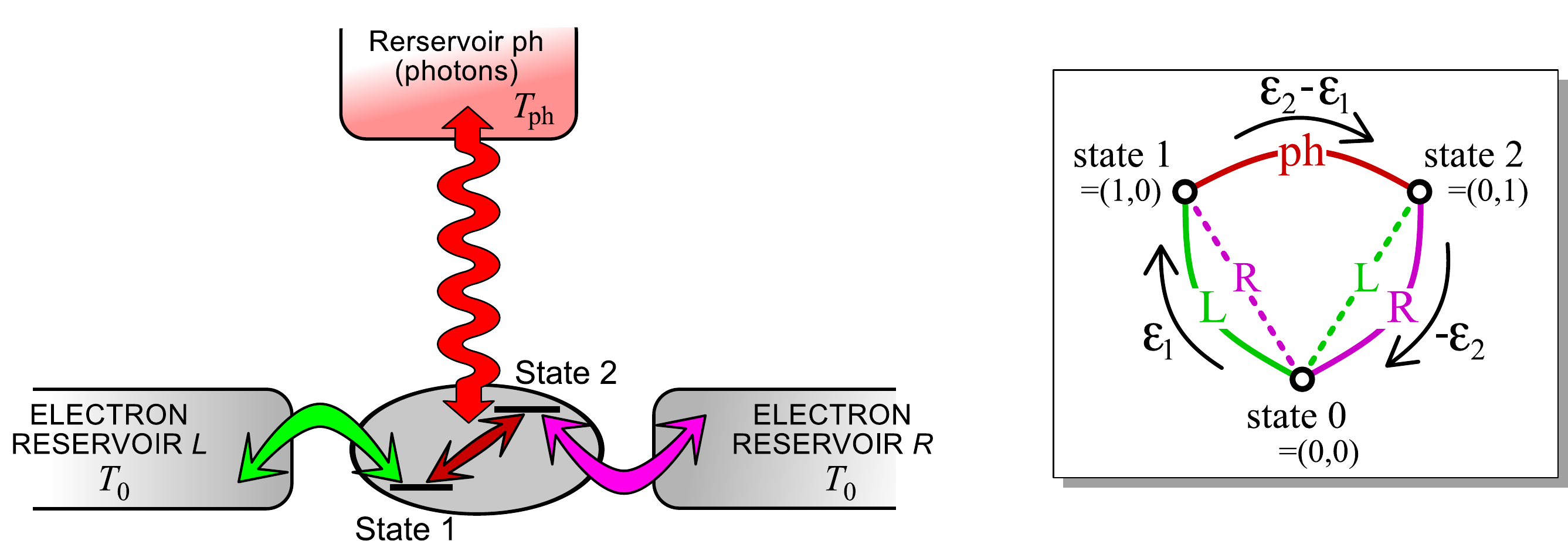}}
\caption{\label{Fig:3-term-sys1}
A system which has been proposed as a heat-engine \cite{rutten09,imry2012}.
It converts heat radiated by the photonic reservoir (at temperature $T_{\rm 
ph}>T_0$)
into electrical power, manifested 
as a charge current between the left and right electronic reservoirs flowing 
against a potential difference.
The inset shows the network associated with the system dynamics, 
the nodes indicating system states, while the bonds indicate transitions 
induced by the coupling to the reservoir indicated (L, R or ph). 
For the system to work well as a heat-engine, the transition rates associated 
with the dashed 
lines must be much smaller than those associated with the solid lines.
The arrows in the inset indicate a heat flow out of reservoir ph, 
which causes an electric current from left to right.  
Next to each arrow we indicate the energy that the reservoir gives to the system 
during that transition.
The same system could also be used as a refrigerator to cool  reservoir ph 
to a temperature $T_{\rm ph} < T_0$, by driving a current between the left and right reservoirs. 
}
\end{figure}

Consider the quantum thermocouple system in Fig.~\ref{Fig:3-term-sys1}, 
suggested in Refs.~\cite{rutten09,imry2012}, 
similar to those in Refs.~\cite{Entin-Wohlman2010a,imry2010,entin2012,sb2012,imryNJP2013,imry2013}. 
In addition to being coupled to two electronic reservoirs (L and R), the quantum system is coupled to a reservoir of
photons (ph).  This reservoir induces transitions between states of the quantum system with the same charge.
If this reservoir is hotter than the electronic reservoirs, it will tend to excite electrons in the quantum system,
so they will leave with more energy than they entered.  
While we will refer to reservoir ph as being a reservoir of photons, it would change nothing in our analysis 
if it were a reservoir of phonons, or of more exotic chargeless excitations (magnons \cite{sb2012}, microwave photons in the circuit itself \cite{ruokola2012}, etc). Models with two phononic baths were considered in Refs.~\cite{pichard2014b,Pichard2016,Lu-Zhou2015,Lu-Wang2016}.
Ref.~\cite{Bergenfeldt13} considered a heat engine \green{which is a sort of hybrid between  those 
in Figs.~\ref{Fig:3-term-sys1} and \ref{Fig:3-term-sys2};
the heat arrives at the heat-engine in the form of hot microwaves flowing though a waveguide (cavity), but those  microwaves are emitted from a hot reservoir of electrons.}

The quantum dot has two 
states; we define state 1 as the lower energy of the two states, so state 1 has 
energy $\eps_1$ and state 2 has energy $\eps_2>\eps_1$. 
We take state 1 to be more strongly coupled to reservoir L and state 2 to be 
more strongly coupled to reservoir R. 
The many-body eigenstates of the system Hamiltonian are given in 
section~\ref{Sect:Qu-Master-Eqn}'s table~\ref{Table:example-H}.
The potential states of this system are $(n_1,n_2)$, where $n_i$ is the 
occupation of state $i$ which can be 0 or 1.  
We assume that we are in a Coulomb blockaded regime, with a strong enough 
$U$, that the two states are never 
occupied at the same time.  Then  the only relevant states are $|0\rangle$,  $|1\rangle$ and  $|2\rangle$,
as shown in the inset of Fig.~\ref{Fig:3-term-sys1}.

%========================================
%========================================

The rate equation for these three states is
\begin{align}
{\rmd  \over \rmd t}
\left( \!\! \begin{array}{c} P_0 \\ P_1  \\ P_2   \end{array} \!\! \right)  = 
\left(\!\! \begin{array}{ccc}
 -\Gamma_{10}-\Gamma_{20} \!\! & \Gamma_{01} & \Gamma_{02}  \\ 
\Gamma_{10} &\!\! -\Gamma_{01}-\Gamma_{21} \!\! & \Gamma_{12} \\
\Gamma_{20} &  \Gamma_{21} & \!\! -\Gamma_{02}-\Gamma_{12} \!\! \\
\end{array} \right)
\left(\!\! \begin{array}{c} P_0 \\ P_1 \\ P_2 \end{array}\!\!\right)\,. 
\label{Eq:master-eqn-3-term-sys1}
\end{align}
Here $\Gamma_{i0}= \Gamma^{\rm (L)}_{i0}+\Gamma^{\rm (R)}_{i0}$ and
 $\Gamma_{0i}= \Gamma^{\rm (L)}_{0i}+\Gamma^{\rm (R)}_{0i}$ for $i\in 1,2$.  
 Following section~\ref{Sect:transition-rates}, the rates involving adding an 
electron to the dot obey
 \begin{align}
\Gamma^{\rm i}_{i 0} = \ {1 \over h} \nu_i \big(\eps_i)  \, 
 f_i\big(\eps_i\big) \, \big|V^{(i)}_i(\eps_i) \big|^2\,,
\end{align}
where $V^{(i)}_i(\eps_i)$ is the coupling of the system state $i$ to the 
state in 
reservoir $i$ with energy $E_i$.
The rates involving an electron leaving the dot, $\Gamma^{\rm i}_{0i}$, 
are then given by Eq.~(\ref{Eq:rate-for-reverse-final}), as discussed in 
section~\ref{Sect:local-detailed}.
 The rates $\Gamma_{21}$ and $\Gamma_{12}$ are for transitions due to the photon 
reservoir.
 The rate $\Gamma_{21}$ involves a photon adding energy to the dot 
($\eps_2>\eps_1$),
 so from section~\ref{Sect:transition-rates} we have
 \begin{align}
\Gamma_{21} \equiv \Gamma^{\rm ph +}_{21}  = {1 \over h} \nu_{\rm ph} 
\big(\eps_2-\eps_1)  \, 
 n_{\rm ph} \big(\eps_2-\eps_1\big) \, \big|V_{\rm (ph)}(\eps_2-\eps_1) \big|^2\,.
\end{align}
where $V_{\rm (ph)}(\omega)$ is the coupling of the system to the photon field at energy $\omega$.
The rate $\Gamma_{12}$ involving the dot losing energy into the photon 
reservoir, is then given by Eq.~(\ref{Eq:rate-for-reverse-final}), as discussed 
in section~\ref{Sect:local-detailed}.

We will show that the machine can absorb heat from the photon reservoir and 
thus generate 
electrical power in the electronic reservoirs, by driving an electrical current 
from reservoir L to reservoir R against a potential difference.  
The two quantities of most interest for a heat engine are the electrical power 
it generates, $P_{\rm gen}$,
and its efficiency $\eta_{\rm eng}$.  
In this case, as reservoir ph is the heat source ($T_{\rm ph}>T_0$), 
the heat-engine's efficiency is given by 
\begin{align}
\eta_{\rm eng} \equiv P_{\rm gen}\big/ \Jheatph\,.
\end{align}
Alternatively, we can use the machine as a refrigerator to cool the photon 
reservoir (ph)
below the temperature of its environment, $T_{\rm ph}<T_0$.  This cooling 
requires
that the electrical current is driven by a potential difference so the power 
absorbed 
by the refrigerator $P_{\rm abs}=-P_{\rm gen} >0$.  
The two quantities of most interest for a refrigerator are its cooling power,
and its coefficient of performance (COP).  
In this case, as reservoir ph is being cooled, the cooling power is the 
\green{the heat current out of reservoir ph, $\Jheatph$.}
The coefficient of performance (efficiency) for this cooling is 
\begin{align}
\eta_{\rm fri} \equiv \Jheatph\big/ P_{\rm abs}\,.
\label{Eq:3-term-sys1-fridge}
\end{align}

Without loss of generality, we will take the zero of energy to be that of the 
electrochemical potential
of reservoir L ($\mu_{\rm L}=0$), and then define $\mu_{\rm R}=\mu$, so that 
$\mu$ 
is the difference in electrochemical potential between reservoir R and L.  

\subsubsection{Results before solving the steady-state equation}
\label{Sect:3-term-sys1-before-solving}

Before explicitly finding the steady-state of the above rate equation, we use 
the results in sections~\ref{Sect:single-loop} and \ref{Sect:rule-Carnot} 
to get information about the system's efficiency, such as whether it can 
achieve Carnot efficiency.
Let us start by considering the case where 
state 1 has negligible coupling to reservoir R,  
while state 2 has negligible coupling to reservoir L, so 
\begin{align}
\Gamma_{10}^{\rm (R)} = \Gamma_{01}^{\rm (R)}  = 
\Gamma_{20}^{\rm (L)} = \Gamma_{02}^{\rm (L)}  = 0.
\label{Eq:3-term-sys1-no-dashed-transitions}
\end{align}   
This means that the transitions marked by dashed lines in the inset of 
Fig.~\ref{Fig:3-term-sys1}
are absent.  Then, there are only two closed primitive trajectories which are not 
self-retracing, 
the first is
\begin{align}
\zeta_1 \ =\
\hskip 2mm \raisebox{6pt}{$0$} 
\hskip -2mm {\xrightarrow{\hspace*{3mm}} } \hskip -2mm
\begin{array}{c} {\scriptstyle {\rm L}} \\ {\bm |} \\ \phantom{{\scriptstyle 
L}} \end{array}  
\hskip 2mm \raisebox{6pt}{$1$} 
\hskip -6.5mm {\xrightarrow{\hspace*{8mm}} } \hskip -2mm
\begin{array}{c} {\scriptstyle {\rm ph}} \\ {\bm |} \\ \phantom{{\scriptstyle 
L}} \end{array} 
\hskip 2mm \raisebox{6pt}{$2$} 
\hskip -6.5mm {\xrightarrow{\hspace*{8mm}} } \hskip -2mm
\begin{array}{c} {\scriptstyle {\rm R}} \\ {\bm |} \\ \phantom{{\scriptstyle 
L}} \end{array} 
\hskip 1mm \raisebox{6pt}{$0$} 
\hskip -5.5mm {\xrightarrow{\hspace*{3mm}} }\,, 
\label{Eq:zeta1-sys1}
\end{align}
while the second is its time-reverse, $\bar{\zeta}_1$.
For the transition 
\hbox{$\raisebox{6pt}{${\scriptstyle 0}$} \hskip 
-2.5mm\xrightarrow{\hspace*{2mm}} \hskip -1.5mm
\raisebox{8pt}{${\scriptstyle {\rm L}}$} \hskip -1.1mm {\bm |} \hskip -1.5mm
\xrightarrow{\hspace*{2mm}} \hskip -2mm \raisebox{6pt}{${\scriptstyle 1}$}$},
the entropy change in reservoir L is $-\eps_1/T_0$.
For the transition 
\hbox{$\raisebox{6pt}{${\scriptstyle 1}$} \hskip 
-2.5mm\xrightarrow{\hspace*{2mm}} \hskip -1.5mm
\raisebox{8pt}{${\scriptstyle {\rm ph}}$} \hskip -1.1mm {\bm |} \hskip -1.5mm
\xrightarrow{\hspace*{2mm}} \hskip -2mm \raisebox{6pt}{${\scriptstyle 2}$}$},
the entropy change in reservoir ph is $-(\eps_2-\eps_1)/T_{\rm ph}$.
For the transition 
\hbox{$\raisebox{6pt}{${\scriptstyle 2}$} \hskip 
-2.5mm\xrightarrow{\hspace*{2mm}} \hskip -1.5mm
\raisebox{8pt}{${\scriptstyle {\rm R}}$} \hskip -1.1mm {\bm |} \hskip -1.5mm
\xrightarrow{\hspace*{2mm}} \hskip -2mm \raisebox{6pt}{${\scriptstyle 0}$}$},
the entropy change in reservoir R is $(\eps_2-\mu)/T_0$.
Thus the total entropy change in the reservoirs associated with trajectory 
$\zeta_1$ 
or $\bar{\zeta}_1$ is
\begin{align}
\Delta\mathscr{S}_{\rm res}(\zeta_1) = - \Delta\mathscr{S}_{\rm res}\left(\bar{\zeta}_1\right) =
{\eps_2-\eps_1-\mu \over T_0} -  {\eps_2-\eps_1 \over T_{\rm ph}} \ .
\end{align}
Using section~\ref{Sect:rule-Carnot}'s rule, we conclude that the steady-state 
will be Carnot efficient if
we choose the electrochemical potential of reservoir R (relative to that of reservoir 
L) to obey
\begin{align}
\mu \ = \   (\eps_2-\eps_1) \left( 1- T_0/T_{\rm ph} 
\right)\,.
\label{Eq:3-term-sys1-mu-for-Carnot}
\end{align}
Since there is only one closed loop in the system, we can also
use sections~\ref{Sect:single-loop} to find the steady-state
efficiencies for heat-engines and refrigerators for arbitrary $\mu$.  
In the context of Eqs.~(\ref{Eq:single-loop-eng},\ref{Eq:single-loop-fridge}), 
the only transition in the loop which contributes to $\Delta Q^{\rm 
(heat)}_{\rm loop}$ (for heat-engines) or
$\Delta Q^{\rm (cold)}_{\rm loop}$ (for refrigerators) is $1 \to 2$, so we have
$\Delta Q^{\rm (heat)}_{\rm loop} =\Delta Q^{\rm (cold)}_{\rm loop} = 
(\eps_2-\eps_1)$. 
The work performed around the loop comes from the transition $2 \to 0$ 
which involves reservoir R (we do not need to take reservoir L into account
because we have defined its electrochemical potential as zero), thus we have
$\Delta W^{\rm gen}_{\rm loop} = -\Delta W^{\rm abs}_{\rm loop}= \mu$. 
Then Eqs.~(\ref{Eq:single-loop-eng},\ref{Eq:single-loop-fridge}) give,
\begin{align}
\eta_{\rm eng} = {\mu \over \eps_2-\eps_1} \ ,\qquad
\eta_{\rm fri} = -\,{ \eps_2-\eps_1 \over \mu} \ ,
\label{Eq:3-term-sys1-eff-single-loop}
\end{align}
where $\mu$ has the same sign as $(\eps_2-\eps_1)$ for the heat-engine,
and the opposite sign for the refrigerator.
These immediately give Carnot efficiencies when we substitute 
in Eq.~(\ref{Eq:3-term-sys1-mu-for-Carnot}).

However, the results in 
Eq.~(\ref{Eq:3-term-sys1-mu-for-Carnot},\ref{Eq:3-term-sys1-eff-single-loop}) 
rely on the fact that we have assumed that the couplings marked by dashed lines 
in the inset of Fig.~\ref{Fig:3-term-sys1} are negligible, as in 
Eq.~(\ref{Eq:3-term-sys1-no-dashed-transitions}).
If we re-introduce these couplings, we see that there is no longer only one 
loop in the network 
of system-states.  There are four other primitive closed 
trajectories which 
are not self-retracing,  
\begin{align}
\zeta_2 \ =\
\hskip 2mm \raisebox{6pt}{$0$} 
\hskip -2mm {\xrightarrow{\hspace*{3mm}} } \hskip -2mm
\begin{array}{c} {\scriptstyle {\rm L}} \\ {\bm |} \\ \phantom{{\scriptstyle 
L}} \end{array}  
\hskip 2mm \raisebox{6pt}{$1$} 
\hskip -6.5mm {\xrightarrow{\hspace*{8mm}} } \hskip -2mm
\begin{array}{c} {\scriptstyle {\rm R}} \\ {\bm |} \\ \phantom{{\scriptstyle 
L}} \end{array} 
\hskip 1mm \raisebox{6pt}{$0$} 
\hskip -5.5mm {\xrightarrow{\hspace*{3mm}} } 
\,,\qquad \qquad 
\zeta_3 \ =\
\hskip 2mm \raisebox{6pt}{$0$} 
\hskip -2mm {\xrightarrow{\hspace*{3mm}} } \hskip -2mm
\begin{array}{c} {\scriptstyle {\rm L}} \\ {\bm |} \\ \phantom{{\scriptstyle 
L}} \end{array}  
\hskip 2mm \raisebox{6pt}{$2$} 
\hskip -6.5mm {\xrightarrow{\hspace*{8mm}} } \hskip -2mm
\begin{array}{c} {\scriptstyle {\rm R}} \\ {\bm |} \\ \phantom{{\scriptstyle 
L}} \end{array} 
\hskip 1mm \raisebox{6pt}{$0$} 
\hskip -5.5mm {\xrightarrow{\hspace*{3mm}} } 
\end{align}
and their time-reverse trajectories, $\bar{\zeta}_2$ and  $\bar{\zeta}_3$.
One might think there are other primitive trajectories, however
we can construct them out of other primitive closed trajectories. For example,
\begin{align}
\hskip 2mm \raisebox{6pt}{$0$} 
\hskip -2mm {\xrightarrow{\hspace*{3mm}} } \hskip -2mm
\begin{array}{c} {\scriptstyle {\rm L}} \\ {\bm |} \\ \phantom{{\scriptstyle 
L}} \end{array}  
\hskip 2mm \raisebox{6pt}{$1$} 
\hskip -6.5mm {\xrightarrow{\hspace*{8mm}} } \hskip -2mm
\begin{array}{c} {\scriptstyle {\rm ph}} \\ {\bm |} \\ \phantom{{\scriptstyle 
L}} \end{array} 
\hskip 2mm \raisebox{6pt}{$2$} 
\hskip -6.5mm {\xrightarrow{\hspace*{8mm}} } \hskip -2mm
\begin{array}{c} {\scriptstyle {\rm L}} \\ {\bm |} \\ \phantom{{\scriptstyle 
L}} \end{array} 
\hskip 1mm \raisebox{6pt}{$0$} 
\hskip -5.5mm {\xrightarrow{\hspace*{3mm}} } 
\end{align} 
looks like it might be a primitive closed trajectory, but we can construct it 
out of
$\zeta_1$ and $\bar{\zeta}_3$.
Following the same logic as for trajectory $\zeta_1$ above, we see that
\begin{align}
\Delta\mathscr{S}_{\rm res}(\zeta_2) =-\Delta\mathscr{S}_{\rm res}(\bar{\zeta}_2) =
\Delta\mathscr{S}_{\rm res}(\zeta_3) =-\Delta\mathscr{S}_{\rm res}(\bar{\zeta}_3) = - \mu\big/T_0 
\ .
\end{align} 
Given Eq.~(\ref{Eq:trajectory-fluct-rel}), this means that for $\mu >0$, 
trajectories $\bar{\zeta}_{2}$ and  $\bar{\zeta}_{3}$ are more probable than 
$\zeta_2$ and $\zeta_3$, respectively. Thus, this set of trajectories leads to a net flow of 
electrons from
a region of high electrochemical potential (reservoir R) to one of low 
electrochemical potential (reservoir L).
As such, they are parasitic processes, reducing the power generation. 
Formally, one can still achieve Carnot efficiency if one chooses the system to 
have 
$\eps_2-\eps_1 \to 0$ and then takes $\mu=0$, so the parasitic back-flow of 
electrons is negligible. Indeed, it is often the case in machines with such 
parasitic processes, that
the dissipation becomes negligible in the limit $\mu \to 0$.

However, this limit $\mu \to 0$ is bad for producing power.   
The reason is that Carnot efficiency always corresponds to vanishing currents
(since it requires reversibility).  If one can achieve this at finite $\mu$, 
then one can get
close to Carnot efficiency for small currents, and the power output will be 
proportional to the small current.  However, if one can only achieve Carnot 
efficiency  at $\mu=0$
then to get non-zero power close to Carnot efficiency one must have a small 
current 
and a small $\mu$,  which means the power output will be the product of two 
small numbers.
Thus, unsurprisingly, the machine with the parasitic dissipation process 
will produce less power at given efficiency that the one without this parasitic 
process.

\subsubsection{Solving the steady-state equation}

Now we return to finding the steady-state properties of the rate equation in 
Eq.~(\ref{Eq:master-eqn-3-term-sys1}).
In the steady-state, the left hand side of this equation is zero, so using the 
fact that
$P_2=1-P_0-P_1$ to eliminate $P_2$, we get a pair of simultaneous equations
\begin{subequations}
\begin{align}
(\Gamma_{10}+\Gamma_{20}+\Gamma_{02})P^{\rm steady}_0 + 
(\Gamma_{02}-\Gamma_{01})P^{\rm steady}_1 &= \Gamma_{02} \ ,
\\
(\Gamma_{12}-\Gamma_{10})P^{\rm steady}_0 +(\Gamma_{01}+\Gamma_{21}+ 
\Gamma_{12})P^{\rm steady}_1&= \Gamma_{12}\ .
\label{Eq:sim-eq-2}
\end{align}
\end{subequations}
The solutions are
\begin{subequations}
\label{Eq:3-term-sys1-occupations}
\begin{align}
P^{\rm steady}_0 \ \ &=\ { \Gamma_{01}\Gamma_{02}+\Gamma_{21}\Gamma_{02}+\Gamma_{01}\Gamma_{12}
\over K} \ ,
\\
P^{\rm steady}_1 \ \ &=\ {  \Gamma_{12}\Gamma_{10}+\Gamma_{12}\Gamma_{20}+\Gamma_{02}\Gamma_{10}
\over K} \ ,
\\
 P^{\rm steady}_2 \ \ &=\ 1-P^{\rm steady}_0-P^{\rm steady}_1  \ =\ {  \Gamma_{01}\Gamma_{20}+\Gamma_{21}\Gamma_{10}+\Gamma_{21}\Gamma_{20}
\over K} \ ,
\end{align}
\end{subequations}
where we have defined
\begin{align}
K \  =& \ \  (\Gamma_{01}+\Gamma_{21}+ 
\Gamma_{12})(\Gamma_{10}+\Gamma_{20}+\Gamma_{02})-(\Gamma_{02}-\Gamma_{01}
)(\Gamma_{12}-\Gamma_{10}) 
\nonumber 
\\
=& \ \ \Gamma_{01}\Gamma_{02}+\Gamma_{21}\Gamma_{02}+\Gamma_{01}\Gamma_{12} 
  +  \Gamma_{12}\Gamma_{10}+\Gamma_{12}\Gamma_{20}+\Gamma_{02}\Gamma_{10}
  + \Gamma_{01}\Gamma_{20}+\Gamma_{21}\Gamma_{10}+\Gamma_{21}\Gamma_{20} \ ,
\end{align}
where the first line is prettier, but the second line shows explicitly that $K$ is positive
(since all $\Gamma$s are positive), ensuring all the above probabilities are positive.
From Eqs.~(\ref{Eq:I^N}-\ref{Eq:J}) we get that
\begin{align}
\JparticleL \ \ &=\ -\JparticleR \ = \  
{\cal I}^{\rm (L)\, steady}_{10} +{\cal I}^{\rm (L)\, steady}_{20}\,,
\label{Eq:I_L-3term-sys1} 
\\
\JheatL \ \ &=\
\JenergyL  \ = \   
\eps_1 {\cal I}^{\rm (L)\, steady}_{10} + \eps_2 {\cal I}^{\rm (L)\, 
steady}_{20}\,,
\label{Eq:J_L-3term-sys1} 
\\
\Jheatph\ \ &= \   \Jenergyph \ =\   ( \eps_2-\eps_1) \,{\cal I}^{\rm 
(L)\, steady}_{21}\,,\label{Eq:J_ph-3term-sys1} 
\end{align}
where ${\cal I}_{ba}^{(i)\, \rm steady}$ is given by 
Eq.~(\ref{Eq:probability-currents})
with the occupation probabilities given by 
Eqs.~(\ref{Eq:3-term-sys1-occupations}). 
Energy conservation gives $\JenergyR= -\JenergyL -  \Jheatph$, so
\begin{align}
\JheatR&= \JenergyR-\mu \JparticleR 
\ =\ \mu \JparticleL - \JheatL - \Jheatph\,.
\label{Eq:3-term-sys1-JR}
\end{align}
The power generated is $P_{\rm gen}= -\mu \JparticleR=\mu  \JparticleL$.
The above results, 
Eqs.~(\ref{Eq:3-term-sys1-occupations}-\ref{Eq:3-term-sys1-JR}), 
constitute a complete solution of the general case of this model.

Returning to the special case discussed in detail above, 
given by Eq.~(\ref{Eq:3-term-sys1-no-dashed-transitions}) in which there is a 
single loop in the state-space.
The particle current is given by Eq.~(\ref{Eq:I_L-3term-sys1}) with ${\cal 
I}_{20}^{\rm (L)}=0$,
its explicit form is thus 
\begin{align}
\JparticleL = -\JparticleR = \Gamma^{\rm (L)}_{10} P_0^{\rm steady} - 
\Gamma^{\rm (L)}_{01} P_1^{\rm steady}\ ,
\label{Eq:particle-current-special-case}
\end{align}
with $P_a^{\rm steady}$ given by Eq.~(\ref{Eq:3-term-sys1-occupations}).
Since $\JparticleL = {\cal I}_{\rm loop}^{\rm steady}$, we can use
Eqs.~(\ref{Eq:I^N+I^E-in-terms-of-I_loop}-\ref{Eq:J-in-terms-of-I_loop}) 
to write all currents in terms of $\JparticleL$.
The energy currents simplify
to $\JenergyL= \eps_1 \JparticleL$ and 
$\JenergyR= \eps_2 \JparticleR =-\eps_2 \JparticleL$. 
Then since energy conservation means that 
$\JenergyL +\JenergyR + \Jheatph = 0$, we have
$\Jheatph = (\eps_2 -\eps_1)  \JparticleL$.
This information complements that in 
Eqs.~(\ref{Eq:3-term-sys1-no-dashed-transitions}-\ref{Eq:3-term-sys1-eff-single-loop}) 
for this special case.

%-------------------------------------------------
\subsection{Machine with three electronic reservoirs}
\label{Sect:3-term-sys2}

%========================================
\begin{figure}
\centerline{\includegraphics[width=0.75\textwidth]{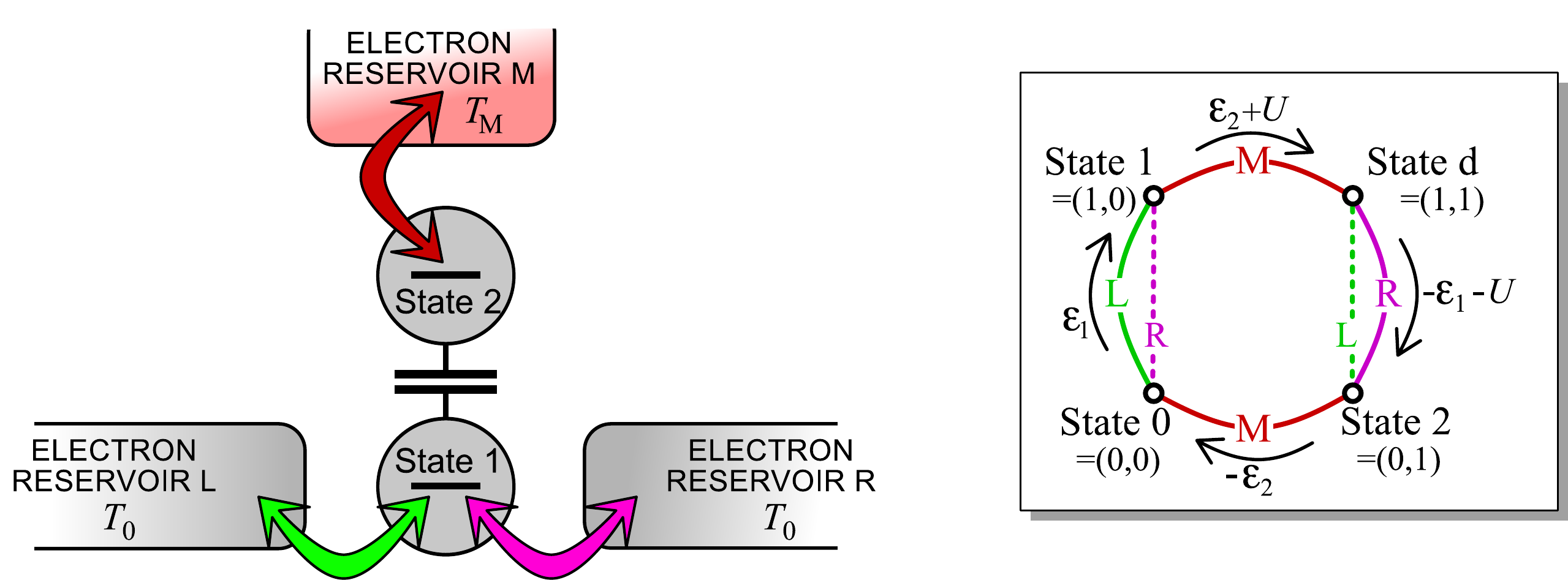}}
\caption{\label{Fig:3-term-sys2}
A system which has been proposed as a heat-engine \cite{Sanchez2011,buttiker2012,Sanchez2013,Esposito-Maxwell-demon}.
It converts heat radiated by the middle electronic reservoir (at temperature 
$T_{\rm M}>T_0$)
into electrical power, manifested 
as a charge current between the left and right electronic reservoirs flowing 
against a potential difference.
The inset shows the network associated with the system dynamics, 
the nodes indicating system states, while the bonds indicate transitions 
induced by the coupling to the reservoir indicated (L, R or M). 
For the system to work well as a heat-engine, the transition rates associated 
with the dashed 
lines must be much smaller than those associated with the solid lines.
The arrows in the inset indicate a heat flow out of reservoir M, 
which causes an electric current from left to right.  
Next to each arrow we indicate the energy that the reservoir gives to the system 
during that transition.
The same system could also be used as a refrigerator to cool  reservoir ph 
to a temperature $T_{\rm ph} < T_0$, by driving a current between the left and right reservoirs. 
}
\end{figure}
%========================================

Consider the system in Fig.~\ref{Fig:3-term-sys2} suggested in 
\colorproofs{Ref.~\cite{Sanchez2011}, see also Refs.~\cite{buttiker2012,Sanchez2013,Esposito-Maxwell-demon},} and recently realized experimentally in Refs.~\cite{Roche15,Hartmann15,Thierschmann15}.
There one electronic reservoir (M) at temperature $T_{\rm M}$ is coupled to a 
dot which is capacitively 
coupled to the rest of the system, thus it cannot exchange charge with the 
other reservoirs, although we will see that it can exchange heat. 
This heat exchange leads to electrical power generation between the left and right reservoirs.
Such experimental quantum dot systems are likely to lose heat directly into their cold environment,
which should be modelled by adding a capacitive coupling to a fourth cold reservoir \cite{wshs2016}.  
We will not consider this heat loss further here, beyond noting that Ref.~\cite{wshs2016} found that such a
quantum system could be designed to out perform any classical system with the same heat loss.

Ref.~\cite{Bergenfeldt13} considered a more complicated system , in which the two dots are replaced by a pair of double-dots sitting at each end of a microwave cavity.  At one end of the microwave cavity is the heat engine which is driven by the microwaves flowing from a hot reservoir of electrons at the other end of the microwave cavity. Such a system could be treated in terms of rate equations in a similar manner to here, 
but one would have to replace the four states in the inset of Fig.~\ref{Fig:3-term-sys2}  
with the relevant eigenstates of the Hamiltonian for the two double-dots coupled to the microwave cavity.

The methods and results discussed in this section
are fairly similar to those in section~\ref{Sect:3-term-sys1-before-solving}.
However, we do not wish to make section~\ref{Sect:3-term-sys1} required reading 
to understand this section, so there will be a certain amount of repetition 
here.
We assume that each dot only has a single energy level, at energies $\eps_1$ 
and $\eps_2$, 
and that the charging energy for each dot is high enough that the neither dots can 
be 
doubly-occupied (for simplicity we also neglect spin). 
We also assume that the capacitative coupling between the dots leads to a charging energy $U$ for occupying the two dots at the same time.
Then the many-body eigenstates of the system are those given in section~\ref{Sect:Qu-Master-Eqn}'s table~\ref{Table:example-H}.
The rate equation for this system is 
\begin{align}
{\rmd  \over \rmd t}
\left( \!\! \begin{array}{c} P_0 \\ P_1  \\ P_2 \\ P_\rmd  \end{array} \!\! 
\right)  = 
\mathbf{\Gamma} \ 
\left( \!\! \begin{array}{c} P_0 \\ P_1  \\ P_2 \\ P_\rmd  \end{array} \!\! 
\right)\,, 
\label{Eq:3term-sys2-master-eqn}
\end{align}
\green{where we now use the matrix form of the rate equation given in Eq.~(\ref{Eq:master-eqn-matrix-form}).}
The four-by-four matrix 
$\mathbf{\Gamma} 
= \mathbf{\Gamma}^{\rm (L)}+ \mathbf{\Gamma}^{\rm (R)}+ \mathbf{\Gamma}^{\rm 
(M)}$ 
is the sum of transitions due to
reservoirs L, R and M.  The transitions due to reservoirs L or R take the form
\begin{align}
\bm{\Gamma}^{\rm (i)} 
&\equiv 
\left(\begin{array}{cccc}
- \Gamma^{\rm (i)}_{10} \phantom{\Big|}\! & 
\Gamma^{\rm (i)}_{01}  
& 0&0 \cr
\Gamma^{\rm (i)}_{10} \phantom{\Big|}\! & 
-\Gamma^{\rm (i)}_{01}  
& 0&0 \cr
0 & 0 & -\Gamma^{\rm (i)}_{\rmd2} \phantom{\Big|}\! & \Gamma^{\rm 
(i)}_{2\rmd} \cr  
0 & 0 & \Gamma^{\rm (i)}_{\rmd2} \phantom{\Big|}\! & -\Gamma^{\rm 
(i)}_{2\rmd} 
\end{array}\right)\,,
\end{align}
for $i =$ L or R.
The transitions due to reservoir M take the form
\begin{align}
\bm{\Gamma}^{\rm (M)} 
&\equiv 
\left(\begin{array}{cccc}
- \Gamma^{\rm (M)}_{20} \phantom{\Big|}\! & 0 &
\Gamma^{\rm (M)}_{02}  & 0 \cr
0 & -\Gamma^{\rm (M)}_{\rmd 1} & 0 & \Gamma^{\rm (M)}_{1\rmd} \cr  
\Gamma^{\rm (M)}_{20} \phantom{\Big|}\! & 0 &
-\Gamma^{\rm (M)}_{02} & 0 \cr
0 & \Gamma^{\rm (M)}_{\rmd1} & 0 & -\Gamma^{\rm (M)}_{1\rmd}  
\end{array}\right)\,,
\end{align}
with the rates obeying Eq.~(\ref{Eq:rate-for-reverse-final}). 
Often, \green{it is sufficient to  consider these rates phenomenological parameters},
however they can be related to the Hamiltonian of the system and environment
as in section~\ref{Sect:transition-rates}. Then the rates are
\begin{subequations}
\label{Eq:3-term-sys2-rates}
\begin{align}
\Gamma^{(i)}_{10} =& \  {1 \over h} \,\nu_i \big(\eps_1)  \, 
 f_i\big(\eps_1\big) \, \big|V_i(\eps_1) \big|^2\,,
\\
\Gamma^{(i)}_{\rmd2} =& \ {1 \over h} \,\nu_i (\eps_1+U) \, 
 f_i\big(\eps_1+U\big)\, \big|V_i (\eps_1+U)\big|^2\,, 
\\
\Gamma^{\rm (M)}_{20} =& \  {1 \over h} \,\nu_{\rm M} \big(\eps_2)  \, 
 f_{\rm M}\big(\eps_2\big) \, \big|V_{\rm M}(\eps_2) \big|^2\,,
\\
\Gamma^{\rm (M)}_{\rmd1} =& \ \ {1 \over h}\, \nu_{\rm M} (\eps_2+U) \, 
 f_{\rm M}\big(\eps_2+U\big)\, \big|V_{\rm M} (\eps_2+U)\big|^2\,, 
\end{align}
\end{subequations}
with the reverse of these rates given by Eq.~(\ref{Eq:rate-for-reverse-final}),
as discussed in section~\ref{Sect:local-detailed}.
Here $V_i(E)$ for $i \in {\rm L,R}$ is the matrix element for an 
electron hopping
between dot 1 and a state in reservoir $i$ with energy $E$, while 
$V_{\rm M}(E)$ is the matrix element for an electron hopping
between dot 2 and a state in reservoir M with energy $E$.
Without loss of generality, we take the zero of energy to be that of the 
electrochemical potential
of reservoir L ($\mu_{\rm L}=0$), and then define $\mu_{\rm R}=\mu$, so that 
$\mu$ 
is the difference in electrochemical potential between reservoir R and L.  
We then note that since reservoir M never exchanges particles with the other 
reservoirs or dot 1,
its electrochemical potential only plays a role with respect to dot 2. 
Thus, without loss of generality, we can measure dot 2 energy $\eps_2$ from the 
electrochemical potential of reservoir M, which is the same as taking $\mu_{\rm M}=0$.

We will show that the machine can absorb heat from the reservoir M and thus 
generate 
electrical power by driving an electrical current from reservoir L to 
reservoir R against a potential difference.  
The two quantities of most interest for a heat engine are 
the electrical power it generates, $P_{\rm gen}$,
and its efficiency $\eta_{\rm eng}$.  
In this case, as reservoir M is the heat source ($T_{\rm M}>T_0$), 
the heat-engine's efficiency is given by 
\begin{align}
\eta_{\rm eng} \equiv P_{\rm gen}\big/ \JheatM\,.
\end{align}
Alternatively, we can use the machine as a refrigerator to cool electron 
reservoir M
below the temperature of its environment ($T_{\rm M}<T_0$).  This cooling 
requires
that the electrical current is driven by a potential difference so the power 
absorbed 
by the refrigerator $P_{\rm abs}=-P_{\rm gen} >0$.  
The two quantities of most interest for a refrigerator are its cooling power,
and its coefficient of performance (COP).  
In this case, as reservoir M is being refrigerated, the cooling power is 
$\JheatM$,
while the coefficient of performance (efficiency) is 
\begin{align}
\eta_{\rm fri} \equiv \JheatM\big/ P_{\rm abs}\ .
\label{Eq:3-term-sys2-fridge}
\end{align}

\subsubsection{Results before solving the steady-state equation}
\label{Sect:3-term-sys2-before-solving}

Before discussing the manner of finding the steady-state solution of the rate 
equation in Eq.~(\ref{Eq:3term-sys2-master-eqn}), 
we use results from section~\ref{Sect:single-loop} and \ref{Sect:rule-Carnot} 
to find the conditions under which this system can achieve Carnot efficiency.

Let us start by considering the case where dot 1 is \green{only} tunnel-coupled to 
reservoir R if dot 2 is full,
and is \green{only} tunnel-coupled to reservoir L if dot 2 is empty;  so
\begin{align}
\Gamma^{\rm (R)}_{10}=\Gamma^{\rm (R)}_{01}
= \Gamma^{\rm (L)}_{d2}=\Gamma^{\rm (L)}_{2d}= 0 \ .
\label{Eq:3-term-sys2-no-dashed-transitions}
\end{align}
This requires that the energy dependence of the tunnel-coupling in 
Eqs.~(\ref{Eq:3-term-sys2-rates}) is
such that $V_{\rm R}(\eps_1)=0$ while $V_{\rm R}(\eps_1+U)$ is finite,
and that  $V_{\rm L}(\eps_1+U)=0$ while $V_{\rm L}(\eps_1)$ is finite.
This corresponds to neglecting the transitions marked by dashed lines in the 
inset of
Fig.~\ref{Fig:3-term-sys2}.
Then there are only two closed primitive trajectories which are not 
self-retracing, 
the first is
\begin{align}
\zeta_1 \ =\
\hskip 2mm \raisebox{6pt}{$0$} 
\hskip -2mm {\xrightarrow{\hspace*{3mm}} } \hskip -2mm
\begin{array}{c} {\scriptstyle {\rm L}} \\ {\bm |} \\ \phantom{{\scriptstyle 
L}} \end{array}  
\hskip 2mm \raisebox{6pt}{$1$} 
\hskip -6.5mm {\xrightarrow{\hspace*{8mm}} } \hskip -2mm
\begin{array}{c} {\scriptstyle {\rm M}} \\ {\bm |} \\ \phantom{{\scriptstyle 
L}} \end{array} 
\hskip 2mm \raisebox{6pt}{$\rmd$} 
\hskip -6.5mm {\xrightarrow{\hspace*{8mm}} } \hskip -2mm
\begin{array}{c} {\scriptstyle {\rm R}} \\ {\bm |} \\ \phantom{{\scriptstyle 
L}} \end{array} 
\hskip 2mm \raisebox{6pt}{$2$} 
\hskip -6.5mm {\xrightarrow{\hspace*{8mm}} } \hskip -2mm
\begin{array}{c} {\scriptstyle {\rm M}} \\ {\bm |} \\ \phantom{{\scriptstyle 
L}} \end{array} 
\hskip 1mm \raisebox{6pt}{$0$} 
\hskip -5.5mm {\xrightarrow{\hspace*{3mm}} } \,,
\end{align}
while the second is its time-reverse, $\bar{\zeta}_1$.
For the transition 
\hbox{$\raisebox{6pt}{${\scriptstyle 0}$} \hskip 
-2.5mm\xrightarrow{\hspace*{2mm}} \hskip -1.5mm
\raisebox{8pt}{${\scriptstyle {\rm L}}$} \hskip -1.1mm {\bm |} \hskip -1.5mm
\xrightarrow{\hspace*{2mm}} \hskip -2mm \raisebox{6pt}{${\scriptstyle 1}$}$},
the entropy change in reservoir L is $-\eps_1/T_0$.
For the transition 
\hbox{$\raisebox{6pt}{${\scriptstyle 1}$} \hskip 
-2.5mm\xrightarrow{\hspace*{2mm}} \hskip -1.5mm
\raisebox{8pt}{${\scriptstyle {\rm M}}$} \hskip -1.1mm {\bm |} \hskip -1.5mm
\xrightarrow{\hspace*{2mm}} \hskip -2mm \raisebox{6pt}{${\scriptstyle \rmd}$}$},
the entropy change in reservoir M is $-(\eps_2+U)/T_{\rm M}$.
For the transition 
\hbox{$\raisebox{6pt}{${\scriptstyle \rmd}$} \hskip 
-2.5mm\xrightarrow{\hspace*{2mm}} \hskip -1.5mm
\raisebox{8pt}{${\scriptstyle {\rm R}}$} \hskip -1.1mm {\bm |} \hskip -1.5mm
\xrightarrow{\hspace*{2mm}} \hskip -2mm \raisebox{6pt}{${\scriptstyle 2}$}$},
the entropy change in reservoir R is $(\eps_1+U-\mu)/T_0$.
For the transition 
\hbox{$\raisebox{6pt}{${\scriptstyle 2}$} \hskip 
-2.5mm\xrightarrow{\hspace*{2mm}} \hskip -1.5mm
\raisebox{8pt}{${\scriptstyle {\rm M}}$} \hskip -1.1mm {\bm |} \hskip -1.5mm
\xrightarrow{\hspace*{2mm}} \hskip -2mm \raisebox{6pt}{${\scriptstyle 0}$}$},
the entropy change in reservoir M is $\eps_2/T_{\rm M}$.
Thus the total entropy change in the reservoirs associated with trajectory 
$\zeta_1$ 
or $\bar{\zeta}_1$ is
\begin{align}
\Delta\mathscr{S}_{\rm res}(\zeta_1) = - \Delta\mathscr{S}_{\rm res}\left(\bar{\zeta}_1\right) =
{U-\mu \over T_0} -  {U \over T_{\rm M}} \ .
\end{align}
Using section~\ref{Sect:rule-Carnot}'s rule, we conclude that the steady-state 
will be Carnot efficient if
we choose the electrochemical potential of reservoir R (relative to that of reservoir 
L) to obey
\begin{align}
\mu  \ =\ U \left( 1- T_0/T_{\rm M} \right) \,. 
\label{Eq:3-term-sys2-mu-for-Carnot}
\end{align}
Since there is only one loop in the system, we can  
use the results in section~\ref{Sect:single-loop} to find the steady-state
efficiencies for heat-engines and refrigerators for arbitrary $\mu$.  
For this, we take Eqs.~(\ref{Eq:single-loop-eng},\ref{Eq:single-loop-fridge}), 
with two transition in the loop contributing to $\Delta Q^{\rm (heat)}_{\rm 
loop}$ \green{(for heat-engines)} or 
$\Delta Q^{\rm (cold)}_{\rm loop}$ \green{(for refrigerators)}, these are 
\hbox{$\raisebox{6pt}{${\scriptstyle 1}$} \hskip 
-2.5mm\xrightarrow{\hspace*{2mm}} \hskip -1mm {\bm |} \hskip -1.5mm
\xrightarrow{\hspace*{2mm}} \hskip -2mm \raisebox{6pt}{${\scriptstyle \rmd}$}$}
and
\hbox{$\raisebox{6pt}{${\scriptstyle 2}$} \hskip 
-2.5mm\xrightarrow{\hspace*{2mm}} \hskip -1mm {\bm |} \hskip -1.5mm
\xrightarrow{\hspace*{2mm}} \hskip -2mm \raisebox{6pt}{${\scriptstyle 0}$}$}.
Thus, we have
$\Delta Q^{\rm (heat)}_{\rm loop} =\Delta Q^{\rm (cold)}_{\rm loop} = -\eps_1 
+(\eps_1+U) =U$. 
The work performed around the loop comes from the transition
\hbox{$\raisebox{6pt}{${\scriptstyle \rmd}$} \hskip 
-2.5mm\xrightarrow{\hspace*{2mm}} \hskip -1mm {\bm |} \hskip -1.5mm
\xrightarrow{\hspace*{2mm}} \hskip -2mm \raisebox{6pt}{${\scriptstyle 2}$}$},
which involves reservoir R (we do not need to take reservoir L into account
because we have defined its electrochemical potential as zero). Thus, we have
$\Delta W^{\rm gen}_{\rm loop} = -\Delta W^{\rm abs}_{\rm loop}= \mu$. 
Then Eqs.~(\ref{Eq:single-loop-eng},\ref{Eq:single-loop-fridge}) give,
\begin{align}
\eta_{\rm eng} = {\mu \over U} \ ,\qquad
\eta_{\rm fri} = -\,{ U \over \mu} \ ,
\label{Eq:3-term-sys2-eff-single-loop}
\end{align}
where $\mu$ is positive for the heat-engine,
and negative for the refrigerator ($U$ is always positive).
These immediately give Carnot efficiencies when we substitute 
in Eq.~(\ref{Eq:3-term-sys2-mu-for-Carnot}).

However, the results in 
Eq.~(\ref{Eq:3-term-sys2-mu-for-Carnot},\ref{Eq:3-term-sys2-eff-single-loop}) 
rely on the fact that we have assumed that the couplings marked by dashed lines 
in the inset of Fig.~\ref{Fig:3-term-sys2} are negligible, as in 
Eq.~(\ref{Eq:3-term-sys2-no-dashed-transitions}).
If we re-introduce these couplings, we see that 
there is no longer a single-loop in the 
network 
of system-states.  There are four other primitive closed trajectories which 
are not self-retracing,  

\begin{align}
\zeta_2 \ =\
\hskip 2mm \raisebox{6pt}{$0$} 
\hskip -2mm {\xrightarrow{\hspace*{3mm}} } \hskip -2mm
\begin{array}{c} {\scriptstyle {\rm L}} \\ {\bm |} \\ \phantom{{\scriptstyle 
L}} \end{array}  
\hskip 2mm \raisebox{6pt}{$1$} 
\hskip -6.5mm {\xrightarrow{\hspace*{8mm}} } \hskip -2mm
\begin{array}{c} {\scriptstyle {\rm R}} \\ {\bm |} \\ \phantom{{\scriptstyle 
L}} \end{array} 
\hskip 1mm \raisebox{6pt}{$0$} 
\hskip -5.5mm {\xrightarrow{\hspace*{3mm}} } 
\,,\qquad \qquad 
\zeta_3 \ =\
\hskip 2mm \raisebox{6pt}{$2$} 
\hskip -2mm {\xrightarrow{\hspace*{3mm}} } \hskip -2mm
\begin{array}{c} {\scriptstyle {\rm L}} \\ {\bm |} \\ \phantom{{\scriptstyle 
L}} \end{array}  
\hskip 2mm \raisebox{6pt}{$\rmd$} 
\hskip -6.5mm {\xrightarrow{\hspace*{8mm}} } \hskip -2mm
\begin{array}{c} {\scriptstyle {\rm R}} \\ {\bm |} \\ \phantom{{\scriptstyle 
L}} \end{array} 
\hskip 1mm \raisebox{6pt}{$2$} 
\hskip -5.5mm {\xrightarrow{\hspace*{3mm}} } 
\end{align}
and their time-reverse trajectories, $\bar{\zeta}_2$ and  $\bar{\zeta}_3$.
One might think there are other primitive trajectories, however
we can construct them out of the above primitive closed trajectories. For 
example,
\begin{align}
\hskip 2mm \raisebox{6pt}{$0$} 
\hskip -2mm {\xrightarrow{\hspace*{3mm}} } \hskip -2mm
\begin{array}{c} {\scriptstyle {\rm L}} \\ {\bm |} \\ \phantom{{\scriptstyle 
L}} \end{array}  
\hskip 2mm \raisebox{6pt}{$1$} 
\hskip -6.5mm {\xrightarrow{\hspace*{8mm}} } \hskip -2mm
\begin{array}{c} {\scriptstyle {\rm M}} \\ {\bm |} \\ \phantom{{\scriptstyle 
L}} \end{array} 
\hskip 2mm \raisebox{6pt}{$\rmd$} 
\hskip -6.5mm {\xrightarrow{\hspace*{8mm}} } \hskip -2mm
\begin{array}{c} {\scriptstyle {\rm L}} \\ {\bm |} \\ \phantom{{\scriptstyle 
L}} \end{array} 
\hskip 2mm \raisebox{6pt}{$2$} 
\hskip -6.5mm {\xrightarrow{\hspace*{8mm}} } \hskip -2mm
\begin{array}{c} {\scriptstyle {\rm M}} \\ {\bm |} \\ \phantom{{\scriptstyle 
L}} \end{array} 
\hskip 1mm \raisebox{6pt}{$0$} 
\hskip -5.5mm {\xrightarrow{\hspace*{3mm}} } 
\end{align}
looks like it might be a primitive closed trajectory, but we can construct it 
out of
$\zeta_1$ and $\bar{\zeta}_3$.

Following the same logic as for trajectory $\zeta_1$ above, we see that
\begin{align}
\Delta\mathscr{S}_{\rm res}(\zeta_2) =-\Delta\mathscr{S}_{\rm res}(\bar{\zeta}_2) =
\Delta\mathscr{S}_{\rm res}(\zeta_3) =-\Delta\mathscr{S}_{\rm res}(\bar{\zeta}_3) = - \mu\big/T_0 
\ .
\end{align} 
Given Eq.~(\ref{Eq:trajectory-fluct-rel}), this means that for $\mu >0$, 
trajectories $\bar{\zeta}_{2}$ and  $\bar{\zeta}_{3}$ are more probable than 
$\zeta_2$ and $\zeta_3$, respectively. Thus, this set of trajectories leads to a net flow  
electrons from
a region of high electrochemical potential (reservoir R) to one of low 
electrochemical potential (reservoir L).
As such, they are parasitic processes, reducing the power generation. 
Formally, one can still achieve Carnot efficiency if one chooses the system to 
have 
$\eps_2-\eps_1 \to 0$ and then takes $\mu=0$, so the parasitic back-flow of 
electrons is negligible. Indeed it is often the case in machines with such 
parasitic processes, that
the dissipation becomes negligible in the limit $\mu \to 0$.
However, this is bad for producing power, for the reasons discussed in the last 
paragraph of section~\ref{Sect:3-term-sys1-before-solving}.

\subsubsection{How to solve the steady-state equation}

Taking the steady-state limit in which the left hand side of the rate 
equation in Eq.~(\ref{Eq:3term-sys2-master-eqn}) is zero,
and then using the fact that $P_\rmd = 1-P_0-P_1-P_2$, we get three 
simultaneous equations, 
\begin{align}
\left(\Gamma_{20} + \Gamma_{10}\right)P^{\rm steady}_0
-\Gamma_{01} P^{\rm steady}_1 - \Gamma_{02} P^{\rm steady}_2 =&0\,,
\nonumber \\
\left(\Gamma_{1\rmd} - \Gamma_{10}\right)P^{\rm steady}_0
+\left(\Gamma_{1\rmd} + \Gamma_{\rmd1} + \Gamma_{01} \right) P^{\rm steady}_1 
+ \Gamma_{1\rmd} P^{\rm steady}_2 =& \Gamma_{1\rmd} \,,
\nonumber
\\
\left(\Gamma_{2\rmd} - \Gamma_{20}\right)P^{\rm steady}_0
+\Gamma_{2\rmd} P^{\rm steady}_1 
+ \left( \Gamma_{2\rmd}+\Gamma_{\rmd2} + \Gamma_{02} \right) P^{\rm steady}_2 
=& \Gamma_{2\rmd} \,,
\nonumber \\
\label{Eq:3-term-sys2-simultaneous}
\end{align}
where 
for compactness we have dropped the reference to the reservoirs in the rates;
however it is easy to see that $\Gamma_{20}\equiv \Gamma_{20}^{\rm (M)}$,
and $\Gamma_{d1}\equiv \Gamma_{d1}^{\rm (M)}$,
while  $\Gamma_{10}\equiv \Gamma_{10}^{\rm (L)}+\Gamma_{10}^{\rm (R)}$,
and $\Gamma_{d2}\equiv \Gamma_{d2}^{\rm (L)}+ \Gamma_{d2}^{\rm (R)}$, and so 
forth.

It is not difficult to solve the above set of simultaneous equations using the 
standard methods,
but it is tedious. The solution are long algebraic expressions, which are hard 
to simplify to something easily comprehensible. 
Thus we do not proceed further with this, if the readers want
the solution, they can evaluate it themselves, or use computer algebra software 
(such as Wolfram's Mathematica) to do so.

From Eqs.~(\ref{Eq:I^N}-\ref{Eq:J}) we get that
\begin{align}
\JparticleL \ =\ -\JparticleR \ = &\ 
{\cal I}^{\rm (L)\, steady}_{10} +{\cal I}^{\rm (L)\, steady}_{d2}\,,
\label{Eq:I_L-3term-sys2} 
\\
\JheatL \ =\
\JenergyL  \ = & \ 
\eps_1 {\cal I}^{\rm (L)\, steady}_{10} + (\eps_1+U) {\cal I}^{\rm (L)\, 
steady}_{d2}\,,
\label{Eq:J_L-3term-sys2} 
\\
\JheatM\ = \   \JenergyM \ =& \ \eps_2 \,{\cal I}^{\rm (L)\, 
steady}_{20}
+ (\eps_2+U) {\cal I}^{\rm (L)\, steady}_{d1}\,,
\label{Eq:J_M-3term-sys2} 
\end{align}
where ${\cal I}_{ba}^{(i)\, \rm steady}$ is given by 
Eq.~(\ref{Eq:probability-currents})
with the occupation probabilities found by solving the simultaneous equations, 
Eq.~(\ref{Eq:3-term-sys2-simultaneous}). 
Energy conservation gives $\JenergyR= -\JenergyL - \JenergyM$, so
\begin{align}
\JheatR&= \JenergyR-\mu \JparticleR 
\ =\ \mu \JparticleL - \JheatL - \JheatM \,.
\label{Eq:3-term-sys2-JR}
\end{align}
The power generated is $P_{\rm gen}= -\mu \JparticleR=\mu \JparticleL$. 
The above results, 
Eqs.~(\ref{Eq:3-term-sys2-simultaneous}-\ref{Eq:3-term-sys2-JR}), 
constitute a formal solution of the general case of this model, however one has 
to 
solve the simultaneous equations, Eqs.~(\ref{Eq:3-term-sys2-simultaneous}), to 
get explicit
formulas for the currents.

Unfortunately, in the special case given by \green{Eq.~(\ref{Eq:3-term-sys2-no-dashed-transitions})}, 
which has a single loop and 
was discussed in detail above, the simultaneous equations are not
simpler to solve than in the general case.  
However, once one has solved them, the formulas for the currents are 
significantly simpler.
The particle current is given by Eq.~(\ref{Eq:I_L-3term-sys2}) with ${\cal 
I}_{d2}^{\rm (L)}=0$,
its explicit form is thus 
\begin{align}
\JparticleL= -\JparticleR= \Gamma^{\rm (L)}_{10} P_0^{\rm steady} - \Gamma^{\rm 
(L)}_{01} P_1^{\rm steady}
\ .
\label{Eq:3-term-sys2-IL-single-loop}
\end{align}
Since $\JparticleL = {\cal I}_{\rm loop}^{\rm steady}$, 
as defined in section~\ref{Sect:single-loop}, 
we can write all currents in terms of  $\JparticleL$ by using Eqs.~(\ref{Eq:I^N+I^E-in-terms-of-I_loop}-\ref{Eq:J-in-terms-of-I_loop}). 
The energy currents simplify
to $\JenergyL= \eps_1 \JparticleL$ and 
$\JenergyR= (\eps_1+U) \JparticleR =-(\eps_1+U) \JparticleL$. 
Then since energy conservation means that 
$\JenergyL +\JenergyR + \JheatR = 0$, we have
$\JheatR = U\,  \JparticleL$.
Thus for this special case, it is sufficient to take $P_0^{\rm steady}$ and 
$P_1^{\rm steady}$ from the solution of the simultaneous equations, 
Eqs.~(\ref{Eq:3-term-sys2-simultaneous}), 
and substitute them into Eq.~(\ref{Eq:3-term-sys2-IL-single-loop}) to get all 
particle, energy and heat currents. 

\subsection{Cooling by heating}
\label{Sect:cooling-by-heating}

While the subject of this review is that of conversion between heat and work, we wish to mention
that one can also use heat directly to do refrigeration (rather than turning the heat into work, and then using that work to do refrigeration).   At the macroscopic scale this is often called an absorption refrigerator.
There has been a lot of work on such {\it cooling by heating} in nanostructures and quantum systems in
recent years \cite{Geusic1959,GevaKosloff94,GevaKosloff96,pekola2007,pekola2011,pekola2012,vandenbroeck2006,vandenbroeck2012,Correa2013,Correa2014, Palao2001,popescu2010,Levy2012a,Levy2012b,popescu2011,popescu2012,fazio2013,eisert2012}.
As mentioned in section~\ref{Sect:Nernst}, a number of works have appeared that implied certain such cooling-by-heating systems could violate Nernst's unattainability principle \cite{kolvar2012-violate-Nernst,Levy2012b,vandenbroeck2012}, 
followed by a number of claims that it is valid 
\cite{Allahverdyan2012-comment,Levy2012-comment,Cleuren-reply2012,Entin-Imry2014-comment,Masanes2014,Freitas2016}.
We will outline cooling-by-heating here, after which readers can study the works on Nernst's principle by themselves.

To perform cooling by heating, a machine must have at least three reservoirs;  reservoirs 0, H and C.
Reservoir 0 is at ambient temperature $T_0$, reservoir H is hotter, $T_{\rm H}>T_0$, and reservoir C is colder,
$T_{\rm C} < T_0$.
The machine then uses the heat flow from reservoir H to reservoir 0 to ``drag'' heat out of reservoir C, even though reservoir C is colder than the other reservoirs. 
There is particular interest in the minimal self-contained machine which can perform such refrigerator.
It was shown \cite{popescu2010,popescu2011,popescu2012} that a refrigerator can consist of three qubits each coupled to a thermal bath.  Ref.~\cite{fazio2013} considered an electronic quantum refrigerator based on four quantum dots in contact with four thermal electronic reservoirs.
Here we outline a system similar to these but with two quantum dots in contact with three reservoirs.

A cooling by heating (cbh) machine's  efficiency is
defined as the heat flow out of reservoir C (the reservoir being refrigerated) divided by the heat flow out of reservoir H (the hot reservoir, whose heat is driving the process);
so its efficiency $\eta_{\rm cbh} = J_{\rm h,C}/J_{\rm h,H}$.
The upper bound on such a machine's efficiency is given by the condition that no entropy is generated, then its efficiency is
\begin{eqnarray}
\eta_{\rm cbh}^{\rm Carnot} = {1- T_0\big/T_{\rm H} \over T_0\big/ T_{\rm C} -1}\,.
\label{Eq:eta_cbh-Carnot}
\end{eqnarray}
It is worth noting that this  Carnot efficiency is exactly the same as the efficiency of a  
Carnot efficient heat engine whose power output all goes into a Carnot efficient refrigerator;
that is to say $\eta_{\rm cbh}^{\rm Carnot} = \eta_{\rm eng}^{\rm Carnot} \eta_{\rm fri}^{\rm Carnot}$.
Thus, in principle, there is no thermodynamic advantage of cooling by heating over 
an ideal heat engine coupled to an ideal refrigerator.
However, in practice, each time we turn heat flows into electricity and back, we can expect sub-Carnot efficiencies,
thus there may well be situations in which cooling by heating achieves higher efficiencies than the available alternatives. 

There are many proposed machines for cooling by heating, however here we limit ourselves to
pointing out that 
\green{the machines considered in sections~\ref{Sect:3-term-sys1} and \ref{Sect:3-term-sys2},  
(shown in Figs.~\ref{Fig:3-term-sys1} and \ref{Fig:3-term-sys2})} 
are capable of doing this.  For example, one could have reservoir L as hot (at temperature $T_{\rm H}$)
and reservoir R as ambient  (at temperature $T_0$,
so heat wants to flow from left to right. This heat flow occurs via an electron flow between reservoirs L and R,
but as we assume the reservoirs are at the same electrochemical potential this electron flow involves no work. 
This heat flow can drag heat out of 
\green{
the other reservoir 
(reservoir ph in Fig.~\ref{Fig:3-term-sys1} or reservoir M in Fig.~\ref{Fig:3-term-sys2})}
even when that reservoir is at a temperature $T_{\rm C}$ colder than the other two reservoirs, $ T_{\rm C} < T_0 < T_{\rm H}$. The system can satisfy the second-law of thermodynamics
because the increase in entropy in reservoir R associated with the heat flow from L to R is larger than
the entropy reduction in the reservoir being cooled.

\green{
Consider the system shown in  Fig.~\ref{Fig:3-term-sys1}
in the ideal case, which obeys Eq.~(\ref{Eq:3-term-sys1-no-dashed-transitions}), 
and for which there are only two primitive trajectories, $\zeta_1$ and  $\bar{\zeta}_1$ (see section~\ref{Sect:3-term-sys1-before-solving}). }
In the set-up which exhibits cooling by heating,  the changes in reservoir  entropy associated with trajectories $\zeta_1$ and  $\bar{\zeta}_1$
are
\begin{align}
\Delta\mathscr{S}_{\rm res}(\zeta_1) \ =\  - \Delta\mathscr{S}_{\rm res}\left(\bar{\zeta}_1\right) \ =\ 
-{\eps_1\over T_{\rm H}} -  {\eps_2-\eps_1 \over T_{\rm C}} + {\eps_2 \over T_{\rm 0}}\ .
\end{align}
Using section~\ref{Sect:rule-Carnot}'s rule, we conclude that the system can achieve Carnot efficiency 
if we tune $\eps_2$ such that $\Delta\mathscr{S}_{\rm res}(\zeta_1)=0$, which occurs when
\begin{eqnarray}
\eps_2 -\eps_1 \ =\   {\eps_1 \ \big(1- T_0\big/T_{\rm H} \big)\over T_0\big/ T_{\rm C} -1}.
\label{Eq:cbh-condition-for-Carnot}
\end{eqnarray}
Since this system is a ``single-loop'' machine, we can use arguments analogous to those in 
section~\ref{Sect:single-loop} to see that
the cooling-by-heating efficiency (as defined above Eq.~(\ref{Eq:eta_cbh-Carnot})), is given by 
\begin{align}
\eta_{\rm cbh}  \ =\  {\eps_2-\eps_1 \over \eps_1}\,.
\end{align}
Thus we immediately see that Eq.~(\ref{Eq:cbh-condition-for-Carnot}) 
does indeed imply the Carnot efficiency in Eq.~(\ref{Eq:eta_cbh-Carnot}).

The method of analysis in section \ref{Sect:single-loop} 
also enables us to calculate all currents; they are given 
in and below Eq.~(\ref{Eq:particle-current-special-case}). 
We see that for any $(\eps_2 -\eps_1)$ with the same sign as $\eps_1$, 
the heat current out of the hot reservoir (reservoir L) 
and cold reservoir (reservoir ph) have the same sign. 
Thus, so long as the hot reservoir is hot enough for the heat current out of that reservoir to be positive,
we know that the system is cooling reservoir ph (the cold reservoir).

Finally, we note that Ref.~\cite{Imry-combined-power+cooling} considered a machine 
coupled to three reservoir, and pointed out that 
one can make it generate power at the same time as it carries out refrigeration.
This is easily seen in the context of the above model, 
where the heat current between the electron reservoirs (L and R) is associated with a
charge current between these reservoirs (again the currents are given in 
and below  Eq.~(\ref{Eq:particle-current-special-case})).
If the machine has $\eps_1 >0$, then cooling by heating is associated with an electron flow from left to right.
Thus, if we raise the electrochemical potential of the right reservoir, the machine will be generating electrical power
at the same time as cooling the cold reservoir of photons.
However, one can see that the  machine's cooling power goes down as its power generation goes up. 
The authors of 
Ref.~\cite{Imry-combined-power+cooling} studied the thermodynamics of such models.
They showed that one can play with the ratio between the two effects to enhance the overall efficiency by tuning the system towards the more efficient process; either power generation (if the cold reservoir is so cold that refrigerating it is inefficient) or refrigeration (if the cold reservoir is close to the ambient temperature).

\subsection{\green{Cold engines and a specific type of Maxwell demon}}

\green{
Instead of having a heat-source as the resource that is used to produce work, one could consider 
a cold-source at temperature $T_{\rm C}$ (below ambient temperature) as the resource.  A rather impractical illustration of this would be if 
one brought a block of ice down from a glacier to somewhere on the equator, so it could be used as a cold-source for power production, (instead of extracting coal and burning it to make a heat-source for power production).
Then what matters is how much work you can get for a given flow of heat into the cold-source,
since that heat flow will eventually deplete the resource (the cold-source).
Hence, a natural definition of efficiency would be power generated, $P_{\rm gen}$ divided by heat current into the cold source, $J_{\rm h,C}$,
\begin{eqnarray}
\eta_{\rm cold\,eng} = {P_{\rm gen} \over J_{\rm h,C}} \ .
\end{eqnarray}
In this case, it is easy to see that the laws of thermodynamics tell us that this efficiency must always be less than
\begin{eqnarray}
\eta^{\rm Carnot}_{\rm cold-eng} = {T_{\rm H} \over T_{\rm C}} -1 \ ,
\end{eqnarray}
where now the ``hot'' reservoir is simply at ambient temperature, which we still call $T_{\rm H}$ because it is 
hotter than the cold source.
This efficiency can be larger than one (much like the coefficient of performance a refrigerator can be greater than one).
Indeed this efficiency goes to infinity in the limit that the cold source's temperature goes to absolute zero. 
This means that the laws of thermodynamics allow a machine to generate work
due to its coupled to a reservoir at absolute zero, even if the heat flow into that reservoir is vanishingly small.
}

\green{
An example of this was considered in Ref.~\cite{Esposito-Maxwell-demon}, which considered a system the same as in our section~\ref{Sect:3-term-sys2} but with a cold reservoir in place of the hot one.
They show that it is a physical implementation of a Maxwell demon. Yet at the same time, it is clear that 
there is no violation of the laws of thermodynamics once one realizes that the implementation of the demon requires a zero-temperature reservoir.  Ref.~\cite{Henriet2015} is a similar work on a different system, which shows that a zero-temperature reservoir (i.e.a reservoir which exhibits vacuum fluctuations but no thermal fluctuations) can break the symmetry between two others reservoirs (left and right) both at a finite temperature, $T_{\rm H}$, 
and thereby cause a net current flow from left to right. 
}

\green{There are, of course, many other types of Maxwell demon, however most of them do not operate in the steady-state 
(unlike \cite{Esposito-Maxwell-demon,Henriet2015}). 
They  are beyond the scope of this review, and we refer the reader to
Refs.~\cite{lloyd97,kieu04,kieu06,quan06,sagawa08,nori09,zhou10,Cottet2017}.
}

%% file: other-steady-state.tex
\section{Other steady-state machines}
\label{Sect:other-steady-state}

\subsection{Heat engine with blowtorch effect}
\label{sec:blowtorch}
\begin{figure}
\begin{center}
\centerline{\includegraphics[width=0.70\columnwidth]{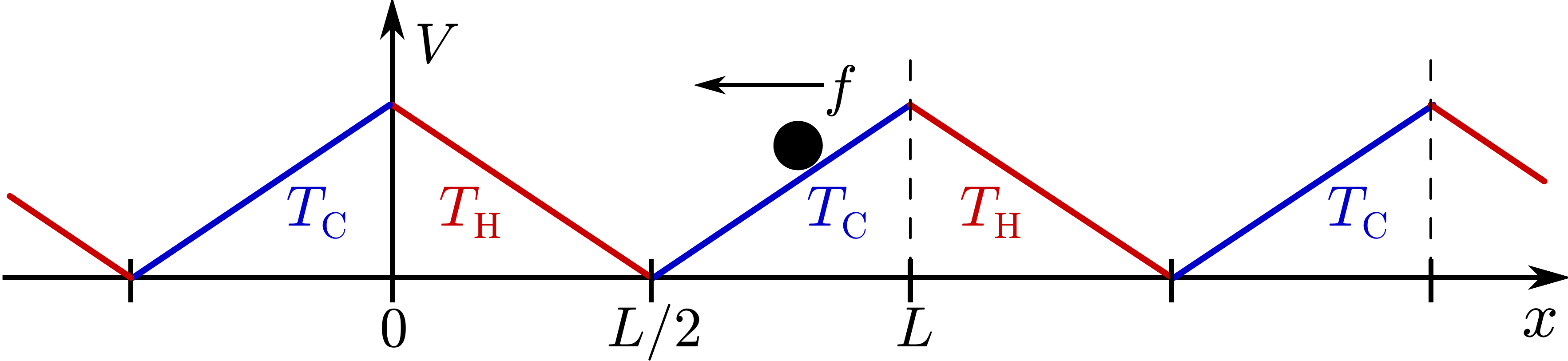}}
\caption{Schematic picture of the 
B\"{u}ttiker-Landauer's heat engine. The temperature is a periodic function of $x$, taking the value 
$T_H$ for $0\le x<L/2$ and $T_C$ for 
$L/2\le x<{L}$}
\label{blowtorch-engine}
\end{center}
\end{figure}

B\"{u}ttiker and Landauer's  motor \cite{buttiker87,landauer88,vankampen} is a rather different example of a heat engine from those treated in the previous chapters. In this example, a particle is trapped in a periodic potential $V(x)$ and subject to a spatially periodic temperature profile. This situation is analyzed using the Langevin dynamics where the particle is alternately in contact, along the spatial coordinate, to thermal baths at different temperatures, see Fig.~\ref{blowtorch-engine}. The equation of motion 
is given by 
\begin{eqnarray}
m \ddot{x} = -\gamma(x) \dot{x} -V'(x)  -  f + \sqrt{2\gamma(x) T(x)} \, \xi (t)  , 
\label{BLdynamics}
\end{eqnarray}
where $\gamma(x)$ is the coefficient of a viscous friction, $f$ is the external force, and $\xi $ is a white Gaussian noise satisfying $\langle \xi (t) \xi(t' )\rangle =\delta (t-t')$. We assume that the potential and temperature depend on the position and are periodic with the period $L$, with the following 
dependence on the position:
\begin{eqnarray}
(T(x), \gamma (x)) =  \left\{
\begin{array}{ll}
(T_H ,   \gamma_{H}), &  0 \le x  < {L\over 2} \, , \\
(T_C ,   \gamma_{C}), &  {L\over 2} \le x < L \, , \\
\end{array}
\right.
\end{eqnarray}
where $T_H>T_C$. A schematic picture for the potential and temperature is presented in Fig.~\ref{blowtorch-engine}.

%%%%%%%

Landauer showed the physical significance of nonuniform temperature in changing the relative 
stability of otherwise locally stable states \cite{landauer88}. He called this phenomenon the
blowtorch effect, since some regions are elevated to higher 
temperatures (the region $0\le x<\frac{L}{2}$ in Fig.~\ref{blowtorch-engine}).
The BL motor has no time-dependent parameters, and hence it is categorized as a steady state engine, 
and may be regarded as an extreme case in the thermoelectric transport where all parts of the system are attached to reservoirs.
 
The energetics and transport properties of the B\"{u}ttiker-Landauer motor have been studied by many authors. 
Periodic temperature with a periodic potential induces a net transport of Brownian particles \cite{buttiker87,landauer88,parrondo02}. In the hot region a Brownian particle can move more easily than in the cold region. Hence, a finite net current is generated. 
The average work \green{generated} by the particle per unit time is 
$\dot{W}=f \langle \dot{x}\rangle$, \green{where the dot indicates a time-derivative, and the average is over all $x$. }
Then the efficiency $\eta =\dot{W}/\dot{Q}_H$, where $\dot{Q}_H$ is the heat supply (per unit time) from the hot region, evaluated as (see \cite{sekimoto}) $\dot{Q}_H = \langle (-\gamma_H \dot{x} + \sqrt{ 2\gamma_H T_H} \xi (t)  ) \dot{x} \rangle$, where the average is taken {\it only} over the hot regions ($0\le x < L/2$, etc).

The overdamped limit for this model is problematic since temperature depends on the position. It has been discussed in the literature that a naive calculation neglecting the inertial term in the Langevin equation (i.e., $\ddot{x}\to 0$ in Eq.~(\ref{BLdynamics})) is not justified \cite{sancho}. The overdamped Langevin equation is instead given by $\gamma (x) \dot{x} = -V' (x) -f + \sqrt{2 T(x) \gamma (x)} \xi (t) - (2 \gamma (x))^{-1}(d/dx)\left[ T(x) \gamma (x) \right]$, and one can derive \cite{sancho} the Fokker-Planck equation
\begin{eqnarray}
{\partial P(x,t) \over \partial t} 
= {\partial \over \partial  x}
\left\{ {1\over \gamma (x) } \left[ V' (x) +  f +{\partial \over \partial x} T(x) \right]
\right\} \, .
\end{eqnarray}
From this equation, one can obtain the net current and show that the efficiency can reach the Carnot efficiency \cite{matsuo.sasa00,asfaw04,asfaw07}. However, it was pointed out that reaching the Carnot efficiency may be problematic due to the abrupt change of temperature at the boundaries between hot and cold regions \cite{astumian99,hs00,ai05,ai06}. Indeed recent first principle calculations using molecular dynamics simulations showed a thermodynamic efficiency much lower than
the Carnot efficiency \cite{benjamin.kawai08}, thus supporting the 
unattainability 
of the Carnot efficiency.

Brownian motors driven by temporal rather than spatial temperature oscillations are discussed in \cite{hanggiPLA}, where the potential has broken spatial symmetry (``ratchet'' potential). The constructive role of Brownian motion for various physical and technological setups is reviewed in \cite{marchesoni}.

\subsection{Photonic heat engines}
\label{sec:photonic}

%%%%%%%
\begin{figure}
\begin{center}
\centerline{\includegraphics[width=0.3\textwidth]{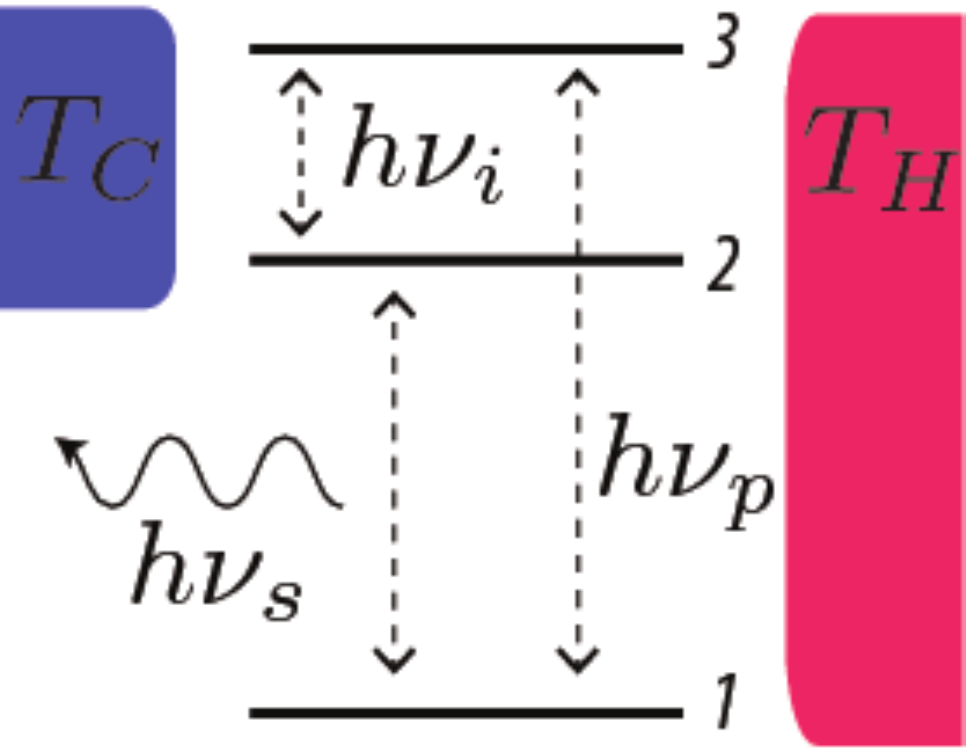}}
\caption{A schematic picture of a three-level maser, which is an ideal model to understand the connection to heat engines.}
\label{3maser}
\end{center}
\end{figure}

Quantum mechanics and thermodynamics have a deep connection, whose investigation started from thermodynamic studies by Planck \cite{planck01} and Einstein \cite{einstein17}. 
The understanding of black-body radiation was a milestone in this context. Nowadays photonic heat engines such as cavity maser systems attract much attention, since quantum effects are anticipated in their working.

Models of lasers and masers can be understood as quantum heat engines in several situations \cite{scovil59}. We follow the idealized model introduced by Scovil and Shulz-DuBois \cite{scovil59} to show the deep connection between the quantum efficiency of the maser and the Carnot cycle. Their system is somewhat similar to the thermoelectric transport in the sense that the model is categorized into the steady state heat engines without time-dependent parameters. We consider an atom with three levels which enters an optical cavity attached to thermal reservoirs,
see the three-level maser depicted in Fig.~\ref{3maser}. The energy gap between states $1$ and $3$ is $h \nu_p$ ($p$ stands for ``pump'') and the transitions between these two states are driven only by the hot reservoir with the temperature $T_H$. The energy gap between states $2$ and $3$ is $h \nu_i$ 
($i$ stands for ``idler'') and the transitions between these states can be induced only by the cold reservoir with the temperature $T_C$. Then state $2$ relaxes to state $1$ by emitting a photon of frequency $\nu_s$ ($\nu_s = \nu_p - \nu_i$, where $s$ stands for ``signal''). For each quantum $h\nu_p$ supplied by the hot
reservoir, an amount of energy equal to $h\nu_i$ goes to the cold reservoir.
Let $n_{i}$ be the population of the $i$-th state. Then, for the maser operation, namely to extract quanta at energy $h \nu_s$, population inversion between $n_1$ and $n_2$ is necessary, i.e., we need $n_2 \gg n_1$. The efficiency of this 
ideal maser setup is defined as the ratio between the extracted energy and 
the energy supplied by the hot reservoir:
\begin{eqnarray}
\eta_M = \nu_s / \nu_p \, .
\end{eqnarray} 
From the Boltzmann factors, we find
\begin{eqnarray}
{n_2 \over n_1} = \exp \left( {h\nu_i \over k_B T_C} - {h \nu_p \over k_B T_H}\right) \, .
\end{eqnarray}
After rearrangement, this becomes
\begin{eqnarray}
{n_2 \over n_1} = \exp \left[ {h\nu_s \over k_B T_C} \left( {\eta_C\over \eta_M} - 1\right) \right] \, .
\end{eqnarray} 
Taking into account the request of population inversion, $n_2 \gg n_1$,
we find the condition for maser action:
\begin{eqnarray}
\eta_M \ll \eta_C \, .
\end{eqnarray}
This shows the deep connection between the maser operation and
thermodynamics.
The Carnot efficiency is obtained at the verge of population inversion, 
namely for $n_2/n_1\to 1$, and in this limit the extracted power vanishes.  

\red{The above description of a three-level maser is 
essentially based on a static quasi-equilibrium viewpoint, where operations
are infinitely slow and the output power vanishes. On the other hand,
engines operate far from the quasi-static limit in order to produce power.
It is therefore necessary to describe finite-time dynamical processes 
and this is possible in simple quantum models where few-level systems are coupled to reservoirs
and the system's dynamics is described via a Markovian (Lindblad) master equation.
These models reproduce generic features of heat engines, in that finite power can
be extracted, but heat leaks to the baths always impose efficiencies smaller than 
the Carnot efficiency, see Ref.~\cite{Kosloff-Levy2014} for a review.} 

Triggered by this pioneering study of Scovil and Shulz-DuBois, a lot of efforts have been devoted to finding quantum effects in photonic quantum heat engines.
%Detailed balance imposed by thermodynamics limits the efficiency 
%of quantum heat engines, including solar cells \cite{shockley61}.
In particular, the role of engineered nonequilibrium 
distributions for the reservoirs 
was investigated \cite{scully02,scully03,scully10,lutz2014,parrondo2016}.
Note that in this case one can overcome the Carnot limit but
this should not be considered surprising since we have nonequilibrium 
distributions for the reservoirs. 
Moreover, the energy cost to engineer 
such distributions should also be taken into account when 
evaluating the overall efficiency of a heat engine.

It is interesting to remark that the photosynthetic reaction center 
has been interpreted as a quantum heat engine \cite{dorfman13}, 
thus suggesting an important intersection between physics and biology.

\subsubsection{\green{Superconductor-based quantum heat-engine and refrigerators}}

\green{
Another promising direction is to consider a Josephson junction coupled to two microwave cavities;
one coupled to a reservoir of hot photons and the other coupled to a reservoir of cold photons.
The temperature difference can be used to perform photon-assisted Cooper pair tunnelling across the Josephson junction against a potential difference (thereby generating electrical work) \cite{Hofer-Clerk2016}.  
Alternatively, one can use the potential difference to drive photon-assisted Cooper-pair tunnelling,
in a manner that extracts photons from the cold photonic reservoir, thereby cooling it down
\cite{Hofer2016b}.  Refs.~\cite{Hofer-Clerk2016,Hofer2016b} used $P(E)$-theory \cite{Ingold-Nazarov} and numerical modelling in the rotating wave approximation
to predict that such a device can reach Carnot efficiency (assuming no flow of photons directly from the hot reservoir to the cold one), but explicitly shows how
the power output vanishes as this efficiency is approached.   
}

%% file: engines.tex
\section{Cyclic thermal machines}

\label{sec:CTM}

\red{So far we have considered steady state (or \emph{autonomous}) heat engines, 
where no time-dependent parameters are involved. 
On the other hand, thermal machines usually discussed in thermodynamics textbooks, such as Carnot and
Otto engines, involve time-dependent parameters for controlling volume, temperature, and so on.
These engines are \emph{nonautonomous}, since they require an external control system.
In this chapter, we will consider cyclic thermal machines, where all parameters return to their original 
position in one period, drawing a cycle in the parameter space. 
We shall discuss general features of power and efficiency in cyclic heat engines,
highlighting the (dis)similarities between steady-state and cyclic heat engines.}

\subsection{Finite-time thermodynamics}
\label{sec:CA}

In the same way as for a steady-state engine,
in a \emph{cyclic thermal machine} operating between 
two (hot and cold) reservoirs at temperatures $T_H$ and $T_C$ $(T_H>T_C)$,
the efficiency $\eta$, defined as the ratio  of the output work
$W$ over the heat $Q_H$ extracted from the hot reservoir,
is bounded by the Carnot efficiency $\eta_C$; so
$\eta = \frac{W}{Q_H} \leq \eta_C = 1-\frac{T_C}{T_H}$.
\red{The Carnot engine achieves the Carnot efficiency for 
a quasi-static transformation which requires infinite time for one cycle
and therefore the extracted power, in this limit, reduces to zero.
Moreover, the total entropy generated in the system plus reservoir is zero, 
hence the process is reversible. The total entropy production per cycle
is sometimes referred to as \emph{dissipation} in heat engines. 
To get finite power, one needs finite-time cycles. As a consequence, 
there is dissipation, and the efficiency is reduced below the Carnot limit. 
It is the purpose of 
finite-time thermodynamics \cite{andresen11} 
to investigate the efficiency as well as performance bounds 
on finite-time, irreversible thermodynamic processes,
addressing the trade-off between efficiency and power.}

In particular, in \emph{endoreversible} thermodynamics
\cite{rubin79,hoffmann97} dissipation is introduced by 
considering finite thermal conductances between heat reservoirs
and the ideal heat engine, namely the engine has no internal dissipation. 
In contrast, in \emph{exoreversible} engines 
no dissipative thermal contacts are 
involved, and irreversibility only arises
due to internal processes. For instance, 
in thermoelectricity the Joule effect is a dissipative internal process.

In this section, we shall discuss in detail the Curzon-Ahlborn endoreversible engine
and compare its efficiency at maximum power with the result obtained for the
Schmiedl-Seifert exoreversible engine (whose detailed discussion will 
be postponed to section~\ref{sec:stochasticSS}, after introduction of 
the necessary tools of stochastic thermodynamics). Both the Curzon-Ahlborn \green{(CA)} and 
the Schmiedl-Seifert \green{(SS)} efficiency at maximum power, $\eta_{CA}$ and
$\eta_{SS}$, feature in 
a model of low-dissipation engines, which highlights the relevance of
asymmetric coupling to the reservoirs. Moreover, the crossover between 
$\eta_{CA}$ and $\eta_{SS}$ can be seen in a model of a thermoelectric device, 
in which dissipation is dominated  
either by thermal contacts with the reservoirs (endoreversible behavior) 
or by internal thermal dissipation (exoreversible behavior).
Finally, we shall briefly discuss the extension of linear response 
formalism for coupled charge and heat flows to driven systems.

\subsection{Endoreversible cyclic engines}
\label{Sect:endoreversible-cyclic}

\begin{figure}
\begin{center}
\centerline{\includegraphics[width=0.55\columnwidth]{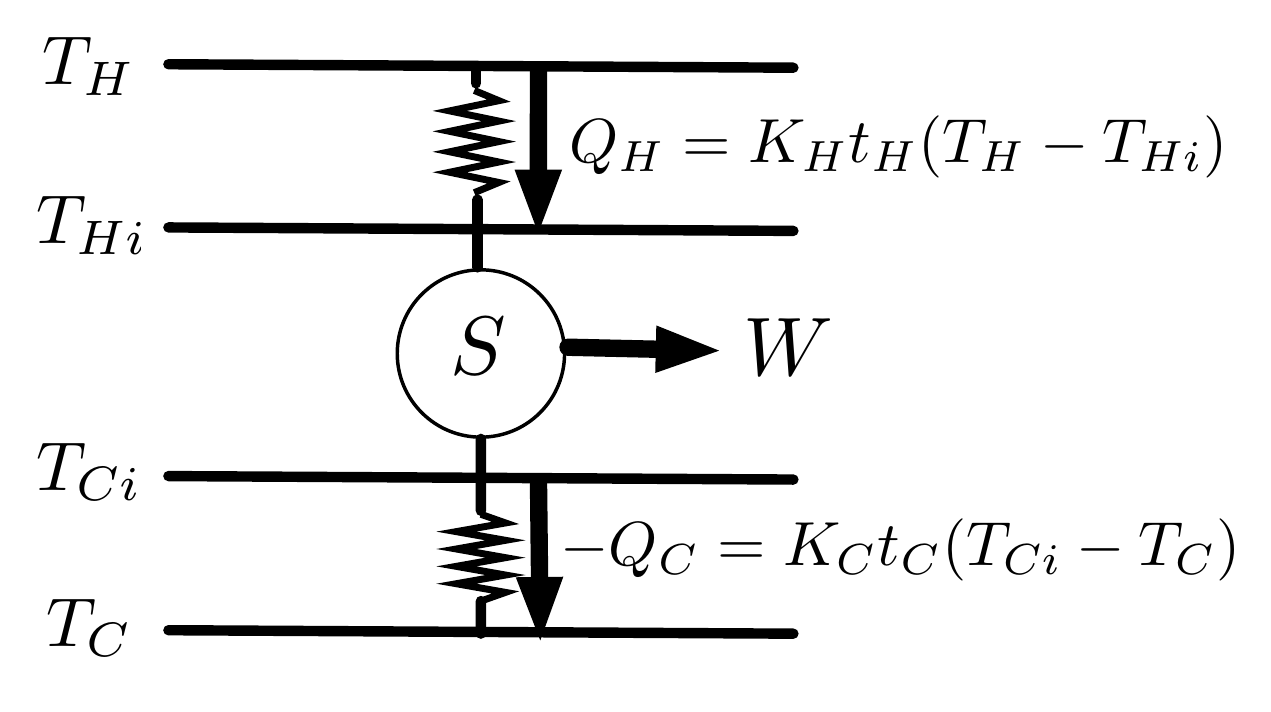}}
\caption{Schematic drawing of the endoreversible engine for
the Curzon-Ahlborn cycle. The two heat baths at temperatures
$T_H$ and $T_C$ are coupled for times $t_H$ and $t_C$ 
to the system $S$ (the working fluid, with output work 
per cycle equal to $W$) by heat conductances 
$K_H$ and $K_C$. The system $S$ is considered as a Carnot 
engine operating between the internal temperatures 
$T_{Hi}$ and $T_{Ci}$ ($T_H>T_{Hi}>T_{Ci}>T_C$).}
\label{fig:curzon}
\end{center}
\end{figure}

The very important concept of efficiency at
maximum power can be conveniently illustrated by means of 
the endoreversible 
{cyclic} Curzon-Ahlborn (CA) engine depicted in 
Fig.~\ref{fig:curzon}.
The Curzon-Ahlborn engine consists of two heat baths at temperatures $T_H$ and 
$T_C$ and a reversible Carnot engine operating between internal temperatures
$T_{Hi}$ and $T_{Ci}$ ($T_H>T_{Hi}>T_{Ci}>T_C$). 
The two processes of heat transfer, from the 
hot reservoir to the system and from the system to the cold 
reservoir, are the only irreversible processes in the Curzon-Ahlborn engine. 
The output work $W$ is the difference between the heat $Q_H$ absorbed
from the hot reservoir and the heat $-Q_C$ ($Q_H>0$, $Q_C<0$) 
evacuated to the cold 
reservoir ($W=Q_H+Q_C$). 
Heat transfers take place during the isothermal strokes
of the Carnot cycle, with the working fluid (the system) at 
internal temperatures
$T_{Hi}$ and $T_{Ci}$. We further assume that 
the rate of heat flow $\dot{Q}_H$  is 
proportional to the temperature difference 
\green{
$T_H-T_{Hi}$ between the hot  reservoir
and the working fluid, and the heat flow $\dot{Q}_C$ is proportional 
to the temperature difference $T_C-T_{Ci}$ between the cold reservoir 
and the working fluid.} Therefore, we need a time $t_H$
to transfer an amount $Q_H$  of heat out of the hot reservoir, so that
\begin{equation}
Q_H= K_H t_H (T_H-T_{Hi}),
\end{equation}
\green{for thermal conductance $K_H$ between the working fluid and the hot reservoir, and
a time $t_C$
to transfer an amount $-Q_C$ of heat into the cold reservoir,} 
\begin{equation}
-Q_C= K_C t_C (T_{Ci}-T_C).
\end{equation}
\green{for thermal conductance $K_C$ between the working fluid and the cold reservoir.}
Finally, we assume that the time spent in the adiabatic strokes of 
the Carnot cycle is negligible compared to the times of the 
isothermal strokes, so that the total time of the cycle is approximately
given by $t=t_H+t_C$. Such \green{an} assumption is justified if the 
relaxation time for the working fluid is \green{short enough that one can operate the adiabatic transformations
as fast as one wishes} \footnote{\label{Footnote:adiabatic} \green{Note that here we use {\it adiabatic} in the thermodynamic sense of the word (a transformation which does not change the working fluid's entropy) rather than in the quantum sense. 
Thus an adiabatic transformation must be slow on the scale of the relaxation rate of
the working fluid.  However, in principle, this relaxation can be arbitrarily fast, and thus 
the adiabatic transformation can also be made arbitrarily fast.}}. 
The output power reads
\begin{equation}
P_{\rm gen}=\frac{W}{t}=\frac{Q_H+Q_C}{t}=\frac{K_Ht_H(T_H-T_{Hi})+
K_Ct_C(T_C-T_{Ci})}{t_H+t_C}.
\end{equation} 
Taking into account that the internal Carnot engine operating between
temperatures $T_{Hi}$ and $T_{Ci}$ has efficiency 
$\eta_{Ci}=1-T_{Ci}/T_{Hi}=1+Q_C/Q_H$ and using the 
relations $Q_H+Q_C=W$ and $t_j=Q_j/[K_j(T_j-T_{ji})]$,
$(j=H,C)$, we can express the power as
\begin{equation}
P_{\rm gen}=\frac{K_H K_C \alpha\beta(T_H-T_C-\alpha-\beta)}{
K_H \alpha T_C + K_C \beta T_H +\alpha\beta(K_H-K_C)},
\label{Eq:P_gen-finite-time-thermodyn}
\end{equation}
where we have defined $\alpha=(T_H-T_{Hi})$ and
$\beta=(T_{Ci}-T_C)$. 
If the working fluid is at the same temperature as 
the reservoir \green{that it is in contact with, then one either has $\alpha=0$ (i.e.~$T_{Hi}=T_H$) 
or  $\beta=0$ (i.e.~$T_{Ci}=T_C$) or both; in all these cases we see from 
Eq.~(\ref{Eq:P_gen-finite-time-thermodyn}) that $P_{\rm gen}$ vanishes. 
Physically, $\alpha=0$ corresponds to the case where the working fluid is at the same temperature 
as the hot reservoir during all the time that they are in contact with each other, and as a result the heat current from the hot reservoir into the working fluid is vanishingly small, 
thus the power generated must be vanishingly small, not matter how efficient the machine is.}

On the other hand, 
\green{if we maximize  the heat flow by maximizing $\alpha$ and $\beta$, we end up taking  $T_{Hi} \to T_{Ci}$ and $(T_H-T_C-\alpha-\beta) \to 0$ in the numerator of Eq.~(\ref{Eq:P_gen-finite-time-thermodyn}). 
Physically, this is the limit where the working fluid is performing a vanishingly small cycle in temperature, which is why
the power generated is again vanishingly small.  Maximum power is clearly between these two extremes.}

By maximizing the power with respect to 
the internal temperatures $T_{Hi}$ and $T_{Ci}$ we obtain
the optimum values 
\begin{equation}
T_{Hi}=c\sqrt{T_H},\;\;T_{Ci}=c\sqrt{T_C},
\quad c\equiv \frac{\sqrt{K_H T_H}+\sqrt{K_C T_C}}{\sqrt{K_H}+\sqrt{K_C}}.
\label{eq:optimumTi}
\end{equation}
These internal temperatures correspond to the maximum power delivered 
by the engine:
\begin{equation}
P_{\rm max}=K_H K_C\left(
\frac{\sqrt{T_H}-\sqrt{T_C}}{\sqrt{K_H}+\sqrt{K_C}}\right)^2.
\end{equation}
From the energy balance $Q_H+Q_C=W$ and from the condition 
$Q_C/Q_H=-T_{Ci}/T_{Hi}$ for the internal Carnot cycle, we obtain 
\begin{equation}
Q_H=\frac{T_{Hi}}{T_{Hi}-T_{Ci}}\,W,
\quad
Q_C=-\frac{T_{Ci}}{T_{Hi}-T_{Ci}}\,W,
\end{equation}
so that the efficiency of the Curzon-Ahlborn engine can be written as 
\begin{equation}
\eta=\frac{Q_H+Q_C}{Q_H}=1-\frac{T_{Hi}}{T_{Ci}}.
\end{equation}
Using the values of $T_{Hi}$ and $T_{Ci}$ from 
(\ref{eq:optimumTi}), we obtain
the efficiency at the maximum power $P_{\rm max}$,
\begin{equation}
\eta_{CA} = 1-\sqrt{\frac{T_H}{T_C}}=1-\sqrt{1-\eta_C}. \label{eq:ca.eff} \ .
\end{equation}
This efficiency is commonly referred to
as the \emph{Curzon-Ahlborn efficiency} \cite{curzon}, even if it already appeared in earlier works  
\cite{Yvon,chambadal,novikov}:
Remarkably, the Curzon-Ahlborn efficiency is independent of the heat conductances
$K_H$ and $K_C$.  

It is interesting to remark that the Curzon-Ahlborn efficiency is
invariant under concatenation \cite{vandenbroeck2005}. 
We consider two thermal machines working in a tandem,
the first one between the hot source at temperature 
$T_H$ and a second heat bath at intermediate temperature $T_i$, the second
one between this latter bath and the cold source at temperature $T_C$. 
The first machine absorbs heat $Q_H$, delivers work 
$W'$ and evacuates heat $|Q_i|=Q_H-W'$, the second machine reuses 
heat $|Q_i|$ and outputs work $W''$.
If both machines function at the Curzon-Ahlborn (CA) efficiency, also the 
overall efficiency $(W'+W'')/Q_H$ is given by 
$\eta_{CA}=1-\sqrt{T_C/T_H}$.  This is the self-concatenation property
as that which is well known for machines working at Carnot efficiency.

The Curzon-Ahlborn efficiency was derived, as described above, for a specific engine. 
It turns out not to be an upper bound for the 
efficiency at maximum power, as shown in several models, see e.g.
\cite{mahler08,izumida_okuda08,izumida_okuda091,izumida_okuda092,ss08,elb09,sb2012}.
Yet $\eta_{CA}$ describes the efficiency of
actual thermal plants reasonably well \cite{curzon,esposito2010}, 
and therefore it 
%the range of validity of
%$\eta_{CA}$ as upper bound 
%for the efficiency at maximum power 
has been widely discussed in the literature, see e.g. 
\cite{vandenbroeck2005,esposito2009,schulman,esposito2010,
linke2010,mahler10,seifert2011,goupil2012a,goupil2012b,shakouri2012} and
\cite{tu12} for a review.
\red{Moreover, the Curzon-Ahlborn efficiency was also derived in models
different from the one proposed by Curzon and Ahlborn, 
for instance for a quantum system (two interacting oscillators) 
coupled to reservoirs and
with the power extracted by an external periodic driving force
\cite{kosloff84} or in an ensemble of quantum oscillators operating 
in an Otto cycle \cite{rezek06}.}
\red{The linear (in $\eta_C$) expansion of the Curzon-Ahlborn bound 
(\ref{eq:ca.eff}), $\eta_{CA}^{(1)}=\eta_C/2$, coincides with the 
exact and universal upper bound for the efficiency at maximum power 
for steady-state systems with 
(i) time-reversal symmetry and (ii) within a regime of linear
response (see section~\ref{sec:ZTmaxpower}).
In the presence of left-right
symmetry in the system (e.g., in thermoelectrics 
the switching of the temperatures
$T_H$ and $T_C$ and of the 
electrochemical potentials $\mu_H$ and $\mu_C$
leads to an inversion of the currents), 
a universal upper bound 
up to quadratic order in the deviation from equilibrium
was derived in Ref.~\cite{esposito2009}.
The obtained result agrees with the expansion of 
$\eta_{CA}$ up to second order in $\eta_C$,
\begin{equation}
\eta_{CA}^{(2)}=
\frac{\eta_C}{2}+\frac{\eta_C^2}{8}.
\label{eq:etaCA2}
\end{equation} 
}
\subsection{Exoreversible cyclic engines}

\label{sec:exoreversible}

Another expression for the efficiency at maximum power was obtained 
by Schmiedl and Seifert
\cite{ss08}, using a model of stochastic cyclic heat engine which
we shall describe in section~\ref{sec:stochastic}. 
This machine is exoreversible,
in that dissipation is fully internal, and the efficiency at maximum power 
is given by
\begin{equation}
\eta_{SS}=\frac{\eta_C}{2-\gamma\eta_C},
\label{eq:etaSS}
\end{equation}
where $\gamma\in[0,1]$ is a parameter related to the ratio of
entropy production during the hot and cold isothermal steps 
of the machine. 
For the symmetric case $\gamma=1/2$. This is for instance the case of
thermoelectricity when internal dissipation is due to Joule heating,
and each end of a thermoelectric device receives half of the produced heat.
It is interesting to remark that for the symmetric case 
$\eta_{CA}$ and $\eta_{SS}$ agree up to second order in $\eta_C$.
%thus confirming that expression (\ref{eq:etaCA2}) is universal in the 
%presence of left-right symmetry.

\subsection{Low-dissipation engines}

The Curzon-Ahlborn efficiency was also derived for the Carnot cycle in the limit
of low and symmetric dissipation 
%An extension of the above results to the class of low-dissipation heat
%engines was reported 
by Esposito \emph{et al.} \cite{esposito2010}.
They considered a Carnot engine which operates under reversible
conditions at the Carnot efficiency when the cycle duration 
becomes infinitely long. In that limit, the system entropy 
increase $\Delta \mathscr{S}=Q_H/T_H$
during the isothermal transformation at the hot temperature
$T_H$ is equal to the system entropy decrease 
$-\Delta \mathscr{S}=Q_C/T_C$ during
the isothermal transformation at the cold temperature $T_C$. 
Hence, there is no overall entropy production 
and the Carnot efficiency $\eta_C=1+Q_C/Q_H=1-T_C/T_H$ is achieved.
Ref.~\cite{esposito2010} considers the weak dissipation regime and 
assumes that the system relaxation is much faster than the times 
$t_H$ and $t_C$ spent in the isothermal strokes, so that the 
overall cycle duration is to a good approximation given by $t_H+t_C$.  
In the low dissipation regime the entropy production is proportional
to $1/t_H$ and $1/t_C$, so that it vanishes in the limit of 
infinite-time cycle where it is supposed that the Carnot efficiency
is recovered. Therefore the amount of heat entering the system 
from the hot (cold) reservoir is, to first order in $1/t_H$ and
$1/t_C$,
\begin{equation}
Q_H=T_H\left(\Delta \mathscr{S}-\frac{\Sigma_H}{t_H}\right),
\quad
Q_C=T_C\left(-\Delta \mathscr{S}-\frac{\Sigma_C}{t_C}\right),
\end{equation}
with $\Sigma_H$ and $\Sigma_C$ coefficients depending on the 
specific implementation. 
The maximum of the output power 
\begin{equation}
P=
%\frac{W}{t_H+t_C}=
\frac{Q_H+Q_C}{t_H+t_C}=
\frac{(T_H-T_C)\Delta \mathscr{S} - T_H\Sigma_H/t_H-
T_C\Sigma_C/t_C}{t_H+t_C}
\end{equation}
is obtained when $\partial P/\partial t_H=\partial P/\partial t_C=0$.
This leads to the efficiency at the maximum output power
\begin{equation}
\eta(P_{\rm max})=
\frac{\eta_C\left(1+\sqrt{\frac{T_C\Sigma_C}{T_H\Sigma_H}}\right)}{\left(1
+\sqrt{\frac{T_C\Sigma_C}{T_H\Sigma_H}}\right)^2+\frac{T_C}{T_H}
\left(1-\frac{\Sigma_C}{\Sigma_H}\right)}.
\label{eq:etas1s2}
\end{equation}
Note that this result was also obtained in the context 
of stochastic thermodynamics by Ref.~\cite{ss08}.
The Curzon-Ahlborn efficiency is recovered for symmetric dissipation,
$\Sigma_H=\Sigma_C$. From (\ref{eq:etas1s2}) we obtain
\begin{equation}
\eta_-=\frac{\eta_C}{2}\le \eta(P_{\rm max}) \le 
\eta_+=\frac{\eta_C}{2-\eta_C},
\label{eq:uppermax}
\end{equation}
with the lower and upper bounds reached in the limits of 
completely asymmetric dissipation, for  
$\Sigma_C/\Sigma_H\to \infty$ and 
$\Sigma_C/\Sigma_H\to 0$, respectively.
\red{Interestingly, the upper bound is obtained when 
dissipation takes place in the hot reservoir. This is an
intuitive result, since heat dissipated to the hot reservoir
can be reused to fuel the heat engine.}
The lower and upper bound coincide in the linear response
regime where $\eta_-=\eta_+=\eta_{CA}=\eta_{SS}=\eta_C/2$ \footnote{Note that the same upper bound as in (\ref{eq:uppermax}) 
was obtained with a different approach by \cite{schulman}.}.
\red{
We note that features of the efficiency at maximum power similar to 
those above discussed for low-dissipation engines are also found 
in a quantum model where the working substance is a single multilevel
particle which undergoes an Otto cycle \cite{Uzdin2014}.}

\subsection{Crossover from endoreversible to exoreversible regime}

The crossover between the endoreversible and the exoreversible regime
was illustrated by Apertet \emph{et al.} \cite{goupil2012a}
in the model of a thermoelectric device. In the ideal case
of no heat leak (open-circuit thermal conductance $K=0$), two 
irreversible sources were taken into account: the internal one
(Joule heating) and the external one (dissipative thermal coupling
to reservoirs). In particular, for the symmetric case (equal thermal
contact conductances, $K_H=K_C$), the Curzon-Ahlborn efficiency $\eta_{CA}$ 
(up to third order in $\eta_C$) is obtained in the endoreversible limit,
when dissipation is dominated by thermal contacts, and the Schmiedl-Seifert
efficiency $\eta_{SS}$ in the exoreversible limit, when dissipation is fully internal.

\subsection{Thermoelectricity for driven systems}
\label{sec:thermodriven}

The linear response formalism can be extended to systems subjected
to a time-dependent driving force $F(t)$, which is applied 
starting from time $t_0$ \cite{hanggi2015}.
In this case, we do not have a steady state but the charge and heat currents
depend on time and are functions of the entire history of the 
applied force, i.e. $J_{e}(t)$ and $J_{h}(t)$ depend on $F(t')$, for all
$t'\in [t_0,t]$.
By linearly expanding the currents at each instant of time
we have
\begin{eqnarray}
\left\{
\begin{array}{l}
J_{e}(t)=\left. J_{e}(t)\right|_{\mathcal{F}_e=0,\mathcal{F}_h=0} 
+ \left( \frac{\partial J_{e} (t)}{\partial  \mathcal{F}_e} 
\right)_{\mathcal{F}_h=0} \mathcal{F}_e
+ \left( \frac{\partial J_{e} (t)}{\partial  \mathcal{F}_h} 
\right)_{\mathcal{F}_e=0} \mathcal{F}_h
\equiv J_{e}^D(t) + L_{ee}[F]\, \mathcal{F}_e + L_{eh}[F]\, \mathcal{F}_h,
\\
\\
J_{h}(t)= \left. J_{h}(t)\right|_{\mathcal{F}_e=0,\mathcal{F}_h=0}
+ \left( \frac{\partial J_{h} (t)}{\partial  \mathcal{F}_e} 
\right)_{\mathcal{F}_h=0} \mathcal{F}_e
+ \left( \frac{\partial J_{h} (t)}{\partial  \mathcal{F}_h} 
\right)_{\mathcal{F}_e=0} \mathcal{F}_h
\equiv J_{h}^D(t) + L_{he}[F]\, \mathcal{F}_e + L_{hh}[F]\, \mathcal{F}_h.
\end{array}
\right.
\label{eq:coupleddriving}
\end{eqnarray}
Here, the coefficients $L_{ab}[F]$ are functionals 
depending on the whole history of the applied force via 
the currents $J_{e}[F]$ and $J_{h}[F]$, and the currents 
$J_{e}^D$ and $J_{h}^D$ at zero thermodynamic forces
are known as \emph{displacement currents}. Note that the 
displacement currents must vanish for an undriven system, 
since we cannot have non-zero steady currents at zero bias
($\mathcal{F}_e=0,\mathcal{F}_h=0$). For undriven systems,
we recover the usual coupled linear transport equations
(\ref{eq:coupledlinear}), with time-independent Onsager 
coefficients $L_{ab}$.

The driving force leads to interesting 
consequences \cite{hanggi2015}. 
The thermodynamic constraints on 
the steady-state Onsager coefficients can be relaxed, i.e. we can have
$L_{eh}[F]\ne L_{he}[F]$ and $\det {\bm L}[F] <0$. 
In such a situation one can have a significant
enhancement of the thermoelectric conversion efficiency, as
shown in a few examples discussed in Ref.~\cite{hanggi2015}. 
%It should be stressed that the overall 
%time-dependent energy conversion efficiency $\eta(t)$ 
%of a driven system must include also the input power $\dot{W}_{in}(t)$,
%with $W_{in}(t)$ work performed
%by the driving force on the system. 
%For thermoelectric power generation,  
%\begin{equation}
%\eta(t)=\frac{\dot{W}(t)}{J_{h}(t)+\dot{W}_{in}(t)},
%\end{equation}
%with $\dot{W}(t)$ output power. 
It should be stressed that the overall
energy conversion efficiency
of a driven system must take into account
as a cost also the input power 
by the driving force. 
($W_{in}(t)$ is work performed
by the driving force on the system).

The thermoelectric analysis of driven systems takes a simple 
and appealing form in the case of adiabatic \emph{ac} 
driving \cite{arrachea2015}. 
In this case, after averaging over one period of the driving, 
one can express the entropy production rate as 
\begin{equation}
\overline{\dot{\mathscr{S}}}=
\overline{J_{e}} \mathcal{F}_e + \overline{J_{h}} \mathcal{F}_h
+\overline{J_\omega} \mathcal{F}_\omega,
\end{equation}
where the overbar denotes time averaging, the average current
$\overline{J_\omega}\equiv \overline{\dot{W}_i}/\hbar\omega$,
with the associated thermodynamic force $\mathcal{F}_\omega=
\hbar\omega/T$, \red{$W_{i}$ and $\omega$ being the work performed
by the driving force on the system and the frequency of the driving,
respectively}. 
The linear response relations between fluxes and thermodynamic
forces then read $\overline{J_a}=\sum_b L_{ab} \mathcal{F}_b$
($a,b=e,h,\omega$),
with the coefficients $L_{ab}$ that satisfy Onsager 
reciprocity relations \cite{arrachea2015}. This theoretical 
framework was applied to quantum motors, quantum generators,
heat engines and heat pumps \cite{arrachea2015}.
\red{For a recent review on energy and heat flow in mesoscopic systems
subjected to periodic driving, see Ref.~\cite{Arrachea_2016}.
}

\subsection{Quantum Carnot engine in the quasi-static limit}
\label{sec:quantum-Carnot-engine}

\red{
\green{The} quantum nature \green{of cyclic} heat engines can emerge from the discreteness of eigenenergies and 
quantum coherence in the dynamics. In this section, we briefly 
discuss effects from the discreteness 
of the eigenenergies, while aspects related to quantum coherence will
be mentioned in section \ref{sec:cycliccoherent}. 
Following Ref.~\cite{quan07}, we clarify the meaning of \green{the}
(quasi-static) isothermal, 
isochoric, and adiabatic processes in quantum engines, and we compare such processes with 
their classical counterpart. These clarifications are crucial when considering 
a quantum mechanical system (the working substance) 
\green{performing} cycles such as Carnot or Otto cycle,
in other words a \emph{quantum Carnot engine} (see e.g. Refs.~\cite{geva92,quan07,eklv10})
or a \emph{quantum Otto engine} 
(see e.g. Refs.~\cite{kosloff96,quan05,rezek06,HRM07,quan07,mahler08}).
\green{We compare these quantum engines with classical ones in which the working substance is}
a classical ideal gas confined within a finite volume.
}

\red{
Suppose that the \green{quantum system's} Hamiltonian is given by 
\begin{eqnarray}
\hat{\cal H} = \sum_{n} E_n \,|n \rangle \langle n | \, ,
\end{eqnarray}
where $| n \rangle$ is the $n$-th eigenstate of the \green{system}, with the 
corresponding \green{eigen-energy $E_n$, where $E_{n+1} \ge E_n$ and without loss of generality 
we can set $E_0=0$}. 
\green{Let us consider a system without coherences, so its density matrix is diagonal in the energy eigenbasis,
then its state is given by the} 
occupation distribution, with probability 
$P_n$ for the $n$-th eigenstate. Then the internal energy $U$ is given by 
$U=\sum_n P_n E_n$, and the entropy $\mathscr{S}$ of the system is given by 
$\mathscr{S}=-\sum_n P_n \ln P_n$.  \green{Note that we can only take this form for the 
entropy, $\mathscr{S}$, because we assume  the system's density matrix is diagonal in the energy eigenbasis.}
We have
$dU=\sum_n(E_ndP_n+P_ndE_n)$ and from the first law of thermodynamics
$d U =  \;\dbar Q - \;\dbar W$,
from which we can identify the heat absorbed from the environment and
the work performed by the system respectively with
\begin{eqnarray}
\;\dbar Q = \sum_n E_n d P_n,\quad 
\;\dbar W = -\sum_{n} P_n d E_n\,. 
\end{eqnarray}
}

\red{
In an isothermal process, 
the system is always in thermodynamic equilibrium with the fixed temperature 
$T$ of the heat reservoir. Then, the density matrix of the system 
at time $t$ is given by the 
canonical distribution
\begin{eqnarray}
\rho_{\rm can} (t) = \frac{1}{Z(t)}\sum_n e^{-E_n(t) / k_B T} 
|n(t)\rangle \langle n(t) | \, , \label{isothermal}
\end{eqnarray}
where $Z(t)=\sum_n e^{-E_n(t) / k_B T}$ is the canonical partition function, 
$|n(t)\rangle$ and $E_n(t)$ are instantaneous system's eigenstates and eigenenergies.
% determined by the control parameters of the engine protocol. 
In the quantum case, temperature is invariant while $U$, $E_n$ and $P_n$ vary,
so that work can be done and heat can be exchanged with the bath
($\;\dbar Q,\;\dbar W\ne 0$). 
In the ideal classical gas, $U$ and $T$ are invariant, while pressure and volume vary.  
}

\red{
A quantum isochoric process has similar properties to those of a classical isochoric process. 
No work is done in this process while heat is exchanged with the heat bath. In the
quantum case, the eigenenergies $E_n$ are invariant while 
the probabilities $P_n$ vary. Hence the entropy 
$\mathscr{S}$ changes until the system reaches thermal equilibrium with the heat bath. 
In a classical isochoric process, pressure and temperature change.
}

\green{
There is a significant difference between a thermodynamically adiabatic process and a quantum adiabatic process. To be thermodynamically adiabatic, a process should not change the system's entropy.  
One way to achieve this is by assuming that the process is adiabatic in the quantum  sense,
in which case the process does not change the distribution of occupation probabilities, 
$d P_n=0$. This implies $\;\dbar Q=0$, while work done 
can still be nonzero.  However, a fast quantum process that interchanges
the occupation probabilities of the levels without changing $\mathscr{S}$ would also be
thermodynamically adiabatic.
}

\red{
Finally, we make a remark on the reversibility condition for the quantum Carnot engine. 
Let us consider the standard four-stroke Carnot cycle in the quasi-static limit,
i.e., 
\begin{itemize}
\item[] (A)$\to$(B) \ = \ isothermal process with temperature $T_H$, 
\item[] (B)$\to$(C) \ = \  quantum adiabatic process, 
\item[] (C)$\to$(D) \ = \ isothermal process with temperature $T_C<T_H$,
\item[] (D)$\to$(A) \ = \ quantum adiabatic process. 
\end{itemize}
From the isothermal property (\ref{isothermal}) 
and quantum adiabatic condition $dP_n=0$, we can readily find  
\begin{eqnarray}
{ P_n (B) \over P_m (B)} = {e^{ - E_n (B) / k_B T_H} \over e^{ - E_m (B) / k_B T_H}} =
{ P_n (C) \over P_m (C)} = {e^{ - E_n (C) / k_B T_C} \over e^{ - E_m (C) / k_B T_C}} \, , 
\end{eqnarray}
where $P_l (X)$ and $E_l (X)$ ($l=m,n$, $X=A,B$) are the occupation probability for the 
$l$-th eigenstate and the $l$-th eigenenergy at point (X), respectively. 
%The above condition must be satisfied for any $n$ and $m$ for achieving the Carnot heat engine. 
From this, the quantum Carnot engine must satisfy
\begin{eqnarray}
E_n (C) - E_m (C) = (T_C / T_H) \left[ E_n (B) - E_m (B) \right] \, .
\end{eqnarray}
}
\red{
This means that \green{to have a reversible (quantum Carnot) heat engine,}
all energy gaps must change by the same ratio in 
\green{the} quantum adiabatic process and this ratio is $T_C / T_H$. 
This is in addition to the condition of
quasi-static transformation.
%even when the protocol is quasi-static limit.
}

\subsection{Cyclic quantum engines and quantum coherence effects}
\label{sec:cycliccoherent}

\red{
To investigate finite-time cyclic quantum heat engines, one has to consider
the dynamics of quantum open systems, coupled to (hot and cold) baths.
Under appropriate assumptions, the dynamics of an open system attached to 
a thermal environment can be analyzed by means of a quantum master equations
of the Lindblad or Redfield type \cite{lindblad,redfield,kubo} discussed in section~\ref{Sect:beyond-master}.
In general such quantum master equation cannot be reduced to classical rate equations 
(as in the analysis of chapters \ref{Sect:Qu-Master-Eqn} and \ref{Sect:master-examples}
for steady-state engines and in the example of section \ref{sec:drivenqd} where a rate
equation for the populations of a quantum dot is used), \green{because} quantum coherent 
superpositions play a role and off-diagonal elements of the system's density matrix 
cannot be neglected. 
}

\red{
We consider the quantum master equation, introduced in the textbooks mentioned in 
section~\ref{Sect:beyond-master}, of the form
\begin{equation}
\frac{d\hat{\rho}}{dt}=-\frac{i}{\hbar}\,[\hat{\cal H},\hat{\rho}] + 
\mathcal{D}(\hat{\rho}), 
\label{eq:masterequation}
\end{equation}
where $\hat{\rho}$ is the density operator describing the state of the 
working substance, governed by the Hamiltonian $\hat{\cal H}$ and coupled to
reservoirs by the dissipator $\mathcal{D}=\mathcal{D}_H+
\mathcal{D}_C$,
where $\mathcal{D}_H$ ($\mathcal{D}_C$) describes the coupling to 
the hot (cold) reservoir. In a generic thermodynamic cycle, 
both $\hat{\cal H}$ and $\mathcal{D}$ depend on time.
Defining the internal energy of the system,
\begin{equation}
U(t)=\langle \hat{\cal H}(t)\rangle = {\rm tr}\left\{\hat{\cal H}(t)\hat{\rho}(t)\right\},
\end{equation}
we take its time-derivative and then use the first law of thermodynamics,
$d U =  \;\dbar Q - \;\dbar W$, 
to obtain \cite{alicki79,kosloff84,GevaKosloff94}
the instantaneous output power
\begin{equation}
P(t)=\dot{W}(t)=
-{\rm tr}\,\left\{ \frac{\partial \hat{\cal H}(t)}{\partial t}\hat{\rho}(t)\right\}
\end{equation}
and the instantaneous heat current absorbed from the environment,
\begin{equation}
\dot{Q}(t)={\rm tr} \,\left\{
\hat{\cal H}(t)  \frac{\partial \hat{\rho}(t)}{\partial t}\right\}.
\end{equation}
It is easy to check from the master equation
(\ref{eq:masterequation}) that
$\dot{Q}(t)=\sum_{k=H,C}\dot{Q}_{k}(t)$,
where
\begin{equation}
\dot{Q}_{k}(t)=
{\rm tr}\left\{ \hat{\cal H}(t) \, \mathcal{D}_{k}\left[\colorproofs{\hat{\rho}(t)}\right]\right\}
\end{equation}
is the instantaneous heat current from bath $k$.
Integrating the power and heat absorbed over one thermodynamic cycle, we
can obtain the output work $W$ and the heat $Q_H$ extracted from the
hot reservoir per cycle, and finally the efficiency   $\eta=W/Q_H$. 
We note that this master equation approach reproduces the Carnot inequality for the efficiency
of any heat engine, $\eta\le\eta_C$ \cite{Spohn78,alicki79}.
}

\red{
The dynamics of many quantum heat engines was analyzed by means of such
quantum master equations.
For instance, Kosloff \cite{kosloff84} showed that two coupled oscillators interacting with hot and cold quantum reservoirs exhibit Curzon-Ahlborn efficiency in the limit of weak coupling. The quantum master equation approach was applied to analyze the performance of heat engines working with spins \cite{geva92,he02,chen02}, harmonic oscillators \cite{lin03-1,amaud02, lin03-2, lin03-3}, and multi-level systems \cite{GevaKosloff94,GevaKosloff96}. Characteristics of the steady state achieved by the iteration of cyclic processes and the monotonic approach to the limit cycle were discussed making use of the quantum conditional entropy \cite{feldmann04}. The unavoidable irreversible loss of power in a heat engine was considered for harmonic systems in the framework of quantum master equation approach \cite{rezek06}.
}

\red{
Other systems studied include quantum heat engines or heat pumps with the working fluid composed 
of non-interacting two-level systems \cite{feldmann00}. 
\green{A refrigerator made of a  pair of periodically driven quantum dots, which cooled an electronic reservoir, was considered in Refs.~\cite{Juergens2013}}.
The concept of ideal quantum heat engine was introduced in cold bosonic atoms confined to a double well potential where thermalization occurs, and the operation of a heat engine with a finite quantum heat bath was proposed \cite{fialko12}. A thermoelectric heat engine with ultracold fermionic atoms was demonstrated, both theoretically and experimentally \cite{Brantut-Grenier-et-al2013}.
}
\red{
Concepts from quantum information theory also provide new insights into the working of quantum heat engines, see e.g. \cite{lloyd97,kieu04,kieu06,quan06,quan07,sagawa08,nori09,zhou10,Cottet2017}, in particular the reviews~\cite{nori09,Vinjanampathy2016,Millen2016,Goold2016}.
}

\red{
Finding signatures and understanding the relevance of quantum coherent superpositions,
quantum correlations and entanglement in heat engines is an intriguing subject. 
However, a complete picture has yet to emerge. 
Quantum thermoelectrics and quantum photonic heat engines
are relevant examples of steady-state engines for exploring this direction. 
In cyclic engines, there exist 
several time-dependent protocols with noncommutability of the Hamiltonian at different times. 
%the external control Hamiltonian does not commute 
%with the internal Hamiltonian. 
%In general, in a cycle a stroke (i.e., an operation 
%that takes place in a certain time segment) does not commute with
%strokes that take place before or after it. 
Purely quantum effects in quantum heat engine were discussed 
in Refs.~\cite{Uzdin2015,uzdin16}:
stationary, two- and four-stroke quantum Otto engines perform equivalently
%in spite of the noncommutability of the time evolution operators, 
when the operator norm of the time-evolution operator is much smaller than the Planck 
constant. This becomes possible for few-level quantum systems.
% and the classical case does not 
%have this properties. 
In Ref. \cite{rezek06}, irreversible loss of power in a heat engine was considered 
for harmonic systems in the framework of the quantum master equation approach and 
the origin of friction was traced back to the noncommutability 
of the kinetic and potential energy of the working substance.
A primary objective in the investigation of quantum heat engines
is to find conditions under which the engine performance can be enhanced 
by quantum mechanical effects.
For instance, it was discussed in Ref.~\cite{scully11}
that radiatively induced quantum coherence in a photonic heat engine can break 
detailed balance and yields lasing without inversion. As a result, one gets more power output.
Work can be significantly boosted also by constructing quantum 
heat engines with collective behavior \cite{uzdin16b}. 
}

%Effects of multilevel systems were investigated \cite{quan05}. 
%A class of quantum heat engines consisting of two subsystems interacting with a work source was studied to maximize the extracted work under various constraints \cite{mahler08}. 

\section{Stochastic heat engines}
\label{sec:machines}
%Since the pioneering work by Sadi Carnot \cite{carnot}, many physical phenomena have been recognized as heat to work conversion. Heat to work conversion can mainly be classified into two categories; the steady state engine such as thermoelectric transport which is the central topic in this review and the cyclic heat engines which contains isothermal and adiabatic processes. These two are different in the sense that the former engines have no time-dependent parameters unlike the latter. It is intriguing to discuss (dis)similarities between them. To this end, we here look mainly at stochastic cyclic heat engines. 

%The Carnot efficiency is achievable at the reversibility condition, which implies quasi-static process where asymptotically vanishing power output is generated. From the practical viewpoint, finite power with high efficiency is desired. This consideration opens the way to the concept of finite-time thermodynamics. Paradigms of finite-time thermodynamic engine are the Carnot or Otto cycles \cite{callen} with a finite time period. 

Recent technological developments allow the realization of finite-time 
thermodynamic devices with high controllability.
In particular, a number of 
stochastic cyclic heat engines have been fabricated of small size 
%and those are scrutinized %under a microscope 
%by virtue of precise measurements 
\cite{steeneken2010,bb2012,ribezzi2015,rosnagel2015,koski2014}. 
 In such cases, the system is not considered to be quantum, but its small size means that thermal fluctuations are significant, making the conversion of heat to work
into a stochastic process.
This is the situation for which stochastic thermodynamics \cite{seifert} was developed.

Section~\ref{Sect:traj} presented stochastic thermodynamics in the context of master equations for quantum system with
a discrete set of states.  Here we will consider it in the context of a systems with 
continuous degrees of freedom, such as those described by a  Langevin equation. 
The fact that the Langevin equation is more complicated than the rate equations discussed in 
section~\ref{Sect:Qu-Master-Eqn} means that the mathematics is more complicated here.
In particular, the fact the Langevin equation is for a continuous degree of freedom means that 
one has to cope with sums of infinite numbers of trajectories (as in a path integral),
rather than discrete trajectories on a network.  However, the basic concepts are exactly the same here as 
in section~\ref{Sect:traj}.

In this section, we start with the basic framework to discuss heat to work 
conversion in stochastic processes and introduce several simple 
stochastic heat engine models. 

%\subsection{SIMPLE PUMP MODELS}
\subsection{Stochastic thermodynamics for a Langevin equation}
\label{sec:stochastic}
Stochastic thermodynamics is a framework to study nonequilibrium thermodynamics in small systems like colloids or biomolecules driven out of equilibrium \cite{sekimoto, seifert}. It describes the energetics of the system of interest surrounded by a thermal environment. Let us consider a system consisting of 
a classical particle and suppose that (i) the time scales of the environment 
and the system are sufficiently separated and (ii) the system's dynamics is well described by the Langevin equation. An important example is that of
a colloidal particle trapped by an external potential, 
where the dynamics is described by the overdamped Langevin equation
\begin{eqnarray}
\dot{x}(t) \ =\  \mu F ( x, \lambda (t) )  + \eta(t) \, ,
\end{eqnarray}
where $x(t)$ is the particle's coordinate, $\mu$ is the mobility, and 
$\eta (t)$ is a Langevin thermal noise satisfying $\langle \eta (t ) \eta (t ' )
\rangle = 2D \delta (t-t') $ with
$D$ being the diffusion constant. The function
$F(x, \lambda (t) )$ is a time-dependent force field, which is given by
\begin{eqnarray}
F(x, \lambda (t)) = - \partial_x V(x, \lambda (t) ) + f\,, 
\end{eqnarray}
where $V(x,\lambda(t))$ is a potential which contains a time-dependent control 
parameter $\lambda(t)$, and $f$ is a nonconservative force which cannot be expressed as the gradient of a potential. The nonconservative force $f$ can drive the system into nonequilibrium states when no time-dependent 
potential is applied. 
%This force is experimentally realizable. 
The diffusion constant $D$ and the mobility $\mu$ are related by the Einstein relation $D=k_B T\mu$, where $T$ is the temperature of the medium surrounding the 
particle. For notational simplicity, we set $k_B=1$ in the rest of this section.

The Langevin dynamics allows a thermodynamic interpretation by applying the energy balance to any individual stochastic trajectory:
\begin{equation}
d U =  \;\dbar Q - \;\dbar W \, ,
\end{equation}
where $\;\dbar Q$ is the amount of heat 
absorption from the thermal environment, $d U$ 
the change of internal energy, and $-\;\dbar W$ the work done 
by the time-dependent potential and by the nonconservative force
(we use the convention that the work is positive 
when it is generated by the system). 
On the coarse-grained time scale where the overdamped Langevin equation is 
valid, inertial effect of the particle is negligible and the particle 
moves by thermal activation.  
%Since the inertial effect is negligible in the overdamped limit,
%the kinetic energy contribution is ignored. 
%Hence, the variation $d U$ of the internal energy
The total energy is then given by the potential term and
the variation $d U$ of the internal energy
is equivalent to the change $d V$ of the potential.
The work done on the particle reads
\begin{equation}
-\;\dbar W = (\partial V/\partial \lambda) \dot{\lambda}dt + f dx,
\end{equation}
hence the amount of heat dissipation is
\begin{equation}
\dbar Q = dV + \;\dbar W  =  -F dx \, .
\end{equation}
The work and heat defined above are the basis for 
investigating the thermodynamic efficiency in stochastic thermodynamics of Langevin systems.
In what follows, we shall describe a few models of stochastic heat engines. 

%%%%%%%
\begin{figure}
\begin{center}
\centerline{\includegraphics[width=0.4\columnwidth]{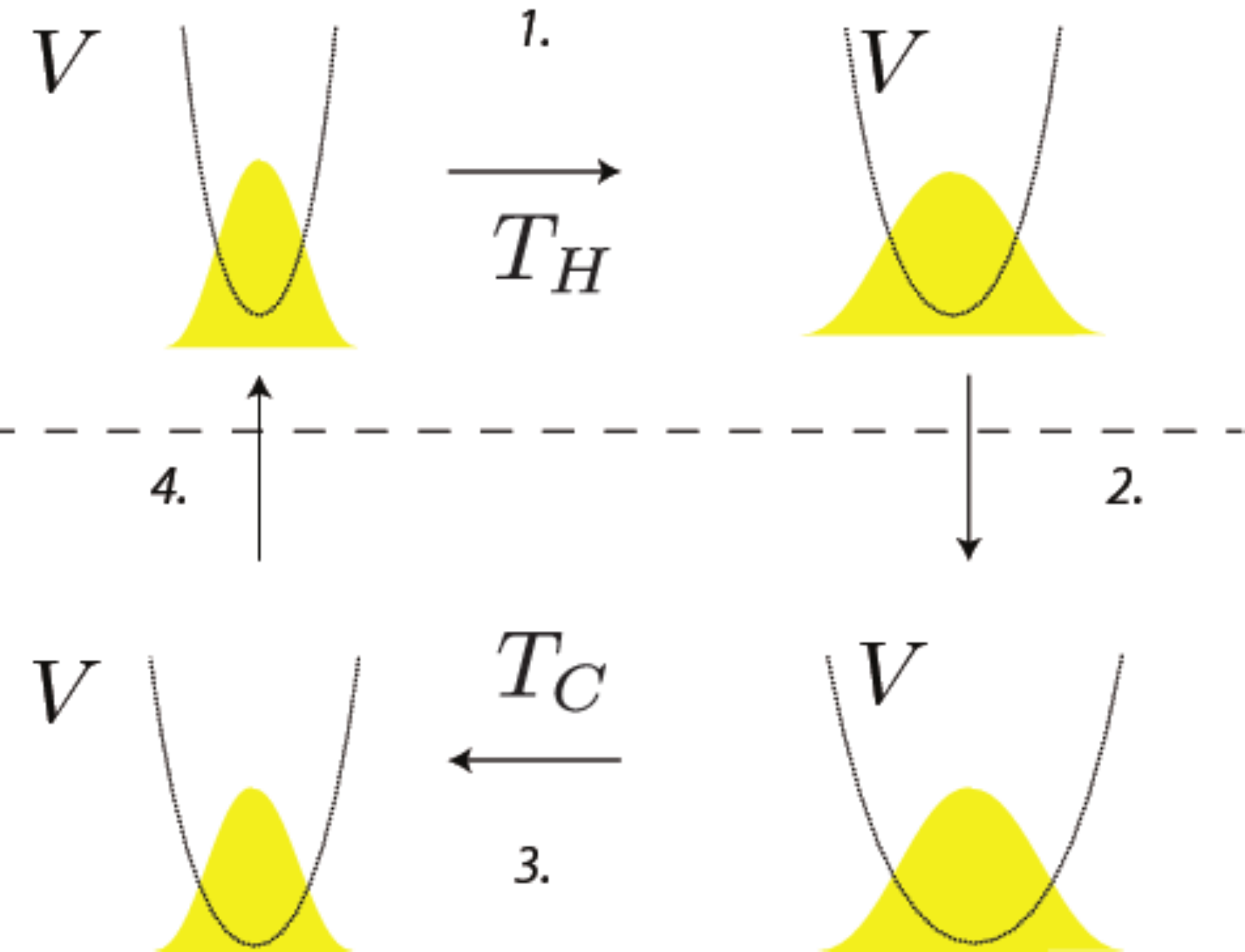}}
\caption{Schematic picture of the stochastic thermodynamic engine in Ref.~\cite{ss08}.
In each plot the curve shows the potential $V$ versus the position $x$, 
the filled region is limited by the curve $p(x)$, representing
the (time-dependent) probability density to find the trapped
particle at $x$.}
\label{ss-engine}
\end{center}
\end{figure}

%%%%%%%
\subsection{Stochastic heat engines I: Schmiedl-Seifert heat engine}
\label{sec:stochasticSS}
Much of thermodynamics was developed with the simple example of the ideal gas contained in a vessel and compressed by a piston. Fundamental and universal laws reveal themselves in this simple model. Hence simple and solvable models are important. Here, we discuss a solvable model, which was introduced by Schmiedl and Seifert \cite{ss08}. Suppose that one particle is trapped by a time-dependent harmonic potential $V(x, \lambda(t) ) = {\lambda (t) x^2 / 2}$ without any non-conservative force. We consider a cycle, depicted in Fig.~\ref{ss-engine}, composed of the following four steps.
\begin{enumerate}
\item Isothermal transition at the hot 
temperature $T_H$ during $0 \le t < t_1$. 
The potential $V(x,\lambda(t))$ changes in time and work is extracted
from the system.
\item \green{An adiabatic transition which should ideally be instantaneous. In other words, we assume 
that adiabatic transitions in the cycle are much faster than the isothermal transitions, 
see footnote~\ref{Footnote:adiabatic} on page~\pageref{Footnote:adiabatic},
% in section~\ref{Sect:endoreversible-cyclic}
from the hot temperature $T_H$ to the cold temperature $T_C$.}
\item Isothermal transition at the cold temperature $T_C$ during the time interval $t_1 \le t < t_1 + t_3$; $V(x,\lambda(t))$ changes in time and work is 
done on the particle.
\item Adiabatic instantaneous transition from the cold temperature $T_C$ to the hot temperature $T_H$.
\end{enumerate}
Let $ Q^{(i)}$ be the amount of heat absorbed from the reservoir in the $i$-th step ($i=1,...,4$). 
Energy conservation means that the work generated $W =  Q^{(1)} + Q^{(3)}$.
Then the thermodynamic efficiency is given by 
\begin{eqnarray}
\eta = { W \over  Q^{(1)}} = 1 + {  Q^{(3)} \over  Q^{(1)}}\, .
\end{eqnarray}
Let $p(x,t)$ be the probability density to find the system (the trapped particle) at position $x$ at time $t$.
The time evolution of the distribution $p(x,t)$ is described by the Fokker-Planck dynamics 
\begin{eqnarray}
{\partial \over \partial t} p(x,t) = \mu \left( \lambda (t) {\partial \over \partial x} x + T {\partial^2 \over \partial x^2 } \right) p(x,t) \, , 
\end{eqnarray}
where $T$ is either $T_H$ or $T_C$ depending on the step $i=1,\cdots, 4$ and $\mu$ is a mobility. The distribution $p(x,t)$ is a Gaussian with zero mean, and hence its variance $\omega (t)=\int dx x^2 p(x,t)$ suffices to describe the time-dependent distribution\footnote{The distribution $p(x,t)$ remains a Gaussian at all times if it is so initially.}.
From the Fokker-Planck equation, we can find the equation 
\begin{eqnarray}
\dot{\omega} (t) = - 2 \mu \lambda (t) \omega (t) + 2 \mu T \, . \label{wdynamics}
\end{eqnarray}
Using the variance $\omega$, the heat absorption between two generic times 
$t_i$ to $t_f$ is calculated as
\begin{eqnarray}
\Delta Q_{t_i \rightarrow t_f} =  \int_{t_i}^{t_f} dt \int dx \dot{p}(x,t) V(x,t) =
-{1\over 4 \mu} \int_{t_i}^{t_f} dt {\dot{\omega}^2 \over \omega } + {T \over 2} \ln {\omega (t_f) \over \omega (t_i ) } \, .
\end{eqnarray}
Similarly, the work done from time $t_i$ to $t_f$ is given by
\begin{eqnarray}
\Delta W_{ t_i\rightarrow t_f}
 =  \int_{t_i}^{t_f} dt \int dx p (x,t) {\partial \over \partial t} V(x,t) 
=
-{1\over 2} \left[ \lambda \omega \right]_{t_i}^{t_f} + {1 \over 4\mu} \int_{t_i}^{t_f} dt {\dot{\omega}^2 \over \omega }
- {T\over 2} \ln {\omega (t_f ) \over \omega (t_i) } \, .
\end{eqnarray}
The maximum work generated is obtained by optimizing 
$\Delta W_{t_i\rightarrow t_f}[\omega]$ with respect to the possible paths of
$\omega(t)$. We obtain the equation 
\begin{eqnarray}
\dot{\omega}^2 - 2 \omega \ddot{\omega} = 0 \, .
\end{eqnarray}
The solution $\omega^{\ast} (t) $ satisfying the boundary 
conditions $\omega (t_i) = \omega_i$ and $\omega (t_f) = \omega_f$ is given by
\begin{eqnarray}
\omega^{\ast} (t) = \left[{ (t - t_f) \sqrt{\omega_i} - (t - t_i)\sqrt{\omega_f} \over t_i - t_f}  \right]^2 \, .
\end{eqnarray}
Correspondingly the function 
$\lambda(t)$ is determined via Eq.~(\ref{wdynamics}). Using the optimum path we calculate the amount of the heat absorbed; $ Q^{(1)}$ and $ Q^{(3)}$. After some manipulation one can find the efficiency as
\begin{eqnarray}
\eta = 1 - {T_C \Delta  \mathscr{S} + A_{\rm irr}/t_3 \over T_H 
\Delta  \mathscr{S} - A_{\rm irr} / t_1} \, ,
\end{eqnarray}
where $\Delta  \mathscr{S} = \ln \sqrt{\omega_b / \omega_a }$ and $A_{\rm irr} = (\sqrt{\omega_b} - \sqrt{\omega_a})^2/\mu$. Here 
$\omega_a = \omega (0)$ and $\omega_b = \omega (t_1)$. We can check that in the quasi-static limit $t_1, t_3 \to\infty$, the Carnot efficiency is recovered. To compute the efficiency at the maximum power, 
one maximizes the power $P$
\begin{eqnarray}
P = {W \over t_1 + t_3} = {1\over t_1 + t_3} \left[ (T_H - T_C) \Delta  \mathscr{S} -A_{\rm irr} \left({1\over t_1} + {1\over t_3}\right) \right]\, ,
\end{eqnarray} 
with respect to $t_1$ and $t_3$. Solving this variational problem, the optimum times $t_1^{\ast}$ and $t_3^{\ast}$ are obtained in terms of $A_{\rm irr}$ and $\Delta  \mathscr{S}$:
\begin{eqnarray}
t_1^{\ast} = t_3^{\ast} = {4 A_{\rm irr} \over \Delta  \mathscr{S} (T_H -T_C)} \, .
\end{eqnarray}
Using these times, one arrives at the efficiency at the maximum power
\begin{eqnarray}
\eta_{SS} = {\eta_C \over 2 - \eta_C/2} \, .
\end{eqnarray}
As already remarked in section~\ref{sec:exoreversible},
this exact result agrees with the \green{Curzon-Ahlborn} efficiency (\ref{eq:ca.eff}) up to 
the second order in $\eta_C$.
\red{A more general treatment \cite{ss08} leads to 
the expression in Eq.~(\ref{eq:etaSS}) for the efficiency at maximum power.}
% CA formula is correct \cite{ss08}. Later, this result was proved in a more general context \cite{esposito2009}. 
%The CA efficiency was also studied for a weakly interacting gas systems where the Carnot cycle is performed within a finite time cycle and the validity of CA efficiency was numerically discussed by means of molecular dynamics simulations \cite{izumida_okuda08,izumida_okuda091,izumida_okuda092}.

A micrometer-sized stochastic engine
using a colloidal particle in a time-dependent harmonic potential
was experimentally realized in \cite{bb2012}.

\subsection{Stochastic heat engines II: Two-level heat engine}
\label{sec:drivenqd}
Among low-dimensional electronic systems, the quantum dot has the potential to provide many kinds of thermodynamic engines \cite{linke2002,elb09,eklv10,linke2010}. Finite-time heat engines can be illustrated by means of quantum-dot systems, where one controls the gate voltage in time to change the on-site energy $\epsilon(t)$ of the dot.
Here we consider the simplest instance, 
we assume only one dot level is close enough the the reservoir's chemical potential to contribute to the 
engine's processes.  We assume that level has a time-dependent energy $\epsilon (t)$, and the reservoir has temperature $T $ and electrochemical potential $\mu(t)$.  
We assume the off-diagonal elements of the density matrix (in the system's energy eigenbasis) are negligible. This implies that quantum coherent superpositions do not play a role and the master equation reduces to a rate equation for the probability 
$p(t)$ of the occupied state in the quantum dot. Then the time evolution is governed by the following master equation:
\begin{eqnarray}
\dot{p}(t) = - \gamma \left[ 1 - f(\beta, \epsilon(t), \mu(t) ) \right]  p(t) + \gamma f(\beta, \epsilon(t), \mu(t) ) \left( 1 - p(t) \right) \, ,
\end{eqnarray}
where $\gamma$ is a rate constant, 
$\beta=1/k_B T$ and the function $f$ is a time-dependent Fermi distribution:
\begin{eqnarray}
 f(\beta, \epsilon(t), \mu(t) )  = {1 \over  e^{\beta ( \epsilon (t) - \mu (t) ) } + 1 } \, .
\end{eqnarray}
From this latter equation it is clear that raising the energy level is 
equivalent to lowering the electrochemical potential, since only 
$\epsilon(t)-\mu(t)$ matter and not $\epsilon(t)$ and $\mu(t)$ separately.
The internal energy $U(t)$ of the quantum-dot system at time 
$t$ is given by 
\begin{equation}
U(t)=[\epsilon(t) -\mu(t)] p(t).
\end{equation}
The rate of change in the internal energy, $\dot{U}$, is the sum of two terms, 
a work flux 
\begin{equation}
-\dot{W}(t)=[\dot{\epsilon}(t) -\dot{\mu}(t)]p(t), 
\end{equation}
and a heat flux
\begin{equation}
\dot{Q}(t)=[\epsilon(t) -\mu(t)]\dot{p}(t)\, .
\end{equation}
\green{We use the convention that $\dot{W}$ is positive when the system is generating work,
and $\dot{Q}(t)$ is positive when the system is absorbing heat.}
Work is done when 
the energy levels are shifted in time, while when an electron enters the 
quantum-dot system at time $t$ 
an amount of heat ${Q}(t)=\epsilon(t)-\mu(t)$ is extracted from the bath.
  
Esposito et al. proposed an exactly solvable model for a quantum-dot heat engine by using the following cycle \cite{eklv10}: % (see Fig.~\ref{dot-engine}) \cite{eklv10}
\begin{enumerate}
\item Isothermal process: The 
quantum dot is in contact with a cold lead at temperature $T_C$ and 
electrochemical 
potential $\mu_C$. The energy level is raised during a finite time $t_1$ 
as $\epsilon (t): \epsilon_0 \to \epsilon_1 ~~(\epsilon_1 > \epsilon_0)$.
\item Adiabatic process: The quantum dot is disconnected from the lead, and the energy level is abruptly lowered
as $\epsilon (t): \epsilon_1 \to \epsilon_2 ~~(\epsilon_1 > \epsilon_2)$. 
Note that, since the quantum dot is isolated during the adiabatic process,
the population of the level does not change.  
\item Isothermal process: The quantum dot is connected to a hot lead with temperature $T_H$ and electrochemical 
potential $\mu_H$. The energy level is lowered during a finite time $t_2$,
$\epsilon (t): \epsilon_2 \to \epsilon_3 ~~(\epsilon_2 > \epsilon_3)$.
\item Adiabatic process: The dot is disconnected, and the energy level abruptly returns to the original value,
$\epsilon (t): \epsilon_3 \to \epsilon_0$.
\end{enumerate}
The period of one cycle is $\tau=t_1 + t_2$, the output power is given by 
\begin{equation}
P={W[p] \over \tau}={Q[p]\over \tau}
={1\over \tau}\int_{0}^{\tau} dt \, \dot{p}(t) \left[ \epsilon(t) - \mu (t)\right], 
\end{equation}
where the net total output work per cycle,
$W[p]=\int_0^\tau dt\,\dot{W}(t)$, and the total absorbed heat,
$Q[p]=\int_0^\tau dt\,\dot{Q}(t)$,
are functionals of the occupation probability $p(t)$. 
Finding the set of parameters that maximize the power may be done with a variational equation. In particular, the \green{Curzon-Ahlborn} efficiency is recovered in the limit of weak dissipation \cite{eklv10}.

\subsection{Onsager matrix in stochastic heat engines}
Thermoelectric transport and stochastic heat engines are categorized into different types of heat to work conversion, since in the thermoelectric transport the power is generated from the steady state electric current, while the work in stochastic heat engines is extracted by using time-dependent thermodynamic protocols. In thermoelectric transport, the Onsager matrix plays a key role in the thermodynamic efficiency in the linear response regime. Here we formulate the Onsager matrix in the stochastic heat engine to discuss the differences between these two types of heat to work conversion.

We follow Brandner \emph{et al.} \cite{brandner15} and discuss the linear response structure by using a general stochastic approach. Let us consider the stochastic heat engine where the heat bath temperature is controlled in time as 
\begin{eqnarray}
T(t)={ T_C T_H \over T_H + (T_C - T_H) \gamma_h (t) } \, ,
\end{eqnarray} 
where $T_H > T_C$ and  the function $\gamma_h (t)$ is a function which takes 
the values $1$ or $0$, so that $T(t)$ takes either $T_H$ or $T_C$. The Hamiltonian is also controlled in time: 
\begin{eqnarray}
H(x,t) = H_0 (x) + \Delta H g_w (x,t) \, .
\end{eqnarray}
Both time-dependent functions $\gamma_h (t)$ and $g_w (x, t)$ are periodic in time with the period ${\cal T}$. We can have in mind as an example the Schmiedl-Seifert engine described in section~\ref{sec:stochasticSS}. We set a small amplitude for the quantities $\Delta T = T_H - T_C $ and $\Delta H$, so that we are within the linear response regime. Then the entropy production rate $\dot{\mathscr{S}}$ is given in the form $\dot{\mathscr{S}}=\mathcal{F}_w J_w + \mathcal{F}_h J_h$, where $\mathcal{F}_w$ and $\mathcal{F}_h$ are affinities: 
\begin{subequations}
\begin{eqnarray}
\mathcal{F}_w = \Delta H/ T\, ,\\
\mathcal{F}_h = \Delta T/T^2 \, ,
\end{eqnarray}
\end{subequations}
where $T=T_C$. The work flux $J_w$ and heat flux into the system $J_{h}$ are respectively defined as
\begin{eqnarray}
J_w &=& {1\over {\cal T} }\int_0^{\cal T} dt \int dx \dot{g}_w (x,t) p_c (x,t) \, , \label{defjw}\\
J_{h} &=& {1\over {\cal T} }\int_0^{\cal T} dt \int dx \gamma_h (t) H( x,t ) \dot{p}_c (x,t) \, \label{defjq} . 
\end{eqnarray}
Here $p_c (x,t)=p_c (x,t +{\cal T})$ is 
the periodic limit to which the time evolution of the probability density
$p(x,t)$ is assumed to converge.
The time evolution of $p(x,t)$  is given by the Fokker-Planck equation: 
\begin{eqnarray}
\dot{p} (x , t) &=& \mathbb{ L} (t) p (x,t) \, , \\
\mathbb{ L} (t)  &=& \mathbb{ L}_0 + \Delta H \mathbb{L}_H (t)  + \Delta T \mathbb{L}_T (t) \, ,
\end{eqnarray}
where the Fokker-Planck operator has been linearized: $\mathbb{L}_0$ is the unperturbed time evolution generator, and $\mathbb{ L}_H (t)$ and $\mathbb{L}_T (t)$ are respectively contributions from time-dependent potential and temperature. %We make a linear response argument using the perturbation method in terms of $\Delta H$ and $\Delta T$. 
We now impose the detailed balance condition to the unperturbed generator, which is the most crucial requirement to get the symmetry in the Onsager matrix:
\begin{eqnarray}
\mathbb{L}_0 p_{\rm eq} = p_{\rm eq} \mathbb{L}_0^{\dagger} \, ,
\end{eqnarray}
where $p_{\rm eq}$ is the equilibrium distribution
of the unperturbed system. 
A standard linear response calculation yields $p_c (x,t)$  
up to the first order:
\begin{eqnarray}
p_c (x,t) = p_{\rm eq} (x) + \sum_{X = H,T} \Delta X \, \int_0^{\infty}
d \tau e^{\mathbb{L}_0 {\tau}} \mathbb{ L}_X (t-\tau) p_{\rm eq} (x)  + O(\Delta^2) \, . \label{steadypc}
\end{eqnarray}
Now we define the Onsager matrix:
\begin{eqnarray}
J_w &=& L_{ww} \mathcal{F}_w + L_{wh} \mathcal{F}_h + O(\Delta^2) \, , \\
J_{h} &=& L_{hw} \mathcal{F}_w + L_{hh} \mathcal{F}_h + O(\Delta^2) \, .
\end{eqnarray}
By using (\ref{defjw}), (\ref{defjq}), and (\ref{steadypc}), one can find the compact expression of the Onsager matrix elements: 
\begin{eqnarray}
L_{\alpha\beta} = L_{\alpha\beta}^{\rm ad} + {1\over k_B}\int_{0}^{\infty} d\tau 
\langle\langle \delta \dot{g}_{\alpha} (0) ; \delta \dot{g}_{\beta} (-\tau ) \rangle \rangle, \, ~~~~(\alpha = w, h)\,,
\end{eqnarray}
where $\delta A \equiv A - \langle A \rangle_{\rm eq}$ with the equilibrium average $\langle A\rangle_{\rm eq}$, $g_h \equiv H_0 \gamma_h$, and the generalized equilibrium correlation function is defined as
\begin{eqnarray}
L_{\alpha \beta}^{\rm ad} &=&  -{1\over k_B {\cal T} } \int_0^{\cal T} dt \int dx\, \delta \dot{g}_{\alpha} (x,t) \, \delta g_{\beta }(x,t)\, p_{\rm eq} (x)\, ,\\
\langle\langle A(t_1); B(t_2) \rangle\rangle  & =&
{1\over {\cal T} } \int_0^{\cal T} dt \int d x \left\{ \begin{array}{ll}
A(x, t_1 + t) e^{\mathbb{ L}_0 (t_1 - t_2)} B(x, t_2 + t) p_{\rm eq} (x), ~~&~~(t_1 \ge t_2)  \\
B(x, t_2 + t) e^{\mathbb{ L}_0 (t_2 - t_1)} A(x, t_1 + t) p_{\rm eq} (x), ~~&~~(t_1 < t_2)  
\end{array} 
\right. \, .
\end{eqnarray}

One intriguing property is that the Onsager matrix elements are in general nonsymmetric unless the above described 
thermodynamic protocol is symmetric under time reversal. In general we can show 
\begin{eqnarray}
L_{\alpha\beta} \left[ H(x,t) , T(t) , {\bm B}\right] 
= 
L_{\beta\alpha} \left[ H(x, - t) , T(- t) , - {\bm B}\right] \, ,
\end{eqnarray}
where the Onsager coefficients are considered functions of the time-dependent Hamiltonian and temperature and of an external magnetic field ${\bm B}$. This nonsymmetric property in the Onsager matrix structure is similar to the thermoelectric transport in the presence of a magnetic field. However, one remark here is that this nonsymmetric property in the stochastic heat engine is present even without any magnetic field, as long as the thermodynamic protocol is not symmetric under time reversal. 

In the stochastic heat engine, the parameters which characterize thermodynamic efficiency are, similarly to what discussed in 
section~\ref{sec:efficiencymagnetic}, 
a generalized figure of merit and the asymmetry between 
the off diagonal Onsager matrix elements.
By algebraic manipulation, one can show an interesting exact bound for the power $P(\eta , \chi)$ for fixed efficiency $\eta$ and fixed ratio between the off-diagonal Onsager matrix elements $\chi=L_{w h} / L_{h w}$:
\begin{eqnarray}
P(\eta , \chi) \le 
\left\{ 
\begin{array}{ll}
4 \bar{P}_0 \bar{\eta} (1 - \bar{\eta}), & |\chi | \ge 1 \, ,\\
\bar{\eta} - \bar{\eta}^2/ \chi^2, & |\chi | < 1 \,  ,
\end{array}
\right.
\end{eqnarray}
where $\bar{\eta}=\eta / \eta_C$ and $\bar{P}_0$ is a model-dependent constant. 
This formula tells us that the Carnot efficiency implies zero power. Isothermal heat engines were discussed in the same framework and the relation between power and work was investigated \cite{vandenbroek2015}.
\red{
Recently a trade-off relation between efficiency and power was proved for 
systems described as Markov process, so that a heat engine with nonvanishing power 
never attains the Carnot efficiency \cite{Shiraishi2016}.}

%% file: conclusions.tex
\section{Concluding remarks}
\label{sec:conclusions}

In this review we have presented  simple and self-contained accounts of some of the main theoretical approaches to the problem of thermoelectric efficiency, and the efficiency of steady-state heat to work conversion in general. Even though the problem has a long history, we believe that the recent theoretical view points described here will be useful. \green{They make a significant contribution to the understanding of the quantum thermodynamics of such steady state quantum machines.}
We expect that they will prove useful in analysing current experiments on  nano-scale systems, 
and hope they will stimulate new generations of experiments. 

Despite the works reviewed here, we believe that the powerful machinery of non-equilibrium statistical mechanics and dynamical system's theory has not yet been fully explored in connection to coupled heat, electric, magnetic or particle transport.   In particular, we believe this machinery will be important in inventing ways of enhancing
heat to work conversion using thermoelectric, thermomagnetic or thermochemical effects. 
An indication of this is the recent realization of the importance of magnetic fields, discussed 
in section~\ref{sec:ZT},
which allow entirely new thermoelectric behaviour by breaking the time-reversal symmetry in the underlying equations of motions.  

The central question which this review identifies is the following:  
What limits do the microscopic dynamics -- for a particular model, or for a particular non-equilibrium steady-state setup -- 
impose on the thermodynamic heat-to-work efficiency?
While the theory for non-interacting quantum systems starts to be well understood, see chapters~\ref{Sect:scattering-theory} and \ref{sec:landauer}, the understanding of general mechanisms connected to interactions \green{is} only begins to emerge.
\green{This is particularly the case when those interactions induce features with no analogue in macroscopic quantum machines, such as quantum coherence and entanglement.} 
One hope is that strongly interacting systems
might be favorable for thermoelectric conversion under suitable conditions, 
for example for the systems close to phase transitions mentioned in chapter~\ref{sec:interacting}.

In all cases, less is known about the physics of heat to work conversion beyond 
the linear-response regime, even though we can expect many 
nanoscale thermoelectric devices to operate far from equilibrium. In that nonlinear regime,
reciprocity relations break down, and there are strong nonlinear effects, such as rectification. 
In this regard, we look forward to new theoretical methods that can treat systems deep in the nonlinear regime,
in situations for which the scattering theory of chapter~\ref{Sect:scatter-nonlin} and the rate equations 
of chapter~\ref{Sect:Qu-Master-Eqn} are not applicable.

Another aspect of the physics of nanostructures currently of great interest, is that of {\it spin caloritronics};
this is the study of situation in which heat is converted into spin-currents (rather than electrical currents).
Many works have recently discussed the spin equivalent of thermoelectric effects, such as spin-Seebeck effects.
We lack the space in this review to do justice to this very active field,
even if the theoretical methods used in the spin caloritronics of nanostructures are often exactly the same as those presented here (scattering theory and master equations).
For the reader interested in spin-caloritronics we suggest starting with the reviews such as Refs.~\cite{spin-caloritronics2012,spin-caloritronics2014}.

Finally, we note that heat to work conversion in the steady-state has the advantage of simplicity (both theoretically and experimentally) over systems that require pumping or driving.  However, this does not mean that we can be sure that steady-state systems are the best route to optimal heat to work conversion.  The types of system considered in sections ~\ref{sec:CTM} and \ref{sec:machines},
may turn out to present advantages over steady-state systems.  Such a possibility is suggested by the fact that, 
as mentioned in section~\ref{sec:thermodriven}, thermodynamic constraints on the steady-state Onsager 
coefficients may be relaxed.
Hence, as our experimental control of the driving of nanoscale system 
improves, it will make sense to consider such systems in more detail.

%% file: appendix-weakly-nonlinear.tex
\section{Evaluation of contributions to the weakly non-linear scattering theory}
\label{Appendix:weakly-nonlinear}

Here we provide technical details to supplement section~\ref{Sect:weakly-nonlinear-microscopic}.
In principle, one just needs to take a Hamiltonian (for the scatterer, its reservoirs and its gates)
and follow the recipe in this appendix to find all ${\cal L}_{\mu\nu\kappa,ijk}$s and hence 
get all currents up to second-order in the thermodynamic forces (bias and temperature difference).
If one could do this exactly, the only approximation in the calculation would be the mean-field approximation that we needed to derive the scattering theory itself.  However, we will see that the recipe is complicated, 
and so one is forced to treat the problem numerically, or to make a set of simplifying assumptions that reduce the problem to one that can be solved analytically (as in Section~\ref{Sect:microsopic-nonlinear-simple}).

\subsection{Transmission functions as a function of the scatterer potential}
\label{Sect:T-as-function-of-U}

 Here we consider calculating   $\big(\rmd  {\cal A}_{ij} \big/ \rmd U_n \big)$, as required 
 in section~\ref{Sect:weakly-nonlinear-microscopic}.
 This can be calculated with whatever theory one used
 to calculate the transmission function in the first place. 
 In practice, this requires enormous work to do without approximation.  
 A reasonable approach is to use the relation between the scattering matrix and the underlying Hamiltonian 
 in Eq.~(\ref{Eq:S-from-H}), 
 with $\hat{\cal H}_{\rm dot}= {1 \over 2m}\hat{p}^2 + \sum_n U_n \hat{x}_n$,
 where the position operator, $\hat{x}_n=|x_n\rangle \langle x_n|$, 
 does not commute with the momentum operator $\hat{p}$.
 However in doing this, one must take the derivative of $\big[E-\hat{\cal H}_{\rm dot}+\rmi \pi \hat{W}\hat{W}^\dagger\big]^{-1}$.
 One should not forget that in general ${\rmd \over \rmd U}\hat{M}^{-1}\neq\hat{M}^{-2} {\rmd \over \rmd U}\hat{M}$
 if $\hat{M}$ is a matrix or operator; instead one must explicitly find $\hat{M}^{-1}$ and take its derivative.  The  $U_n$-dependence of  $\big[E-\hat{H}_{\rm dot}+\rmi \pi \hat{W}\hat{W}^\dagger\big]^{-1}$ will typically be complicated, even though the $U_n$-dependence of  $\big[E-\hat{\cal H}_{\rm dot}+\rmi \pi \hat{W}\hat{W}^\dagger\big]$ is trivial.

\subsection{Characteristic potentials}
\label{Sect:Characteristic-potentials}

Here we consider calculating  $u_{\mu,k} (x)$, as defined in Eq.~(\ref{Eq:u_characteristic}), using the Poisson equation for the nanostructure and all nearby reservoirs, for example inside the region marked by the dashed red ellipse in Fig.~\ref{Fig:nonlinear-example}a.  The Poisson equation reads
 $\nabla^2 U(x) = -q(x)/\eps_0$ where $q(x)$ is the charge at position $x$, and $\eps_0$ is the permittivity of free space.  
Taking the derivative with respect to $ \mathscr{F}_{\mu,k}$ of this Poisson equation gives
\begin{eqnarray}
\nabla^2 u_{\mu,k} (x) = -{1 \over \eps_0} \ 
\left[ \left({\rmd q_{\rm inj}(x) \over  \rmd \mathscr{F}_{\mu,k}}\right)_{\mathscr{F}\to 0} 
+ \left({\rmd q_{\rm pol}(x) \over  \rmd \mathscr{F}_{\mu,k}} \right)_{\mathscr{F}\to 0} \right],
\label{Eq:Poisson1}
\end{eqnarray}
where the equilibrium charge distribution drops out upon taking the derivative,
and the remaining charge distribution can be split in two;  $q_{\rm inj}(x)$ and  $q_{\rm pol}(x)$.
Here $q_{\rm inj}(x)$ is the extra charge directly injected by the leads due to the biases or temperature differences (as represented by the $\mathscr{F}$s), while $q_{\rm pol}(x)$ is the polarization of the charge already in the nanostructure that is induced by the biases or temperature
differences.
The injected charge $q_{\rm inj}(x)$ is the integral over all energies of the local density of states 
at $x$ for electrons arriving from reservoir $k$, multiplied by the occupation probability of that state for given
 $\mathscr{F}_{\mu,k}$ of reservoir $k$, multiplied by $e$.  
 Hence one can define the so-called {\it injectivities} as
  \begin{eqnarray}
D_{\mu,k}(x) \ \equiv \  \left({\rmd q_{\rm inj}(x) \over  \rmd \mathscr{F}_{\mu,k}}\right)_{\mathscr{F}\to 0} 
&=& eT \int \rmd E \ \big(\!-f'(E)\big)\  \left[  e\delta_{\mu { e}} + \delta_{\mu { h}} (E-\mu_1) \right]\ \nu_k (E,x)\, ,
\label{Eq:qinj}
\end{eqnarray}
where the Kronecker $\delta$-functions simply mean that the square bracket is 
$e$ if $\mu={ e}$ (so the thermodynamic force is a bias)
and is 
$(E-\mu_1)$ if $\mu={ h}$ (so the thermodynamic force is a temperature difference).
Here, the local density of states at $x$ for electrons arriving from reservoir $k$,
 can be written as 
 \begin{eqnarray}
 \nu_k (E,x) = - {1 \over 4 \pi i} \sum_j {\rm Tr} \left( 
 {\cal S}_{jk}^\dagger(E) {\rmd {\cal S}_{jk}(E) \over \rmd U(x)} -
 {\cal S}_{jk}(E) {\rmd {\cal S}_{jk}^\dagger(E) \over \rmd U(x)}  \right) \ .
 \label{Eq:nu_k}
 \end{eqnarray}
 The functional derivatives, ${\rmd \big/ \rmd U(x)}$,  can be evaluated on the grid in the manner discussed 
 in \ref{Sect:T-as-function-of-U}.
  Next, the polarization charge is given by $q_{\rm pol}(x) = -e^2\int \rmd^3 y \ \Pi(x,y)\ U(y)$,
where $\Pi(x,y)$ is the Lindhard polarization.
Taking the derivative with respect to $\mathscr{F}_{\mu,k}$
gives 
\begin{eqnarray}
 \left({\rmd q_{\rm pol}(x) \over  \rmd \mathscr{F}_{\mu,k}} \right)_{\mathscr{F}\to 0}  
 = -e^2 \int \rmd^3 y \ \Pi(x,y)\  u_{\mu,k}(y) \, ,
\label{Eq:qpol}
\end{eqnarray}
so that $\left({\rmd q_{\rm pol}(x) \big/  \rmd \mathscr{F}_{\mu,k}} \right)$ is directly related to $u_{\mu,k}(y)$.  
\green{
One might naively guess that temperature changes in the reservoirs or gates do not change their charge
distribution, and so $ \left({\rmd q_{\rm pol}(x) \over  \rmd \mathscr{F}_{\mu,k}} \right)$ would be zero for
$\mu=h$.  In many cases this is a reasonable approximation, however it is rarely strictly true.
For example, whenever the confining potential that defines a gate (or a reservoir) is smooth, increasing the 
temperature of the electrons in that gate (or reservoir) will give some of them the energy to climb that potential a little, and they will thus approach closer to the scatterer.  Thus the scattering potential can be modified by 
changes in gate (or reservoir) temperature.}

To calculate the Lindhard polarization (see for example \cite{Wang-Wang-Guo1999}), we can use
\begin{eqnarray}
\Pi(x,y) = {1 \over 2\pi i} \int \rmd E  f(E) \left(G^{\rm R}(x,y)G^{\rm R}(y,x) -G^{\rm A}(x,y)G^{\rm A}(y,x) \right) \, ,
\end{eqnarray}
where the Green's functions  $G^{\rm R,A}(x,y)$ are evaluated at equilibrium (${\mathscr{F}\to 0}$).
If we take the model in Eq.~(\ref{Eq:S-from-H}),  then
\begin{eqnarray}
G^{\rm R}(x,y) = |x\rangle  \left[{E-\hat{\cal H}_{\rm dot}+i \pi \hat{W}\hat{W}^\dagger}  \right]^{-1} \langle y|,  
\end{eqnarray}
with $G^{\rm A}(x,y)$ having the opposite sign in front of $ \pi \hat{W}\hat{W}^\dagger$. 
If one calculated all the above quantities exactly, one could insert them into Eq.~(\ref{Eq:Poisson1}) 
to get an exact differential equation for $u_{\mu,k} (x)$, which could in turn be solved.
However, in general, one either finds this solution numerically, or one makes a set of simplifying assumptions that reduce the problem to one that can be solved analytically (as in section~\ref{Sect:microsopic-nonlinear-simple}).
In cases where the screening within the scatterer is good, a Thomas-Fermi approximation for the Lindhard function may be sufficient. Then
Eq.~(\ref{Eq:Poisson1}) reduces to \cite{Levinson1989}
\begin{eqnarray}
\nabla^2 u_{\mu,k} (x) - {u_{\rm \mu,k} \over a^2} = -{1 \over \eps_0} \ 
\left({\rmd q_{\rm inj}(x) \over  \rmd \mathscr{F}_{\mu,k}}\right)_{\mathscr{F}\to 0}, 
\label{Eq:Poisson2}
\end{eqnarray}
where $a$ is the Thomas-Fermi screening length, given by
\begin{eqnarray}
{1 \over a^2} &=& {e^2 \over \eps_0} \int \rmd E \ \big(-f'(E)\big) \nu (E,x)\ ,
\end{eqnarray}
where $\nu(E,x)=\sum_k \nu_k(E,x)$ is the local density of states at position $x$.

\subsection{Casting the result in terms of capacitances}

Most author's follow Christen and B\"uttiker \cite{Christen-Buttiker1996a} in presenting the results of such a discretized calculation in terms of capacitances.  The idea is that once one has solved the discretized Poisson equation in the vicinity of the scatterer (e.g.~everywhere inside the dashed red ellipse in Fig.~\ref{Fig:nonlinear-example}a),
one has both the potential $U_n$ and the excess charge $\delta q_n\equiv q_{\rm inj}(x_n)+q_{\rm pol}(x_n)$
at all positions $x_n$ on the grid.  Thus, one can always 
define a capacitance matrix whose $nm$th element satisfies 
\begin{eqnarray}
\delta q_n =\sum_m C_{nm} \ \delta U_m  + \sum_{l \in {\rm res}} \tilde{C}_{n,l}\ eV_l,
\label{Eq:def-C-matrix}
\end{eqnarray}
 where we define 
$\delta U_m\equiv U_n-U_m^{\rm eq}$ as the change in potential compared to its equilibrium value,
$U_m^{\rm eq}$.
The first term on the right of Eq.~(\ref{Eq:def-C-matrix}) is the capacitance between site $n$ and $m$ on the grid,
the second term is the capacitance between site $n$ on the grid and all of reservoir $l$ that is not included within the grid.  This second term is to account for the charge build up in the part of the reservoirs which are not included in the grid; in other words, the part of the reservoirs outside the dashed red ellipse in Fig.~\ref{Fig:nonlinear-example}a.
Hence, $\tilde{C}_{n,l}$ is the effect of the biasing of all of reservoir $l$ outside the dashed ellipse
upon the charge build up at position $x_n$ inside the dashed ellipse.
Taking the derivative with respect to $\mathscr{F}_{\mu,k}$ of Eq.~(\ref{Eq:def-C-matrix}), 
and replacing $\big(\rmd q_n \big/ \rmd \mathscr{F}_{\mu,k} \big)$ by the sum of Eqs.~(\ref{Eq:qinj}) and (\ref{Eq:qpol}) gives
\begin{eqnarray}
\sum_m \left(C_{nm} + e^2\Pi_{nm}\right) \ u_{\mu,k}(x_m) =  -\sum_{l \in {\rm res}} \tilde{C}_{n,l} \delta_{k,l}\delta_{\mu,{ e}}\ +\ {eT \over \eps_0} \int \rmd E \ \big(\!-f'(E)\big)\  \left[  e\delta_{\mu { e}} + \delta_{\mu { h}} (E-\mu_1) \right]\ \nu_k (E,x_n), 
\label{Eq:u-in-terms-of-C-1}
\end{eqnarray}
where we define a discretized polarization matrix ${\bm \Pi}$ whose $nm$th element is $ \Pi_{nm} \equiv (\delta V)\,\Pi(x_n,x_m)$
with $\delta V$ being the volume of each cell in the grid.
The easiest way to solve for $u_{\mu,k}(x_m)$ is to write Eq.~(\ref{Eq:u-in-terms-of-C-1}) as a matrix equation.
For this we define a column vector ${\bm u}_{\mu,k}$ such that its $m$th element is $u_{\mu,k}(x_m)$,
we define a second column vector $\tilde{{\bm C}}_{\mu,k}$ such that its $m$th element is $\sum_l \tilde{C}_{m,l} \delta_{k,l}\delta_{\mu,{ e}}$,
and a third column vector ${\bm D}_{\mu k}$ such that its $n$th element is
\begin{eqnarray}
\left[{\bm D}_{\mu,k}\right]_n = {eT \over \eps_0} \int \rmd E \ \big(\!-f'(E)\big)\  \left[  e\delta_{\mu { e}} + \delta_{\mu { h}} (E-\mu_1) \right]\ \nu_k (E,x_n)\ .
\end{eqnarray}
Then Eq.~(\ref{Eq:u-in-terms-of-C-1}) can be written as $({\bm C}+e^2 {\bm \Pi}) {\bm u}_{\mu,k} = {\bm D}_{\mu,k}-\tilde{\bm C}_{\mu,k}$,
which means that 
\begin{eqnarray}
u_{\mu,k}(x_m) = \left[\left({\bm C}+e^2 {\bm \Pi}\right)^{-1} \ \left({\bm D}_{\mu, k} - \tilde{\bm C}_{\mu, k} \right) \right]_m,
\end{eqnarray}
where $(\cdots)^{-1}$ is the matrix inverse, and $[\cdots]_m$ indicates that we take the $m$th element of the 
vector that is inside the square-brackets.  This gives a simple form for the characteristic potentials, $u_{\mu,k}(x_m)$,
however this simplicity is partially deceptive.  On one hand, one should not forget that to find the matrix ${\bm C}$ 
from first principles we had to completely solve either the Poisson equation in Eq.~(\ref{Eq:Poisson1}), 
or a reasonable approximation of that equation, such as Eq.~(\ref{Eq:Poisson2}). 
On the other hand, capacitances are quantities that can often be measured for a given nanostructure
by A.C. admittance measurements \cite{Christen-Buttiker-AC-PRL,Christen-Buttiker1996b},
thus it is very pretty to be able to write the quantities of importance for thermoelectric transport in terms of such experimental observables.

\subsection{Relations induced by gauge invariance}
\label{Sect:gauge-invariance}

We can use the concept of gauge-invariance (the idea that the physics depends on energy differences, but not the absolute value of energy) to make the following observations.
Since a uniform shift upwards of the potential at all grid points, $U_n$, is the same as a shifting the total energy $E$
downwards,
we have 
\begin{eqnarray}
 -{\cal A}'_{ij} &=& \int \rmd^d x\ \left({\rmd {\cal A}_{ij} \over \rmd U(x)}\right)  
 \ \equiv \ \sum_{n} {\rmd {\cal A}_{ij}\over \rmd U_n} \ ,
\label{Eq:Aprimed-to-functional-derivative}
\end{eqnarray}
where again we define the functional derivative as a sum over the sites on a grid.
Similarly, a shift of all reservoir's electrochemical potentials (including that of all gates) 
by the same amount should not change the physics of any quantity beyond a trivial shift of energy.
Thus, whatever the details of the Poisson equation's solution, the characteristic potentials associated with a change of bias ($\mu={ e}$) on the reservoirs will obey
\cite{Christen-Buttiker1996a,sanchezprb13,Meair-Jacquod2013,Sanchez-Lopez2013,Lopez-Sanchez2014}
\begin{eqnarray}
{1 \over eT_0 }\sum_k  u_{{ e},k}(x_m) = 1
\end{eqnarray}
where the sum is over all reservoirs, including those acting as gates.
Thus, summing Eq.~(\ref{Eq:u-in-terms-of-C-1}) over all reservoirs for $\mu = { e}$ tells 
us that the solution of the Poisson equation will obey
\begin{eqnarray}
\sum_m \left(C_{nm} + e^2\Pi_{nm}\right)  =  -\sum_{l \in {\rm res}} \tilde{C}_{n,l}\ +\ {e^2T_0 \over \eps_0} \int \rmd E \ \big(\!-f'(E)\big)\ \nu (E,x_n), 
\label{Eq:C-relations}
\end{eqnarray}
where $\nu (E,x_n) = \sum_k \nu_k (E,x_n)$ is the local density of states at $x_n$. 
In general, there is no similar relationship for the characteristic potentials associated with temperature differences ($\mu={ h}$), because the physics is not invariant under a shift of all temperatures by the same amount. 

Finally, we note that sometimes Refs.~\cite{Sanchez-Buttiker2005,Buttiker-Sanchez2005,sanchezprb13,Meair-Jacquod2013,Sanchez-Lopez2013,Lopez-Sanchez2014} used Eq.~(\ref{Eq:Aprimed-to-functional-derivative}) to replace ${\cal A}'_{ij}$ by $-\big({\rmd {\cal A}_{ij} \big/ \rmd U_n}\big)$
in Eqs.~(\ref{Eq:nonlinearLs}).  This gives prettier formulas, but is arguably a step in the wrong direction, because 
 ${\cal A}'_{ij}$ is trivial to get from the $E$-dependence of the transmission function,
 while it is very hard work to calculate $\big({\rmd {\cal A}_{ij} \big/ \rmd U(x)}\big)$ for all $x$ 
 (as we saw in \ref{Sect:T-as-function-of-U}).

%% file: appendix-classical-reservoirs.tex
\section{Numerical simulation of classical particle reservoirs}
\label{sec:numerics_c}
In classical systems, a standard model of a particle reservoir is given by the $d$-dimensional ideal gas system with temperature $T$ and electrochemical potential $\mu$.  The ideal gas reservoir is attached to the system of interest via a surface with a finite area $A$. The velocity distribution inside the ideal gas reservoir is given by the Maxwell distribution 
\begin{eqnarray}
f( {\bm v} ) &=& 
\left( {m \over 2\pi k_B T}\right)^{d/2} \exp 
\left( 
- { m |{\bm v }|^2
\over 2 k_B T
}
\right) \, ,~~
\end{eqnarray}
where ${\bm v}$ is a velocity vector ${\bm v}=(v_1 , \cdots , v_d)$ and  $m$ is the mass of a particle. Suppose that the particles enter the system via a surface with a Poisson process and 
the waiting time distribution of injection is parametrized only with the average injection rate $\nu$:
\begin{eqnarray}
P (\tau ) &=& {\nu} \, e^{-\nu \tau} \, . \label{poisson}
\end{eqnarray}
The average injection rate ${\nu}$ is estimated by
\begin{eqnarray}
\nu &=&  A \langle v_1 \rangle n =
A n \int_{0}^{\infty} d v_1 \int_{-\infty}^{\infty} d v_2 \cdots \int_{-\infty}^{\infty} d v_d \, 
v_1  f({\bm v} ) \, , ~~~~ \label{nu}
\end{eqnarray}
where the particles in the reservoir enter the system via an entrance located in the direction $(1,0,\ldots , 0)^T$;
$\langle v_1 \rangle$ is the average of the velocity $v_1$ and $n$ is the density of particles in the reservoir. 
Then numerically, one inject particles according to the Poisson process 
(\ref{poisson}) with a rate (\ref{nu}). 
The velocity of the injected particles is chosen from the distribution
\begin{eqnarray}
p({\bm v}) &=& \sqrt{2\pi m \over k_B T} v_1 f({\bm v}) \, ,
\end{eqnarray}
where $v_1 \in [0,\infty )$ and $\nu_2,...,\nu_d\in(-\infty , \infty )$. 
Once a particle enters into the system, it starts interacting 
with the other particles inside the system.
Whenever a particle of the system crosses the boundary which 
separates the system from one reservoir, it is removed. 

The electrochemical potential is related to the density $n$. 
%Let us consider for concreteness the one-dimensional case. 
To show such a relationship, 
it is convenient to write the grand partition function
\begin{equation}
\Xi=\sum_{N=0}^\infty \frac{1}{N !}
\left\{ \left(\frac{\Lambda}{h}\right)^d e^{\beta \mu}
\int_{-\infty}^{\infty} d v_1 \cdots \int_{-\infty}^{\infty} d v_d
\, m^d \exp \left[-\beta \left(
\frac{1}{2} m |{\bm v }|^2\right)\right]\right\}^N,
\label{eq:grandcanonical}
\end{equation}
with ${\Lambda}^d$ and $N$ volume and number of particles
of the reservoir, respectively~\footnote{It is of course understood
that ${\Lambda}$
is macroscopically large and that the thermodynamic limit is
eventually taken for the reservoir.}, $\beta= 1/k_B T$,
and $h$ the Planck's constant.
We then compute the average
number of particles as
\begin{equation}
\langle N \rangle = \frac{1}{\beta}\frac{\partial}{\partial
\mu} \ln \Xi,
\end{equation}
so that
\begin{equation}
n=\frac{\langle N \rangle}{{\Lambda}^d}=
\frac{e^{\beta \mu}({2\pi m k_B T})^{d/2}}{h^d}.
\label{eq:num5}
\end{equation}
Therefore, we can express the
electrochemical potential of the bath in terms
of the injection rate:
\begin{equation}
\mu = k_B T \ln (\lambda_{T}^d n),
\label{eq:num4}
\end{equation}
where
\begin{equation}
\lambda_{T}=\frac{h}{\sqrt{2\pi m k_B T}}
\end{equation}
is the \emph{de Broglie thermal wave length}.
Note that this relation, even though derived from the grand
partition function of a classical ideal gas, can only be justified
if particles are considered as indistinguishable. The
$1/N!$ term in the grand partition function (\ref{eq:grandcanonical})
is rooted in the above indistinguishability,
of purely quantum mechanical origin~\cite{huang}.

This ideal gas reservoir method is applicable to nonequilibrium situations by applying two particle reservoirs with different temperatures and electrochemical potentials. Coupled heat and matter transport for deterministic classical dynamical systems were discussed in Refs.~\cite{mm2001,mm2003}, which provided the first numerical measurements of the Onsager matrix for interacting chaotic classical gases. The figure of merit for thermoelectric transport was discussed with this method \cite{casati08}, and the divergence of the figure of merit at the thermodynamic limit in low-dimensional systems was reported \cite{casati09,Saito2010,Benenti2013,Benenti2014}. The concept of an ideal gas reservoir was applied to investigate the thermodynamical efficiency in the Nernst effect \cite{stark}.

%% file: appendix-second-to-first-quantization.tex
\section{\green{Example of the many-body basis for a small quantum system}}
\label{Append:second-to-first-quantization}

\green{
As explained in section~\ref{Sect:transition-rates},
to get a rate equation of the type discussed in 
chapters~\ref{Sect:Qu-Master-Eqn} and \ref{Sect:master-examples} from a system's microscopic Hamiltonian,
we must cast that Hamiltonian in terms of its many-body eigenstates.
Here we give an example in which we take a system whose Hamiltonian, ${\cal H}_{\rm s}$, 
is written in terms of creation and annihilation operators, and recast it in terms of its  
many-body eigenstates.
This is a transformation from second quantization to first quantization, when most textbooks on quantum mechanics only discuss the  transformation in the opposite direction. We hope the simple example in this appendix will be enough for the reader to see that the transformation in the direction we need is no more difficult
than that in the textbook direction. }

\green{
Imagine a quantum machine similar to that in Table~\ref{Table:example-H},
with only two fermionic states, which we label state 1 and state 2.
Let us take its Hamiltonian ${\cal H}_{\rm s}$ in second quantization to be
\begin{align}
\hat{H}_{\rm s} \ =\  \epsilon_1 \, \hat{N}_1+\epsilon_2 \, \hat{N}_2 \ -\   \Delta \ \left(\hat{d}^\dagger_1 \hat{d}_2 + \hat{d}^\dagger_2 \hat{d}_1\right) \ +\  U_{12}\, 
\hat{N}_1\hat{N}_2\ , 
\label{Eq:H_s-example-appendix}
\end{align} 
where $\hat{N}_i= \hat{d}_i^\dagger \hat{d}_i$ is the operator that counts the number of electrons in state $i$.
Here the $\Delta$ term corresponds to tunnelling between the two states,
and the $U_{12}$ term to Coulomb repulsion between electrons in the two states.
This Hamiltonian is slightly more complicated than that in Table~\ref{Table:example-H},
because the $\Delta$ term was absent there. 
}

\green{
To write this $\hat{H}_{\rm s}$  in its many-body eigenbasis, we have to first write it is a basis
of many-body states.  In principle one could choose any orthonormal basis, however it is extremely 
natural to choose the basis of many-body states of the form
$| n_1,n_2\rangle$, where $n_i$ is the occupation of fermion state $i$;
the state can be empty ($n_i=0$) or full ($n_i=1$).  
For $M$ fermionic states, there are $N=2^M$ many-body states,
so for two fermionic states, we have $N=4$ many-body states.
The nature of fermionic states means that this basis is orthonormal, 
with $\langle m_1,m_2 | n_1,n_2\rangle = \delta_{m_1,n_1}\delta_{m_2,n_2}$.
Then an arbitrary many-body wavefunction takes the form 
\begin{align}
|\Psi (t) \rangle = &
\psi_{00}(t)\, |0,0\rangle + 
\psi_{10}(t)\, |1,0\rangle + 
\psi_{01}(t)\, |0,1\rangle + 
\psi_{11}(t)\, |1,1\rangle 
\end{align}
Since we are dealing with fermions the order in which we add electrons to states is important
(changing the order will generate minus signs).  Thus let us define
$|1,1\rangle =   \hat{d}^\dagger_2 \hat{d}^\dagger_1 |0,0\rangle = 
- \hat{d}^\dagger_1 \hat{d}^\dagger_2 |0,0\rangle$.
This means that 
$\hat{d}^\dagger_2 |1,0\rangle = |1,1\rangle$ 
but
\begin{eqnarray}
\hat{d}^\dagger_1 |0,1\rangle = -|1,1\rangle,
\label{Eq:minus-sign}
\end{eqnarray}
where the minus sign is a consequence of Fermi statistics.
Let us now write the system state as a vector of these many-body states
\begin{align}
\Psi (t) = \left(\begin{array}{c} 
\psi_{00}(t) \\ \psi_{10}(t) \\  \psi_{01}(t) \\  \psi_{11}(t) 
\end{array}\right)
\end{align}
Then system operators are $4 \times 4$ matrices
acting on this vector of many-body states.
So 
\begin{align}
 \hat{d}^\dagger_1 \ \to&\ 
\left( \begin{array}{ccrc} 0&0&0&0 \\ 1&0&0&0 \\ 0&0&0&0 \\ 0&0&-1&0 \end{array} \right),
\quad
 \hat{d}_1 \ \to\ 
\left( \begin{array}{cccr} 0&1&0&0 \\ 0&0&0&0 \\ 0&0&0&-1 \\ 0&0&0&0 \end{array} \right),
\nonumber \\
 \hat{d}^\dagger_2 \ \to&\ 
\left( \begin{array}{cccc} 0&0&0&0 \\ 0&0&0&0 \\ 1&0&0&0 \\ 0&1&0&0 \end{array} \right),
\quad
 \hat{d}_2 \ \to\ 
\left( \begin{array}{cccc} 0&0&1&0 \\ 0&0&0&1 \\ 0&0&0&0 \\ 0&0&0&0 \end{array} \right).
\label{Eq:d_to_matrix}
\end{align}
Note that the minus signs in $\hat{d}^\dagger_1$ and 
$\hat{d}_1$ originate in the minus sign in Eq.~(\ref{Eq:minus-sign}).  
Then this Hamiltonian as a matrix acting on the above many-body states will read 
\begin{align}
{H}_{\rm s}=
\left( \begin{array}{cccc}
0 \ &\  0 \ & \ 0 & 0 \\
0  \ & \ \epsilon_1 \ & \ -\Delta \  & \ 0 \\
0 & \ -\Delta\ & \ \epsilon_2 \ &\  0 \\
0 & 0 & 0 & \ \epsilon_1+\epsilon_2+U_{12}\ 
\end{array} \right)
\end{align}
In the case where $\Delta=0$, this Hamiltonian is already diagonal, so the many-body eigenstates
are those given in Table~\ref{Table:example-H}, where we defined 
\begin{align}
|0,0\rangle \ \hbox{ as }\,  |0\rangle, \qquad  
|1,0\rangle \ \hbox{ as }\,  |1\rangle, \qquad  
|0,1\rangle \ \hbox{ as }\,  |2\rangle, \qquad \hbox{ and }   
|1,1\rangle \ \hbox{ as }\,  |{\rm d}\rangle, 
\end{align}
}

\green{
However, here we consider the case where $\Delta$ is non-zero, so the matrix form of 
${\cal H}_{\rm s}$ is not diagonal.  In this case, it is not hard to see that it is diagonalized by the matrix
\begin{align}
{\cal U} = 
\left( \begin{array}{cccc}
1 \ &\  0 \ & \ 0 & 0 \\
0  \ & \ \cos (\phi/2) \ & \- \sin (\phi/2) \  & \ 0 \\
0 & \ -\sin (\phi/2)\ \ \  & \ \cos (\phi/2) \ &\  0 \\
0 & 0 & 0 & 1\ 
\end{array} \right)
\end{align}
where we define $\phi = \arctan \left[2\Delta/(\epsilon_1-\epsilon_2)\right]$. 
This can be seen by the fact that
\begin{align}
{\cal H}_{\rm s} = {\cal U}^\dagger\ 
\left( \begin{array}{cccc}
0 \ &\  0 \ & \ 0 & 0 \\
0  \ & \ \epsilon_+ \ & 0  & \ 0 \\
0 & 0  & \ \epsilon_- \ &\  0 \\
0 & 0 & 0 & \ \epsilon_1+\epsilon_2+U_{12}\ 
\end{array} \right)
\ {\cal U}
\label{Eq:H_s-appensix-diag}
\end{align}
where we define $\epsilon_\pm = \half (\epsilon_1+\epsilon_2) \mp \half \sqrt{(\epsilon_1-\epsilon_2)^2 + 4\Delta^2}$.
Thus the four many-body eigenstates are given by 
\begin{align}
 |0,0\rangle \qquad\qquad\qquad \hbox{ with eigenenergy}&=0\ , \\ 
 \cos (\phi/2)\,|1,0\rangle +  \sin (\phi/2)\, |0,1\rangle\qquad \hbox{ with eigenenergy}&=\epsilon_+ \ , \\ 
 \cos (\phi/2)\,|0,1\rangle -  \sin (\phi/2)\, |1,0\rangle\qquad\hbox{ with eigenenergy}&=\epsilon_- \ , \\ 
 |1,1\rangle \qquad\qquad\qquad \hbox{ with eigenenergy}&=\epsilon_1+\epsilon_2+U_{12}\ . 
\end{align}
The eigenstate with energy $\epsilon_+$ is a weighted sum of $|1,0\rangle$ and $|0,1\rangle$,
while eigenstate with energy $\epsilon_-$ is a weighted difference of $|1,0\rangle$ and $|0,1\rangle$;
thus the former is a bonding state, and the latter is the anti-bonding state between the two sites (so $\epsilon_+ <\epsilon_-$).
}

\green{
We now wish to take a Hamiltonian which includes the coupling to the reservoirs, 
such as ${\cal H}_{\rm total}$ in Eqs.~(\ref{Eq:H_total-1}-\ref{Eq:V_ph}),
and recast it in terms of the many-body eigenstates of ${\cal H}_{\rm s}$.
The first step is to rewrite ${\cal H}_{\rm total}$ in terms of many-body matrices,
replacing the system's creation and annihilation operators by the relevant many-body matrices, while leaving the reservoirs' creation and annihilation operators as they are.
If ${\cal H}_{\rm s}$ is given by Eq.~(\ref{Eq:H_s-example-appendix}), then this replacement is that 
given in Eqs.~(\ref{Eq:d_to_matrix}).
The second step is the matrix transformation to the basis in which the matrix ${\cal H}_{\rm s}$ is diagonal.
We see from Eq.~(\ref{Eq:H_s-appensix-diag}) that this transformation 
is done by acting on ${\cal H}_{\rm total}$ to the left with ${\cal U}$ and to the right with ${\cal U}^\dagger$;
this is equivalent to writing the system creation and annihilation operator's as matrices, and then acting on them  to the left with ${\cal U}$ and to the right with ${\cal U}^\dagger$.
For the case where ${\cal H}_{\rm s}$ is given by Eq.~(\ref{Eq:H_s-example-appendix}), 
we can do the two steps in one. 
We can write ${\cal H}_{\rm total}$ directly in the basis of many-body eigenstates of ${\cal H}_{\rm s}$
by making the following substitutions in Eqs.~(\ref{Eq:H_total-1}-\ref{Eq:V_ph}),
\begin{align}
 \hat{d}^\dagger_1 \ \ \to&\ \ 
{\cal U} \ \left( \begin{array}{ccrc} 0&0&0&0 \\ 1&0&0&0 \\ 0&0&0&0 \\ 0&0&-1&0 \end{array} \right) 
\ {\cal U}^\dagger 
\ = \
 \left( \begin{array}{cccc} 0&0&0&0 \\  \cos (\phi/2)&0&0&0 \\ - \sin (\phi/2)&0&0&0 \\ 0&- \sin (\phi/2)&- \cos (\phi/2)&0 \end{array} \right) \ ,
\end{align}
\begin{align}
 \hat{d}_1 \ \ \to&\ \ 
{\cal U} \ \left( \begin{array}{cccr} 0&1&0&0 \\ 0&0&0&0 \\ 0&0&0&-1 \\ 0&0&0&0 \end{array} \right) 
\ {\cal U}^\dagger 
\ = \
 \left( \begin{array}{cccc} 0&\cos (\phi/2) &- \sin (\phi/2) &0 \\  0&0&0&- \sin (\phi/2) \\ 0&0&0&- \cos (\phi/2) \\ 0&0&0&0 \end{array} \right) \ ,
\\
 \hat{d}^\dagger_2 \ \ \to&\ \ 
{\cal U}\ \left( \begin{array}{cccc} 0&0&0&0 \\ 0&0&0&0 \\ 1&0&0&0 \\ 0&1&0&0 \end{array} \right)
\ {\cal U}^\dagger 
\ = \
 \left( \begin{array}{cccc} 0&0&0&0 \\  \sin (\phi/2)&0&0&0 \\ \cos (\phi/2)&0&0&0 \\ 0&\cos (\phi/2)&- \sin (\phi/2)&0 \end{array} \right) \ ,
\\
 \hat{d}_2 \ \ \to&\ \ 
{\cal U} \ \left( \begin{array}{cccc} 0&0&1&0 \\ 0&0&0&1 \\ 0&0&0&0 \\ 0&0&0&0 \end{array} \right)
\ {\cal U}^\dagger 
\ = \
 \left( \begin{array}{cccc} 0&\sin (\phi/2)&\cos (\phi/2)&0 \\  0&0&0&\cos (\phi/2) \\ 0&0&0&- \sin (\phi/2)\ \ \  \\
 0&0&0&0 \end{array} \right) \ .
\end{align}
}